\documentclass[12pt]{report}

\usepackage{a4}
\usepackage{latexsym}
\usepackage{amssymb}

\newcommand{\C}{{\cal C}}

\newcommand{\B}{{\mathcal B}}
\newcommand{\No}{{\mathcal N}}
\newcommand{\PP}{{\mathcal P}}
\newcommand{\D}{{\mathcal D}}
\newcommand{\A}{{\mathcal A}}
\newcommand{\E}{\mbox{$\mathcal E$}}
\newcommand{\X}{{\mathcal X}}
\newcommand{\Y}{{\mathcal Y}}
\newcommand{\Z}{{\mathcal Z}}

\newcommand{\V}{{\mathcal V}}
\newcommand{\F}{{\mathcal F}}
\newcommand{\G}{{\mathcal G}}
\newcommand{\T}{{\mathcal T}}
\newcommand{\I}{{\mathcal I}}
\newcommand{\M}{{\mathcal M}}
\newcommand{\N}{{\mathcal N}}
\newcommand{\BGr}[2][q]{\mbox{$\mathcal{B}_{#1}(#2)$}}

\newcommand{\ZZ}{\mathbb{Z}}

\newcommand{\RR}{\mathbb{R}}
\newcommand{\NN}{\mathbb{N}}

\newcommand{\Kraft}{\mbox{$S$}}

\newcommand{\ltup}[2][l]{\mbox{$\vec{#1}_{#2}$}}

\newcommand{\bfs}[1][l]{\mbox{$\Delta^{#1}_B$}}
\newcommand{\sfs}[1][l]{\mbox{$\Delta^{#1}_S$}}
\newcommand{\pfs}[1][l]{\mbox{$\Delta^{#1}_P$}}

\newcommand{\foi}[3]{\mbox{$(#3_#1)_{#1{\in}\bf{#2}}$}}
\newcommand{\fo}[2][n]{\foi{#1}{\NN}{#2}}

\newcommand{\qed}{
\mbox{\rm\bf\raisebox{-0.5ex}{q.e.d}} }
\newcommand{\mod}{\mbox{$\,\mbox{mod}\, $}}
\newcommand{\num}[1][q]{\mbox{$num_{#1}$}}

\newcommand{\Num}{\mbox{$\,\mbox{Num} $}}
\newcommand{\Sub}{\mbox{$\,\mbox{Sub} $}}

\newtheorem{Def}{Defenition}
\newtheorem{prop}{Proposition}
\newtheorem{lemma}[prop]{Lemma}
\newtheorem{theorem}{Theorem}
\newtheorem{example}{Example}
\newtheorem{conj}{Conjecture}
\newtheorem{cor}{Corollary}
\newtheorem{constr}{Construction}
\newtheorem{claim}{Claim}
\title{}
\author{}

\date{}

\begin{document}

\setlength{\parindent}{0.6cm}
\setlength{\parskip}{1ex}

\title{{\bf On the $\frac{3}{4}\, $-Conjecture for Fix-Free Codes\\ A Survey}}
\author{Holger Schnettler \\[2ex]
Fakult{\"a}t f{\"u}r Mathematik\\
Universit{\"a}t Bielefeld
}

\date{}
\maketitle

\tableofcontents

\newpage

\section*{Introduction}

In this survey we concern ourself with the question, wether there
exists a fix-free code for a given sequence of codeword lengths.
For a given alphabet, we obtain the {\em Kraftsum} of a code, if we divide for every
length the number of
codewords of this length in the code by the total number of all possible words
of this length and then take summation over all codeword lengths which appears
in the code. The same way the Kraftsum of a lengths sequence
$(l_1,\ldots , l_n) $ is given by $\sum_{i=1}^n q^{-l_i} $, where $q$ is the
numbers of letters in the alphabet.
Kraft and McMillan have shown in \cite{kraft} (1956), that there exists a
prefix-free code with codeword lengths of a certain lengths sequence,
if the Kraftsum of the lengths sequence is smaller than or equal to one.
Furthermore they have shown, that the converse also holds
for all (uniquely decipherable) codes.\footnote{In this survey a code means
a set of words, such that any message which is encoded with these words can
be uniquely decoded. Therefore we omit in future the "uniquely decipherable"
and write only "code".}
The question rises, if Kraft's and McMillan's result can be generalized to other
types of codes?
Throughout, we try to give an answer on this question
for the class of fix-free codes. Since any code has Kraftsum smaller than
or equal to one, this answers the question for the second implication
of Kraft-McMillan's theorem. Therefore we pay attention mainly to the first implication.\\

\bigskip

\subsection*{A Kraft-McMillan inequality for fix-free codes}

A {\em fix-free code} is a code, which is prefix-free and
suffix-free, i.e. any codeword of a fix-free code is neither a
prefix, nor a suffix of another codeword. Fix-free codes were
first introduced by Sch{\"u}tzenberg \cite{schuetz}(1956) and
Gilbert and Moore \cite{moore}(1959), where they were called {\em
never-self-synchronizing} codes. A good overview of fix-free code
and some of their properties can be found for example in
\cite{per}. In the literature fix-free codes are also often called
{\em affix-free}, {\em bifix-free} or {\em
reversible-variable-length} (RCLs) codes.

\medskip

Ahlswede, Balkenhol and Khachatrian propose in
\cite{ahlswede}(1996) the conjecture that a Kraftsum of a lengths
sequence smaller than or equal to $\frac{3}{4}$, imply the
existence of a fix-free code with codeword lengths of the
sequence. This is known as the $\frac{3}{4} $-conjecture for
fix-free codes. Ahlswede, Balkenhol and Khachatrian give in
\cite{ahlswede} a justification of this conjecture. Especially
they show that the conjecture holds for $\frac{1}{2} $ in place of
$\frac{3}{4}$. Therefore a formulation of an existence theorem for
fix-free codes in terms of a Kraftinequality similar to the first
implication of Kraft-McMillan theorem is possible. Furthermore
Ahlswede, Balkenhol and Khachatrian prove in \cite{ahlswede}, that
for any number $\gamma $ bigger than $\frac{3}{4} $, there exists
a lengths sequence with Kraftsum smaller than $\gamma $, for which
no corresponding fix-free code exists. Otherwise, there are
fix-free codes with Kraftsum bigger than $\frac{3}{4}$. For
example the set of all words of fixed length $n$ is a fix-free
code with Kraftsum one. This shows that the first implication of
Kraft-McMillans theorem can not hold for fix-free codes with
Kraftsums bigger than $\frac{3}{4}$. Moreover a formulation of
Kraft-McMillan theorem for fix-free codes, in such a way, that both
implications hold for the same upper bound of the Kraftsum, is not
possible. Originally Ahlswede, Balkenhol and Khachatrian examined
only the case of a binary alphabet and a finite codes. However,
Harada and Kobayashi generalized in \cite{harada}(1999) all
results
of \cite{ahlswede} for the case of $q$-ary alphabets and infinite codes.\\

\medskip
Over the last years many attempts were done to prove the
$\frac{3}{4} $-conjecture either for the general case of a $q$-ary alphabet
or at least for the special case of a binary alphabet. All old results which
are related to the $\frac{3}{4}$-conjecture can be found in
\cite{ahlswede}-\cite{zeger}. Most of these results show that the
conjecture holds for some special kinds of lengths sequences or that
a weaker form of the conjecture is true. For example Harada and Kobayashi
show in \cite{harada} the conjecture for two level codes,
in the general case of $q$-ary alphabets or Yekhanin shows that the conjecture
holds for $\frac{5}{8} $ in place of $\frac{3}{4} $ in the case of a binary alphabet.
We survey in this survey all these old results about the $\frac{3}{4}-conjecture $ and furthermore
we obtain some new results, which are mostly generalizations of older
results for the binary case to the case of a $q$-ary alphabet. A collection of
all results can be found in the appendix at the end of this survey. Furthermore
a small summary of this survey can be found at the end of this Introduction.\\

\bigskip
\subsection*{Applications of fix-free codes}

\smallskip
A theorem which shows the existence of a fix-free code for given
codeword lengths and a construction of fix-free codes for a given lengths sequence
is quite important. Commonly variable length prefix-free
codes are used for data compressing. However, fix-free codes have some properties
which make them more favorable for a lot of applications compared with
prefix-free codes.
While fix-free codes are both prefix-free codes and suffix-free codes, it follows
that they are bidirectionally decipherable, whereas prefix-free and suffix-free
codes can be decoded only in one direction.
A string which is encoded with a prefix-free code
can instantaneously be decoded from the beginning toward the end, whereas
a message, encoded by a suffix-free code, can be deciphered backwards, from the
end to the beginning. Therefore the fix-free property ensures, that
messages which are encoded with a fix-free code, can be read from both directions.

\medskip
For example let $\C_1:=\{ 1,00,01 \},\C_2:=\{ 1,10,100 \} $ and
$C_3:=\{ 1,00,010 \} $, then $\C_1 $ is a prefix-free code, $\C_2 $ is a
suffix-free code and $\C_3 $ is a fix-free code. We encode the letters
N,I,A with the codes $\C_1,\C_2 $ and $\C_3 $ respectively, as follows:\\*

\[\begin{array}{|lc|l|r|l|}
\hline
\mbox{Source} & \quad & \C_1 & C_2 & C_3 \\
\hline
\mbox{A} & \longrightarrow & 1  & 1   & 1   \\
\hline
\mbox{I} & \longrightarrow & 01 & 10  & 00  \\
\hline
\mbox{N} & \longrightarrow & 00 & 100 & 010  \\
\hline
\end{array}\]

\smallskip
If the sequence $0001001$ is a message which is encoded with $\C_1 $,
we can decipher the string
step by step from left to right. The first codeword occurring in the string
from the left hand side is $00$. Since $00$ is neither a prefix
of $1$ nor a prefix of $01$, it follows that the message begins with an N.
$01$ is the next codeword of $\C_1 $ which occurs from left to right in the string.
While $01$ is not a prefix of another codeword in $\C_1 $, we obtain as the second
letter $I$. If we proceed in this way, we decode the string $0001001$ as the
message NINA. However, if we try to read the string from right to left, we have
some problems. The first codeword which occurs on the right hand side of the string
is $1$. This can mean, that the message ends with $N$ or $I$, because $1$ is a suffix
of $01$. If we proceed backward we obtain $01$. This gives us the same problem,
because it can mean, that the message ends with $I$ or with $NA$. The next step
backward gives us $001$. This means obviously $NA$. However, this shows that
the string $0001001$ can not decoded codeword by codeword from right to left.

\smallskip
In the same way, a string which is encoded with $\C_2 $ can be
decoded step by step from the end toward the beginning, but it is in general
not possible to decipher such a string by proceeding from left to right.
For example, the string $100101001$ is encoded with $\C_2 $. It can be decoded
as NINA, if we start at the end of the string, go backward to the beginning
and decode directly every codeword when it occurs. If we start on the left hand
side we have the same problem as above. Since $1$ means, that the message
begins with any letter. Since $\C_3 $ is both prefix-free and suffix-free,
we can decode a string which is generated by $\C_3 $ from both sides.
For example $010000101$ can be read from the left-hand side as well as from
the right-hand side as NINA.\\

\medskip
The bidirectional decoding property of fix-free codes is useful for many
applications. For example, a string in a file which is compressed by a fix-free
code, can be searched from both directions or a text which is encoded with a
fix-free code can be decoded from both directions simultaneously.
This reduce the decoding time to half, in comparison
with decoding in one direction only.

\pagebreak
{\samepage
As another example: Suppose, that we have the problem to find a pattern
$*P*$ in a given text which is encoded with some code. $P$ is a string
and $*$ represents an arbitrary string, maybe the empty string, which completes
the string $P$ to a word or a sentence respectively.
If we want to complete the word or the sentence matched by $P$, we have to decode
forward and backward from the position, where $P$ was detected. We can do this,
if all codewords have the same length. However, if we want to
reduce the length of the encoded text, we have to use a variable length
code. Since forward and backward encoding is necessary, the text has to be encoded
with a fix-free code.}

\smallskip
Related to the last example is the
{\em Key Word In Context} (KWIC) display.(see Heaps \cite{heap})
A query for a text consists of one
or several keywords and the location in the text where these words occur.
This is done with a list of pointers for every keyword, which contains all
positions of the appearance of the keyword in the text. A suitable way to present
a query, is to show the context of the appearance of the keywords in the text.
Therefore each of the $k$ words in the text
which appear before and behind the keywords are presented,
where $k$ is a fixed or a variable integer.
This make bidirectional decoding necessary. If the wasteful
way of encoding the text with a fixed-length code should be avoided,
the text has to be encoded with a variable length fix-free code.\\

\medskip
Another advantage of fix-free codes, in comparison with prefix-free codes, is
their higher robustness in the presence of transmission errors. This
is used for example in the development of video and media standards.
Most parts of a video file are commonly encoded with a variable length
prefix-free code (VCL), which minimize or reduce the average codeword length
in comparison with a fixed-length code. Such a code is highly susceptible
to transmission errors. There are two classes of bit errors which can occur,
these are propagating errors and non-propagating errors. A non-propagating error
gives only an incorrect decoding of the codeword in which the error occur.
On the other hand a propagating error causes a loss of synchronization.
In this case the bitstream behind the error will be decoded incorrectly or a
decoding of the resisting bitstream is not possible. In some cases
synchronizing will be reestablished later by itself, but also in this case
often a lot of data is lost. Therefore commonly a frame of a video file
is grouped into several segments. Each two of them are divided by a
synchronization marker, such that a propagating error in one
segment does not cause an erroneous decoding in another segment. Other
kinds of error protection can be used to impose a more reliable code,
if the data is transmitted trough a noisy channel. For example one can encode
the video data with an error correcting code or with a comma-free code.
Another method is to encode the most important parts
of the video data with a more error robust code only.
However, any of these more reliable coding schemes commonly increase
the average codeword length. This defeats the advantage of a careful use of
resources, which is obtained by compressing the video data with a variable length
code. Alternatively somebody can encode the video data with a fix-free code
with the same or at least similar codeword
lengths as the variable length prefix-free code.
In this context a fix-free code is called a {\em reversible-variable-length-code}
(RVLC). If an error burst occurs in a fix-free encoded segment,
the decoder can jump to the synchronization marker at the end of the segment and
decode backward to the error. Thus not all data in a segment behind an error is lost,
if the video file is encoded with a fix-free code. This is shown in the pictures
below. Furthermore it is sometimes possible to locate the position of an error
in a segment by artificially causing additional errors and applying bidirectional
decoding, where the results are compared with the initial decoder output.\\

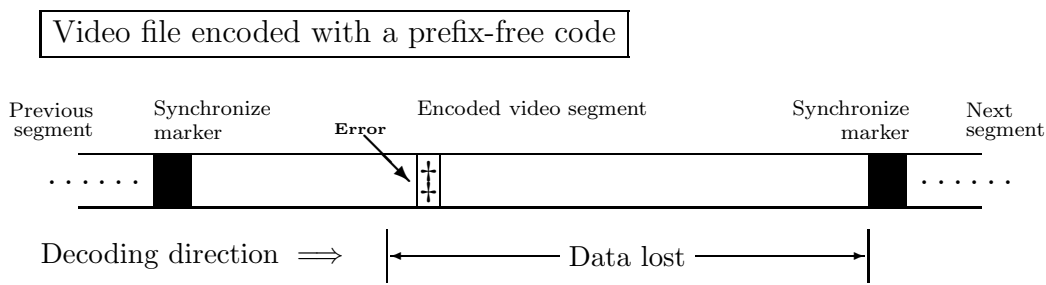
\begin{samepage}
\begin{quote}
\setlength{\unitlength}{1mm}
\begin{picture}(120,60)
\put(0,40){\framebox(78,6){Video file encoded with a prefix-free code}}
\put(20,20){\framebox(90,7){\quad }}
\put(15,19.9){\rule{5mm}{7.25mm}}
\put(110,19.9){\rule{5mm}{7.25mm}}
\put(50,20){\framebox(3,7){\Large $\ddagger $}}
\put(15,19.95){\line(-1,0){10}}
\put(15,27.08){\line(-1,0){10}}
\put(2,19.8){\makebox(10,7){\large $\ldots\ldots $}}
\put(115,19.95){\line(1,0){10}}
\put(115,27.08){\line(1,0){10}}
\put(118,19.8){\makebox(10,7){\large $\ldots\ldots $}}

\put(15,33){\makebox(0,0)[l]{\scriptsize Synchronize}}
\put(15,30){\makebox(0,0)[l]{\scriptsize marker}}

\put(115.5,33){\makebox(0,0)[r]{\scriptsize Synchronize}}
\put(115.5,30){\makebox(0,0)[r]{\scriptsize marker}}

\put(50,33){\makebox(0,0)[l]{\scriptsize Encoded video segment}}

\put(123,33){\makebox(0,0)[l]{\scriptsize Next}}
\put(123,30){\makebox(0,0)[l]{\scriptsize segment}}

\put(7,33){\makebox(0,0)[r]{\scriptsize Previous}}
\put(7,30){\makebox(0,0)[r]{\scriptsize segment}}

\thicklines
\put(43,29.5){\vector(1,-1){6} }
\put(42.5,30.5){\makebox(0,0){\tiny\bf Error}}

\thinlines
\put(46,10){\line(0,1){7}}
\put(110,10){\line(0,1){7}}
\put(46,10){\makebox(64,7){\small Data lost}}

\put(69,13.5){\vector(-1,0){22.5}}
\put(87,13.5){\vector(1,0){22.5}}
\put(0,10){\makebox(0,7)[l]{\small Decoding direction {\bf $\;\Longrightarrow$}}}

\end{picture}\\

\setlength{\unitlength}{1mm}
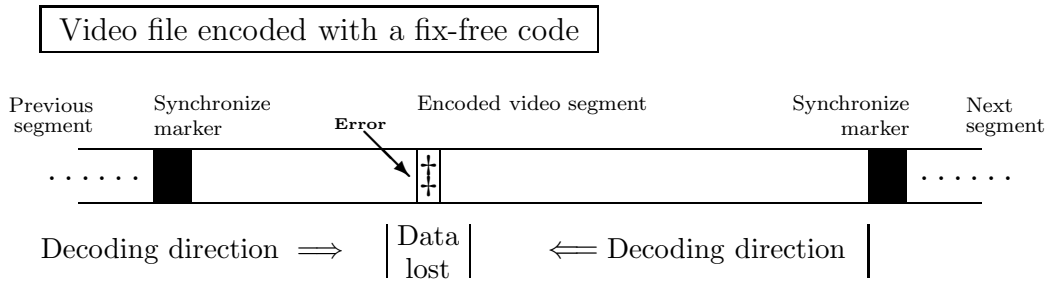
\begin{picture}(120,60)
\put(0,40){\framebox(74,6){Video file encoded with a fix-free code}}
\put(20,20){\framebox(90,7){\quad }}
\put(15,19.9){\rule{5mm}{7.25mm}}
\put(110,19.9){\rule{5mm}{7.25mm}}
\put(50,20){\framebox(3,7){\Large $\ddagger $}}
\put(15,19.95){\line(-1,0){10}}
\put(15,27.08){\line(-1,0){10}}
\put(2,19.8){\makebox(10,7){\large $\ldots\ldots $}}
\put(115,19.95){\line(1,0){10}}
\put(115,27.08){\line(1,0){10}}
\put(118,19.8){\makebox(10,7){\large $\ldots\ldots $}}

\put(15,33){\makebox(0,0)[l]{\scriptsize Synchronize}}
\put(15,30){\makebox(0,0)[l]{\scriptsize marker}}

\put(115.5,33){\makebox(0,0)[r]{\scriptsize Synchronize}}
\put(115.5,30){\makebox(0,0)[r]{\scriptsize marker}}

\put(50,33){\makebox(0,0)[l]{\scriptsize Encoded video segment}}

\put(123,33){\makebox(0,0)[l]{\scriptsize Next}}
\put(123,30){\makebox(0,0)[l]{\scriptsize segment}}

\put(7,33){\makebox(0,0)[r]{\scriptsize Previous}}
\put(7,30){\makebox(0,0)[r]{\scriptsize segment}}

\thicklines
\put(43,29.5){\vector(1,-1){6} }
\put(42.5,30.5){\makebox(0,0){\tiny\bf Error}}

\thinlines
\put(46,10){\line(0,1){7}}
\put(57,10){\line(0,1){7}}
\put(110,10){\line(0,1){7}}

\put(46,10){\makebox(11,7)[t]{\small Data}}
\put(46,10){\makebox(11,7)[b]{\small lost}}

\put(0,10){\makebox(0,7)[l]{\small Decoding direction {\bf $\;\Longrightarrow$}}}
\put(57,10){\makebox(53,7)[r]{
{\bf $\Longleftarrow\;$}\small Decoding direction\hspace{3mm}}}

\end{picture}
\end{quote}
\end{samepage}

\smallskip
\pagebreak
An overview of error handling of fix-free codes and their applications in video
encoding, especially in the video standards H.263 and MPEG-4, can be found in
\cite{tak}-\cite{vil3}. Furthermore in 1999 a data-partition structure based on
reversible variable length codes (fix-free codes), has been adopted
as the addition Annex V to the H.263++ video standard
(see \cite{h263a}, \cite{h263b} and also \cite{vil3}).

\medskip
The most important advantage of variable length codes in comparison with fixed-length
codes, is their low average codeword length for a given source.
A source is a set of finite symbols together with a probability
distribution. For example, one can choose as a source
the Latin alphabet together with the probability distribution which
corresponds to the frequency of the Latin letters in a certain text or
in a certain language. If the symbols in the alphabet are encoded by some
code, the average codeword length is the sum of the codeword
lengths weighted with the probabilities of the source.
If we want to reduce decoding, encoding and transmission time or
memory resources, it is favorable to choose a code with a low average
codeword length. Therefore an optimal code, with respect to a source,
is a code with minimal average
codeword length. Huffman shows in \cite{huff} (1956) that it is possible for
every source,
to choose an optimal code which is prefix-free and that an optimal prefix-free code
is also an optimal code. Furthermore he gave a construction of such prefix-free codes
for a given source. Therefore optimal prefix-free codes are
called {\em Huffman codes}.

\smallskip
Especially Huffman codes are {\em complete}, where
finite complete codes are codes with Kraftsum one. It can also be said, that
the code is a maximal code.\footnote{Take in account that in general for infinite codes,
completeness, maximality and to be code with Kraftsum one are not equivalent
conditions.} Since fix-free codes are especially prefix-free codes, the question
rises, wether there exists a fix-free Huffman code for a given source.
Fraenkel and Klein gave in \cite{klein} (1989) an algorithm which constructs
a fix-free Huffman code for a given source, if there exists one.
Furthermore the existence and properties of complete fix-free codes
are studied extensively in \cite{per}.

\smallskip
\enlargethispage*{14pt}
On the other hand there exists sources, for which no fix-free Huffman codes exist.
An example can be found in \cite{yeung}.
If $(0.7 ,0.1 ,0.1 ,0.1) $ be the probability distribution of
a source, $\{ 00,01,10,11 \} $ is the only complete fix-free code which
corresponds to the source. The average codeword length of this code is $2$, but
$ \{ 0,11,101,1001 \} $ is a fix-free code for the same source with
average codeword length $1.6 $, where the Kraftsum is $\frac{15}{16} $.
Since Huffman codes are complete codes, there does not exist fix-free
Huffman code for the source $(0.7 ,0.1 ,0.1 ,0.1) $. Therefore the
question rises, how we can construct an optimal fix-free code for a given source.
Although such an optimal fix-free code is not an optimal code in general,
the examples above show that some applications make encoding with a fix-free code
necessary or much more favorable than encoding with a prefix-free code.
Since in general an optimal fix-free code is not complete, we have to pay
attention to fix-free codes with Kraftsum smaller than one.

\smallskip
First we could try to answer the question of the existence of a fix-free code
for given lengths. If the $\frac{3}{4} $-conjecture holds, it would answer the
question at least partially in an easy way. However, due to my knowledge
it is not known, wether there are sources with an optimal fix-free code,
which has Kraftsum smaller than or equal to $\frac{3}{4} $.

\smallskip

On the other hand
a proof of the $\frac{3}{4} $-conjecture, will also give an upper bound
for the average codeword length of an optimal fix-free code in the form
of the noiseless coding theorem for prefix-free codes. If the probability
distribution of a source is given by $P=(p_1,\ldots ,p_n) $, the noiseless
coding theorem states, that the average codeword length of a Huffman code for this
source, is bounded by $H(P) $ from below and by $H(P)+1 $ from above.
Where $H(P)$ is the entropy of the source distribution, which is defined
for binary codes  as $H(P)=-\sum_{i=1}^n p_i \log_2 p_i $.
While a fix-free code is also a prefix-free code, we have $H(P) $ also
as a lower bound for the average codeword length of an optimal fix-free code.
Ahlswede, Balkenhol and Khachatrian show in \cite{ahlswede},
that the conjecture holds for $\frac{1}{2} $ instead of $\frac{3}{4}$
and that this imply an
upper bound of $H(P)+2 $ for the average codeword length of the optimal fix-free code.
However, Yekhanin shows in \cite{yek2}
that the (binary) conjecture holds for $\frac{5}{8}$ in place of $\frac{3}{4}$
and this lowers the upper bound of an optimal fix-free code to
$H(P)+4-\log_2 5 $, which is approximately $H(P) +1.678 $.
However it can easily be shown, that the
$\frac{3}{4} $-conjecture would improve this upper bound (for the binary case) to
$H(P)+3+\log_2 3 $, which is approximately $H(P)+1.415 $. The proof of
this and similar statements follows the same line as the proof of the
original noiseless coding theorem for prefix-free codes,
which can be found as an example in \cite{ahlswede2}.
An upper bound for the average code word length of an optimal fix-free
code can also be found in \cite{yeung}.

\medskip
Another way to obtain ``good'' fix-free codes for a given source, is shown  by
Takishima, Wada and Murakami in \cite{tak}(1995) and by Tsai and Wu
in \cite{tsai}(2001).
They gave there algorithms for construction of fix-free codes,
which starts with the lengths of a Huffman code for a given source.
This algorithms was improved by Lakovi{\'c} and
Villasenor in \cite{lak}(2003). The average codeword length of the fix-free codes
constructed by these algorithms for the English alphabet is shown in the tabular below.\\*

\begin{center}
\begin{tabular}{|c|c|c|c|}
\hline
\multicolumn{4}{|c|}{Average codeword length for the English alphabet}\\
\hline
Huffman & Takishima's & Tsai's & Lakovi{\'c}'s \\
code & fix-free code & fix-free code & fix-free code \\
\hline
4.15572 & 4.36068 & 4.30678 & 4.25145 \\
\hline
\end{tabular}\\* \end{center}

\nopagebreak
\noindent
It was not proven, that the
algorithms construct an optimal fix-free code for a given source and it
seems to be, that they do not. However, we pay no more attention
to this algorithms in this survey.

\bigskip
\subsection*{Summary of this survey}\nonumber

\smallskip
In this survey we focus mostly on results which shows the
$\frac{3}{4}$-conjecture for special kinds of lengths sequences or on
results which show that the conjecture holds in a weaker form.
We distinguish between the conjecture for the binary case and the conjecture
for the general $q$-ary case.\\

\medskip
In Chapter 1 we give first an overview and a proof of the original
Kraft-McMillan theorem for prefix-free codes. Then we give a justification
of the $\frac{3}{4}$-conjecture for fix-free codes and examine different
forms of the conjecture and the relations among themselves. Especially
we show for the general $q$-ary case that the conjecture
holds for $\frac{1}{2} $ in place
of $\frac{3}{4} $ and that for every number bigger than $\frac{3}{4} $ the
conjecture can not be hold. These theorems were first shown by
Ahlswede, Balkenhol and Khachatrian in \cite{ahlswede}(1996) for the binary case.
A generalization was shown by Harada and Kobayashi in \cite{harada}(1999).
Finally we study in Chapter 1 the existence of fix-free extensions of a
fix-free code, i.e. we will see, that extensions of
fix-free codes are crucially different to extensions of prfix-free codes.\\

\medskip
Chapter 2 deals with the $\frac{3}{4} $-conjecture in the case of a $q$-ary
alphabet.
We prove three theorems which show that the conjecture
holds for special kinds of lengths sequences. The first theorem
occurs first for the binary case in \cite{ahlswede}(1996) and was generalized in
\cite{harada}(1999). It says, that the conjecture holds, if for two lengths of the
sequence, there is a gap of at least twice time of the smaller length,
where no other codeword length occur. The second theorem in the chapter shows
that the conjecture holds for two level codes and it was proven by Harada and
Kobayashi in \cite{ahlswede}. Finally we show that the
$\frac{3}{4}$-conjecture holds for finite sequences,
if the numbers of codewords on each level is bounded by a term which depends
on $q$ and the smallest codeword length which occurs in the lengths sequence.
This theorem was first shown by Kukorelly and Zeger in \cite{zeger}(2003) for the
binary case. The generalization of this theorem in Chapter 2 to  $q$-ary alphabets,
is one of the new results in this survey.\\

\medskip
Chapter 3 is a long preparation of Chapter 4. While we will
construct fix-free codes from regular subgraphs in the
de Bruijn digraph in Chapter 4, we give in Chapter 3 an introduction to
the $q$-ary, $n$-th level de Bruijn digraph $\BGr{n} $.
Especially we have to know the numbers of vertices, for which there exists
a $k$-regular subgraph in $\BGr{n} $. De Bruijn graphs were introduced by
de Bruijn \cite{bruijn}(1946) and Good \cite{good}(1946) independently.
After a small summary of some basic facts about
digraphs and de Bruijn digraphs, we show that for every number $L$ of vertices
in $\BGr{n}$, there exists a cycle of length $L$ in \BGr{n}. This was shown
independently by Yoeli, Braynt, Heath , Killick, Golomb, Welch and Goldstein
for binary de Bruijn digraps. Lempel generalized this result to the $q$-ary
de Bruijn digraphs. (see for all of these Lempel in \cite{lempel}(1971)).
Especially cycles in \BGr{n} are
$1$-regular subgraphs. Therefore we obtain, that there exist $1$-regular subgraphs
in \BGr{n} for any possible number of vertices. At the end of the chapter,
we try to answer the question of the existence of $k$-regular subgraphs in
\BGr{n} with certain numbers of vertices. We will see that there do not
exist $k$-regular subgraphs in \BGr{n} for vertices numbers smaller than $k^n$
or for vertices numbers between $k^n $ and $k^n-k^{n-1}$. Furthermore we
give some constructions for $k$-regular subgraphs in \BGr{n} with more than
$k^n-k^{n-1}$ vertices. However, we will give no full answer on the question, for
which numbers of vertices there are $k$-regular subgraphs in \BGr{n}.\\

\medskip
In Chapter 4 we pay attention to a theorem which was claimed by Yekhanin
in \cite{yek1}(2001). If the Kraftsum of the first level which occurs
in a lengths sequence together with the Kraftsum of the following level is
bigger than $\frac{1}{2}$, then from  Yekanins theorem follows, that the
$\frac{3}{4}$-conjecture holds. Yekanin claimed this theorem only for the binary case.
However, no full proof of this theorem was published. Therefore we will
give an own proof in Chapter 4, where we follow the proof idea which was proposed
by Yekhanin in \cite{yek1}. Furthermore we give a generalization of the theorem.
For the proof of the theorem and its generalization,
we introduce $\pi $-systems, which are special kinds of fix-free codes
with Kraftsum $\left\lceil\frac{q}{2}\right\rceil q^{-1}$.
Later we show, that $\pi$-systems can be extended to
fix-free codes with Kraftsum smaller than or equal to $\frac{3}{4} $. This
is called the $\pi $-system extension theorem, which we show
in the first section of Chapter 4.
In the second section of Chapter 4 we show, that $\pi $-systems with only
two neighbouring levels and
$L\cdot \left\lceil\frac{q}{2}\right\rceil$ codewords on the first level
exist, if and only if
there exists a $\left\lceil\frac{q}{2}\right\rceil$-regular subgraph of \BGr{n}
with $L$ vertices.
Furthermore we show that arbitrary one level $\pi $-systems exist.
Since there exist
cycles of arbitrary length in $\BGr[2]{n}$, we obtain Yekhanins original theorem
with the $\pi $-system extension theorem. However, in the generalization of Yekhanins
theorem to the $q$-ary case, an extra condition for the existence of
$\left\lceil\frac{q}{2}\right\rceil$-regular subgraph in $\BGr{n}$ occurs.
Moreover we will show another version of all of these theorems, which uses
other bounds than $\frac{3}{4}$ for the Kraftsum. To prove these more
general versions, we work with
$k$-regular subgraphs in \BGr{n} instead of
$\left\lceil\frac{q}{2}\right\rceil$-regular subgraphs in \BGr{n}.
Mainly all of these results are new.
Finally we prove in this chapter some minor new results for very special sequences
by using the $\pi $-extension theorem for $\pi $-systems with more than two levels.\\

\medskip
\pagebreak
Chapter 5 is about the binary version of the $\frac{3}{4}$-conjecture.
It begins with a summary of known results, which are shown only for the binary case.
Then we give a simple construction of binary fix-free codes with the help of
quaternary fix-free codes, by applying this construction to the results
we have obtained in Chapter 2 and Chapter 4, we obtain some new results for
the binary case of the $\frac{3}{4}$-conjecture.
At the end of Chapter 5 we prove
a result which was obtained by Yekanin in \cite{yek2}(2004),
which shows, that the binary conjecture holds,
if we replace in the conjecture $\frac{3}{4}$ by $\frac{5}{8} $.
For this we use some special kinds of fix-free codes, for which
the codewords with the same first letter and the same last letter
are grouped in blocks. The blocks are ordered by the codeword lengths.
Then we try to
apply the technique of Yekhanins prove on the $q$-ary case. This gives us a new
conjecture, which we prove for the ternary case. However, the new conjecture
brings nothing new, because for all $q$ bigger than $2$ we obtain a Kraftsum
smaller than $\frac{1}{2}$. Somebody might only be interested in the special block form
of the fix-free codes, which occurs in the conjecture.\\

\medskip
Finally the appendix contains all known old results and all new results of the survey,
which are related to the $\frac{3}{4}$-conjecture.

\bigskip

\subsubsection*{A new result which is not contained in this survey}

\smallskip
While this survey was in progress, K. Tichler has proven the conjecture
which occurs in the last section of Chapter 5 for arbitrary $q$-ary alphabets.
For a binary alphabet, the conjecture follows from Yekhanins proof in \cite{yek2} of
the $\frac{5}{8}$-version of the $\frac{3}{4}$-conjecture, which can also be found
in the last section of Chapter 5. For a ternary alphabet the conjecture
was first shown by the author of this survey, in the way as it is shown in Chapter 5.
Some months after the author proposed the conjecture in Chapter 5, K. Tichler
gave a counting proof, which shows that the conjecture holds for all
$q$-ary alphabets. This conjecture
gives no new results for the the $\frac{3}{4} $-conjecture for $q$-ary alphabets,
because the fix-free codes in the conjecture have Kraftsums smaller than
$\frac{1}{2} $ for $q>2 $ and the binary case was already shown by Yekhanin.
However, somebody might be interested in the special block form of the fix-free codes
which occurs in the conjecture. Furthermore K. Tichler has proven a variation
of the conjecture in Chapter 5, which shows that for a ternary alphabet
the $\frac{3}{4} $-conjecture holds for some $\gamma_3>\frac{1}{2} $
in place of $\frac{3}{4} $. This is a new result for the
$\frac{3}{4} $-conjecture in the case of ternary alphabets.
Maybe such an variation of the conjecture in Chapter 5 is possible for all $q$.
Since the proofs of K. Tichler are not worked out up to now, they won't be
presented in this survey.\\

\pagebreak
\setlength{\parindent}{1cm}
\bigskip
{\bf Acknowledgement}

\medskip
The author would like to express many thanks to Prof. Dr. Rudolf Ahlswede,
Dr. Christian Deppe, Dr. Haik Mashurian and the anonymous
readers for valuable comments that helped improving the presentation
and also Krisztian Tichler for many helpful notes on Chapter 5.

\setlength{\parindent}{0.4cm}
\newpage

\chapter{The Kraftinequality for fix-free codes}

\smallskip

\section{Notations and Definitions}

\smallskip
Throughout this survey we denote with $\NN $ the set of natural numbers without zero
an with $\NN_0 $ the set of natural numbers with zero. If $\M $ is an arbitrary
set, we write $\PP (M)$ for the powerset of $\M $. This is the set which contains
all subsets of $\M $ as its elements.\\

\smallskip
Let $\A $ be an arbitrary set, which we call an {\em alphabet }.
The elements of $\A $ are called the {\em letters } of the alphabet $\A $. A
{\em word of length $n$ } over the alphabet $\A $ is a finite sequence of length $n$ with
values in $\A $. We write $a_1\ldots a_n \in\A^n $, for a finite sequence.
The empty sequence is called the
{\em empty word } or the {\em word of length $0$ } and is denoted by $e$.\\

\smallskip
For two words $w=w_1\ldots w_n\in\A^n $ and $v=v_1\ldots v_m \in\A^m $, we define
the word $w\cdot v\in\A^{n+m} $ by the concatenation of the two sequences
\[ w\cdot v:= w_1\ldots w_nv_1\ldots v_m ,\]
where we write $wv$ in place of $w\cdot v $. Especially the operation
$\cdot $ is associative and $we =ew =w $ for all $w\in\A^n $ and $n\in\NN_0 $.\\

\smallskip
We denote with $\A^* $ and $\A^+ $ the set of all words on $\A $ with finite length and
all finite words on $\A $ of length bigger than zero, respectively.
\[ \A^* := \bigcup\limits_{n=0}^{\infty } A^n \quad ; \quad
\A^+ :=\bigcup\limits_{n=1 }^{\infty }\A^n =\A^* -\{ e \} \]

\smallskip
A {\em monoid} is a set $\M $ equipped with an associative binary operation\\
$\cdot :\M\times \M\rightarrow \M $ and a neutral element $e\in\M $.
Obviously $(\A^*,\cdot , e) $ is a monoid. Let $(\M,\cdot , e) $ and
$(\N , * , \mbox{\bf 1}) $ be two moniods. A map $\varphi :\M\rightarrow \N $ is called
a {\em monoidhomomorphism}, if $\varphi (e) = \mbox{\bf 1} $ and
$\varphi (u\cdot v) = \varphi (u) *\varphi (v) $ for all $u,v\in\M $. The
monoidhomomorphism $\varphi $ is called an {\em monoidisomorphism}, if $\varphi $
is a bijective map. In this case it follows that the inverse map $\varphi^{-1} $
is also a monoidisomorphism.

\smallskip
For $v,w\in \A^* $ the word $v$ is called a {\em prefix } of the word $w$, if
there exists a word $u\in\A^* $ with $w=vu $. $v$ is called a {\em suffix } of $w$,
if  $w=uv $ for some $u\in\A^* $.

\smallskip
A {\em factor } of $w$ is a subword of $w$. This means
$u\in\A^* $ is a factor of $w$, if there exists words $v,v'\in\A^* $ such that
$w=vuv' $. i.e. $e$ is a prefix, suffix and proper factor of any word in $\A^* $.
Let $\X \subseteq \A^* $ and $w\in\A^* $. A {\em factorization } of $w$ with words in $\X $,
are words $x_1,\ldots ,x_n \in\X $ such that
\[ w= x_1\ldots x_n .\]

\smallskip
For $\X, \Y \subseteq \A^* $ and $v,w\in\A^* $ we define:
\[\begin{array}{lcl}
\X\Y & := &  \{ xy\in\A^* \, |\, x\in\X ,\, y\in\Y \} \,  ,\\
\X^{-1}\Y & := & \{ z\in\A^* \, | \, \exists x\in \X\, , \, \exists y\in\Y \;
\mbox{ with }\; y=xz \} \, , \\
\X\Y^{-1}  & := &  \{ z\in\A^* \, | \, \exists x\in \X\, , \, \exists y\in\Y \;
\mbox{ with }\; y=xz \} \, , \\
w^{-1}\X & := & \{ w \}^{-1}\X \quad ,\quad \X w^{-1} := \X\{ w \}^{-1} \, ,\\
w^{-1}v & := & \{ w \}^{-1} \{ v \} =\{ u\in\A^* | v= wu \} \quad , \quad
wv^{-1} := \{ w \}\{ v \}^{-1} \, ,\\
w^n & := & \underbrace{w\ldots w}_{ \mbox{\scriptsize $n$-times}} \;\mbox{ for }\; n\in\NN \;
\mbox{ and }\; w^0:= e \, , \\[6mm]
\A^-\X & := & \left(\A^+ \right)^{-1}\X \quad , \quad
\X\A^-:=\X  \left(\A^+ \right)^{-1} \, ,\\
\A^{-n}\X & := &  \left(\A^n \right)^{-1}\X \quad , \quad
\X\A^{-n} :=\X \left(\A^n \right)^{-1} \;\mbox{ for }\; n\in\NN \, .

\end{array}\]
Especially $\X(\X^{-1}\Y ) $ is the set of all words in $\Y $ which have a prefix in $\X $,
$(\Y\X^{-1})\X $ is the set of all words in $\Y $ which have a suffix in $\X $,
$\A^-\X $ is the set of all proper suffixes of words in $\X $ and $\X\A^- $ is the set
of all proper prefixes of words in $\X $. Furthermore we obtain for the sets
$\X^n ,\X^* ,\X^+ , \X^-\Y , \Y\X^-  , X^{-n}\Y $ and $\Y\X^{-n } $:

\smallskip
\[
\begin{array}{lcl}
\X^n & = & \{ x_1\ldots x_n\in\A^*\, | \, x_1,\ldots , x_n\in\X \} \ ,\\
\X^* & = &\bigcup\limits_{n=0 }^{\infty }\X^n \quad , \quad
\X^+ = \bigcup\limits_{n=1}^{\infty }\X^n =\X^* -\{ e \} \, , \\
\X^-\Y & = & \left(\X^+ \right)^{-1}\Y \quad , \quad
\Y\X^- = \Y\left(\X^+ \right)^{-1} \, , \\
\X^{-n}\Y & = & \left(\X^{n} \right)^{-1}\Y \quad , \quad
\Y\X^{-n}  =  \Y\left(\X^{n} \right)^{-1} \, .
\end{array}
\]

\smallskip
It is easy to verify, that the following equations hold:
\[\begin{array}{lcl}
\X^{-1}\Y\cap\Y =\emptyset &\Leftrightarrow & \Y\subseteq\X^{-1}\Y  \Leftrightarrow
e\in\X \, ,\\
\X\Y^{-1}\cap\X =\emptyset &\Leftrightarrow & X\subseteq\X\Y^{-1}  \Leftrightarrow
e\in\Y \\
\X^{-1}\Y =\emptyset & \Leftrightarrow &
\mbox{ no word in $\Y $ has a prefix in $\X $} \, ,\\
\Y\X^{-1} =\emptyset &\Leftrightarrow &
\mbox{ no word in $\Y $ has a suffix in $\X $}\, ,\\[2mm]

(\X\Y )^{-1}\Z      & = & \Y^{-1}(\X^{-1}\Z ) =:\X^{-1}\Y^{-1}\Z \, , \\
\X(\Y\Z )^{-1}      & = & (\X\Y^{-1})\Z^{-1} =: \X\Y^{-1}\Z^{-1} \, ,\\
(\X^{-1}\Y )\Z^{-1} & = & (\X\Y^{-1})\Z^{-1} =:\X\Y^{-1}\Z^{-1} \, ,\\
\X^{-1}(\Y\cup\Z )  & = & \X^{-1}\Y \cup \X^{-1}\Z \, ,\\
\X^{-1}(\Y\cap\Z )  & = & \X^{-1}\Y \cap \X^{-1}\Z \, ,\\
\X^{-1}(\Y -\X )    & = & \X^{-1}\Y -\X^{-1}\Z \ .  \\
\end{array}\]
Similar equations holds for $(\X\cup\Y )\Z^{-1} $, $(\X\cap\Y )\Z^{-1} $ and
$(\X-\Y )\Z^{-1} $.\\

\smallskip
For $w\in\A^* $ and $a\in\A $ we denote with $|w| $ the length of the word $w$
and with
$<w|a> $ the number of occurrence of the letter $a$ in $w$. For example let
$\A =\{0,1 \} $ and $w=11010 $, then $|w| =5 $, $<w|0> =2 $ and $<w|1> =3 $.\\

\medskip\noindent
In the rest of this survey we suppose that alphabets are finite sets
with at least two elements. Therefore let $|\A | = q $ for some $q\ge 2 $.\\

\smallskip
Let $(\alpha_l )_{l\in\NN } $ be a sequence of nonnegative integers and
$\C\subseteq\A^+ $. We say the sequence $(\alpha_l )_{l\in\NN } $ {\em fits to }
the set $\C $ or $\C $ fits to $(\alpha_l )_{l\in\NN } $, if
$|\C\cap \A^l | =\alpha_l $ for all $l\in\NN $.

\smallskip
For $\C\subseteq\A^* $ and $n\in\NN_0 $, the {\em Kraftsum } and the
{\em $n$-th level Kraftsum } of $\C $ is defined as:
\[ \Kraft (\C ):= \sum\limits_{l=0}^{\infty } |\C\cap\A^l |q^{-l}\le \infty
\quad , \quad
\Kraft_n (\C ):= \sum\limits_{l=0}^{n } |\C\cap\A^l |q^{-l} <\infty \, .
\]

\medskip
A set $\C \subseteq\A^+ $ is called a {\em code } on the alphabet $\A $,
if every word in $\C^+ $ has a unique factorization of words in $\C $.\footnote{
In information theoretical papers a set with the property above is commonly
called an {\em unique decipherable  }
code and a code is an arbitrary subset of $\A^+ $. }
This means for all $x\in\C^+ $:
\[\begin{array}{ll}
\quad &
x=c_1\ldots c_n =d_1\ldots d_m \;\mbox{ with }\; c_i,d_j\in\C
\;\forall\, 1\le i\le n \, , \, 1\le j \le m \\
\Rightarrow & n=m \;\mbox{ and }\; c_i=c_j \;\forall\, 1\le l\le n=m
\end{array}\]

The next proposition shows, that any message
which is encoded with a code $\C\subseteq\A^+ $, can be decoded uniquely.

\begin{prop}
Let $\C\subseteq\A^+ $ and $\B $ be another alphabet with
$|\B | =|\C | $(, whereas $|\B |=\infty $ should be allowed).\footnote{
Since $\A $ is finite, the set $\B $ is at most countable.}Then $\C $
is a code if and only if there exists a bijection
$\beta :\B^*\leftrightarrow \C^* $, such that
$\beta (uv )=\beta (u)\beta (v) $ for all $u,v\in\B $.
\end{prop}
A proof of the proposition above, can be found for example in \cite{per}.

\smallskip
A code $\C\subseteq\A^+ $ is called a {\em maximal code }, if for
every word $c\in\A^+ - \C $ the set $\C \cup\{ c \} $ is not a code. We call a set
$\D\subseteq \A^+ $ an {\em extension } of the code $\C $, if $\D\supseteq \C $
and $\D $ is a code. $\D $ is called a {\em maximal extension}, if $\D $
is a maximal code.\\

\bigskip

Let $\M $ be an arbitrary set. A binary relation $\preceq $ on $\M $ is called
a {\em (partial) ordering,} if it is reflexive, antisymmetric and transetive.
This means:
\[
\begin{array}{ll@{\, : \;}l}
\mbox{(1)}\quad &\forall\, a\in\M  &\quad a\preceq a \, ,\\
\mbox{(2)}\quad &\forall\, a,b\in\M &\quad a\preceq b, \, b\preceq a\Rightarrow a=b \, ,\\
\mbox{(3)}\quad &\forall\, a,b,c\in\M &\quad a\preceq b, \, b\preceq c\Rightarrow a\preceq c \, .
\end{array} \]
We define $a\prec b $ as:
\[ a\prec b \Leftrightarrow a\preceq b \;\mbox{ and }\; a\neq b \, .\]

\smallskip
The ordering $\preceq $ is called a {\em linear ordering,} if for all $a,b\in\M $
the elements $a,b$ are comparable. This means $a\preceq b $ or $b\prec a $ for all
$a,b\in\M $.\\

\smallskip
Let $\C\subseteq\M $. We call $a\in\C $ a  {\em minimal element} of $\C $, if
$b\not\preceq a $ for all $b\in\C -\{ a \} $. We call $a$ the {\em least element} of
$\C $, if $a\preceq b $ for all $b\in\C $. If there exists a least element in $\C $,
then it is also a minimal element of $\C $ and moreover, there do not exist
other minimal elements in $\C $. In the same way we call $a\in\C $ a
{\em maximal element }of $\C $, if $a\not\preceq b $ for all $b\in\C-\{ a \} $
and $a$ is called the {\em greatest element} of $\C $, if $b\preceq a $ for all
$b\in\C $. If $\preceq $ is a linear ordering on $\C $, then every minimal element of
$\C $ is the unique least element and every maximal element of $\C $ is the unique
greatest element of $\C $.

\smallskip
An ordering $\preceq $ of a set $\M $ is called a {\em well-ordering,} if
it is a linear ordering and if every nonempty subset of $\M $ has a least element.\\

\smallskip
Let $\preceq_1 $ be an ordering of a set $\M_1 $ and $\preceq_2 $ be an ordering
of a set $\M_2 $. We call the orderings $(\M_1 ,\preceq_1  ) $ and
$(\M_2 , \preceq_2) $ {\em isomorph,} if there exists a bijection
$\varphi :\M_1\leftrightarrow \M_2 $ such that $a\preceq_1 b $ if and only if
$\varphi (a)\preceq_2 \varphi (b) $ for all $a,b\in\M_1 $.\\
We call $\varphi $ an {\em isomorphism} between $(\M_1 ,\preceq_1 )$ and
$(\M_2 ,\preceq_2 )$.

\pagebreak
\smallskip\noindent
Let us give two examples:
\begin{example}\upshape
Let $\M $ be an arbitrary set. The subsetrelation $\subseteq $ is a ordering
of $\PP (\M )$, with greatest element $\M $ and least element $\emptyset $.
\end{example}

\begin{example}\upshape
Let $n\in\NN $, then $0< 1<2<\ldots <n-1 $ is a well-ordering of
$\{ 0,\ldots ,n-1 \} $. Furthermore any linear ordering of a set $\M $ with
$|\M | = n$ is isomorphically to this ordering. The ordering
$0< 1 <2 <\ldots $ of $\NN_0 $ is also a well-ordering. This ordering of
the natural numbers is denoted by $\omega $. However, there exist much more
well-orderings and linear orderings of $\NN_0 $ which are not isomorphically
to $\omega $.
Let us define for example
$1\prec 2 \prec 3 \prec \ldots $ and $n \prec 0 $ for all $n\in\NN $. Then this is
a well-ordering of $\NN_0 $ which is not isomorphic to $\omega $.\footnote{
In general every well-ordering is isomorphically to the ordertype of a unique ordinal
number.}
\end{example}

\smallskip
Let $\preceq $ be an ordering of a set $\M $. A subset $\C\subseteq\M $ is called
a {\em chain}, if $\C $ is linear ordered by $\preceq $. $\C $ is called
an {\em antichain } if all elements of $\C $ are incomparable. This means
$a\not\preceq b $ and $b\not\preceq a $ for all $a,b\in\C $ with $a\neq b $.

\smallskip
We call an element $a\in\M $ a {\em lower bound } of $\C $ and an element
$b\in\M $ an {\em upper bound } of $\C $, if
$a\preceq c $ and $ c\preceq b $ for all $c\in\C $. Obviously $a,b\in\C $ if and only
if $a$ is the least element of $\C $ and $b$ is the greatest element of $\C $.

\smallskip
The next lemma is known as Zorn's lemma, which can be found in most books about
set theory (for example \cite{jech}), therefore we omit a proof.

\begin{lemma}[Zorn's lemma]\label{zorn}
Let $\preceq $ be an ordering of a set $\M $. If every chain has an upper bound,
then there exists a maximal element in $\M $.
\end{lemma}
\noindent
It is well known in set theory, that Zorn's lemma is an equivalence of the Axiom of
Choice (see for example \cite{jech}). Therefore proofs which use Zorn's lemma are
none-constructive proofs. As an example for Zorn's lemma we prove, that any code
has a maximal extension.

\begin{prop}
Let $|\A | =q\ge 2 $ and $\C\subseteq\A^+ $ be a code. There exists a maximal code
$\D $ with $\C\subseteq \ D\subseteq \A^+ $.
\end{prop}

\smallskip
\noindent
{\bf Proof: }
Let $\C \subseteq \A^+ $ be a code. We define $\M \subseteq\PP (\A^+ ) $ as
the set of codes extensions of $\C $.
\[ \M := \left\{ \D\subseteq\A^* \, \left| \, \C\subseteq \Y \;\mbox{ and }\;\Y
\;\mbox{ is a code } \right.\right\} \]
Obviously $\M $ is ordered by $\subseteq $ and any maximal element of $\M $
is a maximal extension of $\C $. Therefore it is sufficient to show, that
$\M $ has at least one maximal element.

\smallskip
Let $\M_c \subseteq\M $ be a chain in $\M $ and
$\D' :=\bigcup\limits_{\D \in \M_c }\D $. Then $\C\subseteq \D' $ and
$\D\subseteq \D' $ for all $\D\in\M_c $. Let us assume that $\D' $ is not
a code. Then there exists words $c_1,\ldots ,c_{n+m } \in\D' $ such that:
\begin{equation}\label{zorn1}
c_1\ldots c_n =c_{n+1}\ldots c_{n+m }.
\end{equation}
\noindent
Each of the $c_i$'s is contained in a code $\D_i\in\M_m $. Since $\M_c $
is linear ordered by $\subseteq $, it follows that there exists
$j\in\{ 1,\ldots ,n+m\} $ with
$\D_i\subseteq \D_{j} $ for all $1\le i\le n+m $.
We obtain that $c_1,\ldots ,c_{n+m} \in\D_j $. This is a contradiction, because
$\D_j $ is a code. Thus (\ref{zorn1}) can not hold. This shows that $\D' $
is a code, i.e. $\D'\in\M $.
Therefore $\D' $ is an upper bound of $\M_c $ in $\M $. By Zorn's lemma
follows, that $\M $ has a maximal element.\qed\\

\medskip
Let $\preceq $ be an ordering of a set $\T $. We call $(\T ,\preceq ) $ a
{\em tree }, if for every $a\in\T $ the set $\{ b\in\T | b\prec a \} $ is
well-ordered by $\preceq $ and if $\T $ has a least element, which is
called the {\em root } of the tree $\T $. For a tree $\T $, any chain in $\T $
is well-ordered by $\preceq $.

\smallskip
An element $a\in\T $ is called
a {\em node} of the tree. Furthermore it is called a {\em finite node on the
$l(a)$-th level}, if $|\{ b\in\T | b\prec a \} | =l(a)<\infty $.
If $a\in\T $ is a finite node, then the chain
$\{ b\in\T | b\prec a \} $ is isomorphic to the
well-ordering $0< 1 <2 <\ldots <l(a)-1 $.

\smallskip
Let $l\in\NN_0 $. The $l$-th level of the tree is defined as the set
$\T (l) := \{ a\in\T | l(a) =l \} $. For any $a,b\in\T (l) $ with $a\neq b $
the nodes $a$ and $b$ are incomparable. If $\T (n)=\emptyset $, then
$\T (l) =\emptyset $ for all $l\ge m $.\\

\smallskip
We call the tree $\T $ has {\em height} $h$
for some $h\in\NN $, if $\T (h-1) \neq \emptyset $ and
$\T (l) = \emptyset $ for all $l\ge h $. This means the heights of $\T $ is
the smallest level, which is empty. We write $\T $ has height $\omega $ or
$\T $ is an {\em $\omega $-tree, } if all nodes of $\T $ are finite and
$\T (l )\neq \emptyset $ for all $l\in\NN_0 $. If $\T $ is an $\omega $-tree,
then any chain in $\T $ is either isomorphic to $0< 1 <\ldots < n $ for some
$n\in\NN_0 $ or it is isomorphic to $\omega $.\footnote{For an arbitrary tree
any chain is isomorphic to the ordertype of a (unique) ordinal number and
the ordinal number which is isomorphic to
$\{ b\in\T | b\prec a \} $ is the level of $a$.
The height of $\T $ is the smallest empty level. Furthermore for any
chain the corresponding ordinal is smaller tan or equal to the heights of the tree.}

\smallskip
A {\em branch } of a tree $\T $ is a maximal chain $\C $ in $\T $. This means
$\C $ is a chain and for every $a\in\T-\C $ the set $\C\cup \{ a \} $ is not
a chain. Let $\T $ be a tree with finite heights or an $\omega $-tree and let $\C $ be
a branch in $\T $. We call $l\in\NN $ the length of the branch, if $\C $ is isomorphic
to $0 < 1 <2 \ldots < l-1 $ and we call $\C $ a branch of length $\omega $, if
$\C $ is isomorphic to $\omega $. If $\T $ has height $h\in\NN $, then there
exists a branch of length $h$.

\smallskip
\noindent
The next lemma shows, that this holds also for
$\omega $-trees  which have finite levels.

\begin{lemma}[K\"onig's lemma]\label{koenig}
Let $\T $ be an $\omega $-tree. If any level contains a finite number of
nodes, then there exists a branch of length $\omega $ in $\T $.
\end{lemma}

\smallskip\noindent
The lemma doesn't hold if the levels of $\T $ contain infinite nodes.
An example of such a $\omega $-tree and a proof of the lemma can be found in
\cite{jech}. In the proof of the lemma the Axiom of Choice is used, but the lemma is
not an equivalent of the Axiom of Choice. However, just as the Axiom of choice,
K\"onigs lemma can not be proven
or disproven with the set axioms of Zermalo-Fr\"anklel. This can be found for
example in \cite{choice}. Therefore also proofs
which use K\"onigs's lemma are none-constructive.\\

\smallskip
Let $\A $ be an arbitrary set. For $x,y\in\A^* $ we define
$x \stackrel{p}{\preceq } y $ if $x$ is a prefix of $y$ and
$x \stackrel{s}{\preceq } y $ if $x$ is a suffix of $y$. It is easy to verify
that $(\A^*, \stackrel{p}{\preceq } )$ and $(\A^*, \stackrel{s}{\preceq } ) $
are both $\omega $-trees with root $e$, which we call the prefix-tree and the
suffix-tree, respectively.
Furthermore the $l$-th level of both trees is given by $\A^l $ for all $l\in\NN_0 $.

\pagebreak
\section{Fix-free codes}

A set $\C\subseteq\A^* $ is called {\em prefix-free}, if no word in $\C $ is a prefix
of another word in $\C $ and it is called {\em suffix-free}, if no word in $\C $
is a suffix of another word in $\C $.
The set $\C $ is called {\em fix-free }
or {\em bifix-free }, if it is prefix- and suffix-free.
Since $e$ is a prefix and a suffix of every word in $\A^* $, we obtain, that
$\{ e \} $ is the only prefix- suffix- and fix-free set, which contains $e$
as an element. Therefore we obtain:
\[\begin{array}{lcl}
\C \;\mbox{ is prefix-free }\; & \Leftrightarrow & \C\A^+\cap\C =\emptyset \, ,\\
\C \;\mbox{ is suffix-free }\; & \Leftrightarrow & \A^+\C\cap\C =\emptyset \, .
\end{array}\]
The set $\C $ is called {\em fix-free }
or {\em bifix-free }, if it is prefix- and suffix-free.\\

\smallskip
For an arbitrary set $\C\subseteq\A^* $ the
{\em prefix-, suffix- and bifix-shadow } of $\C $ on the $n$-th level
are defined as:
\[\begin{array}{lclcl}
\pfs[n] (\C ) & := & \bigcup\limits_{l=0 }^{n} (\C\cap\A^l )\A^{n-l}
& \subseteq &\A^n \, ,\\
\sfs[n] (\C ) & := & \bigcup\limits_{l=0 }^{n} \A^{n-l}(\C\cap\A^l )
& \subseteq & \A^n \, ,\\
\bfs[n] (\C ) & := & \pfs[n] (\C )\cup \sfs[n] (\C ) & \subseteq  & \A^n \, .
\end{array}\]

\smallskip
\begin{prop}
Every subset of $\A^* $
which is prefix- suffix- or fix-free and not equal to $\{ e \} $ is also a code.
\end{prop}

\smallskip\noindent
{\bf Proof:}
Let $\C\subseteq \A^+ $ be a prefix-free set and let $\mbox{\bf w}\in\C^+ $,
$x_1,\ldots ,x_n,y_1,\ldots y_m\in\C $, such that
$\mbox{\bf w}=x_1\ldots x_n = y_1\ldots y_m $, where $n\le m $.
Let us assume that there exists $i$ with $x_i\neq y_i $. If we choose $i$ minimal,
we obtain, that either $x_i $ is a prefix of $y_i $ or $y_i $ is a prefix of $x_i $.
This is a contradiction, because $\C $ is prefix-free. Therefore $x_i =y_i $
for all $1\le i \le n $. Furthermore from
$x_1\ldots x_n = y_1\ldots y_m $ and $e\not\in\C $ follows that $n=m$. This shows
that $\C $ is a code. The proof for suffix-free sets follows the same
steps.\qed\\

The next proposition shows how we obtain a prefix-free code from an
arbitrary set $\X\subseteq \A^+ $.

\begin{prop}\label{help0}
Let $\X\subseteq\A^+ $ with $\X\neq \emptyset $ and
$\Y :=\X -\X\A^+ $. Then $\Y\neq\emptyset $ is a prefix-free code
and $\X\A^* =\Y\A^* $.
\end{prop}

\smallskip\noindent
{\bf Proof: }Let $x,z\in\X $, such that $|x|<|z|$ for all $z\in\X $,
then $x\not\in\X\A^+ $.
Hence $\X\neq\emptyset $. From $\X\subseteq\Y $ follows $\Y\A^+\subseteq\X\A^+ $.
Since $\Y =\X-\X\A^+ $, we obtain $\Y\cap\Y\A^+ \subseteq \Y\cap \X\A^+ =\emptyset $.
This shows that $\Y $ is a prefix-free code, because $e\not\in\X $.

\smallskip
Obviously $\Y\A^*\subseteq\X\A^* $ holds. We show the other direction.
Let $x\in\X $. If $x\in\Y $, then $x\in\Y\A^* $. Otherwise
$x\in\X\A^+$, whence $x=x_1z_1 $ for some $x_1\in\X ,z\in\A^+ $. Especially
we obtain $|x_1|< |x| $. By induction on the length of $x$ it follows, that
$x=x_nz_1\ldots z_n $ for some $z_1,\ldots ,z_n \in\A^+ $ and $x_n\in\Y $.
It follows that $x\in\Y\A^+ $. Therefore we obtain $\X\subseteq\Y\A^* $.
Since $\X\A^*\subseteq\Y\A^*\A^* =\Y\A^* $, this shows $\X\A^* =\Y\A^* $.\qed\\

\medskip
A prefix-free code $\C\subseteq\A^+ $ is  called a {\em maximal prefix-free code}
if for every $c\in\A^+ -\C $ the set $\C\cup\{ c \} $ is not prefix-free.
In the same way {\em maximal suffix-free codes } and {\em maximal fix-free codes }
are defined. The question rises, if a maximal fix-free code is also a maximal
code? Indeed in general, this is not the case for infinite codes. However,
it is true for finite codes.\\

\smallskip
A set $\X\subseteq\A^* $ is called {\em dense}, if

\[ \A^* w\A^* \cap \X \neq\emptyset \quad\mbox{ for all }\; w\in\A^* \, .\]
The set $\X $ is called {\em thin}, if $\X $ is not a dense set.
This means $\X $ is a thin set if and only if there exists a word $w\in\A^* $
such that $\A^* w\A^* \cap \X =\emptyset $.

\smallskip
\begin{prop}\label{thin1} Every finite set is a thin set, as well.
\end{prop}
\smallskip\noindent
{\bf Proof:}
Let $\X\subseteq\A^* $ be a finite set.
Since $\X $ is finite, there exists a word $w\in\A^* $ such that
$|w|> |x| $ for all $x\in\X $. It follows that $\A^*w\A^*\cap \X =\emptyset $.\qed\\

\smallskip
The next two examples shows, that dense fix-free codes and infinite
thin fix-free codes exist.

\begin{example}\label{dyck}\upshape
Let $\A :=\{ 0,1 \} $ and
\[ \D := \left\{ w\in\A^+ \left| <w|0>=<w|1> \;\mbox{ and }\;
<u|0>\neq<u|1> \;\mbox{ for all }\; u\in w\A^{-} \right.\right\} \, .\]
This means, if  $w\in\D $, then the number of $0$'s in the word $w$ is
equal to the number of $1$'s in the word, but $<u|0>\neq <u|1> $
for any proper prefix $u$ of $w$. Obviously $\D $ is a prefix-free code.
Let us assume that there exists $w, w'\in\D $ such that $w'$ is a proper
suffix of $w$. Then there exists a word $u\in\A^+ $ with $w=uw' $.
We obtain:
\[ \begin{array}{lcl}
<w|0> & = & <u|0> + <w'|0> = <u|0> + <w'|1> \\
\quad & \neq & <u|1> + <w'|1> = <w|1> \, .
\end{array}\]
This is a contradiction. Therefore $\D $ is a fix-free code. $\D $ is called
the {\em binary Dyckcode}.

Let $w\in\A^+ $, then $0^{2<w|1>}w1^{|w|} \in\A^*w\A^* $. We obtain:

\[\begin{array}{lcl}
\left<0^{2<w|1>}w1^{|w|}\left| 0\right.\right> & = &
2 <w|1> + <w|0> = |w| + <w|1> \\
\quad & = & \left<0^{2<w|1>}w1^{|w|}\left| 1\right.\right> \, .
\end{array}\]
This shows that $\A^*w\A^* \cap\D \neq \emptyset $ for all $w\in\A^* $.
Therefore $\D $ is a dense fix-free code. Furthermore
$\D $ is maximal prefix-free and maximal suffix-free.
Let $w\in\A^+ $. We show that $w$ has a prefix in $\D $ or that
there exists a word in $\D $ with prefix $w$.

{\em Case 1:} $<u|0> =<u|1> $ for some prefix $u$ of $w$.\\
Let $u$ be the prefix of $w$ with $<u|0> =<u|1> $ and minimal length.
Then $u$ is in $\D $.

{\em Case 2:} $<u|0> \neq <u|1> $ for any prefix $u$ of $w$.\\
We can suppose that $\big< w \big| 0 \big> \, < \, \big< w|1\big> $.
Let $n := <w|1> - <w|0> $. Then
$w0^n $ is a word in $\D $ and has $w$ as a prefix.

\noindent
This shows that $\D $ is maximal prefix-free. The prove that $\D $
is maximal suffix-free follows the same steps.

\end{example}

\medskip
\begin{example}\label{thin2}\upshape
Let $\A =\{ 0,1 \} $ and $C:= \{ 10^n1 \, |\, n\ge 1 \} \cup \{ 0 \} $.
Obviously $\C $ is a fix-free code. Furthermore we obtain:
\[ \A^*11\A^* \cap \C = \emptyset\, . \]
This shows that $\C $ is an infinite thin fix-free code.
We obtain for the Kraftsum of $\C$:
\[ \Kraft(\C ) =\sum\limits_{n=1}^{\infty }\left(\frac{1}{2} \right)^n =
\frac{1}{1-\frac{1}{2}}-1 =1 \]
\end{example}

\medskip
For thin codes the following theorem holds:
\begin{samepage}
\begin{theorem}\label{finitemaximal}
Let $\C\subseteq \A^+ $ be a thin set.
\begin{enumerate}
\renewcommand{\labelenumi}{(\roman{enumi})}
\item $\C $ is a maximal prefix-free code
$\Leftrightarrow \C $ is prefix-free and a maximal code.

\item $\C $ is a maximal suffix-free code
$\Leftrightarrow \C $ is suffix-free and a maximal code.

\item
\begin{tabular}[t]{ll}
\quad & $\C $ is a maximal fix-free code. \\
$\Leftrightarrow $ & $\C $ is fix-free and a maximal code.\\
$\Leftrightarrow  $ & $\C $ is a maximal prefix-free and a maximal suffix-free code.\\
\end{tabular}
\end{enumerate}
\end{theorem}
\end{samepage}
A proof of the theorem can be found in \cite{per}. Furthermore we will
prove the theorem for finite prefix-free codes at the end of the next section.
However, in general the theorem does not hold in for
infinite codes as the next example shows.

\pagebreak
\begin{example}\upshape
Let $\A := \{ 0,1 \} $ and
\[ X:= \{ u10^{|u|} \, |\, u\in\A^* \} \, .\]
It is easy to verify, that $\X $ is a suffix-free code, but not a prefix-free code.
For example $1\in\X $ is a prefix of $110\in\X $.
>From Proposition~\ref{help0}
follows that
$\Y := \X - \X\A^+ $ is a prefix-free code.
Since $\X $ is not a prefix-free code it follows, that $\Y\neq \X $.
Let $w\in\A^+ $. Then  we have $w10^{|w|} \in w\A^*\cap\X $. It follows that:
\[ \emptyset \neq w\A^*\cap\X \subseteq \A^*w\A^*\cap\X \quad\mbox{ for all }\;
w\in\A^* \, .\]
Especially $\X $ is a dense code and $w\A^*\cap\X\A^*\neq \emptyset $ for all
$w\in\A^* $. By Proposition~\ref{help0} we have $\X\A^* =\Y\A^* $.
Therefore we obtain:
\[ w\A^*\cap \Y\A^* \neq\emptyset \quad\mbox{ for all }\; w\in\A^* \, .\]

\noindent
The equation above means, that for any word $w\in\A^* $ the word $w$
is a prefix of a word in $\Y $ or that there exists a word in $\Y $ which
is a prefix of $w$. Therefore the code $\Y $ is a maximal prefix-free code.
Indeed $\Y $ is not a maximal code, because $\X\neq \Y $ and $\Y\subseteq\X $.
Furthermore $\Y $ is a maximal fix-free code, since $\X $ is suffix-free.
However, obviously $\Y $ is not a maximal suffix-free code.\\
\end{example}

\bigskip
The next lemma gives us the relationship between the Kraftsum of
a prefix- suffix- or fix-free code, respectively and its shadow on the
$n$-th level.
For $x,y\in\A^* $ and $n\in\NN_0 $ we define the sets
$I_n(x,y)\subseteq\A^n $ and $I(x,y)\subseteq \A^* $ as:
\[\begin{array}{lcl}
I_n (x,y) & := & \{ z\in\A^n |\mbox{\footnotesize $x$ is prefix of $z$ and
$y$ is suffix of $z$ } \} =x\A^{n-|x|}\cap \A^{n-|y|}y\, , \\
I(x,y) & := & \bigcup\limits_{n=0 }^{\infty } I_n(x,y) =x\A^*\cap \A^*y\, .
\end{array}\]

\pagebreak
\begin{lemma} \label{shadow}
Let $|\A | =q\ge 2 $, $ x,y\in\A^* $ and $ \X\subseteq\A^* $. \\
\renewcommand{\labelenumi}{(\roman{enumi})}
\begin{enumerate}
\item For any $N\in\NN $ we have:\\
\begin{tabular}[t]{lcl}
$\X $ is prefix-free & $\Leftrightarrow $ & $ |\pfs[n] (\X )| =q^n\cdot\Kraft_n (\X )
\quad\forall\, n\ge N $\\
$\X $ is suffix-free & $\Leftrightarrow $ & $ |\sfs[n] (\X )| =q^n\cdot\Kraft_n (\X )
\quad\forall\, n\ge N $\\
$\X $ is fix-free & $\Leftrightarrow $ & $
|\sfs[n] (\X )|=|\pfs[n] (\X )| =q^n\cdot\Kraft_n (\X )
\quad\forall\, n\ge N $\\

\end{tabular}

\item If $\X\subseteq \bigcup\limits_{l=0}^{N }\A^l $ for some $N\in\NN_0 $ and
$n\ge N $ then:\\
\begin{tabular}{lcl}
$\X $ is prefix-free & $\Leftrightarrow $ & $ |\pfs[n] (\X )| =q^n\cdot\Kraft (\X )$\\
$\X $ is suffix-free & $\Leftrightarrow $ & $ |\sfs[n] (\X )| =q^n\cdot\Kraft (\X )$\\
$\X $ is fix-free & $\Leftrightarrow $ & $
|\sfs[n] (\X )|=|\pfs[n] (\X )| =q^n\cdot\Kraft (\X )$\\

\end{tabular}

\item {\small $ \big| \bfs[n] (x) \big|= \left\{ \begin{array}{ll}
0 & \mbox{for } \quad |x|>n \\
2\cdot q^{n-|x|}- q^{n-2|x|} & \mbox{for }\quad 2|x|\le n \\[2mm]
2\cdot q^{n-|x|}-1 &
\mbox{for }\quad 2|x|>n \mbox{ and }\\
\quad & x_{n-|x|+1}\ldots x_{|x|}=x_{1}\ldots x_{2|x|-n} \\[2mm]

2\cdot q^{n-|x|} &
\mbox{for }\quad  2|x|>n \mbox{ and }\\
\quad & x_{n-|x|+1}\ldots x_{|x|}\neq x_{1}\ldots x_{2|x|-n} \\

\end{array}
\right. $ }

\item {\small $ \big| I_n ( x,y ) \big| =\left\{ \begin{array}{lll}
0 & \mbox{for } & n< |x| \mbox{ or } n < |y| \\[2mm]

0 & \mbox{for } & n\ge |x|,|y|, n\le |x|+|y| \mbox{ and }\\
\quad & \quad & x_{n-|y|+1}\ldots x_{|x|+|y|-n+1} \neq y_{1}\ldots y_{|x|+|y|-n}\\[2mm]

1 & \mbox{for } & n\ge |x|,|y|, n\le |x|+|y| \mbox{ and } \\
\quad & \quad & x_{n-|y|+1}\ldots x_{|x|+|y|-n+1} = y_{1}\ldots y_{|x|+|y|-n}\\[2mm]
q^{n-|x|-|y|} & \mbox{for } & n\ge |x| + |y|
\end{array}\right. $ }

\item If $\X $ is fix-free:\\
\begin{eqnarray*}
\big|\bfs[n]{(\X )} \big| & = &  \big|\pfs[n]{(\X )} \big| + \big|\sfs[n]{(\X )}
\big| -
\big|\pfs[n]{(\X )} \cap \sfs[n]{(\X )}  \big| \\
 \quad  & = & 2\cdot \Kraft_n (\X ) -
\sum\limits_{x,y\in\X ,|x|,|y|\le n} |I_n (x,y) | \\
\end{eqnarray*}

\end{enumerate}
\end{lemma}

\noindent
{\bf Proof:}\quad\\
\begin{enumerate}
\renewcommand{\labelenumi}{(\roman{enumi})}
\item Let $\X\subseteq\A^* $ be prefix-free and $n\ge N $.
If $x,y\in\X $ with $x\neq y $
and $|x|,|y|\le n $, then the sets $x\A^{n-|x|} $ and $y\A^{n-|y|} $ are disjoint.
It follows that the sets $(\X\cap\A^l)\A^{n-l} $ and
$(\X\cap\A^k)\A^{n-k} $ are also disjoint for $l<k\le n $. Therefore we obtain:

\[\begin{array}{lcl} |\pfs[n] (\X ) | & = &
\big| \bigcup\limits_{l=0 }^n (\X\cap\A^l)\A^{n-l}\big| =
\sum\limits_{l=0}^n |\X\cap\A^l |\cdot |\A^{n-l} | \\
\quad & = &
q^n \cdot \sum\limits_{l=0}^n |\X\cap\A^l | q^{-l } =q^n\cdot \Kraft_n (\X ) \, .
\end{array} \]

\smallskip
Let $\X\subseteq\A^* $ and $x,y \in\X $ with $x\neq y $ such that $x$ is a prefix
of $y$. Let $n:=\max \{ N,|y| \} $, then
$\emptyset\neq y\A^{n-|y| }\subseteq (\X\cap \A^{|x|} )\A^{n-|x|}\,\cap\,
(\X\cap \A^{|y|} )\A^{n-|y|} $. Therefore it follows:

\[\begin{array}{lcl} |\pfs[n] (\X ) | & = &
\big| \bigcup\limits_{l=0 }^n (\X\cap\A^l)\A^{n-l}\big| <
q^n \cdot \sum\limits_{l=0}^n |\X\cap\A^l | q^{-l } =q^n\cdot \Kraft_n (\X ) \, .
\end{array} \]

This shows (i) for prefix-free sets. The proof for suffix-free sets follows
the same steps.

\item Let $\X\subseteq\bigcup\limits_{l=0}^N \A^l $. Since
$\Kraft_n (\X ) =\Kraft (\X ) $ for all $n\ge N $, part (ii) follows from part (i).

\item Since $ \{ x \} $ is a fix-free set for all $x\in\A^* $, part (iii) follows from
(iv) and (v).

\item Let $|x|> n $ or $ |y|> n $. Obviously there does not exist a
word of length $n$,
which has $x$ as a prefix and $y$ as a suffix. Therefore we obtain
$|I_n (x,y) |=0 $ for $\max\{ |x|,|y| \} > n $.

\smallskip
If $|x|+|y| \ge n $ and $z\in\A^n $, such that $x$ is a prefix of $z$ and $y$ is
a suffix of $z$, the picture below shows that\\
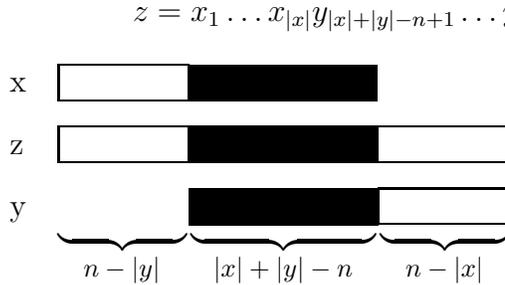
\[ z= x_1\ldots x_{|x|}y_{|x|+|y|-n+1}\ldots y_{|y|} =
x_1\ldots x_{n-|y|}y_{1}\ldots y_{|y| } \, .\]

\begin{tabular}{ll}
x & \framebox{\parbox[b][2.5mm][c]{1.5cm}{\footnotesize \quad  }}\rule[-1.2mm]{2.5cm}{4.8mm} \\[3mm]
z & \framebox{\parbox[b][2.5mm][c]{1.5cm}{\footnotesize \quad  }}\rule[-1.2mm]{2.5cm}{4.9mm}\framebox{\parbox[b][2.5mm][c]{1.5cm}{\footnotesize \quad  }} \\[3mm]
y & $\mbox{\parbox[b][2.5mm][c]{1.75cm}{\footnotesize \quad  }}\rule[-1.2mm]{2.5cm}{4.9mm}\framebox{\parbox[b][2.5mm][c]{1.5cm}{\footnotesize \quad  }} $ \\[-3mm]
\quad & $\underbrace{\hspace*{1.7cm}}_{\mbox{\footnotesize $n-|y| $}}\underbrace{\hspace*{2.45cm}}_{\mbox{\footnotesize $|x|+|y|-n $}}\underbrace{\hspace*{1.7cm}}_{\mbox{\footnotesize $n-|x| $}} $

\end{tabular}\\

In this case follows $|I_n (x,y)|=1 $.\\

\smallskip
If $|x|+|y|\le n $, then we obtain $I_n (x,y) =x\A^{n-|x|-|y|}y $.
In this case  follows $|I_n (x,y) | =q^{n-|x|-|y|} $  .

\item Let $\X $ fix-free. While
$\bfs[n](\X ) = \pfs[n](\X )\cup \sfs[n](\X ) $, we obtain
\[|\bfs[n](\X )| = |\pfs[n](\X )|+|\sfs[n](\X )|-|\pfs[n](\X )\cap \sfs[n](\X )| \]
By (i) follows $\Kraft_n (\X )q^n =|\pfs[n](\X ) | =|\sfs[n](\X ) |$. Furthermore
we have:
\[ \pfs[n](\X )\cap \sfs[n](\X ) = \bigcup\limits_{\mbox{\scriptsize
$\begin{array}{l} x,y\in\X \\ |x|,|y|\le n \end{array}$ }} I_n(x,y) . \]
Since $\X $ is fix-free, no word in $\A^n $ has two different prefixes in $\X $
or two different suffixes in $\X $. It follows:
\[ I_n (x,y)\cap I_n(x',y')=\emptyset \quad \forall\; x,y,x',y'\in\X \;\mbox{ with }\;
(x,y)\neq (x',y') .\]
Therefore we obtain
\[  |\pfs[n](\X )\cap \sfs[n](\X )| = \sum\limits_{\mbox{\scriptsize
$\begin{array}{l} x,y\in\X \\ |x|,|y|\le n \end{array}$ }} I_n(x,y) \, .
\mbox{\qed }\]
\\
\end{enumerate}

\smallskip
Let $(\X_n )_n\in\NN $ be a sequence of sets with $X_n\in\A^* $ for all $n\in\NN $.
We write $X_n\uparrow \X $ if $ X_1\subseteq X_2\subseteq X_3\subseteq \ldots $
and $X=\bigcup\limits_{n=1}^{\infty } \X_n $. And we write
$\X_n\downarrow \X $ if $ X_1\supseteq X_2\supseteq X_3\supseteq \ldots $
and $X=\bigcap\limits_{n=1}^{\infty } \X_n $.\\

\begin{prop}\label{help0}
If $\X_n $ is prefix-free for all $n\in\NN $ and $\X_n\uparrow\X $, then
$\X $ is prefix-free, too.
\end{prop}
\smallskip\noindent
{\bf Proof:} Let us suppose, that there exists $x,y\in\X $ such that
$x$ is a prefix of $y$. Then there exists an $n\in\NN $, such that $x,y\in\X_n $.
This is a contradiction, because $\X_n $ is prefix-free.\qed\\

\noindent
Obviously the proposition holds also for suffix-free and fix-free sets. \\

\smallskip
We finish this section with two lemmas, which deals with the
construction of fix-free codes and which we will use in Chapter 2.

\pagebreak
\begin{lemma}\label{help1}
Let $\X ,\Y \subseteq\A^* $ be fix-free sets. Then the set $\X\Y $ is also fix-free.
\end{lemma}
Furthermore the lemma above holds also for suffix-free and prefix-free sets.\\

\smallskip\noindent
{\bf Proof:}
Obviously the lemma holds for $\X=\{e \} $ or $\Y =\{ e \} $.
Let $\X,\Y \subseteq\A^+ $ be prefix-free codes. Let us assume that
$xy $ is a prefix of $x'y' $, where $x,x'\in\X $, $y,y'\in\Y $ and $xy\neq x'y' $.
It follows that either $x$ is a proper prefix of $x'$ , $x'$ is a proper prefix
of $x$ or $x=x'$. Since $\X $ is a prefix-free code, we obtain that $x=x'$,
but then $y$ is a prefix of $y'$. This is a contradiction because also $\Y $
is prefix-free. Therefore $\X\Y$ is prefix-free. The proof for suffix-free follows
the same way.\qed\\

\smallskip
\begin{lemma}\label{help2}
Let $\X \subseteq \bigcup\limits_{l=0}^{n-1}\A^l  $, $\Y ,\Z\subseteq \A^n $ be
such that $\X\cup\Y $ is fix-free. If $\X'\subseteq \X $ such that:
\begin{enumerate}
\renewcommand{\labelenumi}{(\arabic{enumi})}
\item every word in $\Z $ has a prefix in $\X' $ or no prefix in $\X $,
\item every word in $\Z $ has a suffix in $\X' $ or no suffix in $\X $,
\end{enumerate}
then the set $(\X-\X' )\cup\Y \cup\Z $ is fix-free.
\end{lemma}

\smallskip\noindent
{\bf Proof:} By symmetry, it is sufficient to prove that (1) implies that
$(\X-\X' )\cup\Y \cup\Z $ is prefix-free.  Obviously the lemma holds, if
$\X' =\emptyset $ or $\X=\X' $. If $e\in\X $ then $\X=\{ e\} $ and therefore
$\X' =\emptyset $ or $\X=\X' $. Let $\emptyset \neq \X' \neq \X $ and suppose that
$(\X-\X' )\cup\Y \cup\Z $ is not prefix-free. Then there exists $z\in\Z $ and
$x\in\X-\X' $ such that $x$ is a prefix of $z$. By (1) follows, that $z$
has a prefix $x'$ in $\X' $. Since $x\neq x' $ and since both words $x$ and $x'$
are prefixes of $z$, it follows that either $x$ is a proper prefix of $x'$ or
$x'$ is a proper prefix of $x$. This is a contradiction, because $x,x'\in\X $
and $\X $ is prefix-free.\qed\\

\newpage
\smallskip
\section{The Kraftinequality for prefix-free codes}
In this section we will show the Kraft-McMillan inequality for prefix-free codes
(see McMillan \cite{kraft}) and related results which can be
found for example in \cite{per}.
\begin{Def}
\hspace*{1cm}\\
A Map $\pi:\, \A^* \longrightarrow \RR^{\ge 0 }$ is called a Bernoulli Distribution on $\A^*$ if:\\
\renewcommand{\labelenumi}{(\arabic{enumi})}
\begin{enumerate}
\item $\pi (xy)=\pi (x)\pi (y)\quad \forall\, x,y\in \A^*$ \item $\pi (e)=1$ \item
$\sum\limits_{x\in \A}\,\pi (a)=1$
\end{enumerate}
$\pi$ is called positive if $\pi (a)\neq 0 \quad \forall\, a\in \A$
\end{Def}

\smallskip\noindent
>From (1) and (2) follows $\pi (x)=\prod\limits_{k=1}^n\pi (x_k) \mbox{ for }
x=x_1\ldots x_n$ Therefore $\pi$ is unique determined by its  values on $\A$.
If $\pi$ is positive, we obtain $\pi (x)\neq 0$ for all $x \in A^*$.
It follows by
(1) and (2), that a positive Bernoulli distribution is a monoidhomomorphism from
$\A^*$ into $(\RR^{>0},1,\cdot)$. For an arbitrary
Bernoulli distribution $\pi$ and $n\in\NN_0$ we obtain:
\begin{equation} \label{formelA}
\sum\limits_{x\in\A^n}\,\pi (x)=\sum\limits_{x\in\A^{n-1}}\pi
(x)\underbrace{\sum\limits_{x\in \A}\,\pi (a)}_{=1}= \sum\limits_{x\in\A^{n-1}}\,\pi
(x)=\,\ldots\,=1.
\end{equation}
Thus $\pi_{\big|A^n}$ is a probability distribution on $\A^n$.\\

\medskip
Let $\M $ be an arbitrary set. A {\em measure } on $\M $ is a map
$\pi :\PP (\M)\longrightarrow \RR^{\ge 0}\cup \{\infty \} $ with the properties:
\begin{enumerate}
\renewcommand{\labelenumi}{(\arabic{enumi})}
\item $\pi (\emptyset ) =0 $,
\item If $\X_1,\X_2,\ldots \subseteq\M $ are pairwise disjoint, then
$\pi \bigg( \bigcup\limits_{n=1}^{\infty } \X_n \bigg) =
\sum\limits_{n=1}^{\infty }\pi (\X_n ) $. \\
{\em ($\sigma $-additivity) }
\end{enumerate}
Furthermore, $\X\subseteq Y\subseteq \M $ satisfy the inequality
$\pi (\X )\le \pi (\Y )$
and if $\pi (\Y )<\infty $, then $\pi (\Y-\X )=\pi (\Y )-\pi (\X )$.

\pagebreak
\begin{prop}Let $|\A | =q\ge 2 $, $\X\subseteq\A^* $ and
$\pi :\A^*\rightarrow \RR^{\ge 0 } $ be a Bernoulli
distribution.
If we define
$\pi (X):=\sum\limits_{x \in X}\pi(x)\,$, then
$\pi:\,\mathcal{P}\left(\A^*\right)\longrightarrow \RR^+\cup\{\infty\}$ is a measure
on $\mathcal{P}\left(\A^*\right)$.
\end{prop}

\smallskip\noindent
{\bf Proof: }Obviously $\pi (\emptyset )=0$.
Let $\X_1,\X_2,\ldots \subseteq\M $ pairwise disjoint, then:
\[ \pi\big(\bigcup\limits_{n\in\NN}^{.}\X_n\big)=
\sum\limits_{x\in\bigcup\limits_{n\in\NN}^{.}\X_n}\pi (x)=
\sum\limits_{n=0}^{\infty}\sum\limits_{x\in X_n} \pi
(x)=\sum\limits_{n=0}^{\infty}\pi (X_n)\]
This shows the $\sigma$-additivity of $\pi $. \qed\\

\smallskip\noindent
While $\pi$ is a measure on $\mathcal{P}\left(\A^*\right)$,
it has the following properties:\\

\begin{equation}\begin{array}{l}
\mbox{For each sequence \fo{\X} in $\mathcal{P}\left(\A^*\right)$ with
$\X^n\uparrow\X$ the equation\hspace{2cm}}\\
\mbox{$\lim\limits_{n\to\infty}\pi\left(\X_n\right)=\pi\left(\X\right)$ holds. This means
$\pi$ is continues from below.}
\end{array}\end{equation}

\begin{equation}\begin{array}{l}
\mbox{For each sequence \fo{\X} in $\mathcal{P}\left(\A^*\right)$ with $\X^n\downarrow\X$ and
$\X_n <\infty$ for at\hspace{0.2cm}}\\
\mbox{least one $n\in\NN$, the equation
$\lim\limits_{n\to\infty}\pi\left(\X_n\right)=\pi\left(\X\right)$ holds. This means}\\
\mbox{
$\pi$ is continues from above.}
\end{array}\end{equation}

\begin{equation}\begin{array}{l}
\mbox{The inequality $\pi\big(\!\!\bigcup\limits_{n\in\NN}\X_n\big)
\le\sum\limits_{n=0}^{\infty}\pi\left(\X_n\right)$
holds for all sequences\hspace{2.8cm}}\\
\mbox{\fo{\X} in $\mathcal{P}\left(\A^*\right)$. }
\end{array}\end{equation}

\noindent
A proof of the properties above can be found for example in \cite{bauer}.\\

\smallskip
The next example shows that the Kraftsum $\Kraft :\PP (\A^* )\rightarrow \RR^{\ge 0}
\cup \{ \infty \} $ can be obtained from a positive Bernoulli distribution.
Especially the map $\Kraft $ is measure on $\PP (\A^*) $.\\

\pagebreak
\begin{example}\label{kraftbernex} \upshape
Let $|\A | =q\ge 2 $. We define $\pi_K :\A^*\rightarrow \RR^+ $ as the
(unique) positive
Bernoulli distribution, given by the uniform distribution on $\A $. This means
$\pi_K (a):=\frac{1}{q}$ for all $a\in\A $. It follows
$\pi_K (x) =\pi_K (x_1)\ldots \pi_K(x_{|x|} ) =q^{-|x| } $ for all $x\in\A^+ $ with
$x=x_1\ldots x_{|x|} $ and $x_1,\ldots ,x_{|x|}\in\A $. Let $\X\subseteq\A^* $, then
we obtain:
\[
\pi_K(\X)=\sum\limits_{x\in\mathcal{X}}q^{-|x|}=
\sum\limits_{l=0}^{\infty}\sum\limits_{x\in\mathcal{X}\cap\mathcal{A}^l}q^{-l}=
\sum\limits_{l=0}^{\infty}|\X\cap\A^l | \cdot q^{-l} =\Kraft (\X )
\]
\end{example}

\medskip
\begin{lemma}\label{kleiner}
Let $\C\subseteq\A^+$ be a code and $\pi$ be a Bernoulli distribution. Then \\
$\pi (\C )\le 1$. Especially $\Kraft (\C ) \le 1 $ for every code $\C \subseteq\A^+ $.

\end{lemma}

\smallskip\noindent
{\bf Proof: }
Let $\C\subseteq\A^+$ be a code and $\pi$ be a Bernoulli Distribution. We claim
\begin{equation}\label{bern1lem1}
\pi (\C^n)=\pi (\C )^n\;\mbox{ for }\;n\in\NN \, .
\end{equation}
Let $c_1,\ldots ,c_n ,c'_1,\ldots ,c'_n \in\C $ such that $c_i\neq c'_i $ for some
$i$.
Since $\C $ is a code, we have $\{c_1\cdot\ldots\cdot
c_n\}\cap\{c'_1\cdot\ldots\cdot c'_n \}=\emptyset$. With
the $\sigma $-additivity of $\pi $, follows:

\begin{eqnarray*}
\pi (\C^n )&=&\pi\Big(\bigcup\limits^{.}_{c_1,\ldots,c_n\in\C}\left\{c_1\cdot\ldots\cdot c_n\right\}\Big)\\
&\stackrel{\tiny\sigma\mbox{-Add.}}{=}&
\sum\limits_{c_1,\ldots,c_n\in\C}\pi (c_1)\cdot\ldots\cdot\pi(c_n)\\
&=&\underbrace{\sum\limits_{c_1\in\C}\pi (c_1)}_{=\pi (\C
)}\ldots\underbrace{\sum\limits_{c_1\in\C}\pi (c_n)}_{=\pi (\C )} =\pi (\C )^n \, .
\end{eqnarray*}

\noindent
Next we claim:
\begin{equation}\label{bern1lem2}
\mbox{If $\left|\C\right|<\infty$ then $\pi \left(\C\right)\le 1$} \, .
\end{equation}

\noindent
Let us suppose that $\pi (\C )=1+\epsilon$ for some $\epsilon >0$.
While $\C $ is finite, there exists $k\in\NN$ with
$\C\subseteq\A\stackrel{.}{\cup}
\ldots\stackrel{.}{\cup}A^k$. Then
$\C^n\subseteq \A\stackrel{.}{\cup}
\ldots\stackrel{.}{\cup}A^{n\cdot k}$. It follows with the measure properties of
$\pi$:\\[0.5ex]
$(1+\epsilon )^n=
\pi\!\left(\C\right)^{n}\parbox[t][2ex][t]{3ex}{$=$\\{\scriptsize\raisebox{1.5ex}{(\ref{bern1lem1})}}}
\pi\!\left(\C^n\right)\hspace{-2ex}
\parbox[t][3.5ex][t]{5.5ex}{\centering $\le$\\[-1ex]{\scriptsize $\pi$ is}\\[-1.5ex]{\scriptsize measure}}
\hspace{-1ex} \pi\bigg(\bigcup\limits_{l=1}^{n\cdot k}\A^l\bigg) \hspace{-0.5ex}
\parbox[t][3.5ex][t]{5.5ex}{\centering $=$\\[-1.5ex]{\scriptsize $\sigma$-Add.}}
\hspace{-0.1ex} \sum\limits_{l=1}^{n\cdot k} \underbrace{\pi\big(\A^l\big)}_{=1}
=n\cdot k$ for all $n\in\NN$.\\[0.5ex]
This is a contradiction, because
for $\epsilon>0, k\in\NN$ there exists an $N\in\NN$ with
$(1+\epsilon )^n > n\cdot k \quad\forall\, n\ge N$. This proves (\ref{bern1lem2}).

\noindent
We claim:
\begin{equation}
\mbox{$\pi \left(\C\right)\le 1$ is also true for $\left|\C\right|=\infty$} \, .
\end{equation}
For $n\in\NN$ let $\C_n:=\C\cap\bigg(\bigcup\limits_{k=0}^n\A^k\bigg)$, then
$\C_n\uparrow\C$ and $\left|\C_n\right|<\infty$. From (\ref{bern1lem2}) follows
$\pi\left(\C_n\right)\le 1 $ for all $ n\in\NN$ and since every measure is
continues from bellow, we conclude:

\[ \pi\left(\C\right)=\pi\!\!\left(\lim\limits_{x\to\infty}C_n\right)=\lim\limits_{x\to\infty}\pi\left(\C_n\right)\le
1\, .\]

\noindent
This shows the lemma. Furthermore from Example \ref{kraftbernex} follows, that
$\Kraft (\C )\le 1 $ for every code $\C\subseteq\A^+ $. \qed\\

\medskip
The next theorem shows, that for prefix-free codes and $\pi = \Kraft $, also
the converse of Lemma \ref{kleiner} holds.
\begin{theorem}[Kraft and McMillan \cite{kraft}]\label{kraft}
Let $\left|\A\right|=q \ge 2$
and \fo[l]{\alpha} be a sequence of nonnegative integers.
There exists a prefix-free code $\C\subseteq\A^+ $ which fits to
\fo[l]{\alpha}, if and only if $\sum\limits_{l=1}^{\infty }\alpha_lq^{-l }\le 1 $.
\end{theorem}
Furthermore the theorem holds also for suffix-free codes,
in place of prefix-free codes.

\smallskip\noindent
{\bf Proof:}
If $\C \subseteq\A^+ $ is a prefix-free code which fits \fo[l]{\alpha}, then from
Lemma \ref{kleiner} follows, that
$\sum\limits_{l=1}^{\infty }\alpha_lq^{-l } =\Kraft (\C ) \le 1 $.
Let \fo[l]{\alpha} be a sequence of nonnegative integers such that
$0< \sum\limits_{l=0}^{\infty}\alpha_lq^{-l}\le 1$. Since $1\ge\alpha_1\cdot q^{-1} $,
it follows that $|\A |=q\ge \alpha_1$. Therefore we can choose a set $\C_1\subseteq\A$ with
$\left|\C_1\right|=\alpha_1$. Obviously this set is prefix-free.\\

\noindent
Let $\C_n $ be a prefix-free set such that $\C_n\subseteq\bigcup\limits_{l=1}^n\A^l$
and $\alpha_l=\left|\C_n\cap\A^l\right| $ for all \\
$ 1\le l\le n$. By Lemma~\ref{shadow} (ii) follows:
\begin{equation}\label{formelC}
\left|\pfs[{n+1}]\left(\C_n\right)\right|=q^{n+1}\Kraft (\C_n )
=q^{n+1}\cdot\sum\limits_{l=1}^n\alpha_l\cdot q^{-l}.
\end{equation}
Since $\sum\limits_{l=1}^{\infty}\alpha_lq^{-l}\le 1$, we obtain:
\begin{equation}\label{formelD}
\alpha_{n+1}\cdot
q^{-(n+1)}\le\sum\limits_{l=n+1}^{\infty}\alpha_l q^{-l}
\le1-\sum\limits_{l=1}^n\alpha_l\cdot q^{-l} = 1-\Kraft (\C_n ) \, .
\end{equation}
By (\ref{formelC}) follows:
\begin{displaymath}
\alpha_{n+1}\le q^{n+1} -q^{n+1}\Kraft (\C_n ) =
\left|\A^{n+1}\right|-\left|\pfs[{n+1}]\left(\C_n\right)\right|.
\end{displaymath}
Thus we we can choose $\alpha_{n+1} $ codewords
$c_1,\ldots,c_{\alpha_{n+1}}\in\A^{n+1}$, which are not in the prefix shadow
of $\C_n$.\footnote{If we order the finite alphabet in some arbitrary linear
ordering, we can order the words in $\A^{n+1} $ by the lexicographical (well-)ordering
, which is forced by the linear ordering of $\A $. Then we can choose
the words $c_1,\ldots ,c_{\alpha_{n+1}} $ in ascending order and avoid
in this way some dubious choice principles. Especially the words which are chosen
in the next step are all bigger than the words which were chosen before.}
Then the set
$\C_{n+1}:=\C_n\cup\left\{c_1,\ldots,c_{\alpha_{n+1}}\right\}$ is prefix-free and
fits to $\alpha_1,\ldots ,\alpha_{n+1} $.

\smallskip\noindent
By induction we obtain prefix-free sets $\C_1\subseteq \C_2\subseteq \ldots $, such
that $\C_n $ fits to $(\alpha_1,\ldots ,\alpha_{n} )$ for all $n\in\NN $.
Since $\C_n\uparrow \C $, it follows that $\C:=\bigcup\limits_{n=1}^{\infty }\C_n $
is a prefix-free code which fits to $(\alpha_l )_{l\in\NN }$. This shows the
theorem for prefix-free codes, the proof for suffix-free codes
follows the same way.\qed\\

\medskip
Let us now examine the relationship between maximal codes and Bernoulli
distributions on $\A^* $.
As a result of Lemma~\ref{kleiner} it is easy to give a necessary condition for
maximal codes.

\begin{prop}\label{max1}
Let $|\A |=q\ge 2 $ and $\C\subseteq \A^* $ be a code.
If there exists a positive Bernoulli distribution $\pi $ on $\A^* $
with $\pi (\C ) =1 $, then $\C $ is a maximal code.
\end{prop}

\smallskip\noindent
{\bf Proof: } Let $\pi (\C ) =1 $ for some positive Bernoulli
distribution $\pi $ on $\A^* $. Let us further assume, that $\C $ is not
a maximal code. Then there exists a word $w\in\A^+-\C $ such that
$\C\cup\{ w \} $ is a code. Since $\pi $ is positive, we obtain that
$\pi (w) > 0 $. With Lemma~\ref{kleiner} we obtain the contradiction:
\[ 1\le \pi \left( C\cup\{ w \} \right) = \pi (\C )+\pi (w) =
1+\pi (w) > 1 \, .\]
Therefore $\C $ is a maximal code.\qed\\

\smallskip
For prefix-free codes
the following converse of the proposition above holds.

\begin{prop}\label{max2}
Let $|\A |=q\ge 2 $ and $\C\subseteq \A^* $ be a finite prefix-free code.
$\C $ is a maximal prefix-free code if and only if $\Kraft (\C )=1 $.
if and only if $\C $ is a maximal code.
\end{prop}
Furthermore the lemma above holds also for suffix-free codes.

\smallskip\noindent
{\bf Proof:} Let $\C \subseteq\A^* $
be a finite prefix-free code with $\Kraft (\C ) =1 $.
Since the map $\Kraft $ is a positive Bernoulli distribution on $\A^* $,
it follows from Proposition~\ref{max1}, that $\C $ is maximal.

Let $\C\subseteq\bigcup\limits_{l=1}^n $
be a finite prefix-free code with $\Kraft (\C )< 1 $.
>From Lemma~\ref{shadow} follows that
$|\pfs[n] (\C )| =q^{n}\Kraft (\C ) < q^n =|\A^n | $. We conclude that there
exists a word $w\in\A^n $, which is not in the prefix-shadow of $\C $, i.e.
$w$ has not a prefix in $\C $. Obviously the set $\C \cup \{ w \} $ is a prefix-free
code. Therefore $\C $ is not a maximal prefix-free code.
This shows that $\Kraft (\C ) =1 $ if $\C $ is maximal prefix-free.

Obviously any code which is maximal and prefix-free is also a maximal prefix-free
code. Thus we have shown:

\noindent
$\C $ is maximal prefix-free $\Rightarrow \Kraft (\C )=1 \Rightarrow \C $
is maximal $\Rightarrow \C $ is maximal prefix-free.\qed\\

\smallskip
The proposition above shows Theorem~\ref{finitemaximal} (i) for finite prefix-free
codes. However the next theorem gives us a more general reversal of
Proposition~\ref{max1} for thin codes.

\begin{theorem}
Let $\C\subseteq\A^* $ be a thin code. Then the following properties are all
equivalent.
\begin{enumerate}
\renewcommand{\labelenumi}{(\roman{enumi})}
\item $\C $ is maximal.
\item $\pi (\C )=1 $ for every positive Bernoulli distribution $\pi $ on $\A^* $.
\item There exists a positive Bernoulli distribution $\pi $ on $\A^* $
with $\pi (\C )=1 $.
\end{enumerate}
\end{theorem}

\smallskip\noindent
A proof of the theorem above can be found in \cite{per}. For dense codes
the theorem above is in general wrong. For example, it is shown in
\cite{per} that although the Dyckcode $\D $ in Example~\ref{dyck} has
Kraftsum one, $\pi (\D )< 1$ for any other positive Bernoulli
distribution $\pi $.
On the other hand $\D $ is a maximal code, because of $\Kraft (\D )=1 $.

\pagebreak
\section{Kraftsums of fix-free codes and the $\frac{3}{4}$-conjecture}
One might ask the question, wether Kraft's Theorem \ref{kraft} holds  for fix-free
codes, as well. We will answer this question in general with no. However, the first part
of Theorem~\ref{kraft} which is Lemma~\ref{kleiner} for $\pi =\Kraft $, holds for
all codes, i.e. for fix-free codes. If $\C := \A^n $ for some $n\in\NN $
then $\C $ is a (maximal) fix-free code with $\Kraft (\C )=1 $. This shows, that
$1$ is the smallest number $\gamma $, such that $\Kraft (\C ) \le \gamma $ for
all fix-free codes $\C\subseteq\A^* $.
Furthermore Lemma \ref{34lemma2} below
in this section, shows that for any $0< \gamma \le 1 $ there exists a fix-free code
with Kraftsum equal to $\gamma $.
Other construction of fix-free codes with Kraftsum $1$ ,
which are especially maximal fix-free codes, can be found in \cite{per} and
\cite{gil}. The question rises, if there exists a number $0< \gamma \le 1 $
for which the other direction of Theorem \ref{kraft} holds for fix-free codes.
More precisely: Does there exist a number $\gamma $
such that $\sum\limits_{l=1}^{\infty } \alpha_l q^{-l} \le \gamma $ imply the
existence of a fix-free code which fits to $(\alpha_l)_{l\in\NN }$
and which is the smallest
possible $\gamma $? In \cite{ahlswede}  Ahlswede, Balkenhol and Khachatrian
gave the conjecture below for binary codes and finite sequences, which was
generalized by Harada and Kobayashi in \cite{harada} to arbitrary (finite) alphabets
and infinite sequences in the form given below.

\smallskip
\begin{conj}[$\frac{3}{4}$-conjecture]\label{34conj1}
Let $|\A | =q \ge 2 $ and $(\alpha_l )_{l\in\NN } $ be a sequence of
nonnegative integers, then
$\sum\limits_{l=1}^{\infty } \alpha_l q^{-l}\le \frac{3}{4} $ implies
the existence of a fix-free $\C\subseteq\A^* $ which fits to
$(\alpha_l )_{l\in\NN } $.
\end{conj}

\smallskip
The next lemma shows that for every number bigger than $\frac{3}{4} $ the conjecture
above can not hold. The lemma was first showed by
Ahlswede, Balkenhol and Khachatrian in \cite{ahlswede}
for binary codes and finite sequences and it was generalized for arbitrary (finite)
alphabets and infinite sequences by Harada and Kobayashi in \cite{harada}.

\begin{lemma}\label{34lemma1}
Let $|\A | =q\ge 2 $.
For every $\varepsilon > 0 $, there exists
a finite sequence $(\alpha_1,\ldots ,\alpha_n )\in\NN_0^n $ with
$ \sum\limits_{l=1}^{n }\alpha_lq^{-l}\le \frac{3}{4} +\varepsilon $,
such that for every fix-free code \\
$\X \subseteq \bigcup\limits_{l=1}^{n}\A^l $,
there exists some $1\le l\le n $ with  $|\X \cap \A^l | < \alpha_l $.
\end{lemma}

\smallskip\noindent
{\bf Proof:}
Let $\left|\A\right|=q\ge 2 $. It is sufficient to show the
lemma for $0< \varepsilon<\frac{1}{4}$.

\noindent
Let $m\in\NN$ with $\varepsilon\cdot q^m>2$. We can choose
$\alpha_m\in\NN $ such that\\
$q^m<2\alpha_m<q^m+\epsilon\cdot q^m=q^m(1+\varepsilon)$ holds. It follows:
\begin{equation}\label{bsp1gl1}
\frac{1}{2}<\alpha_m\cdot q^{-m}<\frac{1}{2}+\frac{\epsilon}{2}.
\end{equation}
Let $n\in\NN$, such that $n\ge 2m$ and $2\varepsilon q^n>4$. Then we can
choose a number $\alpha_n\in\NN$ with
\[ q^n<4\alpha_n<q^n+2\epsilon
q^n=q^n(1+2\epsilon )\, .\]
We obtain for $\alpha_n $:
\begin{equation}\label{bsp1gl2}
\frac{1}{4}<\alpha_n\cdot q^{-n}<\frac{1}{4}+\frac{\varepsilon}{2}.
\end{equation}
If we define $\alpha_l=0 $ for all $l\in\{1,\ldots
,m-1,m+1,\ldots ,2m,\ldots ,n-1\}$ we obtain for
$\alpha_1,\ldots \alpha_n$ the desired property
\begin{equation}\label{bsp1gl3}
\frac{3}{4}<\sum\limits_{l=1}^{n}\alpha_lq^{-l}<\frac{3}{4}+\varepsilon.
\end{equation}
Let $\C\subseteq\A^n$ be a set with $\left|\C\right|=\alpha_m$. We obtain:
\begin{equation}\label{bsp1gl4}
\left|\bfs[n]\left(\C\right)\right|=\left|\pfs[n]\left(\C\right)\cup
\sfs[n]\left(\C\right)\right|=
\left|\pfs[n]\left(\C\right)\right|+\left|\sfs[n]\left(\C\right)\right|-
\left|\pfs[n]\left(\C\right)\cap \sfs[n]\left(\C\right)\right| \, .
\end{equation}
While $\C $ is a one-level set, $\C$ is fix-free. Therefore it follows:
\[ \left|\pfs[n]\left(\C\right)\right|=
\left|\sfs[n]\left(\C\right)\right|=\sum\limits_{x\in\C}q^{n-m}
=\alpha_m\cdot q^{n-m}\, . \]
Since $n\ge 2m$, from Lemma \ref{shadow} follows:
\[ \pfs[n]\left(\C\right)\cap \sfs[n]\left(\C\right)=\C\A^{n-2m}\C \, .\]
It follows:
\[ \left|\pfs[n]\left(\C\right)\cap
\sfs[n]\left(\C\right)\right|=\left|\C\right|^2\cdot \left|\A^{n-2m}\right|=
\alpha_m^2\cdot q^{n-2m} \, . \]
By (\ref{bsp1gl4}) follows:\\
\begin{eqnarray*}
\left|\bfs[n]\left(\C\right)\right|& = &2\alpha_mq^{n-m}-\alpha_m^2q^{n-2m}\\
\Rightarrow \hspace{5mm} \frac{\left|\bfs[n]\left(\C\right)\right|+\alpha_n}{q^n}& =
&2\frac{\alpha_m}{q^m}-\left(\frac{\alpha_m}{q^m}\right)^2+
\frac{\alpha_n}{q^n}\\
& = &\frac{\alpha_n}{q^n}+1-\left(1-\frac{\alpha_m}{q^m}\right)^2\\
\quad & \begin{array}{c} > \\ \mbox{\scriptsize (\ref{bsp1gl1}), (\ref{bsp1gl2})}
\end{array} &
\frac{1}{4}+1-\left(1-\frac{1}{2}\right)^2 = 1\\[1.5ex]
\Rightarrow \hspace{5mm} \left|\bfs[n]\left(\C\right)\right|+\alpha_n & > & q^n=\left|\A^n\right| \, .\\
\end{eqnarray*}
It follows that $\D\cap |\bfs[n] (\C ) | \neq \emptyset $ for every
set $\D\subseteq\A^n$ with $\left|\D\right|=\alpha_n$.
We conclude that for such $\D$'s the set
$\C\cup\D$ is not a fix-free code.
Since $\C$ was chosen arbitrarily, this shows that there exist no fix-free code
$\X $ with $\left|\X\cap\A^l\right|=\alpha_l $ for all $1\le l\le n$.\qed\\

\smallskip
Let $\alpha_1,\ldots ,\alpha_n $ be as in the proof above,
$\gamma :=\frac{3}{4}+\varepsilon\in ]\frac{3}{4},1]$ and\\
$s:=\frac{3}{4}+\epsilon -\left(\alpha_mq^{-m}+ \alpha_nq^{-n}\right)\in
[0;\epsilon[$.
There exists a sequence $(\beta_k)_{k\in\NN }$ such that
$s=\sum\limits_{k=1}^{\infty}\beta_kq^{-k}$ and $\beta_k\in\{0,\ldots ,(q-1)\}$
for all $k\in\NN $. If we define $\tilde{\alpha }_l := \alpha_l +\beta_l $
for $1\le l\le n $ and $\tilde{\alpha }_l := \beta_l $ for $l> n $, then we obtain
that there exists no fix-free code which fits to $(\tilde{\alpha_l})_{l\in\NN }$.
Furthermore, if it is possible to write $\gamma $ as a finite Kraftsum, then
also the sequence $(\tilde{\alpha_l})_{l\in\NN }$ is finite. This shows:

\begin{cor}\label{34lem1b}
Let $|\A |=q\ge 2 $ and $\frac{3}{4} < \gamma $. There exists
a sequence $(\alpha_l )_{l\in\NN } $ of nonnegative integers with
$\gamma =\sum\limits_{l=1}^{\infty } \alpha_l q^{-l} $ such that for every
fix-free code $\X\subseteq\A^* $ there exists an $l\in\NN $ with
$|\X\cap \A^l | < \alpha_l $.
\end{cor}

\smallskip
However the next lemma shows that for any $0< \gamma \le 1 $ there exists
a fix-free code with Kraftsum $\gamma $.

\begin{lemma}\label{34lemma2}
Let $|\A |=q \ge 2 $. For every $0< \gamma \le 1 $ there exists a fix-free code
$\X\subseteq\A^+ $ with $\Kraft (\X )=\gamma $.
\end{lemma}

\smallskip\noindent
{\bf Proof: } If $\gamma =1 $ then for every $n\in\NN $ the fix-free code
$C:=\A^n $ has Kraftsum $1$. Furthermore it is sufficient to show the lemma for
$\A =\{ 0,\ldots ,q-1 \} $. Let $0 < \gamma < 1 $. There exists a
sequence $(\beta_l )_{l\in\NN }$ with
$\gamma =\sum\limits_{l=1}^{\infty}\beta_lq^{-l},$  $\beta_l\in\{0,\ldots
,q-1\}$ for all $ l\in\NN$. Furthermore the sequence is unique if we assume that
for any $n\in\NN $ there exists an $l\ge n $ with $\beta_l\neq q-1$.
Let $\C_1:=\left\{0,\ldots ,\beta_1-1\right\}$, if $\beta_1\ge
1$ and $C_1:=\{ 0 \} $ if $\beta_1=0 $. Furthermore we define for $l\ge 2 $:
\[ \D:=\A-\C_1 \;\mbox{ and }\; \C_l:=\D\C_1^{l-2}\D\subseteq\A^n
 \, . \]

\noindent
Obviously the set $\C := \bigcup\limits_{l=1}^{\infty } \C_l$ is a fix-free code,
where $\C\cap \A^l =\C_l $ for all $l\in\NN $.
For $l\ge 2 $, the number of codewords on the $l$-th level of $\C $ is given by:
\[ \left|\C\cap\A^l\right|=\left|\C_l\right|
=\left|\D\right|\cdot\left|\C_1^{l-2}\right|\cdot\left|\D\right|=
(q-\beta_1)^2\beta_1^{(l-2)}\;\mbox{ for }\; \beta_1\neq 0 \, ,\]
\[ \left|\C\cap\A^l\right|=(q-1)^2 \;\mbox{ for }\; \beta_1 =0 \, .\]

\smallskip\noindent
{\em Case 1:} $\beta_1\neq 0 $  and $\beta_2\le(q-\beta_1)^2$\\
An easy derivation shows that for $\beta_1\neq 0 $ the
global minimum of \\
$f(q):=\beta_1(q-\beta_1)^2-q+1$ is given by
$q_{\min}=\frac{1}{2\beta_1}+\beta_1$ and that there exists no other local minima.
If $0<\beta_1<q $ and $q\in\NN $, it follows that:
\[ \begin{array}{ll}
\quad &  0\le f(\beta_1)\le\beta_1(q-\beta_1)^2-q+1\, , \\
\Rightarrow & q-1\le\beta_1(q-\beta_1)^2 \, , \\
\Rightarrow & \beta_l\le q-1\le\beta_1(q-\beta_1)^l=\left|\C\cap\A^l\right|
\,\forall\,l\ge 3 \end{array}\, .\]
Therefore we obtain
$\left|\C\cap\A^n\right|\ge \beta_l $ for all $l\ge 1$.
Thus we can choose a subset
$\X$ of $\C$ with $\left|\X\cap\A^n\right|=\beta_l$ for all $l\in\NN$. This
gives us a fix-free Code with Kraftsum~r.\\

\smallskip\noindent
{\em Case 2:} $\beta_2 \ge 2 $ and $\beta_2>(q-\beta_1)^2$. \\
We define $\X_1:=C_1$ and as long as
\[ 0<\sum\limits_{l=2}^n
\bigg(q^{(n-l)}\beta_k-q^{(n-l)}(q-\beta_1)^2\beta_1^{(l-2)}\bigg) \]
we define $\X_n:=\C_n$.
Since $\beta_1\ge 2$ and $q-1\ge\beta_l\,\forall n\in \NN$,
there exists an $n\ge 2 $ such that the sum above is smaller than or
equal to zero.
Let N be the smallest of such numbers. It follows:
\[\begin{array}{lcl}
\left|\C_N\right| & = & (q-\beta_1)^2\beta_1^{(N-2)}\\
\quad & \ge &
\beta_N+q\sum\limits_{l=2}^{N-1}\bigg(q^{(N-1-l)}\beta_l\big)-q^{(N-1-l)}
(q-\beta_1)^2\beta_1^{(l-2)}\bigg) >0
\end{array}\]
Thus we can choose
$\beta_N + q\sum\limits_{l=2}^{N-1}\bigg(q^{(N-1-l)}\beta_k\big)
-q^{(N-1-l)}(q-\beta_1)^2\beta_1^{(l-2)}\bigg)$
words from $\C_N$ to obtain an $\X_N\subseteq\C_N $.
With this definition of $\X_1,\ldots ,\X_n$ we obtain:
\begin{eqnarray*}
\sum\limits_{l=1}^N\left|\X_l\right|q^{-l}&=&\frac{\beta_1}{q}+
\sum\limits_{l=2}^{N-1}(q-\beta_1)^2\beta_1^{(l-2)}q^{-l}
+\sum\limits_{l=2}^{N-1}\bigg(q^{-l}\beta_l-q^{-l}(q-\beta_1)^2\beta_1^{(l-2)}\bigg)
+\beta_N \\
&=&\sum\limits_{l=1}^{N}\beta_lq^{-l}\\
\end{eqnarray*}
In the same way as in Case 1, it follows, that $\beta_l\le q-1 \le
\beta_1^{l-2}(q-\beta_1)^2=\left|\C_k\right|$ for $l\ge 3$. Therefore we can choose
for every $l> N$ a set $\X_l\subseteq\C_l$ with $\left|\X_l\right|=\beta_l$.
It follows that $\X :=\bigcup\limits_{l=1}^{\infty } \X_l \subseteq \C $ is fix-free
and $\Kraft (\X )=\sum\limits_{l=1}^{\infty }\beta_l q^{-l} =\gamma $.

\smallskip\noindent
{\em Case 3:} $\beta_1\le 1$.
We obtain $\beta_l\le q-1\le (q-1)^2=\left|\C_n\right|\,\forall l\ge2$. Therefore we can
choose for every $l\in\NN $ a set $\X_l\subseteq \C_l $ with $|\X_l|=\beta_l $.
For $\X :=\bigcup\limits_{l=1}^{\infty }\X_l $ we obtain that $\X $ is fix-free and
and $\Kraft (\X ) =\sum\limits_{l=1}^{\infty }\beta_l q^{-l} =\gamma $.\qed\\

\pagebreak
Let $|\A |=q\ge 2 $. For a number $\gamma\in\RR $ we define the following properties:\\

{\em

\begin{equation}\label{34pro1}\begin{array}{l}
\mbox{For any sequence $(\alpha_l )_{l\in\NN} $ of nonnegative integers with
$\sum\limits_{l=1}^{\infty }\alpha_lq^{-l}\le \gamma \, $,}\\
\mbox{there exists a fix-free
$\C\subseteq\A^* $ which fits to $(\alpha_l )_{l\in\NN} $. }
\end{array}\end{equation}

\quad\\

\begin{equation}\label{34pro2}\begin{array}{l}
\mbox{For any sequence $(\alpha_l )_{l\in\NN} $ of nonnegative integers with
$\sum\limits_{l=1}^{\infty }\alpha_lq^{-l} < \gamma \, $,}\\
\mbox{there exists a fix-free
$\C\subseteq\A^* $ which fits to $(\alpha_l )_{l\in\NN} $. }
\end{array}\end{equation}

\quad\\

\begin{equation}\label{34pro3}\begin{array}{l}
\mbox{For any $n\in\NN $ and finite sequence $(\alpha_1,\ldots ,\alpha_n ) $
of nonnegative \hspace{1cm}}\\
\mbox{integers with $\sum\limits_{l=1}^{n }\alpha_lq^{-l}\le \gamma \, $, there exists a
fix-free $\C\subseteq\A^* $ which}\\
\mbox{fits to $(\alpha_1,\ldots ,\alpha_n ) $. }
\end{array}\end{equation}

\quad\\

\begin{equation}\label{34pro4}\begin{array}{l}
\mbox{For any $n\in\NN $ and finite sequence $(\alpha_1,\ldots ,\alpha_n ) $
of nonnegative\hspace{1.2cm}}\\
\mbox{integers with $\sum\limits_{l=1}^{n }\alpha_lq^{-l}< \gamma \, $, there exists a
fix-free $\C\subseteq\A^* $ which}\\
\mbox{fits to $(\alpha_1,\ldots ,\alpha_n ) $. }
\end{array}\end{equation}

\quad\\

\begin{equation}\label{34pro5}\begin{array}{l}
\mbox{For any sequence $(\alpha_l )_{l\in\NN} $ of nonnegative integers
with the properties\hspace{0.2cm}}\\
\mbox{that for every $n\in\NN $ there exists an $l\ge n $
with $\alpha_l\neq 0 $ and}\\
\mbox{$\sum\limits_{l=1}^{\infty }\alpha_lq^{-l} < \gamma \, $,
there exists a fix-free
$\C\subseteq\A^* $ which fits to $(\alpha_l )_{l\in\NN} $. }
\end{array}\end{equation}
\quad\\
}
\pagebreak
\noindent
The next proposition shows the relation between the different properties above:

\begin{prop}\label{34prop}\quad\\
\begin{enumerate}

\renewcommand{\labelenumi}{(\roman{enumi})}
\item For $\gamma\in\RR $ we have:\\
\[ (\ref{34pro1})\Leftrightarrow (\ref{34pro3})\Rightarrow
(\ref{34pro2})\Leftrightarrow (\ref{34pro4})\Leftrightarrow (\ref{34pro5}) \]

\item If there exists an $\gamma $ with one of the property above we obtain:
\[ \sup_{\mbox{\tiny $\gamma $ has (\ref{34pro1})} }  \gamma
= \sup_{\mbox{\tiny $\gamma $ has (\ref{34pro2})} }  \gamma
=\sup_{\mbox{\tiny $\gamma $ has (\ref{34pro3})} }  \gamma
=\sup_{\mbox{\tiny $\gamma $ has (\ref{34pro4})} }  \gamma
=\sup_{\mbox{\tiny $\gamma $ has (\ref{34pro5})} }  \gamma \]
and the suprema above have the properties (\ref{34pro2}), (\ref{34pro4}) and
(\ref{34pro5}).
\end{enumerate}
\end{prop}

\smallskip\noindent
{\bf Proof: }\\
{\bf (\ref{34pro2}) $ \Rightarrow $ (\ref{34pro4}): } This holds obviously.\\

\noindent
{\bf (\ref{34pro4}) $ \Rightarrow $ (\ref{34pro5}): }
Let $\gamma\in (0,1] $ be a real number with property (\ref{34pro4}).
Let $(\alpha_l)_{l\in\NN } $ be a sequence of nonnegative integers
such that for every $n\in\NN $ there exists an $l\ge n $ with
$\alpha_l\neq 0 $ and $\sum\limits_{l=1}^{\infty }\alpha_lq^{-l} \le \gamma $.
It follows:
\begin{equation}\label{treegl1}
\sum\limits_{l=1}^n\alpha_l q^{-l} <\gamma \quad \forall\, n\in\NN \, .
\end{equation}
While $\gamma $ has property (\ref{34pro4}), it follows that
\begin{equation}\label{treegl2}
\mbox{for all $n\in\NN $ there exists a fix-free $\D\subseteq\A^* $ which fits
to $(\alpha_1,\ldots \alpha_n ) $.}
\end{equation}
>From the property of the sequence $(\alpha_l)_{l\in\NN } $ follows, that
there exists\\
$n_1< n_2 <n_3 <\ldots $ such that $\alpha_{n_l }\neq 0 $ for all
$l\in\NN $ and $\alpha_{n} =0 $ for all\\
$n\not\in\{ n_l |l\in\NN \} $.
We define for $l\in\NN $:
\[\begin{array}{lcl}
\T (l) & := & \{ \D\subseteq\A^* | \D \;\mbox{ is fix-free and fits to }\;
(\alpha_1,\ldots ,\alpha_{n_l}) \}  \\
\T & := & \bigcup\limits_{l=1}^{\infty } \T (l) \cup \{ \emptyset \}\\
\quad & = & \{ \D\subseteq\A^* | \D \;\mbox{ is fix-free and fits to }\;
(\alpha_1,\ldots ,\alpha_{n}) \;\mbox{ for some }\; n\in\NN_0 \}
\end{array}\]
Obviously $(\T,\subseteq ) $ is a tree, where the $l$-th level is given by
$\T (l) $, i.e. every node in $\T $ is a finite node. $\T $ is an
$\omega $-tree, because by (\ref{treegl2}) follows that $\T (l) \neq \emptyset $
for all $l\in\NN $. Furthermore $|\T (l) |<\infty $ for all $l\in\NN $,
because every $\D\in\T (l) $ is a subset of the finite set
$\bigcup\limits_{i=0 }^{n_l }\A^i $. From K\"onigs's Lemma~\ref{koenig}
follows, that there exists an infinite branch in $\T $. This means, there exists
$\D_1\subseteq \D_2 \subset \ldots $ with $\D_l \in\T (l) $ for all $l\in\NN $.
Especially $\D_l $ is fix-free and fits to $(\alpha_1,\ldots ,\alpha_{n_l}) $
for all $l\in\NN $. Let $\C:=\bigcup\limits_{l=1}^{\infty }\D_l $, then
$\C $ is fix-free, because $\D_l\uparrow \C $ and obviously $\C $ fits
to $(\alpha_l)_{l\in\NN } $.\\

\smallskip\noindent
{\bf (\ref{34pro5}) $ \Rightarrow $ (\ref{34pro2}): } Let $0< \gamma \le 1 $
be a real  number which  fulfill (\ref{34pro5}). Then also (\ref{34pro4}) holds for
$\gamma $.
Let $(\alpha_l )_{l\in\NN } $
be a sequence of nonnegative integers with
$\sum\limits_{l=1}^{\infty }\alpha_l q^{-l } <\gamma $. Either the sequence
has the property in (\ref{34pro5}) or there exists an $n\in\NN $ such that
$\alpha_l = 0 $ for all $l\ge n $. In the first case the existence
of a fix-free code which fits to the sequence follows from (\ref{34pro5}) and
in the second case the existence of the fix-free code follows from (\ref{34pro4}).\\

\smallskip\noindent
{\bf (\ref{34pro1}) $ \Rightarrow $ (\ref{34pro3}) $\Rightarrow $
(\ref{34pro4}): } This holds obviously.\\

\smallskip\noindent
{\bf (\ref{34pro3}) $ \Rightarrow $ (\ref{34pro1}): } Let $0< \gamma \le 1 $
such that (\ref{34pro3}) holds, then ,as shown above, also (\ref{34pro5}) holds
for $\gamma $. Let $(\alpha_l )_{l\in\NN } $
be a sequence of nonnegative integers with
$\sum\limits_{l=1}^{\infty }\alpha_l q^{-l } \le\gamma $. If there exists
an $n\in\NN $ such that $\alpha_l = 0 $ for all $l\ge $, there exists a
fix-free code which fits the sequence by (\ref{34pro3}). Otherwise for
every $n\in\NN $ there exists an $l\ge n $ with $\alpha_l \neq 0 $ and from
(\ref{34pro5}) follows that there exists a fix-free code which fits to
$(\alpha_l )_{l\in\NN } $.\\

\smallskip\noindent
This shows part (i). Part (ii) follows from part (i).\qed\\

\medskip
The next lemma shows that there exist a $\gamma $  which fulfill
(\ref{34pro1}) holds. The lemma was first proven by Ahlswede, Balkenhol and Khachatrian
in \cite{ahlswede} for binary codes and finite sequences.
Harada and Kobayashi gave in \cite{harada} a proof of the lemma for arbitrary
(finite) alphabets and infinite sequences.

\begin{lemma}\label{einhalb}Let $|\A |=q\ge 2 $ and
\fo[l]{\alpha} be a sequence of nonnegative integers. If
$\sum\limits_{l=1}^{\infty}\alpha_lq^{-l}\le\frac{1}{2}$ , then there exists a
fix-free $\C\subseteq\A^*$ which fits to \fo[l]{\alpha}.
\end{lemma}

\smallskip\noindent
{\bf Proof: }(The proof is very similar to the proof of theorem \ref{kraft})\\

\noindent
By the condition we obtain $\alpha_1\frac{1}{q}\le \frac{1}{2}$. Thus
$\alpha_1\le\frac{q}{2}<\left|\A\right|$ and we can choose a set
$\C_1\subseteq\A$ with
$\left|\C_1\right|=\alpha_1$. Obviously $\C_1$ is fix-free.\\

\smallskip\noindent
Let $\C_n\subseteq\bigcup\limits_{k=0}^n\A^k$ be a fix-free set with
$\left|\C_n\cap\A^l\right|=\alpha_l $ for all $ 1\le l\le n$.
By Lemma~\ref{shadow} we have:
\begin{equation}\label{gl1}
\left|\pfs[n+1]\left(\C_n\right)\right|=\left|\sfs[n+1]\left(\C_n\right)\right|=\sum\limits_{l=1}^{n}\alpha_lq^{n+1-l}\le\frac{1}{2}
\end{equation}
While $\sum\limits_{l=1}^{n+1}\alpha_lq^{-l}\le
\sum\limits_{l=1}^{\infty}\alpha_lq^{-l}\le\frac{1}{2}\, $, it follows:
\[ \left|\A^{n+1}\right|=q^{n+1}\ge
2\alpha_{n+1}+2\sum\limits_{l=1}^{n}\alpha_lq^{n+1-l} \, .\]
Thus we obtain:
\begin{eqnarray*}
\left|\A^{n+1}\right|-2\alpha_{n+1}& \ge & 2\sum\limits_{l=1}^{n}\alpha_lq^{n+1-l}\\
\quad & \begin{array}{c} \ge \\ \mbox{\footnotesize (\ref{gl1}) }\end{array}
 & 2\left|\pfs[n+1]\left(\C_n\right)\right|\\
\quad & \begin{array}{c} \ge \\ \mbox{\footnotesize (\ref{gl1}) } \end{array}
& \left|\pfs[n+1]\left(\C_n\right)\right|+\left|\sfs[n+1]\left(\C_n\right)\right|
-\left|\pfs[n+1]\left(\C_n\right)\cap\sfs[n+1]\left(\C_n\right)\right|\\
& = &\left|\bfs[n+1]\left(\C_n\right)\right| \, .
\end{eqnarray*}
It follows that
$ \left|\A^{n+1}\right|\ge\left|\bfs[n+1]\left(\C_n\right)\right|+\alpha_{n+1}$.\\

\noindent
Therefore we can choose $\alpha_{n+1}$ words $c_1,\ldots
,c_{\alpha_{n+1}}\in\A^{n+1}$, which are not in the $(n+1)$-th level bifix-shadow
of $\C_n $.
Thus $\C_{n+1}:=C_n\cup\left\{c_1,\ldots
,c_{\alpha_{n+1}}\right\}$ is fix-free and fits $(\alpha_1,\ldots ,\alpha_{n+1} )$.\\

\noindent
Let $\C:=\bigcup\limits_{l=1}^{\infty }\C_l $, then $\C $ fits to $(\alpha_l)_{l\in\NN }$.
and since
$\C_l\uparrow \C $, the code $\C $ is fix-free.\qed\\

\medskip
>From Lemma \ref{34lemma1} and Lemma \ref{einhalb} follows, that there exists
a $\gamma \in [\frac{1}{2},\frac{3}{4}] $ which fulfill (\ref{34pro1})
and that for every $\gamma > \frac{3}{4} $ property (\ref{34pro1})
does not hold. This gives us the conjecture:

\begin{conj}\label{34conj2}
\[ \sup\limits_{\mbox{\tiny $\gamma $ fullfil (\ref{34pro1})}}\gamma =\frac{3}{4} \]
\end{conj}

\smallskip\noindent
However, the conjecture above is weaker than the Conjecture
\ref{34conj1}, since from Proposition \ref{34prop} follows that Conjecture
\ref{34conj2} is equivalent to:
\begin{equation}\label{34conj3}\begin{array}{l}
\mbox{For any sequence $(\alpha_l)_{l\in\NN }$ of nonnegative integers
with $\sum\limits_{l=1}^{\infty } \alpha_l q^{-l} <\frac{3}{4} \, $,}\\
\mbox{there exists a fix-free code which fits to $(\alpha_l)_{l\in\NN }$.}
\end{array}
\end{equation}

\section{Extensions of fix-free codes}
\smallskip
Let $P$ be a property defined for sequences of nonnegative integers. We call
$P$ an {\em extension property} for sequences if
\begin{enumerate}
\renewcommand{\labelenumi}{(\arabic{enumi})}
\item for any finite sequence $(\alpha_1,\ldots ,\alpha_n ) $ which fulfill $P$ also
$(\alpha_1,\ldots ,\alpha_{n-1} )$ fulfill $P$,

\item for any infinite sequence $(\alpha_l )_{l\in\NN } $ for which $P$ holds, also
$(\alpha_1,\ldots ,\alpha_n ) $ fulfill $P$ for all $n\in\NN $.

\end{enumerate}

\smallskip\noindent
We call $P$ an {\em $\sigma $-extension} property, if further on,
$(\alpha_1 ,\ldots , \alpha_n)$ has property $P$ for all $n\in\NN$ for
a sequence $(\alpha_l )_{l\in\NN } $ imply that also
$(\alpha_l )_{l\in\NN } $ fulfill $P$.\\

\smallskip
Let $\M\subseteq \PP (\A^* ) $. We call $\M $ an {\em extension class}, if the
following properties hold for $\M$:

\begin{enumerate}
\renewcommand{\labelenumi}{(\arabic{enumi})}

\item $\emptyset\in\M $,

\item if there exists $\C\in\M $ which fits to a finite sequence
$(\alpha_1,\ldots ,\alpha_n )$ then there exists a set $\D\in\M $ which fits
to $(\alpha_1,\ldots ,\alpha_{n-1} )$,

\item if there exists a set $\C\in\M $ which fits to a sequence $(\alpha_l )_{l\in\NN } $
then for every $n\in\NN $ there exists a set $\C_n\in\M $ which fits to the finite
sequence $(\alpha_1,\ldots ,\alpha_n ) $.

\end{enumerate}

\smallskip\noindent
Furthermore we call $\M $ an {\em $\sigma $-extension class}, if
$\bigcup\limits_{l\in\NN }\C_n \in\M $
for every ascending set sequence
$\C_1\subseteq \C_2\subseteq \ldots $ with $C_1,\C_2, \ldots\in\M $.\\

\medskip
For example the classes of prefix-, suffix- and fix-free sets are all
$\sigma $-extension classes. Let $0< \gamma \le 1 $. For sequences of nonnegative
integers. We define the properties
$P_{\gamma }$ and $P_{<\gamma } $ as follows:
\[ P_{\gamma } : \quad \sum\limits_{l=1}^{\infty } \alpha_l q^{-l }
\le \gamma \quad\mbox{ and }\quad
P_{<\gamma } :\quad \sum\limits_{l=1}^{\infty }
\alpha_l q^{-l } < \gamma \,.\]
Obviously $P_{\gamma } $ and $P_{<\gamma } $ are $\sigma $-extension properties.\\

\pagebreak
We denote with $\I (P) , \F (P) ,\M (P ) $ and $ \M_F ( P) $ the following sets:

\begin{quote}
$\I (P) $ is defined as the set of all sequences which have property $P$.\\

$\F (P) $ is defined as the set of all finite sequences which fulfill $P$.\\

$\M (P) $ is defined as the class of all sets in $\M $ which fits to a sequence
in $I (P)$.\\

$\M_f (P) $ is defined as the class of all sets in $\M $ which fits to a sequence
in $\F (P) $.\\

\end{quote}

\noindent
Obviously $\M_f (P) \subseteq \M (P)$.
If $P$ is an extension property and $\M $ an extension class, it is easy to verify
that $\M_f (P)$ and $\M (P)$ are extension classes. Furthermore if $\M $ is an
$\sigma $-extension class and $P$ a $\sigma $-extension property,
then $\M (P) $ is an $\sigma $-extension class, as well.\\

\medskip
Let $P$ an extension property. We call an extension class $\M $ a
{\em $P$-simple extension class} if for $\M $ the following property holds:

\begin{equation}\label{simext1}
\mbox{\parbox{12cm}{\em
{\bf Simple extension property:} Let $(\alpha_1,\ldots, \alpha_n)\in\F (P) $.
If there exists a set in $\M $ which fits to $(\alpha_1,\ldots ,\alpha_n) $,
then for every $\C\in\M $ which fits to $(\alpha_1,\ldots ,\alpha_{n-1}) $
there exists an extension in $\M $ which fits to
$(\alpha_1,\ldots ,\alpha_n )$, i.e. the extension and $\C $ are a sets
in $\M_f (P) $.}}
\end{equation}

\noindent
If $\M $ fulfill (\ref{simext1}) for all sequences of nonnegative integers,
then we call $\M $ a
simple extension class. Property (\ref{simext1}) means, that for a finite set
$C\in\M $, the existence of an extension in $\M $ which fits to
a sequence in $\F (P) $, does not depend on the words contained in $\C $,
but on the values of $|\C\cap \A^l | $ for $l\in\NN $.
Therefore the following simple strategy is possible,
for finding a set in a $P$-simple extension class $\M $ which
fits to a sequence $(\alpha_1,\ldots ,\alpha_n )\in\F (P) $:

\begin{enumerate}
\renewcommand{\labelenumi}{\arabic{enumi}.}
\item Choose an arbitrary set $\C_1\subseteq \A $ with $\C_1\in\M $ and
$|\C_1 | =\alpha _1 $.
\item If a set $\C_l \in\M $ which fits to $(\alpha_1,\ldots ,\alpha_l ) $
is already constructed, then choose $\C_{l+1} $ as an arbitrary extension
of $\C_l $ in $\M $ which fits to $(\alpha_1,\ldots ,\alpha_l ) $.
\end{enumerate}

\noindent
If there exists at least one set in $\M $ which fits to
$(\alpha_1, \ldots ,\alpha_n) $, then from the simple extension property follows,
that the construction above gives us after $n$ steps a set $\C_n \in\M $ which
fits to $(\alpha_1, \ldots ,\alpha_n) $.
Furthermore, if $\M $ is a $P$-simple $\sigma $-extension class
for some $\sigma $-extension property $P$, then there exists
a set in $\M $ which fits to a sequence $(\alpha_l)_{l\in\NN } $
in $\I (P) $ if and only if
the construction above doesn't stop. In this case the set
$\C:=\bigcup\limits_{l=1}^{\infty}\C_l $ is a set in $\M $ which
fits to $(\alpha_l)_{l\in\NN } $.\\

\medskip
Since the cardinality of the prefix-shadow on the $(n+1)$-th level of
a prefix-free set $\C\subseteq\bigcup\limits_{l=0}^n \A^l $ only depends on
the Kraftsum of $\C $, it follows that the class of prefix-free sets is
a simple $\sigma $-extension class. However, for a fix-free set
$\C\subseteq\bigcup\limits_{l=0}^n \A^l $ the bifix-shadow on the
$(n+1)$-th level is given by:
\[ |\bfs[n+1] (\C )| =2|\pfs[n+1] (\C )| -\sum\limits_{x,y\in\C } I_{n+1} (x,y)
=2|\pfs[n+1] (\C )|-|\pfs[n+1] (\C )\cap\sfs[n+1] (\C )|. \]
In general the sum $\sum\limits_{x,y\in\C } I_{n+1} (x,y) $ depends on the codewords
contained in $\C $ and does not depends on the codeword lengths only.

\medskip
The next example shows that, the class of fix-free sets is not
a simple extension class.
\begin{example}\upshape Let $\A =\{ 0,1 \} $ and
$\alpha_1=0, \alpha_2 =1 , \alpha_3=2, \alpha_4=4 , \alpha_l=0 $ for $l\ge
5 $. For the Kratsum we obtain:
\[\sum_{l=1}^{\infty}\alpha_l\cdot\frac{1}{2}^l=\frac{1}{4}+
\frac{2}{8} +\frac{4}{16}= \frac{3}{4} \]

\noindent
$\D :=\{ 00,101,110 \} $ is a fix-free code which fits to
$(\alpha_1,\alpha_2,\alpha_3) $.
Since \\
$\big| \A_4-\bfs[4](\D) \big|=\big|\{ 1111,0111,1001 \}\big|
=3<4=\alpha_4 $, it follows that there does not exist a fix-free
$\C\supseteq\D $ which fits to $(\alpha_1,\alpha_2,\alpha_3,\alpha_4 )$.

\noindent
Indeed $\D' := \{ 10,000,111,1100,0011,0101,1101 \} $ is a fix-free code which
fits to $(\alpha_1,\alpha_2,\alpha_3,\alpha_4 )$.
\end{example}

\smallskip
Moreover the example above shows, that the class of fix-free sets, is also
not a $P_{\frac{3}{4}}$-simple extension class. The question arise if there
exists extension properties $P$ such the class of fix-free codes is a
$P$-extension class? The proof of Lemma \ref{einhalb} shows for example
that the the class of fix-free sets is a $P_{\frac{1}{2}}$-simple extension class.
While for fix-free codes with Kraftsum smaller than or equal to $\frac{1}{2} $,
there are for an extension, such less codewords necessary, that it is possible
to ignore the value of $|\pfs[n+1] (\C )\cap\sfs[n+1] (\C )| $.

\pagebreak
Another example is the following property $P_{\frac{3}{4}}^* $ for
sequences $(\alpha_l )_{l\in\NN } $ of  nonnegative integers:
\[P_{\frac{3}{4}}^*\quad :\quad
\sum\limits_{l=1}^{\infty } \alpha_l q^{-l} \le \frac{3}{4} \quad\mbox{ and }\quad
\alpha_l\neq 0 \Rightarrow \alpha_{l+1}=\alpha_{l+2}=\ldots =\alpha_{2l-1} =0
\]
Obviously $P_{\frac{3}{4}}^* $ is an $\sigma $-extension property. Furthermore
in the proof of Theorem \ref{qfall1} at the beginning of the next chapter, we
will show that the class of fix-free sets is a $P_{\frac{3}{4}}^* $-simple
extension class. We might ask for the supremum
of numbers $\gamma $ for which the class of fix-free sets is a
$P_{\gamma }$-simple extension class. For the supremum $\gamma^* $ follows that
the class of fix-free sets is a $P_{<\gamma^*} $-simple extension class.

\medskip
The next
example shows that the class of fix-free codes is not a $P_{<\frac{3}{4}}$-simple
extension class. Therefore it follows that $\gamma^*<\frac{3}{4} $.

\begin{example}\upshape
Let $\A :=\{ 0,1\} $ and  $\alpha_1=\alpha_2=0,
\alpha_3=4,\alpha_4=1,\alpha_5=5 ,\alpha_l=0 $ for $l\ge 6 $.
We obtain for the Kraftsum of the sequence:
\[\sum_{l=1}^{\infty}\alpha_l\cdot\frac{1}{2}^l=
\frac{4}{8}+\frac{1}{16}+\frac{5}{32}=\frac{23}{32} <\frac{3}{4} \,.\]

\noindent
The set $\D :=\{ 000,111,011,001,0101 \} $ is a fix-free code and fits to
$(\alpha_1,\alpha_2,\alpha_3,\alpha_4 )$. \\

\smallskip
\begin{center}
\begin{tabular}{|l|c|c||l|c|c||l|c|c|}\hline
Level 5 & P & S & Level 5 & P & S & Level 5 & P & S \\ \hline
00000 & x     & x     & 01011 & x     & x     & {\bf 10110} & \quad & \quad \\
00001 & x     & x     & 01100 & x     & \quad & 10111 & \quad & x     \\
00010 & x     & \quad & 01101 & x     & \quad & 11000 & \quad & x     \\
00011 & x     & x     & 01110 & x     & \quad & 11001 & \quad & x     \\
00100 & x     & \quad & 01111 & x     & \quad & {\bf 11010} & \quad & \quad \\
00101 & x     & x     & 10000 & \quad & x     & 11011 & \quad & x     \\
00110 & x     & \quad & 10001 & \quad & x     & 11100 & x     & \quad \\
00111 & x     & x     & {\bf 10010} & \quad & \quad & 11101 & x     & \quad \\
01000 & \quad & x     & 10011 & \quad & x     & 11110 & x     & \quad \\
01001 & \quad & x     & {\bf 10100} & \quad & \quad & 11111 & x     & x     \\
01010 & x     & \quad & 10101 & \quad & x     & \quad & \quad & \quad \\ \hline
\end{tabular}
\end{center}

\smallskip\noindent
The tabular above shows that:
\[ \big| \A^5 -\bfs[5](\D ) \big| = \big| \{
10010,10100,10110,11010 \} \big| =4 < 5=\alpha_5 \, .\]
It follows that there does not exist a fix-free code $\D $ with $\C \supseteq \D $
which fits to $(\alpha_1,\alpha_2,\alpha_3,\alpha_4,\alpha_5) $.
However there exist fix-free codes which fits to
$(\alpha_1,\alpha_2,\alpha_3,\alpha_4,\alpha_5) $. As an example:
\[ \{ 000,001,010,011,1110,10100,10101,10110,10111,11111 \} \]
\end{example}

\pagebreak
Since the class of fix-sets is a $P_{\frac{1}{2}}$-simple extension class, we
give the following conjecture:

\begin{conj}\label{extconj}
\[ \frac{1}{2} = \max \left\{ \gamma \, \left|\,
\mbox{the class of fix-free codes is
a $P_{\gamma } $-simple extension class} \right.\right\} \]
\end{conj}

\smallskip
Instead of searching for properties $P$ of sequences for which the class
of fix-free sets is a $P$-simple extension class, one might try to find
a subclass $\M $ of fix-free sets such that $\M $ is a simple extension
or a $P_{\frac{3}{4}}$-simple extension class. However in this survey we don't
pay attention to this problem.

\smallskip
Let $\alpha_1,\ldots ,\alpha_{n+1} \in\NN_0 $ and $\D\subseteq\A^* $ be a fix-free
set which fits to $(\alpha_1,\ldots ,\alpha_n ) $. By Lemma~\ref{shadow} follows:
\[ |\bfs[n+1] (\D )| =2q^{n+1}\Kraft (\D ) -|\pfs[n+1](\D )\cap \sfs[n+1] (\D )| \, . \]
Therefore the existence of an extension of $\D $ which fits to
$(\alpha_1,\ldots ,\alpha_{n+1}) $ depends on the value of
$|\pfs[n+1](\D )\cap \sfs[n+1] (\D )| $ and $(\alpha_1,\ldots ,\alpha_{n+1}) $.
The next lemma shows, for which values of $|\pfs[n+1](\D )\cap \sfs[n+1] (\D )| $
an extension is possible, if the Kraftsum of the sequence is smaller than or
equal to $\frac{3}{4} $. The following lemma can be found in \cite{harada}.

\begin{lemma}\label{ext} Let $|\A |=q \ge 2 $, $n>k $,
$\alpha_1,\ldots ,\alpha_k\in\NN_0 $, $\alpha_{k+1}=\ldots =\alpha_{n-1}=0 $
and $\alpha_{n}\in\NN $ such that
$\sum\limits_{l=0}^{n} \alpha_l\cdot q^{-l} \le \frac{3}{4} $.
Let $\D $ be a fix-free set which fits to
$(\alpha_1,\ldots , \alpha_k )$.

\renewcommand{\labelenumi}{(\roman{enumi}):}
\begin{enumerate}

\item If
$\quad\frac{\big| \pfs[n] (\D )\cap \sfs[n] (\D ) \big|}{q^n} \ge \left\{
\begin{array}{ll}
\frac{\pfs[n] (\D )}{q^n}+\frac{\lfloor \frac{3}{4} q^{n} \rfloor }{q^n} -1 & \mbox{ if q is odd}\\
\frac{\pfs[n] (\D )}{q^n} -\frac{1}{4} & \mbox{ if q is even}
\end{array}\right. \; $, \\[2mm]
then there exists a fix-free extension $\C\supseteq \D $ which fits to
$(\alpha_1,\ldots ,\alpha_{n+k}) $.

\item If $\quad \frac{\big| \pfs[n] (\D )\cap \sfs[n] (\D ) \big|}{q^n} \ge \bigg(
\frac{\pfs[n] (\D )}{q^n} \bigg)^2 \; $,
then there exists a fix-free extension $\C\supseteq \D $ which fits to
$(\alpha_1,\ldots ,\alpha_{n+k}) $.

\end{enumerate}
\end{lemma}

\smallskip\noindent
{\bf Proof:}
\renewcommand{\labelenumi}{(\roman{enumi}):}
\begin{enumerate}

\item Let $(\alpha_1,\ldots ,\alpha_{n+k} ), \D $ be as in the Lemma.
For even $q$ and $n\ge 2 $ we have
$\frac{\lfloor \frac{3}{4} q^n \rfloor }{q^n} -1 =-\frac{1}{4} $. Therefore
it is sufficient to show that

\pagebreak
\begin{equation}\label{extgl1} \frac{\big| \pfs[n] (\D )\cap \sfs[n] (\D ) \big|}{q^n}
\ge \frac{\pfs[n] (\D )}{q^n}+\frac{\lfloor \frac{3}{4} q^n \rfloor }{q^n} -1
\end{equation}
imply the existence of a fix-free Code $\C \supseteq\D $ which fits to
$(\alpha_1,\ldots,\alpha_n) $.\\
We obtain from the conditions of $(\alpha_1,\ldots ,\alpha_n ) $ :
\[\begin{array}{ll}
\quad & \frac{3}{4}\ge\sum\limits_{l=1}^n\alpha_lq^{-l} =\alpha_nq^{-n}
+\sum\limits_{l=1}^k\alpha_lq^{-l} \, ,\\
\Rightarrow & \frac{3}{4}q^n\ge \alpha_n + \sum\limits_{l=1}^k \alpha_lq^{n-l}
=\alpha_n +\big| \pfs[n] (\D ) \big| \in\NN \, ,\\
\Rightarrow &
\lfloor \frac{3}{4} q^n \rfloor \ge \alpha_n +\big| \pfs[n] (\D ) \big| \, .
\end{array}\]
By (\ref{extgl1}) it follows:
\[
\big| \pfs[n] (\D )\cap \sfs[n] (\D ) \big| \ge 2\big| \pfs[n] (\D ) \big| +\alpha_n
- q^n \, .
\]
While $\big| \bfs[n] (\D ) \big| = 2\big| \pfs[n] (\D ) \big| -\big|
\pfs[n] (\D )\cap \sfs[n] (\D ) \big| $
(by Lemma~\ref{shadow}) and\\
$q^n=\big| \A^n \big| $, we conclude:
\[
\big| \A^n \big| - \big| \bfs[n] (\D ) \big| \ge \alpha_n \, .
\]
Thus we can choose $\alpha_n $ different words $c_1,\ldots ,c_{\alpha_n}\in\A^n $ of
length $n$, which are not in the bifix-shadow of $\D $. Then
$\C:=\D\cup\{c_1,\ldots ,c_{\alpha_n} \} $ is a fix-free Code with the desiered
properties.

\item The function $f(x):=x^2 $ is convex. Therefore we have:
\begin{equation}\label{extgl3}
x^2\ge f'(\frac{1}{2} )(x-\frac{1}{2} )+f(\frac{1}{2} )=x-\frac{1}{4}\ge x +
\frac{\lfloor \frac{3}{4} q^n \rfloor }{q^n} -1
\end{equation}
If
$\frac{\big| \pfs[n] (\D )\cap \sfs[n] (\D ) \big|}{q^n} \ge \bigg( \frac{\pfs[n] (\D
)}{q^n} \bigg)^2 $, then the existence of a fix-free extension $\C\supseteq\D $
which fits to $(\alpha_1,\ldots ,\alpha_{n})$ follows by (i) and (\ref{extgl3})
for $x=\frac{\pfs[n] (\D)}{q^n} $.\qed
\end{enumerate}

\smallskip\noindent
The proof of the lemma shows, that the condition in (i) imply
the condition in (ii).\\

\pagebreak
There is another difference between fix-free codes and prefix-free codes
which was mentioned in \cite{yeung}:\\

\medskip\noindent
We call a finite sequence $\vec{l_n} :=(l_1,\ldots l_n)\in\NN^n $ a lengths
sequence, if\\
$l_1\le l_2\le \ldots \le l_n $. A set $\C\subseteq\A^+ $ fits to
the lengths sequence $\vec{l_n} $ if $l_1,\ldots ,l_n $ are the lengths of
the words in $\C $. This means there exists a word $c_i \in\C $ with
$|c_i|=l_i $ for every $1\le i \le n $ and $\C =\{ c_1,\ldots ,c_n \} $.
If $\alpha_l :=|\C \cap\A^l | $ for all $l\in\NN $, then $\alpha_l $ is
the number of occurrence of $l$ in the lengths sequence $\vec{l_n} $.
We call the sequence $(\alpha_l )_{l\in\NN } $ the sequence which corresponds
to the lengths sequence $\vec{l_n} $. It follows that:
\[ \sum\limits_{i=1}^n q^{-l_i} =\sum\limits_{l=1}^{l_n}\alpha_l q^{-l} \quad\mbox{ and }\quad
\alpha_l =0 \;\mbox{ for all }\; l>l_n \]
Let $\vec{l_n}=(l_1,\ldots ,l_n ) ,\vec{l'_n} =(l'_1,\ldots ,l'_n ) $ be lengths
sequences. We write $\vec{l'_n}\ge \vec{l_n} $ if $l'_i\ge l_i $ for all
$1\le i\le n $.

\medskip
Let $\C\subseteq\A^+ $ be a prefix-free code with lengths sequence
$\vec{l_n} $ and $\vec{l'_n} $ be another lengths sequence with
$\vec{l'_n}\ge \vec{l_n} $. Since $\C $ is prefix-free, it follows:
\[ 1\ge \Kraft (\C ) = \sum\limits_{i=1}^n q^{-l_i} \ge
\sum\limits_{i=1}^n q^{-l'_i} \, .\]
>From Theorem \ref{kraft} follows, that there exists a prefix-free code $\C' $
which fits to the length sequence $\vec{l_n} $. Furthermore $\C' $ can be chosen
in such a way, that $\C' $ lies in the prefix-shadow $\C\A^* $ of $\C $.
Let $\C =\{ c_1,\ldots ,c_n \} $ with $|c_i|=l_i $ for all $1\le i\le n $.
We obtain $\C' $ by replacing of every $c_i $ by a word
$c'_i\in c_i\A^{l'_i-l_i}\subseteq \pfs[l'_i] (\C ) $. Then the set
$C' := \{ c'_1,\ldots ,c'_n \} $ is a prefix-free code which fits to
$\vec{l'_n }$ and $\C'\subseteq
\C\A^* =\bigcup\limits_{l=0}^{\infty }\pfs[l] (\C ) $.

\begin{prop}\label{extlang}
Let $\C\subseteq\A^+ $ a prefix-free code which fits to a length sequence
$\vec{l_n} $ and let $\vec{l'_n}$ another lengths sequence with
$\vec{l'_n}\ge \vec{l_n} $.
\begin{enumerate}
\renewcommand{\labelenumi}{(\roman{enumi})}
\item There exists a prefix-free Code $\C' $ which fits to $\vec{l'_n} $.
\item $\C' $ can be chosen in such a way that $\C'\subseteq\C\A^* $.
\end{enumerate}
\end{prop}

The question rises, wether the proposition above is also true for fix-free codes?
The next example shows, that this is not the case.

\pagebreak
\begin{example}
\upshape Let $\A =\{ 0,1 \} $, $\C :=\{ 0,11,101,1001 \} $ and
$\ltup[l']{n} :=(1,2,4,4) $. $\C $ is a fix-free
code with lengths sequence $ \ltup[l]{4} =(1,2,3,4) $. We
have $\ltup[l']{n}\ge \ltup[l]{n} $.
Let us assume that there exists a fix-free code
$\C' =\{ c'_1,c'_2,c'_3,c'_4 \} $ with $c'_i={l'}_i $ for all $1\le i \le 4 $.\\

\noindent
{\em Case $c'_1= 0 $ :} Since $\C' $ is fix-free it follows that $c'_2= 11 $.
Then it is easy to verify that the $1001$ is the  only word in $\A^4 $
which is not in the
bifix-shadow of $\{ c'_1,c'_2 \}$. This is a contradiction,
because $\C' $ contains two words of length $4$.\\

\noindent
{\em Case $c_1=1 $ :}It follows that $c'_2=00 $. Then $0110 $ is only
word of length $4$
which is not in the bifix-shadow of $\{ c_1,c_2 \} $. This is
once again a contradiction.\\

\noindent
This shows that there does not exists a fix-free code which fits to
$\vec{l'_4} $.
For the Kraftsum of $\vec{l_n} $ we obtain:
\[ \sum\limits_{i=1}^4 2^{-l_i } =\frac{13}{16}>\frac{3}{4} \quad\mbox{ and }\quad
\sum\limits_{i=1}^4 2^{-l'_i } =\frac{7}{8}>\frac{3}{4} \, .\]
\end{example}

\bigskip
If the $\frac{3}{4} $-conjecture holds, then obviously Proposition~\ref{extlang} (i)
holds fix-free codes with Kraftsum smaller than or equal to $\frac{3}{4} $.
In general Proposition~\ref{extlang} (i) holds for fix-free codes
with Kraftsum smaller than or equal to $\gamma$, if (\ref{34pro3}) holds
for $\gamma $. On the other hand, if we assume that the $\frac{3}{4} $-conjecture hold
, the next example shows that Proposition~\ref{extlang} (ii)
does not hold for fix-free codes with Kraftsum smaller than or equal to
$\frac{3}{4} $.

\begin{example}
\upshape  Let $\A=\{ 0,1 \} $ and $\C := \{ 011,110,010,1001 \} $. Then $\C $ is a
fix-free code with lengths $\ltup[l]{4} =(3,3,3,4) $ and Kraftsum $\frac{7}{16}
<\frac{3}{4} $. Let $\ltup[l']{4} :=(3,3,3,5) $, then $\ltup[l]{n}\le \ltup[l']{n} $.
Every word in $\bfs[5] (1001) = \{ 11001,01001, 10010,10011 \} $ has at least
one word in $\{ 011,110,010 \} =\D\cap A^3 $ as a suffix or as a prefix. It follows
that there does not exist a fix-free Code $\C'\subseteq\C\A^* $ with lengths
sequence $\ltup[l']{4} $. On the other hand there exists a fix-free
Code with lengths sequence $\ltup[l']{4} $. For example
$ \C' := \{ 011,110,010,10001 \} $.
\end{example}

\newpage
\chapter{The $\frac{3}{4}$-conjecture for $q$-ary fix-free codes}
This chapter is about the cases, where the $\frac{3}{4}$-conjecture
can be shown for an arbitrary finite alphabet $\A $. We show first
two theorems from Ahlswede, Balkenhol and Khachatrian \cite{ahlswede}
and Harada and Kobayashi \cite{harada}
which stated, that the $\frac{3}{4}$-conjecture holds for sequences
with $2k\le \inf \{ l\, |\, \alpha_l\neq 0 \, , l>k\}$ for all $k\in\NN $
and that the $\frac{3}{4}$-conjecture holds for two level fix-free codes.
Finally we give a generalization of a theorem
from Kukorelly and Zeger \cite{zeger}, which was shown for the binary case
originally. This theorem shows, that the $\frac{3}{4}$-conjecture
holds for finite codes, if the number of codewords on each level,
expect the maximal level, is bounded by term which depends on the minimal level.
The generalization of this theorem for $q$-ary alphabets is one of the
new results in this survey.\\

\smallskip
The next theorem shows that $\frac{3}{4} $-conjecture holds for sequences with
property $P^*_{\frac{3}{4} } $. It was first shown by
Ahlswede, Balkenhol and Khachatrian in \cite{ahlswede} for binary codes
and finite sequences. In \cite{harada} Harada and Kobayashi generalized
the theorem to the form given below for arbitrary finite alphabets and
infinite sequences.

\begin{theorem}[Ahlswede, Balkenhol and Khachatrian]\label{qfall1}
Let $|\A |=q\ge 2 $ and $\fo{\alpha } $ be a sequence of nonnegative integers.
If the sequence has the properties
\[ \sum\limits_{l=1}^{\infty } \alpha_l q^{-l}\le \frac{3}{4}
\;\mbox{ and }\;
2k\le \inf \{ l\, |\, \alpha_l\neq 0 \, , l>k\}
\;\mbox{ for all }\; k\in\NN \;\mbox{ with }\; \alpha_k\neq 0\, ,
\]
then there exists a fix-free code $\C\subseteq\A^* $, which fits to $\fo{\alpha } $.
\end{theorem}

\noindent
Furthermore from the proof of the theorem follows, that the class of fix-free
sets is a $P^*_{\frac{3}{4} } $-simple extension class.\\

\pagebreak
\noindent
{\bf Proof:} (by induction)\\

\begin{enumerate}
\renewcommand{\labelenumi}{(\roman{enumi})}
\item Let $k_1>0 $ the smallest number with $\alpha_{k_1}\neq 0 $. It follows that:

\[\frac{3}{4} \ge \sum\limits_{l=1}^{k_1}\alpha_lq^{-l} =\alpha_{k_1}q^{-k_1}
\Rightarrow \big|\A^{k_1} \big|\ge \frac{3}{4} q^{k_1} =\alpha_{k_1} \]
Therefore we can choose
a set $\C_1 \subseteq \A^{k_1} $ with $\big| \C_1|=\alpha_{k_1} $.
Obviously $\C_1 $ is fix-free code which fits to $(\alpha_1,\ldots ,\alpha_{k_1}) $.

\item {\bf $\bf k_i \rightarrow k_{i+1} $:}\\
Let $k_i\in\NN $ such that $\alpha_{k_i}\neq 0 $ and $\C_i $ be a fix-free code
which fits to $(\alpha_1,\ldots ,\alpha_{k_i})$.
Let $k_{i+1}:=\inf \{\, l\in\NN \, |\, \alpha_l>0, l>k_i\}\le \infty $.
If $k_{i+1} =\infty $, the set $\C :=\C_i $
is a fix-free code which fits to $\fo{\alpha }$. Therefore let us suppose
that $k_{i+1}<\infty $. By the conditions of the theorem  follows  $2k_{i}\le
k_{i+1} $. Therefore we obtain $|x|+|y|\le k_{i+1} $
for all $x,y\in\C_i $. From Lemma \ref{shadow} (iii) follows:\\

\begin{displaymath}
\big| I_{k_{i+1}} (x,y) \big| = q^{k_{i+1} -|x|-|y|} \;\mbox{ for all }\; x,y\in\C_i\, .
\end{displaymath}

While $|x|< k_{i+1} $ for all $ x\in\C_i $,  from Lemma \ref{shadow}  follows:\\

\[\begin{array}{lcl}
\big|\pfs[k_{i+1}]{(\C_i )}\cap \sfs[k_{i+1}]{(\C_i )} \big| & = &
\sum\limits_{x,y\in\C_i}\big| I_{k_{i+1}} (x,y) \big|
 =  \sum\limits_{x,y\in\C_i}q^{k_{i+1} -|x|-|y|}\\
& = & q^{k_{i+1}}\sum\limits_{l_1,l_2=1}^{k_i}\alpha_{l_1}\alpha_{l_2} q^{-l_1-l_2}
=  q^{k_{i+1}}\bigg(\sum\limits_{l=1}^{k_i} \alpha_lq^{-l} \bigg)^2\, .
\end{array}\]

Since $\big| \pfs[k_{i+1}] (\C_i ) \big| = \sum\limits_{x\in\C_1}
q^{k_{i+1}-|x|}= q^{k_{i+1}}\cdot \sum\limits_{l=1}^{k_i} \alpha_lq^{-l} $, we obtain:

\begin{displaymath}
\frac{\big|\pfs[k_{i+1}]{(\C_i )}\cap \sfs[k_{i+1}]{(\C_i )} \big|}{q^{k_{i+1}}}=
\bigg( \frac{\big| \pfs[k_{i+1}] (\C_i ) \big|}{q^{k_{i+1}}}\bigg)^2 \, .
\end{displaymath}

Furthermore we have $\sum\limits_{l=1}^{k_{i+1}}\alpha_lq^{-l}\le \frac{3}{4} $.
This shows, that the conditions of Lemma \ref{ext} (ii) hold.
Therefore it follows that there exists a fix-free
extension $\C_{i+1} $ of $\C_i $ which fits to
$(\alpha_1,\ldots ,\alpha_{k_i} )$.

\pagebreak
\item If there exists  $i\in\NN $ such that $k_i =\infty $, then
$\C:= C_{i} $ is a fix-free code which fits to $\fo{\alpha } $. If
for every $l\in\NN $ there exists a $k> l $ with $\alpha_k\neq 0 $, then the
procedure above doesn't stop. In this case the set
$\C :=\bigcup\limits_{i=1}^{\infty }\C_i $ is a fix-free code which fits to
$\fo{\alpha } $, because $\C_i\uparrow \C $.\qed\\
\end{enumerate}

\medskip
We have shown in the last section of Chapter 1, that it is in general not possible to
obtain a fix-free code $\C $ which fits to a sequence $(\alpha_1,\ldots ,\alpha_n) $
with Kraftsum smaller or equal
$\frac{3}{4} $, by the following procedure.
Choose a set $C_1\subseteq\A^1$ with $\big|C_1 \big| =\alpha_1 $.
Then extend $\C_1 $ to a fix-free set $\C_2 $ which fits to
$(\alpha_1,\alpha_2 )$, after this extend $\C_3 $ to a fix-free set $\C_3 $
which fits to $(\alpha_1 ,\alpha_2 ,\alpha_3 )$ etc. . Although this works fine
for the case in the theorem above, the next example shows, that this is
even not possible for a two level fix-free code.\\

\begin{example} \upshape
Let $\A:=\{ 0,1 \} $ and $\alpha_1=\alpha_2 =0 ,\alpha_3=\alpha_4 =4 $,
$\alpha_l=0 $ for $l>4 $. We obtain$\sum\limits_{l=0}^{\infty}\alpha_lq^{-l}
=\frac{3}{4} $. If we choose $C_1=\{ 001,101 ,110 ,111 \} $, then $\C_1 $ fits
to $(\alpha_1,\ldots ,\alpha_3 ) $.
The tabular below shows that
$|\A^4| -|\bfs[4] (\C_1 ) | = 3 $. Therefore it is not possible to extend
$\C_1 $ to a fix-free code which fits to $(\alpha_1,\ldots ,\alpha_4 )$.
\smallskip
\begin{center}
\begin{tabular}{|l|c|c||l|c|c||l|c|c|}\hline
Level 4 & P & S & Level 4 & P & S & Level 4 & P & S \\ \hline
{\bf 0000} & \quad & \quad & 0110 & \quad & x     & 1100 & x     & \quad \\
0001 & \quad & x     & 0111 & \quad & x     & 1101 & x     & x     \\
0010 & x     & \quad & {\bf 1000} & \quad & \quad & 1110 & x     & x     \\
0011 & x     & \quad & 1001 & \quad & x     & 1111 & x     & x     \\
{\bf 0100} & \quad & \quad & 1010 & x     & \quad & \quad & \quad &\quad \\
0101 & \quad & x     & 1011 & x   & \quad & \quad & \quad & \quad \\ \hline

\end{tabular}
\end{center}
\quad\\
\end{example}

\smallskip
The proof of the next theorem shows, how to choose the first level of a two level
fix-free code, if the Kraftsum of the code is smaller than or equal to
$\frac{3}{4} $. The theorem was shown by Harada and Kobayashi in
\cite{harada} and shows that the $\frac{3}{4} $-conjecture holds for two
level fix-free codes.\\

\pagebreak
\begin{Def} Let $q\ge 2 $ and $\A  = \{ 0,\ldots ,q-1 \}  $.
We define the map  \\
$\num :\A^+ \rightarrow \NN $ as:
\[ \num (x) = \sum\limits_{l=0}^{|x|-1}x_{|x|-l}q^{l}
\;\mbox{ for }\; x=x_1\ldots x_{|x|}\in\A^+ \, .\]
\end{Def}

\smallskip
In the following we identify a finite alphabet $\A $ with $\{ 0,\ldots ,q- \} $, if
$|\A |=q\ge 2 $. Obviously the function $\num_{\big|\A^l} $ is
a one-to-one map onto $\{ 0,1,\ldots ,q^l-1\} $ for every
$l\in\NN $.\\

\begin{prop} \label{num}
Let $|\A |=q\ge 2  $ and $x,y\in\A^+ $ with $|x|\le |y| $.\\
\begin{enumerate}
\item x is a suffix of y $\Leftrightarrow \num (y)\mod \, q^{|x|} =\num (x) $ \\
\item x is prefix of y $\Leftrightarrow \num (x)\cdot q^{|y|-|x| }\le \num (y) <
\big( \num (x) +1 \big)q^{|y|-|x|}$\, .
\end{enumerate}
\end{prop}

\quad\\
\smallskip
\begin{theorem}[Harada and Kobayashi]\label{qfall2}
Let $q\ge 2 ,\A  = \{ 0,\ldots , q-1 \}$, \\
$m<n$ and $\fo{\alpha }$ be a sequence of
of nonnegative integers with $\alpha_l =0 $ for all $l\not\in\{ m,n \} $.
If $\alpha_m q^{-m} +\alpha_n q^{-n} \le \frac{3}{4} $, then

\begin{enumerate}
\item there exists a fix-free Code $\C \subseteq\A^+ $ which fits to
$\fo{\alpha } $.
\item If we choose $C_1 =\{ x\in\A^m \, |\, 0\le\num (x) \le \alpha_m-1 \,
\}$, then there exists a fix-free extension of $\C_1 $ which fits to
$\fo{\alpha } $.
\end{enumerate}
\end{theorem}

\smallskip\noindent
{\bf Proof: } From $\frac{3}{4}\ge \alpha_m q^{-m} +\alpha_n q^{-n} $ follows that
$\alpha_m \le q^m =\big| \A^m \big| $. Thus we can define :

\begin{quote}
$\C_1 :=\{ x\in\A^m \, |\, 0\le\num (x) < \alpha_m \, \} \subseteq \A^m $.
\end{quote}

\noindent
If we take in account that $\num_{\big|\A^m} $ is a one-to-one map onto
$ \{0, \ldots , q^m-1 \} $, we obtain $\big| \C_1 \big| =\alpha_m $.
Let $ s:=\big\lfloor \frac{\alpha_m q^{n-m}}{q^m} \big\rfloor $.
We define:

\[\begin{array}{lcl}
T_{t} & := &\{ y\in\A^n \, |\, (t-1)q^m \le \num (y) < tq^m \,\}
\;\mbox{ for all }\; 1\le t \le s \;\mbox{ and}\\
T_{s+1} & := & \{ y\in\A^n \, |\, s\cdot q^m\le \num (y) <\alpha_m q^{n-m} \,\}\, .
\end{array}\]

\noindent
While $\alpha_mq^{n-m} \le \frac{3}{4} q^n \le \big| \A^n \big| $ and
since $\num_{\big|\A^n} $ is a one-to-one map onto \\
$ \{0, \ldots , q^n-1 \} $, it follows that:

\[\begin{array}{lcl}
\big| T_{t} \big| & = & q^m $ for all $1\le t\le s \;\mbox{ and } \\
\big| T_{s+1} \big| & = & \alpha_m q^{n-m}- q^m\big\lfloor \frac{\alpha_m q^{n-m}}{q^m} \big\rfloor \le q^m \, .
\end{array}\]

\noindent
By Proposition \ref{num} (ii) follows:

\[ \bigcup_{t=1}^{s+1} T_t = \{ y\in\A^n \, |\, 0\le \num (y) < \alpha_m \, \} = \pfs[n] {(\C_1 )}  \, .\]

\noindent
The $T_i $'s are pairwise disjoint.
Therefore $T_1, \ldots T_s ,T_{s+1} $ is a partition of $\pfs[n]{(\C_1 )} $\\
Because of Proposition \ref{num} (i)
we we obtain for every $\X\subseteq A^m $ and $1\le t \le q^{n-m}$:\\

\begin{eqnarray*}
& &\{ y\in\A^n \, |\, (t-1)q^m \le \num (y) < tq^m \,\} \cap \sfs[n]{(\X )} \\
& = &\{ y\in\A^n \, |\, x\in\X, \num (y)=(t-1)q^m + \num (x) \,\} \, .\\
\end{eqnarray*}

\noindent
>From the definition of $\C_1 ,T_1,\ldots ,T_{s+1} $ and
$ \big| T_{s+1} \big| < q^m $ follows:

\begin{displaymath}
T_{t} \cap \sfs[n]{(\C_1 )} =\{ y\in\A^n \, |\,(t-1)q^m\le \num (y)<(t-1)q^m +
\alpha_m  \,\}  \mbox{ for } 1\le t\le s\, .
\end{displaymath}

\begin{equation} \label{gl0}
T_{s+1} \cap \sfs[n]{(\C_1 )} = \left\{ \begin{array}{ll} \{ y\in\A^n \, |\,s\cdot
q^m\le \num (y)< s\cdot q^m + \alpha_m \,\}
& \mbox{ for } \big| T_{s+1} \big| \ge \alpha_m \\
T_{s+1} & \mbox{ for } \big| T_{s+1} \big| < \alpha_m
\end{array}\right.
\end{equation}

\noindent
Since the $ T_i $'s are a partition of $ \pfs[n]{(\C_1 )} $, we have \\

\[ \big| \pfs[n]{(\C_1 )} \cap \sfs[n]{(\C_1 )} \big|=\sum\limits_{i=1}^{s+1} \big| T_i \cap \sfs[n]{(\C_1 )} \big| \, .\]

\noindent
By (\ref{gl0}) follows:

\begin{equation} \label{gl1}
\begin{array}{rcl}
\small
\big| \pfs[n]{(\C_1 )} \cap \sfs[n]{(\C_1 )} \big| & = & s\cdot\alpha_m + \big| T_{s+1} \cap \sfs[n]{(\C_1 )} \big|\\
\quad & = &\big\lfloor \frac{\alpha_m q^{n-m}}{q^m} \big\rfloor\alpha_m + \left\{
\begin{array}{ll}
\alpha_m & \mbox{if } \big| T_{s+1} \big| \ge \alpha_m \\
\alpha_m q^{n-m}- q^m\big\lfloor \frac{\alpha_m q^{n-m}}{q^m} \big\rfloor & \mbox{if }
 \big| T_{s+1} \big| < \alpha_m
\end{array}\right.
\small
\end{array}
\end{equation}

\noindent
While $\frac{3}{4}\ge \alpha_m q^{-m} +\alpha_n q^{-n} $, from Lemma \ref{ext}
(ii) follows, that it is sufficient for the existence of a set
$\C_2\subseteq\A^n $ with $\big| \C_2 \big| =\alpha_n $ and
$\C:= \C_1\cup\C_2 $ is fix-free, to show
that the following inequality holds.

\begin{equation} \label{ungl1}
\big| \pfs[n] (\C_1 ) \cap \sfs[n] (\C_1 ) \big| \ge \frac{\big| \pfs[n] (\C_1 )
\big|^2 }{q^n}
\end{equation}

\noindent
We show the above inequality by distinguishing two cases.
This will finish the proof.\\

\smallskip\noindent
{\em Case 1:} $\big| T_{s+1}\big|\ge\alpha_m $\\

\smallskip\noindent
By equation (\ref{gl1} ) we have:\\

\begin{equation} \label{gl2}
\big| \pfs[n]{(\C_1 )} \cap \sfs[n]{(\C_1 )} \big|=\big\lfloor \frac{\alpha_m
q^{n-m}}{q^m} \big\rfloor\alpha_m +\alpha_m =\big\lfloor \frac{\alpha_m
q^{n-m}}{q^m} +1 \big\rfloor\alpha_m \ge\frac{\alpha_m^2q^{n-m}}{q^m} \, .
\end{equation}

\noindent
Since $\big| \pfs[n] (\C_1 ) \big| =\alpha_m q^{n-m} $, we have

\begin{equation} \label{gl3}
\frac{\big| \pfs[n] (\C_1 ) \big|^2 }{q^n}=\frac{\alpha_m^2q^{2n-2m}}{q^n} =
\frac{\alpha_m^2 q^{n-m}}{q^m}
\end{equation}

\noindent
By equations (\ref{gl2}) and (\ref{gl3}), it follows that the desired inequality
(\ref{ungl1}) holds.\\

\smallskip\noindent
{\em Case 2:} $\big| T_{s+1}\big| <\alpha_m $ \\
\smallskip\noindent
In this case we have:

\[
\begin{array}{clcl}
\quad & \frac{\alpha_mq^{n-m}}{q^m} & \ge &
\big\lfloor \frac{\alpha_m q^{n-m}}{q^m}\big\rfloor \\[2mm]

\begin{array}{c}\Leftrightarrow \\\mbox{\scriptsize $\alpha_m-q^m\le 0$}\end{array} &
(\alpha_m -q^m)\frac{\alpha_m q^{n-m}}{q^m} & \le &
(\alpha_m -q^m)\big\lfloor \frac{\alpha_m q^{n-m}}{q^m}\big\rfloor  \\[4mm]

\Leftrightarrow
& \frac{\alpha_m^2q^{n-m}}{q^m} & \le & \alpha_m\big\lfloor
\frac{\alpha_m q^{n-m}}{q^m}\big\rfloor + \alpha_m q^{n-m}
- q^m \big\lfloor \frac{\alpha_m q^{n-m}}{q^m}\big\rfloor \\[2mm]

\begin{array}{c}\Rightarrow \\ \mbox{\scriptsize (\ref{gl1}) } \end{array} &
\frac{\alpha_m^2q^{n-m}}{q^m} & \le &
\big| \pfs[n]{(\C_1 )} \cap \sfs[n]{(\C_1 )} \big| \\[2mm]

\begin{array}{c}\Rightarrow \\ \mbox{\scriptsize (\ref{gl3}) } \end{array} &
\frac{\big|\pfs[n] (\C_1 ) \big|^2 }{q^n} & \le &
\big| \pfs[n]{(\C_1 )} \cap \sfs[n]{(\C_1 )}\big|
\end{array} \, .
\]
Thus the inequality  (\ref{ungl1}) holds in this case, as well.\qed\\

\pagebreak
In \cite{zeger} Kukorelly and Zeger show the $\frac{3}{4}$-conjecture for binary
codes and finite sequences, if the number of codewords on each level which is
smaller than the maximal level is limited by $2^{l_{min} -2} $, where $l_{min}$
is the first nonempty level of the code.

\begin{theorem}[Kukorelly and Zeger]\label{qfall3}
Let $A =\{ 0,1 \} $, $\fo{\alpha } $ be a sequence of nonnegative integers
with $\sum\limits_{l=1}^{\infty }\alpha_l \left(\frac{1}{2}\right)^l \le\frac{3}{4}\, $,
$l_{min} :=\min \{ l  | \alpha_l > 0 \} \, $ and \\
$\, l_{max} :=\mbox{sup}\, \{ l \in\NN | \alpha_l > 0 \} \le \infty $.
If $\; l_{min} \ge 2 $, $l_{max} <\infty $ and $\alpha_l \le 2^{l_{min} -2} $ for all
$l\neq l_{max} $,
then there exists a fix-free code $\C\subseteq\{ 0,1 \}^* $ which fits to
$\fo{\alpha }$.
\end{theorem}

\smallskip
We prove a generalization of the theorem above for arbitrary finite alphabets.
This is one of the new results in this survey. However the proof of
the generalization is similar to the proof of the binary case given in
\cite{zeger}, if the binary alphabet $ \{ 0, 1 \} $ is replaced by
$\{ X, Y \} $, where $\X ,\Y $ is a partition of the alphabet $\A $
with $|\X | = \left\lfloor \frac{\A }{2 } \right\rfloor $ and
$|\Y | = \left\lceil \frac{\A }{2 } \right\rceil $.\\

\begin{theorem}\label{qfall4}
Let  $|\A | = q \ge 2$, $\fo[l]{\alpha } $
be a sequence of nonnegative integers with
$\sum\limits_{l=l_{min } }^{l_{max}} \alpha_l q^{-l} \le \frac{3}{4} $
and $\;l_{min} :=\min \{ l\, |\alpha_l \ge 0 \} \,$,\\
$l_{max}:=\mbox{sup}\,\{ l\, |\alpha_l\ge \} \le \infty $.
If $l_{min}\ge 2 \,$, $l_{max}<\infty $ and
$\alpha_l \le  q^{l_{min}-2} \big\lfloor \frac{q}{2} \big\rfloor^2 \big\lceil
\frac{q}{2} \big\rceil^{l-l_{min}} $ for all $l \neq l_{max} $, then there exists a
fix-free Code $\C\subseteq \A^* $ which fits to $\fo[l]{\alpha } $.
\end{theorem}

\noindent
Theorem~\ref{qfall3} follows from Theorem~\ref{qfall4} for $q=2 $.

\smallskip\noindent
{\bf Proof:} If $\; l_{max} \le l_{min} +1 $ then we need only a  one level or
two level code. In this case the theorem follows from Theorem~\ref{qfall2}.
Thus we can assume that
$l_{max} \ge l_{min} +2 $. In the proof we first
construct a fix-free Code $\C_0 $ such that\\
$|\C_0 \cap \A^l | = q^{l_{min}-2}
\big\lfloor \frac{q}{2} \big\rfloor^2 \big\lceil \frac{q}{2} \big\rceil^{l-l_{min}}
$ for all $l_{min} \le l < l_{max} $ and $\sum\limits_{l=l_{min } }^{l_{max}}|\C_0
\cap \A^l | q^{-l} = \frac{3}{4} $.

\noindent
Then in four steps we delete $q^{l_{min}-2}
\big\lfloor \frac{q}{2} \big\rfloor^2 \big\lceil \frac{q}{2} \big\rceil^{l-l_{min}}
-\alpha_l $ codewords from each level $l_{min} \le l < l_{max} $, replace each of
this codeword with more than $q^{l_{max} -l} $ new codewords on the $l_{max} $-th
level, and show with Lemma~\ref{help2}, that this new code $\C $ is also fix-free. Then
the Kraftsum of this new Code is bigger or equal  $\frac{3}{4} $, \\
$|\C \cap \A^l | =\alpha_l \;\forall\, l_{min} \le l < l_{max} $ and
$|\C \cap \A^{l_{max}} | \ge \alpha_{l_{max}} $.
To obtain the desired Code we have only to delete some codewords on
the $l_{max}$-th level.\\

\pagebreak
Let $\X , \Y  $ be a partition of the alphabet $\A $ into two parts with
$|\X |= \big\lfloor \frac{q}{2} \big\rfloor $ and $|\Y |=
\big\lceil \frac{q}{2} \big\rceil $. We define:

\begin{eqnarray*}
\B_0 & := & \{ x_1 y x_2 | x_1,x_2 \in\X ,\, y\in \Y^i ,\, 0\le i \le l_{max} -l_{min} -1 \} \\
\D_1 & := & \Y\A^{l_{max}-l_{min}}\Y\A^{l_{min}-2} \subseteq \A^{l_{max}} \\
\D_2 & := & \X\Y^{l_{max}-l_{min}}\A^{l_{min}-1} \subseteq \A^{l_{max}} \\
\B & := & \B_0\A^{l_{min}-2} \subseteq \bigcup\limits_{l=l_{min}}^{l_{max}-1} \A^l \\
\C_0 & := & \B \cup D_1 \cup D_2
\end{eqnarray*}
$\B $ is fix-free, because $\B_0 $ is fix-free. Obviously no codeword in $\B $
is a prefix or suffix from a word in $\D_1 \cup \D_2 $.
Thus $\C_0 $ is fix-free, as well.\\

\noindent
We have:

\begin{eqnarray}
|\C_0 \cap A^l|= |\B \cap A^l| & = & q^{l_{min}-2}\bigg\lfloor \frac{q}{2}
\bigg\rfloor^2 \bigg\lceil \frac{q}{2} \bigg\rceil^{l-l_{min}}
\quad\;\mbox{ for } l_{min} \le l < l_{max} ,\\
| \D_1 | & = & q^{l_{max} -2} \bigg\lceil \frac{q}{2} \bigg\rceil^2 , \nonumber\\
| \D_2 | & = & \bigg\lfloor \frac{q}{2} \bigg\rfloor \bigg\lceil \frac{q}{2}
\bigg\rceil^{l_{max}-l_{min}} q^{l_{min}-1} . \nonumber
\end{eqnarray}

\noindent
It follows:
\begin{eqnarray*}
S(\D_1 ) & = &|\D_1 | q^{-l_{max}}  =   \bigg( \frac{1}{q}\bigg)^2 \bigg\lceil \frac{q}{2} \bigg\rceil^2 ,\\
S(\D_2 ) & = & |\D_2 | q^{-l_{max}} =  \bigg\lfloor \frac{q}{2} \bigg\rfloor
\bigg\lceil \frac{q}{2} \bigg\rceil^{l_{max}-l_{min}} q^{l_{min}-l_{max}-1} =
\frac{1}{q} \bigg\lfloor \frac{q}{2} \bigg\rfloor
\left( \frac{1}{q} \bigg\lceil \frac{q}{2} \bigg\rceil\right)^{l_{max} -l_{min}}  ,\\
S(\B ) & = & \sum\limits_{l=l_{min}}^{l_{max}-1}|\B \cap A^l|q^{-l}  = q^{l_{min}-2
} \bigg\lfloor \frac{q}{2} \bigg\rfloor^2 \sum\limits_{l=l_{min}}^{l_{max}-1}
\bigg\lceil \frac{q}{2} \bigg\rceil^{l-l_{min}} \bigg( \frac{1}{q} \bigg)^l ,\\
 & = & \left( \frac{1}{q} \bigg\lfloor \frac{q}{2} \bigg\rfloor\right)^2\cdot\,
 \sum\limits_{l=0}^{\scriptsize l_{max}-l_{min}-1}
\left(\bigg\lceil \frac{q}{2} \bigg\rceil  \frac{1}{q} \right)^l ,\\
 & = &\left( \frac{1}{q} \bigg\lfloor \frac{q}{2} \bigg\rfloor\right)^2\cdot\,
 \frac{1-\left(\big\lceil\frac{q}{2}\big\rceil \frac{1}{q}\right)^{l_{max}-l_{min}}  }{1-\big\lceil\frac{q}{2}\big\rceil\frac{1}{q}}. \\
\end{eqnarray*}

\pagebreak\noindent
We obtain from the last equation:

\[S(\B ) = \frac{1}{q} \bigg\lfloor \frac{q}{2} \bigg\rfloor -  \frac{1}{q} \bigg\lfloor \frac{q}{2} \bigg\rfloor
\left( \bigg\lceil \frac{q}{2} \bigg\rceil  \frac{1}{q}\right)^{l_{max}-l_{min}}
=\frac{1}{q} \bigg\lfloor \frac{q}{2} \bigg\rfloor -S (\D_2 ) \, .\]

\noindent
The sets $\B ,\D_1 \mbox{ and } \D_2 $ are disjoint and so
together with the equations above follows:

\begin{equation} \label{ksumgl1}
S(\C_0 ) = S(\B ) + S(\D_1 ) +S(\D_2 ) =\frac{1}{q} \bigg\lfloor \frac{q}{2}
\bigg\rfloor + \bigg( \frac{1}{q}\bigg)^2 \bigg\lceil \frac{q}{2} \bigg\rceil^2 \, .
\end{equation}

\noindent
We claim
\[ \frac{1}{q} \bigg\lfloor \frac{q}{2} \bigg\rfloor +
\bigg( \frac{1}{q}\bigg)^2 \bigg\lceil \frac{q}{2} \bigg\rceil^2 \ge \frac{3}{4} \, .\]

\noindent
If $q$ is even, we have $\frac{1}{q} \big\lfloor \frac{q}{2} \big\rfloor +
\big( \frac{1}{q}\big)^2 \big\lceil \frac{q}{2} \big\rceil^2 = \frac{1}{2} +
\big(\frac{1}{2}\big)^2 =\frac{3}{4} $.\\

\noindent
If $q$ is odd, we have $ q=2p+1 $ for some $p\in\NN $ and we obtain:
\begin{eqnarray*}
\quad & \quad & \frac{1}{q} \bigg\lfloor \frac{q}{2} \bigg\rfloor+
\bigg( \frac{1}{q}\bigg)^2 \bigg\lceil \frac{q}{2} \bigg\rceil^2\ge \frac{3}{4}\\
 & \Leftrightarrow & \frac{p}{2p+1} +\left( \frac{p+1}{2p+1} \right)^2\ge \frac{3}{4} \\
 & \Leftrightarrow & (2p+1)p+ (p+1)^2\ge \frac{3}{4}(2p+1)^2\\
 & \Leftrightarrow & 3p^2+3p+1 \ge 3p^2 +3p +\frac{3}{4} \, .
\end{eqnarray*}
This holds for all $p\in\NN $. With (\ref{ksumgl1}) follows:

\begin{equation}
S(\C_0 ) =\sum\limits_{l=l_{min}}^{l_{max}} |\C_0 \cap \A^l | q^{-l} \ge \frac{3}{4} \,.
\end{equation}

\smallskip
Let  $\E \subseteq \B  $ be a set which contains $q^{l_{min}-2}\bigg\lfloor
\frac{q}{2} \bigg\rfloor^2 \bigg\lceil \frac{q}{2} \bigg\rceil^{l-l_{min}} -\alpha_l
$ codewords of length $l$ for each $l_{min}\le l < l_{max} $.
Furthermore let $F \subseteq
\A^{l_{max}}-( \D_1 \cup \D_2 )$ be an arbitrary set of  at least
\[ \sum\limits_{l=l_{min}}^{l_{max}-1} \bigg(q^{l_{min}-2}\bigg\lfloor \frac{q}{2} \bigg\rfloor^2 \bigg\lceil \frac{q}{2} \bigg\rceil^{l-l_{min}} -\alpha_l  \bigg)q^{l_{max}-l} \]
codewords. If we remove the words of $\E $ from $\C_0 $ and add the words of $\F
$ we obtain the set:
\[ \C := (\B -\E ) \cup (\D_1\stackrel{.}{\cup }\D_2 \stackrel{.}{\cup } \F ) =(\C_0 -\E)\stackrel{.}{\cup } \F \, .\]

\pagebreak
\noindent
For this set we have:

\[ \left| \C \cap A^l \right| = \alpha_l \mbox{ for } l_{min} \le l < l_{max} \, .\]

\begin{eqnarray*}
 S(\C ) & = & S(\C_0 ) - S(\E ) +S(\F ) \\
 & \ge &S( \C_0) -
\sum\limits_{l=l_{min}}^{l_{max}-1 }q^{-l}\bigg( q^{l_{min}-2}\bigg\lfloor \frac{q}{2} \bigg\rfloor^2 \bigg\lceil \frac{q}{2} \bigg\rceil^{l-l_{min}} -\alpha_l \bigg) \\
 & \quad & +q^{-l_{max}} \cdot \sum\limits_{l=l_{min}}^{l_{max}-1} \bigg(q^{l_{min}-2}\bigg\lfloor \frac{q}{2} \bigg\rfloor^2 \bigg\lceil \frac{q}{2} \bigg\rceil^{l-l_{min}} -\alpha_l  \bigg)q^{l_{max}-l} \\
 & = &  S(\C_0 ) \ge \frac{3}{4}\, .
\end{eqnarray*}

\noindent
We obtain a
set which fits to $(\alpha_l)_{l\in\NN } $, by deleting some words of length $ l_{max}
$ from $\C $, because $\sum\limits_{l=0}^{\infty } \alpha_l q^{-l}
=\sum\limits_{l=l_{min}}^{l_{max}-1} \alpha_l q^{-l} \le \frac{3}{4} $\\

\smallskip
To complete the proof we have to show, that we can choose $\E $
and $\F $ such that $\C $ is fix-free. To show this, we use Lemma~\ref{help2}, which
says, that $\C $ is fix-free if the following two conditions holds.
\begin{enumerate}
\renewcommand{\labelenumi}{(\roman{enumi})}
\item Each word in $\F $ has a prefix in $\E $ or no prefix in $\C_0 $.
\item Each word in $\F $ has a suffix in $\E $ or no suffix in $\C_0 $.
\end{enumerate}

\smallskip
We construct the sets $\F $ and $\E $ in three steps:

\begin{enumerate}
\renewcommand{\labelenumi}{\arabic{enumi}.}
\renewcommand{\labelenumii}{\alph{enumii})}

\item For each $ l_{min} \le l \le l_{max} - l_{min} +1 $ we include in
$\E_1\subseteq \B  $ all\\
$q^{l_{min}-2}\bigg\lfloor \frac{q}{2} \bigg\rfloor^2
\bigg\lceil \frac{q}{2} \bigg\rceil^{l-l_{min}} -\alpha_l $ words of the form:
\[ x_1yx_2w \, ,\;\mbox{ where }\; x_1,x_2\in\X ,\, y\in \Y^{l-l_{min} } ,\, w\in\A^{l_{min}-2}.\]
For each of these words, we include in $\F_1 \subseteq \A^{l_{max} } - (\D_1 \cup
\D_2 ) $:
\begin{enumerate}
\item the $\lceil \frac{q}{2} \rceil q^{l_{max}-l-1} $ words of the sets:

\begin{equation}
 x_1yx_2w\A^{l_{max}-l_{min}-l+1}\Y\A^{l_{min}-2}\subseteq \A^{l_{max}}
\end{equation}

Each of these words has a prefix in $\E_1 $, but they have no suffix in $\B $,
because for every word in $\E_1 $ the $(l_{min}-1) $-th letter
from the left-hand side is an element of $\Y $.
Furthermore each word of $\F_1 $ has a prefix of $x_1yx_2\in\C_0 $
Thus $\F_1 $ is disjoint from $\D_1\cup\D_2 $, since $\C $ is prefix-free.

\item choose  $\lfloor \frac{q}{2} \rfloor q^{l_{max}-l-1} $ arbitrary words of
the sets:

\begin{equation}
\Y\A^{l_{max}-l-1}x_1yx_2w \subseteq \A^{l_{max}}
\end{equation}
Each of these words have a suffix in $\E_1 $, but they have no prefix in $\B $, because
they begin with a letter in $\Y $.
Since $\C_0 $ is suffix-free, none of these words are in $\D_1\cup \D_2 $
and of course also disjoint from the other part of $\F_1 $.

\end{enumerate}
Thus for the sets $\E_1, \F_1 $ the above conditions of Lemma~\ref{help2}
holds and we obtain:
\begin{equation} \label{F1}
|\F_1 | = \sum\limits_{l=l_{min}}^{l_{max}-l_{min}+1}q^{l_{max}-l}\bigg(
q^{l_{min}-2}\bigg\lfloor \frac{q}{2} \bigg\rfloor^2 \bigg\lceil \frac{q}{2}
\bigg\rceil^{l-l_{min}} -\alpha_l\bigg)\, .
\end{equation}

\item For each $l_{max} -l_{min}+2 \le l < l_{max} $ and $\alpha_l \ge
q^{l_{min}-3}\bigg\lfloor \frac{q}{2} \bigg\rfloor^3 \bigg\lceil \frac{q}{2}
\bigg\rceil^{l-l_{min}} $ we include in $\E_2 \subseteq \B $ any
$q^{l_{min}-2}\bigg\lfloor \frac{q}{2} \bigg\rfloor^2 \bigg\lceil \frac{q}{2}
\bigg\rceil^{l-l_{min}} -\alpha_l $ words of the form:

\begin{equation}
\begin{array}{l}
x_1yx_2w \in\B \mbox{ with } x_1,x_2\in\X ,\, y\in\Y^{l-l_{min}} \;\mbox{ and } \\
w\in\A^{l_{max}-l-1}\Y\A^{l-(l_{max} -l_{min}+2) }\subseteq \A^{l_{min}-2 }
\end{array}
\end{equation}
The letters at the $(l_{max} -l_{min} +2) $-th
position of these words of $\B $ are in $\Y $ .
For each possible $l$ there are $q^{l_{min}-3}\bigg\lfloor
\frac{q}{2} \bigg\rfloor^2 \bigg\lceil \frac{q}{2}\bigg\rceil^{l-l_{min}+1} $ such
words and from the condition for $\alpha_l $ follows:

\begin{eqnarray*}
q^{l_{min}-2}\bigg\lfloor \frac{q}{2} \bigg\rfloor^2 \bigg\lceil \frac{q}{2} \bigg\rceil^{l-l_{min}} -\alpha_l & \le & q^{l_{min}-2}\bigg\lfloor \frac{q}{2} \bigg\rfloor^2 \bigg\lceil \frac{q}{2}\bigg\rceil^{l-l_{min}}-q^{l_{min}-3}\bigg\lfloor \frac{q}{2} \bigg\rfloor^3 \bigg\lceil \frac{q}{2} \bigg\rceil^{l-l_{min}}\\
 & = & q^{l_{min}-3}\bigg\lfloor \frac{q}{2} \bigg\rfloor^2 \bigg\lceil \frac{q}{2} \bigg\rceil^{l-l_{min}}\cdot\bigg(q- \bigg\lfloor \frac{q}{2}\bigg\rfloor\bigg) \\
 & = & q^{l_{min}-3}\bigg\lfloor \frac{q}{2} \bigg\rfloor^2 \bigg\lceil
 \frac{q}{2}\bigg\rceil^{l-l_{min}+1}
\end{eqnarray*}

Therefore we can include enough words in $\E_2 $.\\
For each of this words we include in $\F_2 \subseteq \A^{l_{max}} - (\D_1 \cup
\D_2\cup \F_1 ) $:
\begin{enumerate}
\item the $\bigg\lceil\frac{q}{2} \bigg\rceil q^{l_{max} -l -1} $ words of the
set:

\begin{equation}
\Y\A^{l_{max}-l-1}x_1yx_2w \subseteq \A^{l_{max}}
\end{equation}
These words have a suffix in $\E_2 $,but they have no prefix in $\B $,
because they begin with a letter in $\Y $.
Moreover they are neither contained in $\F_1 $ nor they are contained in
$\D_1 \cup \D_2 $, because $\C_0 $ is suffix-free.

\item choose any $\bigg\lfloor\frac{q}{2} \bigg\rfloor q^{l_{max} -l -1} $ from
the set:
\begin{equation}
x_1yx_2w\A^{l_{max}-l} \subseteq \A^{l_{max} }
\end{equation}
These words have a prefix in $\E_2 $, but they have no suffix in $\B $, because they
have a letter at the $(l_{max}-l_{min}+2) $-th position which is an element of $\Y $
and therefore ends with with a word in $\Y\A^{l_{min}-2} $,
whereas all codewords in $\B $ ends with a word in $\X\A^{l_{min}-2} $.
Furthermore they are
not contained in $\D_1 \cup \D_2 $, because $\C_0 $ is prefix-free and
obviously they are also not contained in $\F_1 $.

\end{enumerate}

Thus for the sets $\E_2 $ and $ \F_2 $ the conditions of the lemma holds. For every
possible $l$ the number of codewords in $\F_2 $ is:

\begin{eqnarray*}
 & \quad & \bigg(q^{l_{min}-2} \bigg\lfloor \frac{q}{2} \bigg\rfloor^2 \bigg\lceil \frac{q}{2} \bigg\rceil^{l-l_{min}} -\alpha_l \bigg) \cdot \bigg( \bigg\lceil\frac{q}{2} \bigg\rceil q^{l_{max} -l -1} +\bigg\lfloor\frac{q}{2} \bigg\rfloor q^{l_{max} -l -1} \bigg) \\
 & = &\bigg(q^{l_{min}-2} \bigg\lfloor \frac{q}{2} \bigg\rfloor^2 \bigg\lceil
 \frac{q}{2} \bigg\rceil^{l-l_{min}} -\alpha_l \bigg) \cdot q^{l_{max}-l}\, .
\end{eqnarray*}
>From this follows with $\beta_l:=q^{l_{min}-3}\bigg\lfloor \frac{q}{2}
\bigg\rfloor^3 \bigg\lceil \frac{q}{2} \bigg\rceil^{l-l_{min}} $:

\begin{equation} \label{F2}
|\F_2 | = \sum\limits_{\stackrel{ l=l_{max}-l_{min}+2 }{\alpha_l \ge \beta_l
}}^{l_{max}-1}\bigg( q^{l_{min}-2}\bigg\lfloor \frac{q}{2} \bigg\rfloor^2
\bigg\lceil \frac{q}{2} \bigg\rceil^{l-l_{min}} -\alpha_l \bigg)\cdot q^{l_{max}-l }\, .
\end{equation}

\item For each $l_{max} -l_{min}+2 \le l < l_{max} $ and $\alpha_l <
q^{l_{min}-3}\bigg\lfloor \frac{q}{2} \bigg\rfloor^3 \bigg\lceil \frac{q}{2}
\bigg\rceil^{l-l_{min}} $ we include in $\E_3 \subseteq \B \quad$
$q^{l_{min}-3}\bigg\lfloor \frac{q}{2} \bigg\rfloor^2 \bigg\lceil \frac{q}{2}
\bigg\rceil^{l-l_{min}+1} $ codewords of the form:

\begin{equation} \label{form1}
\begin{array}{l}
x_1yx_2w \in\B \mbox{ with } x_1,x_2\in\X ,\, y\in\Y^{l-l_{min}} \;\mbox{ and }\\
w\in\A^{l_{max}-l-1}\Y\A^{l-(l_{max} -l_{min}+2) }\subseteq \A^{l_{min}-2 }
\end{array}
\end{equation}
All these words are contained in $\B\cap\A^l $ and therefore
the letter at the
$(l_{max}-l_{min}+2) $-th position is in $\Y $\\

Furthermore we include in $\E_3 $
any $q^{l_{min}-3}\bigg\lfloor \frac{q}{2} \bigg\rfloor^3 \bigg\lceil \frac{q}{2}
\bigg\rceil^{l-l_{min}} -\alpha_l $ words of the form:

\begin{equation} \label{form2}
\begin{array}{l}
x_1yx_2w \in\B \mbox{ with } x_1,x_2\in\X ,\, y\in\Y^{l-l_{min}} \;\mbox{ and }\\
w\in\A^{l_{max}-l-1}\X\A^{l-(l_{max} -l_{min}+2) }\subseteq \A^{l_{min}-2 }
\end{array}
\end{equation}

Each of these words have at the $(l_{max}-l_{min}+2)$-th position a letter in $\X $.

For each possible $l$ the number of codewords in $\E_3 $ of length l is:

\begin{eqnarray*}
 & \quad  &\bigg( q^{l_{min}-3}\bigg\lfloor \frac{q}{2} \bigg\rfloor^3 \bigg\lceil \frac{q}{2} \bigg\rceil^{l-l_{min}} -\alpha_l\bigg) +q^{l_{min}-3}\bigg\lfloor \frac{q}{2} \bigg\rfloor^2 \bigg\lceil \frac{q}{2} \bigg\rceil^{l-l_{min}+1} \\
 & = &q^{l_{min}-3}\bigg\lfloor \frac{q}{2} \bigg\rfloor^2 \bigg\lceil \frac{q}{2} \bigg\rceil^{l-l_{min}}\bigg(\bigg\lfloor \frac{q}{2} \bigg\rfloor + \bigg\lceil \frac{q}{2} \bigg\rceil \bigg)-\alpha_l \\
  & = & q^{l_{min}-2}\bigg\lfloor \frac{q}{2} \bigg\rfloor^2 \bigg\lceil \frac{q}{2}
  \bigg\rceil^{l-l_{min}}-\alpha_l
\end{eqnarray*}

codewords in $\E_3 $.\\

For each word in $\E_3 $ of the form (\ref{form1}) or (\ref{form2})  we include in
$\F_3 $:

\begin{itemize}

\item[a)] the $\bigg\lceil\frac{q}{2} \bigg\rceil q^{l_{max} -l -1} $ words of the
set:

\begin{equation}
\Y\A^{l_{max}-l-1}x_1yx_2w \subseteq \A^{l_{max}}
\end{equation}

These words have a suffix in $\E_3 $, but do not have a prefix in $\B $,
because the first letter is an element of $\Y $.
Obviously they are not contained in $\F_1 \cup \F_2 $ and
they are also not contained in $\D_1 \cup \D_2 $, because $\C_0 $ is suffix-free.

\end{itemize}

For every word in $\E_3 $ of the form (\ref{form1}) we include in $\F_3 $:
\begin{itemize}
\item[b)] the $q^{l_{max}-l} $ words of the set:
\begin{equation}
x_1yx_2w\A^{l_{max}-l} \subseteq \A^{l_{max} }
\end{equation}
These words have a prefix in $\E_3 $, but do not have a suffix in $\B $, because they
have a letter at the $(l_{max}-l_{min}+2)$-th position which is in $\Y $ and
therefore they have a suffix in $ \Y\A^{l_{min}-2} $ whereas
all codewords in $\B $ have a suffix in $\X\A^{l_{min}-2} $.
They are not in $\D_1 \cup \D_2 $, because $\C_0 $ is prefix-free
and obviously they are also not contained in $\F_1\cup\F_2 $.
\end{itemize}

Therefore $\E_3 ,\F_3 $ fulfill the condition of the lemma and\\
$\F_3 \subseteq \A^{l_{max}} -(\D_1\cup\D_2\cup\F_1\cup \F_2 ) $.\\

\pagebreak
For every possible $l$ the number of codewords of length $l$ in $\F_3 $ is equal to

\begin{eqnarray*}
 & \quad & q^{l_{min}-3}\bigg\lfloor \frac{q}{2} \bigg\rfloor^2 \bigg\lceil \frac{q}{2} \bigg\rceil^{l-l_{min}+1}\cdot\bigg(\bigg\lceil\frac{q}{2} \bigg\rceil q^{l_{max} -l -1} + q^{l_{max}-l} \bigg)\\
 & \quad & +\bigg(q^{l_{min}-3}\bigg\lfloor \frac{q}{2} \bigg\rfloor^3 \bigg\lceil \frac{q}{2} \bigg\rceil^{l-l_{min}} -\alpha_l\bigg)\cdot\bigg\lceil\frac{q}{2} \bigg\rceil q^{l_{max} -l -1}\\
 & = & q^{l_{max}-l}\cdot\bigg( q^{l_{min}-2}\bigg\lfloor \frac{q}{2} \bigg\rfloor^2 \bigg\lceil \frac{q}{2} \bigg\rceil^{l-l_{min}}\cdot\bigg( \frac{1}{q^2}\bigg\lceil\frac{q}{2} \bigg\rceil^2 + \frac{1}{q}\bigg\lceil\frac{q}{2} \bigg\rceil + \frac{1}{q^2}\bigg\lceil\frac{q}{2} \bigg\rceil \bigg\lfloor \frac{q}{2} \bigg\rfloor \bigg) -\alpha_l\frac{1}{q}\bigg\lceil\frac{q}{2} \bigg\rceil  \bigg)\\
 & = & q^{l_{max}-l}\cdot\bigg( q^{l_{min}-2}\bigg\lfloor \frac{q}{2} \bigg\rfloor^2 \bigg\lceil \frac{q}{2} \bigg\rceil^{l-l_{min}}\cdot\bigg(\frac{2}{q}\bigg\lceil\frac{q}{2} \bigg\rceil\bigg) -\alpha_l\frac{1}{q}\bigg\lceil\frac{q}{2} \bigg\rceil  \bigg) \\
 & \ge & q^{l_{max}-l}\cdot\bigg( q^{l_{min}-2}\bigg\lfloor \frac{q}{2}
 \bigg\rfloor^2 \bigg\lceil \frac{q}{2} \bigg\rceil^{l-l_{min}} -\alpha_l \bigg) \, .
\end{eqnarray*}

\noindent
With $\beta_l:=q^{l_{min}-3}\bigg\lfloor \frac{q}{2}
\bigg\rfloor^3 \bigg\lceil \frac{q}{2} \bigg\rceil^{l-l_{min}} $ follows:

\begin{equation} \label{F3}
|\F_3 | \ge \sum\limits_{\stackrel{ l=l_{max}-l_{min}+2 }{\alpha_l <\beta_l
}}^{l_{max}-1}\bigg( q^{l_{min}-2}\bigg\lfloor \frac{q}{2} \bigg\rfloor^2
\bigg\lceil \frac{q}{2} \bigg\rceil^{l-l_{min}} -\alpha_l \bigg)\cdot q^{l_{max}-l} \, .
\end{equation}

\end{enumerate}

\noindent
Let $\E := \E_1\cup\E_2\cup\E_3 \subseteq \B $ and $\F := \F_1\cup\F_2\cup\F_3
\subseteq \A^{l_{max}} - (\D_1\cup\D_2)$ then from Lemma~\ref{help3} it follows that $\C
:= (\C_0 -\E ) \cup \F $ is fix-free. Moreover we have $\big| \C \cap \A^l \big| =
\big| (\B \cap \A^l) \big| - \big| (\E \cap \A^l) \big|=\alpha_l $ for all $l_{min}
\le l < l_{max} $. Since $\F_1 , \F_2  ,\F_3  $ are disjoint, by (\ref{F1}),
(\ref{F2}) and (\ref{F3}) follows:
\[
 |\F |  =  |\F_1|+|\F_2 | +|\F_3 | \ge  \sum\limits_{l=l_{min}}^{l_{max}-1} \bigg(
 q^{l_{min}-2}\bigg\lfloor \frac{q}{2} \bigg\rfloor^2 \bigg\lceil \frac{q}{2}
 \bigg\rceil^{l-l_{min}} -\alpha_l \bigg)\cdot q^{l_{max}-l}
\]
As described above, we obtain  $ S(\C ) \ge S(\C_0 ) \ge \frac{3}{4} $ and
because of $\sum\limits_{l=l_{min}}^{l_{max} } \alpha_l \cdot q^{-l} \le \frac{3}{4}
$ we obtain a fix-free code which fits $(\alpha_l )_{l\in\NN } $ by deleting some
codewords of length $l_{max} $ from $\C $. \qed

\newpage

\chapter{The de Bruijn digraphs \BGr[q]{n}}

In this chapter we examine the existence of certain cycles and regular
subgraphs of the de Bruijn digraphs.  In Chapter 4 we will generate
fix-free codes with $k\cdot L $ codewords on
the first nonempty level with the help of
$k$-regular subgraphs with $L$ vertices of de Bruijn digraps. Therefore
it is important to know for which numbers $L$ of vertices such subgraphs exist.
The de Bruijn digraph of span n over an alphabet $\A $
contains all $\A $-words of length $n$ as its vertices and for every word $w\in\A^n $
the successors of $w$ are given by the words which are contained in the set $\A^{-1}w\A $.
de Bruijn digraphs were first constructed by de Bruijn \cite{bruijn}(1946) and
independently by Good \cite{good}(1946), while examining
the existence of binary cyclic sequences of length $2^n $ containing $2^n $
different subwords of length $n$. Such sequences are called
a (binary) de Bruijn sequence and they correspond with Hamilton circuits
of the de Bruijn digraph of span $n$.
One might ask, wether such sequences exist and how much of them exist
for certain values of $n\in\NN $? We will come back to this problem in Section 3 of
this chapter. de Bruijn digraphs have a lot of applications. For example,
they are used for computer network building (see for example \cite{cayl}).
However, in this chapter we focus on the question, wether there exists
a $k$-regular subgraph in the de Bruijn graph of span $n$ for a given
number of vertices.
We begin with an Introduction of digraphs. Then we give
an overview of some basic facts of de Bruijn digraphs. In the third
section of this chapter we show, that there are cycles of arbitrary length
in the $q$-ary de Bruijn digraph. This result was obtained independently by
Yoeli, Bryant, Heath, Killik, Golomb, Welch and Goldstein for the binary case
(\cite{lempel} and \cite{gol1}). Lempel generalized this result
to $q$-ary de Bruijn graphs in \cite{lempel}. Since cycles are connected one regular digraphs,
this shows that there are $1$-regular subgraphs of the de Bruijn digraph of span
$n$, for every given number of vertices. Finally the last section of this
chapter is dealing with the study of
$k$-regular subgraphs of the de Bruijn digraph, i.e.
we obtain that there do not exist $k$-regular subgraphs for any number of vertices.

\pagebreak

\section{Introduction of digraphs}

\subsection*{Some basics about digraphs}

\subsubsection{Definition of digraphs}

Let $\V ,I $ be arbitrary sets and $\E\subseteq \V\times\V\times I $. We call
$\Gamma := (\V,\E )$ the {\em digraph} with vertices in $\V $ and
edges in $\E $. An element $v\in\V $ is called a {\em vertex} of $\Gamma $
and an element $e=(v_1,v_2,i)\in\E $ is called an {\em (directed) edge}
in $\Gamma $, where $e$ runs from the vertex $v_1 $ to the vertex $v_2 $.
For this we write $v_1\stackrel{e}{\rightarrow } v_2 $ or
$v_1\stackrel{i}{\rightarrow } v_2 $. A {\em loop} in $\Gamma $ is
an edge $v\stackrel{e}{\rightarrow } v $ for which the terminal vertex
is equal to the initial vertex.
We call an edge $e$ in $\Gamma $,
{\em incident to} a vertex $v\in\V $, if
$u\stackrel{e}{\rightarrow } v $ for some $u\in\V $. We call $e$
{\em incident from} $v$, if
$v\stackrel{e}{\rightarrow } u $ for some $u\in\V $ and $e$ is
{\em incident at} $v$, if $e$ is incident to $v$ or incident from $v$.
A vertex $v\in\V $ is called an {\em isolated vertex} in $\Gamma $, if
there does not exist an edge which is incident at $v$.
If $|\V | ,| I| < \infty $, then $\Gamma $ is called a {\em finite digraph}.
If $|I|=1 $, then $\Gamma $ is called a digraph without multiple edges.
In this case, we suppose
that $\E\subseteq\V\times\V $ and we write $v_1\rightarrow v_2 $
for the edge $e=(v_1,v_2)\in\E $.
All graphs which occurs in this survey
are digraphs, therefore we use digraph, directed graph and graph simultaneously.

\subsubsection{Subgraphs}
We call a graph $\tilde{\Gamma }=(\tilde{\V } ,\tilde{\E } ) $ a
{\em subgraph } of a graph $\Gamma =(\V ,\E ) $, if $\tilde{\V } \subseteq \V $ and
$\tilde{\E } \subseteq \E $,for this we write $\tilde{\Gamma }
\subseteq \Gamma $. $\tilde{\Gamma }$ is called a {\em spanning
subgraph} of $\Gamma $, if $\tilde{\Gamma } $ has the same vertex
set as $\Gamma $.\\

Let $\Gamma_1:=(\V_1 ,\E_1 ) \, $,
$\Gamma_2:=(\V_2 ,\E_2 ) \, $ be graphs and  $\Lambda :=(\V_3,\E_3 )$
be a subgraph of $\Gamma_1 $. The union and intersection of $\Gamma_1 $
with $\Gamma_2 $ and the complement of $\Lambda $ in $\Gamma $ is
defined as the graphs:
\begin{eqnarray*}
\Gamma_1\cup\Gamma_2 & := & \big( \V_1\cup\V_2 ,\E_1\cup\E_2 \big) \, ,\\
\Gamma_1\cap\Gamma_2 & := & \big( \V_1\cap\V_2 ,\E_1\cap\E_2 \big) \, ,\\
\Lambda^c & := & \big( \V , \E_1 -\E_2 \big) \quad \mbox{ whereas
} \Lambda^c \subseteq \Gamma_1 \, .
\end{eqnarray*}

\pagebreak
\subsubsection{Graph isomorphism}
Let $\Gamma =(\V_1,\E_1),\Lambda =(\V_2,\E_2) $ be two graphs,
where $\E_1\subseteq\V_1\times\V_1\times I_1 $ and
$\E_2\subseteq\V_2\times\V_2\times I_2 $.
We call $\Gamma $ and $\Lambda $ {\em isomorph} graphs and write
$\Gamma\cong \Lambda $, if there exists a bijective map
$\phi :\V_1\leftrightarrow \V_2 $ such that
\[ \big| \{ i\in I_1 | (v_1,v_2, i)\in\E_1 \} \big| =
\big| \{ j\in I_2 | (\, \phi (v_1) ,\phi( v_2), j \,)\in\E_2 \} \big|\quad
\mbox{ for all }\; v_!,v_2\in\V_1 \, .\]
This means for graphs without multiple edges, that $(v_1,v_2)\in\E_1 $ holds
if and only if $(\, \phi (v_1) ,\phi(v_2)\,)\in\E_2 $ hold.
The map $\phi $ is called a {\em graph isomorphism}.

\subsubsection{Vertex degree}
Let $\Gamma :=(\V ,\E ) $ be a finite graph and $v$ be a vertex in $\Gamma $.
We denote with $d_i(v) $ the {\em indegree} and with $d_o(v) $ the {\em
outdegree} of the vertex $ v $. This is the total number of edges
which are incident to $ v $ respectively incident from $ v $.
We write the vertex $v$ in $\Gamma $  has {\em degree} $d(v)$, if the total
number of edges which are incident at $v $ is $d$ and for this we
write $d(v)=d $. It follows that
\[ d(v)=d_i(v)+d_o(v)-\mbox{\em numbers of loops at }v \, . \]

We call $\Gamma $ a {\em $q$-regular } digraph, if
$d_i(v)=d_o(v) =q $ for all $v\in\Gamma $. This means that for every
vertex $v $ there exists exactly $q$ edges of $\Gamma $
with initial vertex $v $.
and $q $ edges of $\Gamma $ with terminal vertex $v$.
If $\Gamma $ is a $q$-regular graph with $L$ vertices, $\Gamma $
contains $qL$ edges. Vice versa, if $\Gamma $ is $q$-regular with $M$ edges,
$\Gamma $ contains $\frac{M}{q} $ vertices.\\

A digraph $\Gamma :=(\V ,\E ) $ is called an {\em Euler graph}, if
for every vertex $v\in\V $ we have $d_i(v) =d_o(v) $. Obviously
every regular graph is also an Euler graph.

\subsubsection{Walks in a graph}
Let $\Gamma =(\V ,\E )$ be a digraph.
A {\em walk } $P $ in $\Gamma $ of length n, from $v_1
$ to $v_{n+1} $, is a sequence $v_1e_1v_2\ldots v_ne_nv_{n+1} $
with $v_1,\ldots ,v_{n+1} \in\V $ and $v_i\stackrel{e_i}{\rightarrow} v_{i+1} $
is an edge of $\Gamma $
for all $1\le i \le n $. For this we write:
\[v_1
\stackrel{P}{\rightarrow } v_{n+1}  \;\mbox{ or }\;
 v_1\stackrel{e_1}{\rightarrow } v_2\stackrel{e_2}{\rightarrow }
\ldots \stackrel{e_n}{\rightarrow } v_{n+1}  \, .\]
Particulary a length $0$ walk is a single vertex of $\Gamma $.
We denote with $\left| P \right| =n $ the length of $P$.
If $|P|> 0 $, the walk $P$ is uniquely defined by its sequence of
edges $e_1\ldots e_n $.If moreover $\Gamma $ is a graph without multiple
edges, $P$ is also uniquely determined by the sequence of vertices
$v_1\ldots v_{n+1} $.
In this case we cease sometimes the edges over the
arrows in the notation above.\\

The walk $P$ is called a {\em closed walk}, if $v_1 = v_{n+1}$.
$P$ is called a {\em path}, if $P$ runs through every edge
one time at the most. This means $e_i\neq e_j $ for all $i\neq j $.
$P$ is called a {\em cycle}, if it is a closed walk
and runs trough every vertex different to the start vertex
not more than one time. This means $v_1 = v_{n+1} $ and $v_i\neq v_j $
for all $1\le i < j \le n$.
Furthermore $P$ is called a {\em simple path} if every vertex in $P$
occurs exactly one time or if $P$ is a cycle. This means that $P$ is a
simple path if $v_i\neq v_j $ for all $i\neq j $ or if $P$ is a cycle.
Obviously any simple path is also a path.

\noindent
The associated graph $\PP =(\V_P ,\E_p )$ of the walk $P$ is given by:
\[ \V_P :=\{ v_1,\ldots ,v_{n+1} \} \;\mbox{ and }\;
\E_P :=\{ e_1,\ldots ,e_n \} \, .\]
We obtain the following relations between a walk $P$ and its
associated subgraph:
\begin{quote}
\begin{tabular}[t]{lcl}
$P$ is a closed path & $\Leftrightarrow $ & $\PP $ is a connected Euler graph, \\
$P$ is a cycle & $\Leftrightarrow $ & $\PP $ is a connected $1$-regular graph.
\end{tabular}
\end{quote}
If $P$ is a simple path, but not a cycle, then every
vertex in $\PP $ expect $v_{n+1} $ has a unique successor vertex in $\PP $ and
every vertex in $\PP $ expect $v_1 $ has a unique antecessor vertex in $\PP $.
Therefore a simple path which is not a cycle is uniquely determined by its associated
graph. Furthermore if the starting respectively the end vertex in a
cycle is not important, we can interpret a cycle also as its associated
one-regular subgraph.
Therefore, if the context is clear, we don't distinguish between a simple pathes
its associated subgraphs.\\

We call a closed path $E$ in $\Gamma $ an {\em Euler circuit} of $\Gamma $, if $E$
runs through every edge of $\Gamma $ exactly one time. A cycle $C$
is called a {\em Hamilton circuit } if C runs through every vertex
of $\Gamma $ exactly
one time.\\

Let $P_1=v_1e_1\ldots e_nv_{n+1} $ be a walk of length $n$ and
$P_2=\tilde{v}_1\tilde{e}_1\ldots \tilde{e}_n\tilde{v}_{m+1} $ be
a walk of length $m$.
If the end vertex of $P_1$ is equal to the
starting vertex of $P_2 $, we define $P_1P_2 $ as the length $(n+m)$
walk given by the concatenation of the two pathes:
\[ P_1P_2:=v_1e_1\ldots e_n\tilde{v}_1\tilde{e}_1\ldots
\tilde{e}_n\tilde{v}_{m+1} \, .\]

\pagebreak
\subsubsection{Factors of a graph}
Let $\Gamma =(\V ,\E )$ be a graph and $\Lambda =(\tilde{\V } ,\tilde{\E })$
be a finite subgraph of $\Gamma $. The subgraph $\Lambda $ is called a
{\em $q$-factor} of $\Gamma $, if $\Lambda $ is a $q$-regular graph and a
spanning subgraph in $\Gamma $. This means $\tilde{\V } =\V $ and
$d_o(v) =d_i(v) =q$ for all vertices $v$ in $\Lambda $. The next proposition
is obviously.

\begin{prop}\label{fac1} Let
$\Gamma =(\V ,\E )$ be a graph without multiple
edges and $\Lambda =(\V  ,\tilde{\E })$ be a finite
spanning subgraph of $\Gamma $.
The following conditions are equivalent.
\begin{enumerate}
\renewcommand{\labelenumi}{(\roman{enumi})}
\item $\Gamma $ is a $1$-factor in $\Gamma $.
\item $\Lambda $ is the union of vertex disjoint cycles.
\item There exists a bijective map $\phi :\V \leftrightarrow \V $, such that
\[  (v,u)\in\tilde{\E } \Leftrightarrow u =\phi (v) \]
holds for all $u,v\in\V $.
\end{enumerate}
\end{prop}
Furthermore every $q$-regular graph is the union of vertex disjoint cycles.

\subsection*{Connected graphs, connected components and Euler graphs}
Let $\Gamma :=(\V ,\E )$ be a graph.
A vertex $v\in\V$ is called {\em reachable }
from another vertex $u\in\V $, if there exists a walk from u to v. This
is the same as to say that there exists a simple path from u to
v. The {\em distance} between $u$ and $v$ in $\Gamma $ is given by
\[ dist(u,v) := \min \{\left| P\right| \, |\, u\stackrel{P}{\rightarrow } v
 \mbox{ $P$ is simple path } \} \, ,\]
if $v$ is reachable from $u$ and $dist(u,v):=\infty $ otherwise.
Let
\[ \Gamma^n (v):=\{ u\in\V | \mbox{there is a
simple path of length $n $ from $v$ to $u$} \} \]
be the set of all
vertices which have distance $n$ from $v\in\V $ in $\Gamma $.
Furthermore we define
\[\,^n\Gamma (v) :=\{ u\in\V
| \mbox{there is a simple path of length $n $ from $u$ to $v$} \, . \}
\]
Then $\,^n\Gamma (v) $ is the set of all vertices,
from which the distance to $v$ is $n$.
We call the vertices in $\Gamma^1(v) $ the {\em successors}
of $v$ in $\Gamma $ and the vertices in $\,^1\Gamma (v) $ the
{\em antecessors} of v. Let $A\subseteq\V $ be an arbitrary set of vertices.
Then we define the successor set of $A $ as $\Gamma^n(A) :=\bigcup_{v\in A} \Gamma^n (v) $
and the antecessor set of $A $ as
$\,^n\Gamma (A) = \bigcup_{v\in A}\,^n\Gamma (v) $.
Instead of $\Gamma^1(v)$ and $\Gamma^1(A) $ we write commonly
$\Gamma (v) $ and $ \Gamma (A) $ respectively.
By this we have
\[ \left| \Gamma (v) \right| =d_o(v) \;\mbox{ and }\;
\left| \,^1\Gamma (v) \right| =d_i(v) \]
It follows that $\Gamma $ is q-regular if and only if
$|\Gamma (v)| = |\,^1\Gamma (v)| =q $ for all $v\in\V $.\\

\smallskip
We call $\Gamma $ {\em (strongly) connected}, if  $u$ is reachable from
$v$ for all $u,v\in\V $. In the most books about graph theory, a digraph
is called connected, if the underlying undirected graph is connected
and a digraph is called strongly connected if it fulfill the condition above.
Since we pay no attention to undirected graphs in this survey, we cease
the ``strongly''. This means that in this survey a connected graph $\Gamma $ is
a digraph for which there exists a simple path between every two vertices
of $\Gamma $.\\

\smallskip
Let $\Gamma :=(\V ,\E ) $ be a graph. $\Gamma $ can be {\em split into cycles }
if there exists a set $\C $ of edge disjoint cycles, such that $\Gamma $ is the edge
disjoint union of the cycles in $\C $. This means two different
cycles in $\C $ have no common edge and the cycles in $\C $ covers
$\Gamma $, where we understand an isolated vertex as a cycle of length 0.
We call the set $\C $ a {\em cycle splitting} of $\Gamma $.

\begin{prop}\label{splitting} Let $\Gamma =(\V ,\E )$ be a finite graph.Then
$\Gamma $ is an Euler graph if and only if $\Gamma $ can be split into
cycles.
\end{prop}

\smallskip\noindent
{\bf Proof:} Let $\Gamma $ be a finite Euler graph. We show that
there exists at least one cycle in $\Gamma $.
We choose an arbitrary vertex $v_1\in\V $,
if $v_1 $ is isolated, we have found a cycle of length 0.
If $v_1 $ is not isolated, than there exists at least one
edge which is incident at $v_1 $. Since  $d_i(v_1)=d_o(v_1) \ge 1 $, there exists also an
edge $e_1 $  with initial vertex $v_1 $.
Let $v_2 $ be the terminal vertex of $e_1 $, then $v_1e_1v_2 $ is a
simple path in $\Gamma $. If $v_1 =v_2 $ we have found a cycle.
Let $v_1e_1v_2\ldots e_nv_{n+1} $ be a simple path of length $n$, which is not a
cycle. We have $d_i(v_{n+1} )\ge 1 $,
because $\Gamma $ is an Euler graph there exists an edge $e_{n+1} $
incident from $v_{n+1} $. Let $v_{n+2} $ the terminal vertex of
$e_{n+1} $. If $v_{n+2} \neq v_i $ for all $1\le i \le (n+1) $ we obtain
that $v_1e_1v_2\ldots v_{n+1}e_{n+1}v_{n+2} $ is a simple path of length $(n+1)$
which is not a cycle. If we continue with this procedure, it follows that
at some point $v_{n+2} =v_i $ for some $1\le i\le n+1 $, because
$\Gamma $ has only a finite number of vertices.
Since $v_1e_1\ldots e_{n}v_{n+1} $ is a simple path,we obtain that
$v_ie_i\ldots e_nv_{n+1}e_{n+1}v_{n+2} $ is a cycle.
Therefore every finite Euler graph obtain at least one cycle. If
we delete in $\Gamma $ the edges of the cycle, we obtain a new graph
$\Gamma $ which is also an Euler graph. While $\Gamma $ is finite,
we obtain by induction a finite numbers of edge disjoint cycles
which covers $\Gamma $. This shows that there exists a cycle
splitting for $\Gamma $.

\pagebreak
\noindent
Thus let $\Gamma $ be a finite graph and $\C $ be a cycle splitting
of $\Gamma $. Let $v$ be a none
isolated vertex of $\Gamma $. Then there exists cycles
$C_1,\ldots,C_m \in\C $, such that every edge incident at v is contained in
one of the cycles. While a cycle is
1-regular and the cycles in $\C $ are edge disjoint, it follows that
$d_i(v)=d_o(v)=m $. This shows that $\Gamma $ is an Euler graph. \qed\\

\medskip
Let $\Gamma =(\V ,\E )$ be a finite Euler graph and  $u,v \in\V $. If $u$ is
reachable from $v$, then there exists a simple walk
\[ v_1e_1v_2\ldots v_ne_nv_{n+1} \mbox{ with } v_1=v \mbox{ and } v_{n+1}=u \, .\]
Let $\C :=\{ C_1,\ldots ,C_k \} $ be a cycle splitting of $\Gamma $.
Then there exists for every $i\in\{ 1,\ldots , n \} $ a (unique) cycle
$C\in\C $, such that $e_i $ is an edge in $C $. Let us denote with
with $P_i $ the simple path, which is obtained by deleting $e_i $
in $C $. It follows that $P_nP_{n-1}\ldots P_1 $ is a walk from $u $
to $v$. Therefore $v$ is also reachable from $u$. This shows:

\begin{prop} \label{connection}
Let $\Gamma =(\V ,\E ) $ be a finite Euler graph. Then we have for every $u,v\in\V $:
\[ u \mbox{ is reachable from } v \Leftrightarrow v \mbox{ is reachable from } u \, . \]
\end{prop}

\smallskip
If $u$ is reachable from $v$ and $w$ is
reachable from $u$ then obviously $w$ is reachable from $v$.
This shows that the relation "reachable" is an equivalent
relation on the vertex set of a finite Euler graphs.
Let $\{ \V_1,\ldots , \V_m \} $ be the
partition of the vertex set $\V $ of $\Gamma $ given by this
equivalent relation. i.e. the $\V_i$´s are the equivalent classes.
Let $\E_i $ be the edge sets given by
\[ (u,v)\in\E_i \mbox{ iff } u,v\in\V_i \quad i\in\{ 1,\ldots , m \} , . \]
Then $\{ \E_1 ,\ldots ,\E_m\} $ is a partition of the edge set $\E $ of
$\Gamma $. Furthermore it follows that
$\Gamma_i :=\big( \V_i ,\E_i \big) $ is a connected Euler subgraph
of $\Gamma $ for all $1\le i\le m $ and that there does not exists connections
in $\Gamma $  between two different of this subgraphs.\\

\smallskip
Let $\Gamma =(\V ,\E )$ be an arbitrary graph.
We call a collection $\Gamma_1 ,\ldots \Gamma_m \subseteq\Gamma$ of subgraphs a
{\em decomposition into the connectivity components } of $\Gamma $,
if the following properties hold:
\begin{enumerate}
\renewcommand{\labelenumi}{(\arabic{enumi})}
\item $\Gamma =\Gamma_1\cup\ldots\cup\Gamma_m $,
\item all $\Gamma_i$'s are connected graphs,
\item there does not exist connections in $\Gamma $  between the subgraphs
$\Gamma_1,\ldots ,\Gamma_m $.
Especially for every vertex $u$ in $\Gamma_i $ and vertex
$v$ in $\Gamma_j$ $u$ isn't reachable from $v$ in $\Gamma $ and vice versa.
\end{enumerate}
\pagebreak
\noindent
Obviously such a decomposition is uniquely. We have shown above, that
every finite Euler graph has a decomposition into connected components.
Furthermore every connected component of an Euler graph is by itself
an Euler graph.\\

\medskip
For finite Euler graphs the well known theorem from Euler holds:
\begin{theorem}[Euler]\label{euler}
Let $\Gamma $ be a finite graph without isolated vertices. Then
\[ \Gamma \mbox{ has an Euler circuit } \Leftrightarrow
\Gamma \mbox{ is a connected Euler graph }\, ,\]
i.e. for every regular connected finite graph exists an Euler circuit.
\end{theorem}

\smallskip
\noindent
{\bf Proof:} Let $\Gamma =(\V ,\E ) $ be a finite graph without isolated
vertices and $E$ be an
Euler circuit for $\Gamma $. Let $u,v\in\V $. Since $u,v$ are none-isolated vertices
and $E$ runs through every edge of $\Gamma $ one time it follows that
$u$ and $v$ occurs at least once in $E$. Let us suppose that
$u$ occurs before $v$ in the sequence $E$, then $v$ is reachable
from $u$. While $E$ is a closed walk we obtain, that $u$ is also
reachable from $v$. Therefore $\Gamma $ is a connected graph.
If $v$ occurs in $E$ $m$-times, then there exist exactly $m$ edges in $\Gamma$
with initial vertex $v$ and $m$ edges with terminal vertex $v$, because
$E$ is a closed walk which runs through every edge of $\Gamma $ exactly
one time. It follows that $\Gamma $ is an Euler graph.\\

\smallskip\noindent
Thus let $\Gamma $ be a finite connected Euler
graph without isolated vertices. We proof by induction on the
number of edges of $\Gamma $, that there exists an Euler circuit in
$\Gamma $. If $\Gamma $ has only one edge, then $\Gamma $ consists
of a single vertex and a loop at this vertex.
In this case, the loop is an Euler circuit for $\Gamma $.
Let us assume, that for $n>1 $, every finite
connected Euler graph without isolated vertices and $k$ edges has an
Euler circuit, if $k<n$.
Let $\Gamma $ be a finite connected
Euler graph with $n$ edges and without isolated vertices.
By Proposition~\ref{splitting} follows,
that there exists a cycle $C$ (with length bigger than $0$) in $\Gamma $.
If we delete the edges of this cycle in $\Gamma $,
we obtain a new graph $\tilde{\Gamma } $ which is also an Euler
graph, but $\tilde{\Gamma } $ has less than k edges.
In general, this graph is not a connected graph , but every connectivity component
of $\tilde{\Gamma } $ is a
connected Euler graph with less than k edges.
By the induction hypothesis follows, that every connectivity component of $\tilde{\Gamma }$
has an Euler circuit. While $\Gamma $ is a connected graph,
every connectivity component of $\tilde{\Gamma } $ has at least one vertex
on the cycle $C $.
We obtain an Euler circuit for $\Gamma $, by travelling along the cycle.
If we come to a vertex $v$
in a connectivity component of $\tilde{\Gamma }$ which we haven't visited
before, we stop and run through the corresponding Euler circuit of the
connectivity component. After one round we come back to $v$ and continue
to travel around the cycle $C$. If we finish one round in $C$, we
had visited every edge in $\Gamma $ exactly one time. This gives
us an Euler circuit of $\Gamma $.\qed\\
\pagebreak

\subsection*{Factors of $q$-regular digraphs}
Let us recall Hall's matching theorem.
A graph $\Lambda =(\tilde{\V },\tilde{\E } ) $ is called a
{\em bipartite digraph}, if there exists a partition $\V_1 ,\V_2 $ of
$\V $, such that all edges in $\Lambda $ have their initial vertex in
$\V_1 $ and their terminal vertex in $\V_2 $. Furthermore we allow multiple
edges. A {\em matching set} in $\Lambda $ is an
edge set $M \subseteq \E $, such that any two different edges in $M$
have neither the same initial vertex nor the same terminal vertex.
A {\em complete matching} in $\Lambda $ is a matching set $M$, such that
for every $v\in\V_1 $ there exists an edge in $M$ with initial vertex $v$.

\begin{theorem}[Hall's matching theorem]\label{hall1}
Let $\Lambda =(\V_1\cup\V_2 ,\tilde{\E } ) $ be a finite
bipartite digraph.
There exists a complete matching set $ M\subseteq \E $ for $\Gamma $ if and
only if $ |\Gamma (A) |\ge |A| $  for all $A\subseteq \V_1 $
\end{theorem}

\noindent
A proof of the theorem above should be found in nearly every book
about graph theory or combinatorics, for example \cite{lint1}.

\begin{cor} \label{hall2}
Let $\Gamma = (\V,\E )$ be a finite $q$-regular digraph.
Then there exists a $1$-factor in $\Gamma $.
\end{cor}

\smallskip\noindent
{\bf Proof:} Let $\Gamma :=(\V ,\E ) $ be a finite $q$-regular graph, with
$\E\subseteq\V\times\V\times I $, $|I|<\infty $.
We define a bipartite digraph $\Lambda =(\V_1\cup\V_2 , \tilde{\E } ) $
as follows.
Let $\V_1 $ and $\V_2 $ be two different duplicates of $\V $.
This means $\V_1 := \V\times \{ 1 \} $ and $ \V_2 := \V \times \{ 2 \} $.
We define the edge set $\tilde{\E }\subseteq\V_1\times\V_2\times I $ as follows.
\begin{quote}
For $u,v\in\V ,i\in I $ the edge
$(v,1)\stackrel{i}{\rightarrow } (u,2) $ exists in $\Lambda $ if and only if
$v\stackrel{i}{\rightarrow } u $ is an edge in $\Gamma $.
\end{quote}
Obviously $\Lambda $ is a finite bipartite digraph. Since $\Gamma $ is $q$-regular
it follows that
\[
d_o (v) =q =d_i(u) \;\mbox{ for all }\; v\in\V_1 ,u\in\V_2 \, .
\]
Let $A\subseteq\V_1 $. It follows that there are $q|A|$ edges with initial
vertex in $A$ and terminal vertex in $\Lambda (A) $, but that there exists
totally $q|\Lambda (A)| $ edges with terminal vertex in $\Lambda (A) $.
Therefore we obtain $|A|\le |\Lambda (A) |$ for all $A\subseteq\V_1 $.
>From Hall's matching theorem follows, that there exists a complete
matching set $M\subseteq\V_1 $ for $\Lambda $. This means, for
every $v\in\V $, there exists unique $u\in\V ,i\in I $, such that
$(v,1)\stackrel{i}{\rightarrow } (u,2) $ is an edge in $M$.
With $|\V_1|=|\V |=|\V_2| $ we conclude, that also for every
$u\in\V $, there are unique $v\in\V ,i\in I $, such that
$(v,1)\stackrel{i}{\rightarrow } (u,2) $ is an edge in $M$.
Thus we obtain a $1$-factor of $\Gamma $, if we identify
the edges in $M$ with its corresponding edges in $\E $.\qed\\

\pagebreak

Let $\Gamma =(\V ,\E ) $ be a $q$-regular digraph and
$\Lambda = (\V ,\E' ) $ be a $k$-factor of $\Gamma $.
Obviously the subgraph $\Lambda^c = (\V , \E -\E' )\subseteq\Gamma $
is a $(q-k)$-factor of $\Gamma $, i.e. $\Lambda^c $ is $(q-k)$-regular.
Vice versa, let $\Lambda_1 $ be a
$k_1$-factor of a graph $\Gamma $ and $\Lambda_2 $ be a $k_2$-factor
of $\Gamma $. If $\Lambda_1 $ and $\Lambda_2$ are edge disjoint, then
$\Lambda_1\cup\Lambda_2 $ is a $(k_1+k_2)$-factor of $\Gamma $.
Therefore we obtain the following proposition with the help of Corollary~\ref{hall2}.

\begin{prop}\label{factor} Let $\Gamma :=(\V,\E )$ be a finite q-regular graph.
\begin{enumerate}
\renewcommand{\labelenumi}{(\roman{enumi})}
\item There exists a $k$-factor of $\Gamma $, for every $1\le k\le q $.

\item Let $\Lambda $ be a $k$-factor of $\Gamma $ and $1\le m\le k $.
A subgraph $\tilde{\Lambda } \subseteq\Lambda $ is a $m$-factor
of $\Lambda $ if and only if $\tilde{\Lambda }$ is a $m$-factor
of $\Gamma $.

\item Let $k_1,\ldots,k_m \in\NN $, such that $k_1+\ldots +k_m \le q $.
Then there exists edge disjoints factors $\Lambda_1 ,\ldots ,\Lambda_k $
of $\Gamma $, with $\Lambda_i $ is a $k_i$-factor for all $1\le i\le m $.
If $k_1+\ldots + k_m = q $, then $\Gamma $ is the edge disjoint union of the
$\Lambda_i $'s. Especially there exists an edge disjoint decomposition
of $\Gamma $ into $q$ $1$-factors.

\item If $\Lambda $ is a $m$-factor of $\Gamma $ and $1\le k\le m $, then
there exists a $k$-factor $\tilde{\Lambda } $
of $\Gamma $ with $\tilde{\Lambda }\subseteq\Lambda  $.

\item If $\Lambda $ is a $k$-factor of $\Gamma $ and $1\le k\le m\le q  $, then
there exists a $m$-factor $\tilde{\Lambda } $
of $\Gamma $ with $\Lambda \subseteq \tilde{\Lambda } $.

\end{enumerate}
\end{prop}

\noindent
Moreover part (ii) and (iv) in the proposition above holds for
any digraph $\Gamma $.

\subsection*{Linegraphs}
Let  $\Gamma = (\V , \E ) $ be a graph. The {\em linegraph } $L\Gamma
:= (\V_1 ,\E_1 ) $ is defined as:
\begin{enumerate}
\item $\V_1 :=\E $. This means, the vertices of $L\Gamma $ are
the edges in $\Gamma $.
\item Let $e_1,e_2\in\V_1 $ be two vertices of $L\Gamma $.
There exists an edge from $e_1 $ to $e_2 $ in $L\Gamma $
if and only if $e_1e_2$ is a walk in $\Gamma $ (of length 2).
\[ \E_1 := \big\{  (e_1 ,e_2)\in\E^2  | \mbox{ the terminal vertex of
$e_1$ is the initial vertex of $e_2$} \big\} \]
\end{enumerate}
This means, that the edges in $L\Gamma $ are the walks of length 2 in
$\Gamma $. We observe that $L\Gamma $ has no multiple edges.\\

\pagebreak
We define the {\em k-iterated linegraph } of $\Gamma $ recursively as:
\[ L_{k+1}\Gamma = L(L_k\Gamma ) \mbox{ with } L_0\Gamma := \Gamma \, .\]
By an easy induction we obtain:
\[ L_{k+m}\Gamma =L_k(L_m\Gamma ) =L_m(L_k\Gamma ) \quad\mbox{ for } k,m\in\NN_0 \, . \]
By induction it is easy to verify, that the vertices of
$L_k\Gamma $ can be labelled with the walks in $\Gamma $ of length $k$
and that the edges of $L_k\Gamma $ can be labelled with the walks in
$\Gamma $ of length $(k+1) $, in the following way:\\

\begin{quote}
Let $u,u'$ be two vertices in $L_k\Gamma $, where $u$ should be labelled
with the walk $P=v_1e_1v_2\ldots v_ne_nv_{n+1} $ in $\Gamma $
and $u'$ should be labelled with the walk
$P'=v'_1e'_1v'_2\ldots v'_ne'_nv'_{n+1} $ in $\Gamma $.
Then there exists an edge from $u$ to $u'$ in $L_k\Gamma $ if and only if
$ v_2e_2v_3\ldots v_ne_nv_{n+1} =v'_1e'_1v'_2\ldots v'_{n-1}e'_{n-1}v'_{n} \, . $
Furthermore this edge is labelled with the walk of length $(k+1)$ given by:
\[ v_1e_1v_2\ldots v_{n+1}e_nv'ne'_nv'_{n+1} =
v_1e_1v'_1e'_1v'_2\ldots v'_{n}e'_{n}v'_{n+1} \,. \]
\end{quote}
\noindent
More precisely:

\noindent
Let us understand walks as sequences of edges. We define:
\begin{eqnarray*}
 \V_k & := & \{ P | P=e_1\ldots e_k  \mbox{ walk of length k in } \Gamma \}\, , \\
 \E_k & := & \big\{
(e_1\ldots e_k ,e_2\ldots e_{k+1})\subseteq \E^k\times \E^k
\big|\,
 e_1\ldots e_{k+1} \mbox{\footnotesize is a length $ (k+1)$  walk in $ \Gamma $} \big\}\, .
\end{eqnarray*}
Then an easy induction proof shows, that $L_k\Gamma \cong (\V_k ,\E_k ) $
and that
\[ (e_1,\ldots e_k ,e_2\ldots e_{k+1} )\in\E_k \leftrightarrow
e_1\ldots e_{k+1} \mbox{ length k+1 walk in } \Gamma \]
gives us a one-to-one relation between the walks
of length $(k+1)$ in $\Gamma $ and the edges of $L_k \Gamma $.\\

\smallskip
If $\Lambda $ is a subgraph of $\Gamma $, then obviously  $L_k\Lambda $
is a subgraph of $L_k\Gamma $. Furthermore for Linegraphs the
following proposition holds. Since the proposition is mostly obviously,
we omit a proof.

\pagebreak
\begin{prop}\label{linegraph1}
Let $\Gamma := (\V ,\E ) $ be a graph and $L_k\Gamma  =(\V_k ,\E_k )$
be the $k$-iterated linegraph of $\Gamma $.
\begin{enumerate}
\renewcommand{\labelenumi}{(\roman{enumi})}
\item $L_k\Gamma $ has no multiple edges and no isolated vertices
for all $k\ge 1 $.

\item $e\in\E $ is a loop in $\Gamma $ if and
only if there exists a loop at the vertex $e$ in $L\Gamma $, i.e. the
number of loops in $L_k\Gamma $ is equal to the number of loops in
$\Gamma $.

\item If $\Gamma $ is $q$-regular then also $L_k\Gamma $
is q-regular, as well. Moreover, if $\Gamma $ is finite,
then $\big| \V_k \big|=q^k \big| \V_k \big| $

\item $\Gamma $ is a connected graph if and only
if $\L_k\Gamma $ is a connected graph for some $k\in\NN_0 $, i.e. $L_k\Gamma $
is a connected graph for all $k\in\NN_0 $, if $\Gamma $
is a connected graph.

\item If $\Lambda $ is a $q$-regular subgraph of
$\Gamma $ with $p$ vertices, then $\L_k\Lambda $ is a $q$-regular
subgraph of $L_k\Gamma $ with $q^kp $ vertices.

\item If $\Lambda_1 $ ,$\Lambda_2 $ are edge disjoint subgraphs of $\Gamma $,
then $L_k\Lambda_1 $ and $L_k\Lambda_2 $ are vertex disjoint
subgraphs of $L_k\Gamma $ for all $k\ge 1 $.

\item Let $\C $ be a cycle of length $k$ in $\Gamma $, then
$L_k\C $ is also a cycle of length $k$ in $L_k\Gamma $.

\item Let $\Gamma $ be $q$-regular and
$\Lambda_1 ,\ldots ,\Lambda_q $ be edge disjoint $1$-factors of $\Gamma $.
Then $L\Lambda_1 ,\ldots ,L\Lambda_q $ are vertex disjoint and
$\big(L\Lambda_1\cup\ldots\cup L\Lambda_q \big) $ is a $1$-factor of $L\Gamma $.

\end{enumerate}
\end{prop}

\bigskip

Let $\Gamma =(\V ,\E ) $ be a graph and
$ P=e_1\ldots e_n $ be a walk of length $n$ in $\Gamma $ with
$e_i\in\E \;\forall\, 1\le i\le n $.
We denote with $LP $ the {\em linewalk } walk of length
$(n-1)$ in the linegraph $L\Gamma $ which is given by:
\[LP := (e_1,e_2)\ldots (e_{n-1},e_n)\, ,\qquad
\mbox{whereas the $ (e_i,e_{i+1}) $ are all edges
in $L\Gamma $.} \]
Obviously $P$ is a walk of length $(n-1)$ in $L\Gamma $.
Furthermore we define recursively $L_k P $ as:
\[ L_{k+1}P :=L\big(L_k P\big ) \, ,\quad L_0P:=P \quad
\mbox{ for all }\; 1\le k\le n \, .\]
Obviously $L_kP $ is a walk of length $(n-k) $ in $L\Gamma $ for all
$1\le k\le n$. Especially $L_nP$ is a single vertex in $\L_n\Gamma $.\\

\pagebreak
If $P $ is a closed walk, then $e_ne_1 $ is a walk of length 2 in $\Gamma $.
Therefore we can define for closed walks, the {\em closed linewalk} $\hat{L}P $ as:
\[ \hat{L}P :=(e_1,e_2)\ldots (e_{n-1},e_n)(e_n,e_1) \, . \]
Obviously $\hat{L}P $ is a closed walk in $\L\Gamma $ of length $n$ and
$\hat{L}P = (LP)(e_n,e_1) $.
We define recursively:
\[ \hat{L}_{k+1}P :=L\big(\hat{L}_k P\big ) \, ,\quad \hat{L}_0P:=P \]
It follows that $\hat{L}_kP $ is a closed walk in $L_k\Gamma $ of length $n$.\\

\noindent
We obtain the following proposition, which is obviously for the most part.

\begin{prop}\label{linegraph2} Let $\Gamma :(\V ,\E )$ be a graph.
\begin{enumerate}
\renewcommand{\labelenumi}{(\roman{enumi})}

\item If $P_1,P_2 $ are two edge disjoints walks in $\Gamma $,
then $LP_1 $ and $LP_2 $ are vertex disjoints walks in $L\Gamma $.
The same holds for closed walks and $\hat{L}$.

\item If $P$ is a path in $\Gamma $,
then $LP $ is a simple path in $LP $

\item If $P$ is a closed path (of length n ) in $\Gamma $,
then $\hat{L}_kP $ is a cycle (of length n ) for all $k\ge 1 $

\item If $P$ is a cycle in
$\Gamma $ with corresponding subgraph $\PP $, then $L_k\PP $ is the
the corresponding subgraph of $\hat{L}_kP $ in $L_k\Gamma $.

\end{enumerate}
\end{prop}

\smallskip
Let $\Gamma =(\V, \E ) $ be a graph and $\Lambda $ be a finite
Euler subgraph in $\Gamma $ with $n$ edges. There exists an
Euler circuit $E$ for $\Lambda $. Especially $E$ is a closed path
of length $n$ in $\Gamma $. From part (iii) of the proposition above follows
that $\hat{L}_k P $ is a cycle of length $n$ in
$L_k\Lambda \subseteq L_k\Gamma $ for all $k\in\NN $.
Especially $\hat{L}_k P $ is a Hamilton circuit of $L_k\Lambda $.\\

\smallskip
Let $P' $ be a cycle of length $n$
in $L\Gamma $. Since $L\Gamma $ has no multiple edges,
we can understand $P'$ as a sequences of vertices in $L\Gamma $:
\[ P=e_1\ldots e_n e_1 \; \mbox{ with }\; e_1,\ldots ,e_n \;\mbox{ vertices in }\;
L\Gamma \, .\]
While the vertices in $L\Gamma $ are the edges of $\Gamma $, we can understand
the sequence $e_1\ldots e_n$ of edges in $\Gamma $ as a walk of length $n$ in
$\Gamma $. Furthermore it follows that $e_1\ldots e_n $ is a closed path
in $\Gamma $, because $P$ is a cycle in $L\Gamma $. If we denote
with $P' $ the closed path $e_1\ldots e_n $ in $\Gamma $, then follows
that $\hat{L}P' =P $. Furthermore
we define $\hat{L}_{*} P $ as the subgraph in $\Gamma $ which
corresponds to the closed path $P'$. Since $P$ is a closed path of length $n$,
we obtain that $\hat{L}_{*} P $ is an Euler subgraph in $\Gamma $ with $n$
edges. Let $E$ be an Euler circuit for an Euler subgraph
$\Lambda\subseteq \Gamma $ with $n$ edges. We have already shown above, that
$\hat{L}E $ is a cycle in $L\Gamma $ of length $n$.
Therefore $\hat{L}_{*}\hat{L} E$ is defined.
Furthermore it is easy to verify, that $\hat{L}_{*}\hat{L} E = \Gamma $.
Therefore $\hat{L}_{*} $ maps the cycles of length $n$ in $L\Gamma $
onto the Euler subgraphs in $\Gamma $ with $n$ vertices.
Let $\Lambda $ be a Euler subgraph in $\Gamma $ with $n$ vertices and
edge set $\E_{\Lambda }$. It follows that there exists
a cycle of length $n$ in $L\Gamma $ if and only if there exists
an Euler subgraph in $\Gamma $ with $n$ vertices.
Furthermore we have:
\begin{equation}\label{linehat1}
\hat{L}^{-1}_{*}\Lambda =\{ P=e_1\ldots e_ne_1 \, | \,
e_1\ldots e_n\in\E^n_{\Lambda } \;\mbox{ is a closed path } \}
\end{equation}

\newpage

\section{The de Bruijn digraph \BGr[\A ]{n}}

\subsection*{Definition of de Bruijn digraphs}
Let $\A $ be an arbitrary alphabet, where we allow infinite alphabets.
We define for $n\in\NN $ the {\em $n $-th level de Bruijn
Graph} $\BGr[\A ]{n}=(\V ,\E ) $ as follows:
\begin{enumerate}
\renewcommand{\labelenumi}{\arabic{enumi}.}
\item The vertices of \BGr[\A ]{n} are the words over $\A $ of length $n$.
This means $\V :=\A^n $.

\item Let $w,w'\in\A^n $. There is an edge from $w$ to $w'$ in \BGr[\A ]{n}
if and only if the letters of $w$ at the $(i+1)$-th position is equal to
the letter of $w'$ at the $i$-th position for all $1\le i \le n-1$.
Therefore the edge set of \BGr[\A ]{n} is given by:
\[ \E:= \{ (au,ub)\in\A^n\times\A^n | a,b\in\A , u\in\A^{n-1} \}
=\bigcup\limits_{w\in\A^n} \left( w,\A^{-1}w\A \right) \, . \]
This means $w_1\ldots w_n \rightarrow w_2\ldots w_{n+1} $ is an edge in
$\BGr[\A ]{n}=(\V ,\E ) $ for all $w_1,\ldots ,w_{n+1}\in\A $.
\end{enumerate}
Obviously \BGr[\A ]{n} has no multiple edges for $n\in\NN $.
If n=0, the graph \BGr[\A ]{0} is defined as the multiple edge
digraph which has the empty sequence $e\in\A^0 $ as its only vertex and for every
$a\in\A $ there exists a loop $ e \stackrel{a }{\rightarrow } e $
in \BGr[\A ]{0}.\\

If $\A = \{ 0,1,\ldots ,q \} $, then we will write \BGr{n} in place of
\BGr[\A ]{n}. If $|\A |= q $ for some arbitrary finite alphabet, then
we can understand \BGr[\A ]{n} also as the graph \BGr{n}, because
a bijective map between $\A $ and $\{ 0,1,\ldots ,q \} $ gives us
a isomorphism  between \BGr[\A ]{n} and \BGr{n}. Especially we obtain
$\BGr[\A ]{n}\cong \BGr{n} $.\\


Let $u\in\A^{n-1}\,, \, a,b\in\A $ for some arbitrary finite or infinite
alphabet $\A $.
We will write for the edge from $au $
to $ub $ in \BGr[\A ]{n} sometimes $au\stackrel{b}{\rightarrow } ub $.
Furthermore we obtain that $u\A $ as the set of successors of the vertex
$au $ and $\A u $ is the set of antecessors of the vertex
$ub$. If $\X\subseteq\A^n $ is a set of vertices of \BGr[\A ]{n}, then it follows that
\[\mbox{\begin{tabular}{ll}
$\A^{-1}\X\A $ & is the set of successor of vertices in $\X $,\\
$\A\X\A^{-1} $ & is the set of antecessors of vertices in $\X $.\\
\end{tabular}} \]
Let $|\A | =q<\infty $. Obviously \BGr[\A ]{0} is a $q$-regular graph.
Let $n\in\NN $ and $w\in\A^n $. It follows that
\[ d_i(w) =|\A w \A^{-1} | = |\A | =q =|\A | = |\A^{-1}w\A | =d_o(w) \, .\]
Therefore \BGr[\A ]{n} is a $q$-regular graph for all $n\in\NN $.
Especially \BGr{n} is a $q$-regular graph. Since $\A^n $ is the
vertex set of \BGr[\A ]{n}, it follows that \BGr[\A ]{n} has
$q^{n+1} $ edges.

\pagebreak
Let $\A $ be an arbitrary alphabet and $w,v\in\A^n $, where
$w=w_1\ldots w_n $, \\
$v=v_1\ldots v_n $ with $w_i,v_i\in\A $ for all $i$.
We obtain a walk of length $n$ from $w$ to $v$ by
\[ w\stackrel{v_1}{\rightarrow }w_2\ldots w_nv_1 \stackrel{v_2}{\rightarrow }
 w_3\ldots w_nv_1v_2 \stackrel{v_3}{\rightarrow } \ldots
\stackrel{v_{n-1}}{\rightarrow } w_nv_1\ldots v_{n-1}
\stackrel{v_n}{\rightarrow } v \, .\]
This shows that for any two vertices $w,v$ in \BGr[\A ]{n} there is
a walk of length $n$ from $w$ to $v$. Especially this shows that
\BGr[\A ]{n} is a connected graph.\\

\medskip\noindent
The pictures below show the graphs \BGr[2]{0} -
\BGr[2]{3}  and \BGr[3]{2}.\\

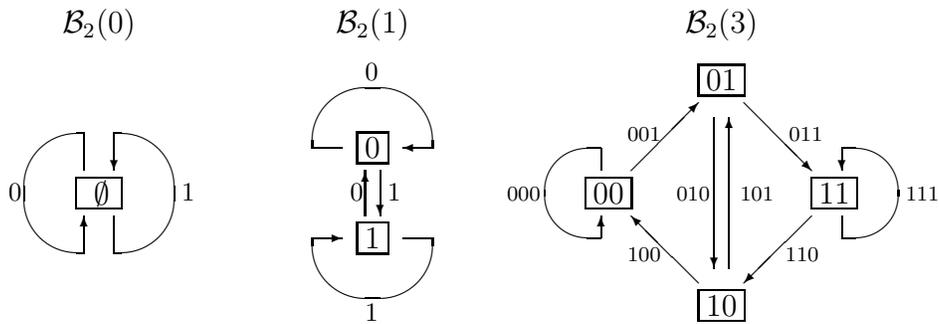
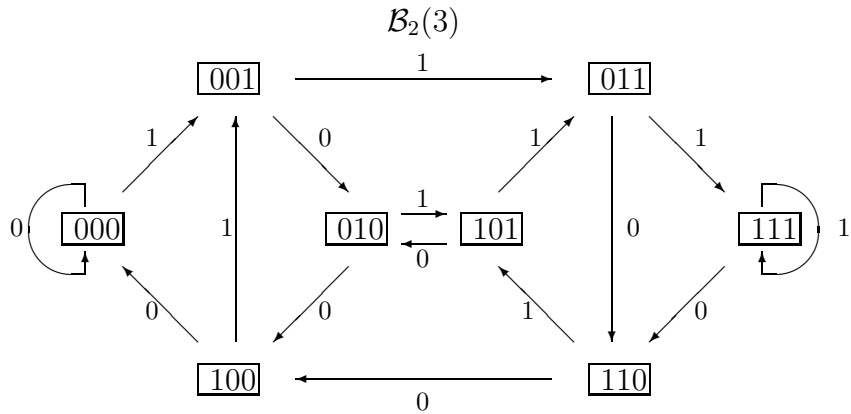
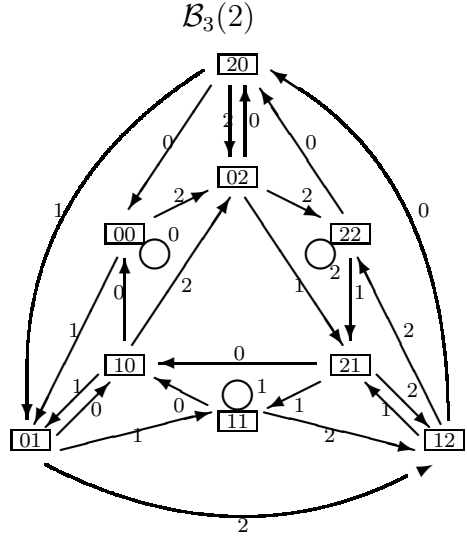
\begin{center}
\setlength{\unitlength}{1mm}
\parbox[b]{4cm}{\begin{center} \BGr[2]{0} \\
\begin{picture}(40,40)
\put(17,18){\framebox(6,4){$\,\emptyset $}}
\put(18,20){\oval(16,16)[l]} \put(18,12){\vector(0,1){5}}
\put(18,28){\line(0,-1){5}} \put(22,20){\oval(16,16)[r]}
\put(22,12){\line(0,1){5}} \put(22,28){\vector(0,-1){5}}
\put(8,19){\footnotesize 0} \put(31,19){\footnotesize 1}
\end{picture}
\end{center}}
\parbox[b]{3cm}{\begin{center} \BGr[2]{1} \\
\begin{picture}(30,40)

\put(13,12){\framebox(4,4){1}} \put(13,24){\framebox(4,4){0}}

\put(15,14){\oval(16,16)[b]} \put(7,14){\vector(1,0){4}}
\put(23,14){\line(-1,0){4}} \put(14,3){\footnotesize 1}

\put(15,26){\oval(16,16)[t]} \put(7,26){\line(1,0){4}}
\put(23,26){\vector(-1,0){4}} \put(14,35){\footnotesize 0}

\put(14,17){\vector(0,1){6}} \put(12,19){\footnotesize 0}

\put(16,23){\vector(0,-1){6}} \put(17,19){\footnotesize 1}
\end{picture}
\end{center}}
\parbox[b]{6cm}{\begin{center} \BGr[2]{3} \\
\begin{picture}(60,40)
\put(27,33){\framebox(6,4){01}} \put(27,3){\framebox(6,4){10}}
\put(42,18){\framebox(6,4){11}} \put(12,18){\framebox(6,4){00}}

\put(46,20){\oval(15,12)[r]} \put(46,26){\vector(0,-1){3}}
\put(46,14){\line(0,1){3}} \put(54.5,19){\scriptsize 111}

\put(14,20){\oval(15,12)[l]} \put(14,26){\line(0,-1){3}}
\put(14,14){\vector(0,1){3}} \put(1.5,19){\scriptsize 000}

\put(42,17){\vector(-1,-1){9}} \put(38.5,11){\scriptsize 110}

\put(33,32){\vector(1,-1){9}} \put(39,27){\scriptsize 011}

\put(18,23){\vector(1,1){9}} \put(17.5,27){\scriptsize 001}

\put(27,8){\vector(-1,1){9}} \put(17.5,11){\scriptsize 100}

\put(29,30){\vector(0,-1){20}} \put(24,19){\scriptsize 010}

\put(31,10){\vector(0,1){20}} \put(32.5,19){\scriptsize 101}

\end{picture}
\end{center}} \\
\parbox[b]{12cm}{\begin{center} \BGr[2]{3} \\
\begin{picture}(120,50)
\put(12,23){\framebox(8,4){$\; 000$}}
\put(30,43){\framebox(8,4){$\; 001$}}
\put(82,43){\framebox(8,4){$\; 011$}}
\put(82,3){\framebox(8,4){$\; 110$}} \put(30,3){\framebox(8,4){$\;
100$}} \put(47,23){\framebox(8,4){$\; 010$}}
\put(65,23){\framebox(8,4){$\; 101$}}
\put(102,23){\framebox(8,4){$\; 111$}}

\put(30,10){\vector(-1,1){10}} \put(20,30){\vector(1,1){10}}
\put(43,45){\vector(1,0){34}} \put(90,40){\vector(1,-1){10}}
\put(100,20){\vector(-1,-1){10}} \put(77,5){\vector(-1,0){34}}

\put(35,10){\vector(0,1){30}} \put(85,40){\vector(0,-1){30}}

\put(40,40){\vector(1,-1){10}} \put(50,20){\vector(-1,-1){10}}
\put(70,30){\vector(1,1){10}} \put(80,10){\vector(-1,1){10}}

\put(57,27){\vector(1,0){6}} \put(63,23){\vector(-1,0){6}}

\put(15,25){\oval(15,12)[l]} \put(15,31){\line(0,-1){3}}
\put(15,19){\vector(0,1){3}}

\put(105,25){\oval(15,12)[r]} \put(105,31){\line(0,-1){3}}
\put(105,19){\vector(0,1){3}}

\put(5,24){\footnotesize 0} \put(115,24){\footnotesize 1}
\put(23,36){\footnotesize 1} \put(23,13){\footnotesize 0}
\put(96,36){\footnotesize 1} \put(96,13){\footnotesize 0}
\put(33,24){\footnotesize 1} \put(87,24){\footnotesize 0}
\put(46,36){\footnotesize 0} \put(46,13){\footnotesize 0}
\put(74,36){\footnotesize 1} \put(73,13){\footnotesize 1}
\put(59,28){\footnotesize 1} \put(59,20){\footnotesize 0}
\put(59,1){\footnotesize 0} \put(59,46){\footnotesize 1}
\end{picture}
\end{center}}\\
\setlength{\unitlength}{0.5mm}
\parbox[b]{7cm}{\begin{center} \BGr[3]{2} \\
\begin{picture}(140,140)
\put(70,35){\framebox(10,5){\scriptsize 11}}
\put(40,50){\framebox(10,5){\scriptsize10}}
\put(40,85){\framebox(10,5){\scriptsize 00}}
\put(70,100){\framebox(10,5){\scriptsize 02}}
\put(100,85){\framebox(10,5){\scriptsize 22}}
\put(100,50){\framebox(10,5){\scriptsize 21}}
\put(15,30){\framebox(10,5){\scriptsize 01}}
\put(70,130){\framebox(10,5){\scriptsize 20}}
\put(125,30){\framebox(10,5){\scriptsize 12}} \thicklines
\put(45,59){\vector(0,1){22}} \put(42,70){\scriptsize 0}

\put(53,92){\vector (2,1){14}} \put(58,96){\scriptsize 2}

\put(83,99){\vector (2,-1){14}} \put(92,96){\scriptsize 2}

\put(105,81){\vector (0,-1){22}} \put(106,70){\scriptsize 1}

\put(97,48){\vector (-2,-1){14}} \put(90,40){\scriptsize 1}

\put(67,42){\vector (-2,1){14}} \put(58,40){\scriptsize 0}

\put(96,53){\vector (-1,0){42}} \put(74,54){\scriptsize 0}

\put(47,58){\vector (2,3){26}} \put(60,72){\scriptsize 2}

\put(77,97){\vector (2,-3){26}} \put(90,72){\scriptsize 1}

\put(38,50){\vector (-1,-1){14}} \put(31,45){\scriptsize 1}

\put(27,34){\vector (1,1){14}} \put(36,39){\scriptsize 0}

\put(123,34){\vector (-1,1){14}} \put(113,39){\scriptsize 1}

\put(112,50){\vector (1,-1){14}} \put(120,44){\scriptsize 2}

\put(77,108){\vector (0,1){19}} \put(78,116){\scriptsize 0}

\put(73,127){\vector (0,-1){19}} \put(71,116){\scriptsize 2}

\put(68,127){\vector (-2,-3){22}} \put(55,110){\scriptsize 0}

\put(103,93){\vector (-2,3){22}} \put(93,110){\scriptsize 0}

\put(43,81){\vector (-1,-2){22}} \put(30,60){\scriptsize 1}

\put(129,37){\vector (-1,2){22}} \put(119,60){\scriptsize 2}

\put(28,30){\vector (4,1){40}} \put(47,32){\scriptsize 1}

\put(82,40){\vector (4,-1){40}} \put(98,32){\scriptsize 2}

\qbezier(19,49)(18,100)(66,131) \put(19,49){\vector (0,-1){11}}
\put(26,92){\scriptsize 1}

\qbezier(131,38)(131,100)(91.5,126) \put(92,125.5){\vector
(-3,2){8}} \put(123,92){\scriptsize 0}

\qbezier(24,26)(75,0)(120,23) \put(120,22.5){\vector(2,1){7} }
\put(75,8){\scriptsize 2}

\put(53,82){\circle{8}} \put(56.5,85.5){\scriptsize 0}
\put(97,82){\circle{8}} \put(99.5,75.5){\scriptsize 2}
\put(75,44.5){\circle{8}} \put(80,45){\scriptsize 1}
\end{picture}
\end{center} }
\end{center}

\medskip
In the picture of \BGr[2]{3} the edges are labelled with the
words of length $3$. In general it is possible to label the the edges
in \BGr[\A ]{n} with the words in $\A^{n+1}$. If $n=0$ we can choose
an arbitrary bijection between the edges of \BGr[\A ]{0} and $\A $.
Thus let $n\in\NN $ and $\BGr[\A ]{n} =(\A^n,\E )$.
Let $w_1,\ldots ,w_{n+1} \in\A $,
we obtain a one-to-one relation between the edges in \BGr[\A ]{n} and
the set $\A^{n+1} $ of words of length $(n+1)$ by:

\begin{equation}\label{edge1}
w_1\ldots w_{n+1}\in\A^{n+1} \longleftrightarrow \big( w_1\ldots w_n
\stackrel{w_{n+1}}{\rightarrow } w_2\ldots w_{n+1} \big)\in\E
\end{equation}
This means that for $u\in\A^{n-1} ,a,b\in\A $ the edge
$au\rightarrow ub $ corresponds to the word $aub\in\A^{n+1} $.
In this way we understand edges in \BGr[\A ]{n} as words over $\A $ of length
$(n+1)$ and vice versa.
If we talk about words of length $(n+1)$ as edges of \BGr[\A ]{n}
it is always meant in the way described above.\\

\medskip
Let $e=aub\in\A^{n+1} $ be an edge in \BGr[\A ]{n} and $e=aub $
with $a,b\in\A ,u\in\A^{n-1}$. Then $au$ is the initial vertex of $e$
and $ub$ is the terminal vertex of $e$. Let $\E\subseteq\A^{n+1} $
a set of edges of \BGr[\A ]{n}. Then it follows that
\[\mbox{\begin{tabular}[t]{ll}
$\E\A^{-1} $ & is the set of initial vertices of edges in $\E $,\\
$\A^{-1}\E $ & is the set of terminal vertices of edges in $\E $.
\end{tabular}}\]

\pagebreak
\subsection*{Walks in \BGr[\A ]{q}}
Let $e_1,e_2\in\A^{n+1} $, with $e_1=aw $ and $e_2 =w'b $.
We can understand $e_1,e_2 $ as edges in \BGr[\A ]{n}, as vertices in
\BGr[\A ]{n+1} or as vertices in $L\BGr[\A ]{n} $.
It follows that:
\begin{quote}
\begin{tabular}{ll}
$\quad $ & $(e_1,e_2)$ is an edge in $L\BGr[\A ]{n} $\\
$\Leftrightarrow $ & The sequence $e_1e_2$ of edges in \BGr[\A ]{n}
is a walk of length $2$.\\
$\Leftrightarrow $ & The terminal vertex of the edge $e_1$ in \BGr[\A ]{n}
is equal to the\\
$\quad $ & initial vertex of the edge $e_2 $ in \BGr[\A ]{n}.\\
$\Leftrightarrow $ & $w=w'$\\
$\Leftrightarrow $  & $e_1\stackrel{b}{\rightarrow } e_2 $ is an edge in \BGr[\A ]{n+1}.\\
\end{tabular}\end{quote}
Therefore we obtain $L\BGr[\A ]{n} \cong \BGr[\A ]{n+1} $. By induction follows:
\begin{equation}\label{linedebr}
L_k\BGr[\A ]{n} \cong \BGr[\A ]{n+k} \quad\mbox{for all }
n,k\in\NN_0\, .
\end{equation}
The vertex set of \BGr[\A ]{n+k} is $\A^{n+k }$ and the vertices
of $L_k\BGr[\A ]{n}$ are the walks of length $k$ in \BGr[\A ]{n}.
It follows that there exists a one-to-one relation between $\A^{n+k} $
and walks of length $k$ in \BGr[\A ]{n}. Therefore we can interpret
words of length $(n+k)$ as walks of length $k$ in \BGr[\A ]{n}.\\

\smallskip
Let $w=w_1\ldots w_{n+k}\in\A^{n+k} $, with $w_1,\ldots ,w_{n+k} \in\A $.
An easy inductive proof shows, that the walk in \BGr[\A ]{n} which
corresponds to $w$ is given by:

\begin{equation}\label{walkword}
 w_1\ldots w_n \stackrel{w_{n+1}}{\rightarrow }
w_2\ldots w_{n+1} \stackrel{w_{n+2}}{\rightarrow } \ldots
\stackrel{w_{n+k}}{\rightarrow } w_{k+1}\ldots w_{n+k} \, .
\end{equation}

\smallskip\noindent
In common we don't make a distinction between a word $w\in\A^{n+k}$ and
its corresponding walk of length $k$ in \BGr[\A ]{n}.
With (\ref{walkword}) the next proposition is obviously.

\pagebreak
\begin{prop}\label{cycleprop1}
Let $P$ be a length $k$ walk in \BGr[\A ]{n} and
$w=w_1\ldots w_{k+n}\in\A^{n+k} $ the corresponding word, where
$w_1,\ldots ,w_{k+n}\in\A $.

\begin{enumerate}
\renewcommand{\labelenumi}{(\roman{enumi})}
\item $P$ is a closed walk  if and only if $ w_1\ldots w_n =
w_{k+1}\ldots w_{n+k} $.

\item $P$ is a path if and only if  $w$
has $k$ different subwords of length $(n+1) $
This means $w_{i+1}\ldots w_{i+n+1} \neq
w_{j+1}\ldots w_{j+n+1} $ for all $0\le i < j \le k-1 $.

\item $P $ is a simple path, but is not a cycle, if and only if $w$
has $k+1$ different subwords of length $n$. This means
$w_{i+1}\ldots w_{i+n} \neq
w_{j+1}\ldots w_{j+n} $ for all $0\le i < j \le k $

\item $P$ is a cycle if and only if the word $w_1\ldots w_{n+k-1} $
has $k$ different subwords of length $n$ and $w_1\ldots w_n =
w_{k+1}\ldots w_{n+k} $.

\item Let $0\le m \le n+k $ and $P_m $ be the
walk of length length $(n+k-m) $ in \BGr[\A ]{n+m} which corresponds to $w$.
If we identify $L_m\BGr[\A ]{n} $ with $\BGr[\A ]{n+m} $, it follows that
$L_mP=P_m $.

\item Let $P $ be
a closed walk, $m\in\NN_0 $.Let $p\in\NN_0 ,0\le l<(n+k) $ such that
$m=p(n+k)+l $.
Then the word $(w_1\ldots w_{n+k})^{p+1}w_1\ldots w_l \in\A^{n+k+m} $ corresponds
to the walk $\hat{L}_mP $ of length $k$ in \BGr[\A ]{n+k}, where we identify
$L_m\BGr[\A ]{n} $ with $\BGr[\A ]{n+m} $.
\end{enumerate}
\end{prop}

\bigskip

\subsection*{Cyclic sequences and closed walks in \BGr[\A ]{n}}

\smallskip
For a word $w\in\A^n $ with $w=w_0\ldots w_{n-1} $, $w_i\in\A ,\;\forall 0\le i<n $,
the {\em cyclic sequence of length $n$}
\[ [w] =[w_0\ldots w_{n-1}] \]
is defined as the map $w_l :\ZZ \rightarrow \A $ given by
\[ w_l :=w_{l \mbox{\scriptsize $\,\mod\, $}n} \quad \forall\, l\in\ZZ \quad ;\quad
[w] = \ldots w_0\ldots w_{n-1}w_0\ldots w_{n-1}\ldots  \quad . \]

\medskip
If we work with cyclic $n$ length sequences, we take all subscripts $\mod n $
without writing this explicitly every time. This means we write $w_l $ in place of
$w_{l \mbox{\scriptsize $\,\mod\, $}n} $ for all $l\in\ZZ $. \\

\pagebreak\noindent
For $t\in\ZZ ,n\in\NN $ and $u\in\A^n $ we define:
\[\begin{array}{lcl}
\Num_w (u) & := &
 \left| \{ l\in\NN_0 | u=w_l\ldots w_{l+n-1}\, ,\, 0\le l < n\} \right|\, , \\
\Sub_w (n) & := & \{ u\in\A^n | u=w_l\ldots w_{l+n-1} \, ,\, l\in\ZZ \}\, , \\

[ w ]_t & := & [ w_t\ldots w_{t+n-1} ]\, ,
\end{array}\]
then $\Sub_w (n)\subseteq\A^n $ is the set of subwords of length n in the
cyclic sequence $[w]$, $\Num_w (u) $ is the total number of occurrence
of the word $u$ as a subword in $[w]$ and $[w]_t $ is the
the $t$-shift of the cyclic sequence $[w]$. Obviously $[w]_t =[w]_s $ for
$s= t\mod n $ and $[w]=[w]_t $ if $t=sn $ for some $s\in\ZZ $.\\

\medskip
We call a word $w\in\A^* $ {\em primitive }, if it is not the power of some other
word in $\A^* $. This means
\[ w\neq u^n \quad \forall\, u\in\A^*-\{ w\} ,\, n\in\NN\, . \]
A cyclic sequence $[w] $ is called primitive, if $w\in\A^+ $ is primitive.

\begin{prop}\label{primitive}
Let $w\in\A^n $ for some $n\in\NN $. The following conditions are equivalent.\\
\begin{enumerate}
\renewcommand{\labelenumi}{(\alph{enumi})}
\item $w$ is primitive.
\item $[w]_t$ is primitive for all $t\in\ZZ $.
\item $[w]\neq[w]_t $ for all $1\le t< n $
\item $|\Sub_w (n) | = n $
\end{enumerate}
\end{prop}
We omit a proof of this proposition.\\

\medskip
Let us show, that there exists a one-to-one relation between
closed pathes of length k in \BGr{n} and cyclic sequences of length k.
Let $v=v_0\ldots v_{n+k-1}\in\A^{n+k} $ with $v_i\in\A $ for all
$0\le i\le n+k-1$ .Furthermore let $p\in\NN_0 $ and $l\in\{ 0,\ldots ,k-1\} $
be the unique numbers with $(n-1)=pk+l $.

\[ \begin{array}{ll}
\quad & v_0\ldots v_{n+k-1} \mbox{ denotes a closed walk of length $k$ in \BGr{n} }  \\
\Leftrightarrow & v_0\ldots v_{n-1}= v_k\ldots v_{n+k-1}  \\
\Leftrightarrow &\begin{array}[t]{cl} \quad & (v_0\ldots
v_{k-1})(v_k\ldots v_{2k-1})\ldots (v_{(p-1)k}\ldots v_{pk-1})
(v_{pk}\ldots v_{pk+l}) \\
= & (v_k\ldots v_{ 2k-1 })\ldots (v_{ (p-1)k }\ldots v_{ pk-1
})(v_{ pk }\ldots v_{ (p+1)k-1} ) (v_{(p+1)k}\ldots v_{(p+1)k+l})
\end{array} \\
\Leftrightarrow & \left\{\begin{array}{rclcl}
v_0\ldots v_{k-1} & = & v_0\ldots v_{k-1} & = & v_{ik}\ldots v_{(i+1)k-1} \\
v_0\ldots v_l & = & v_0\ldots v_l & = & v_{ik}\ldots v_{ik +l }
\end{array}\right.
\quad \forall i\in\{ 0,\ldots , p \}\\
\Leftrightarrow & v=v_{\mbox{\scriptsize $0 \mod k $}}
v_{\mbox{\scriptsize $1 \mod k $}}\ldots v_{\mbox{\scriptsize $n+k-1 \mod k $}}
\end{array}\]

\pagebreak\noindent
This shows, that we can interpret the closed walk $P$ and its corresponding
word $v$ as the cyclic sequence $[v_0\ldots v_{k-1}]$.
Thus there exists a one-to-one relation between walks of length $k$
in $\BGr[\A ]{n} $, cyclic sequences of length $k$ and
words $v=v_0\ldots v_{n+k-1} \in\A^{n+k} $
with $v_0\ldots v_{n-1} = v_{k}\ldots v_{k+n-1} $, as it is shown below:\\

\begin{center}
{\large $ [ \, v \, ] $ }\\
{\footnotesize $v=v_0\ldots v_{k-1}\in\A^k$ }\\[2mm]
{\Large$ \updownarrow $ } \\[2mm]
{\large $v_0\ldots v_{k+n-1} $}\\
{\footnotesize $v_i\in\A\;\forall\, i\, ,\; v_j=v_{j+k}
\;\forall\, 0\le k <n$}\\[2mm]
{\Large $ \updownarrow $ } \\[2mm]
$P:= v_0\ldots v_{n-1}
\stackrel{v_n}{\rightarrow } v_1\ldots v_{n-1}
\stackrel{v_{n+1}}{\rightarrow }\ldots
\stackrel{v_{n+k-1}}{\rightarrow } v_k\ldots v_{n+n-1}  $\\
{\footnotesize $P$ is a closed walk of length $k$ in \BGr[\A ]{n}}
\end{center}
If we talk about cyclic sequences as closed walks,
closed pathes or cycles, it is meant in this way.\\

Let $t\in\ZZ, w\in\A^{k} $, $ u\in\A^{n} $ be a
vertex in \BGr[\A ]{n},
$v\in\A^{n+1}$ be an edge of \BGr[\A ]{n} and $P$ be the closed walk of
length $k$ in \BGr[\A ]{n} which corresponds to $[w]$. Then we obtain:\\

\begin{quote}
$\Sub_w (n) $  is the set of vertices in P.\\[3mm]
$\Sub_w (n+1) $  is the set of edges in P.\\[3mm]
The walk $P$ pass the vertex $u$ $\Num_w (u) $ times.\\[3mm]
The walk $P$ runs $\Num_w (v) $ times through the edge $v$.\\[3mm]
The closed walk $[w]_t $ differs from $P$ only in the starting vertex.
Especially the starting vertex of $[w]_t $ lays $t$ steps forwards in $P$
if $t\ge 0 $ and $t$ steps backwards in $P$ if $t<0 $.\\
\end{quote}

\pagebreak

\begin{prop}\label{cycleprop2}
Let $w\in\A^k $ for some $k\in\NN $.
\begin{enumerate}
\renewcommand{\labelenumi}{(\roman{enumi})}

\item For all $n\in\NN_0 $ we have:
\[ \begin{array}{ll}
\quad & [w] \;\mbox{ is a (closed) path in \BGr{n} (of length $k$)} \\
\Leftrightarrow &
\big| \Sub_w (n+1) \big| =k \\
\Leftrightarrow & \Num_w(u)\in\{ 0,1 \} \quad \forall\,
u\in\A^{n+1}
\end{array} \]

\item For all $n\in\NN_0 $ we have:
\[ \begin{array}{ll}
\quad & [w] \;\mbox{ is a cycle in \BGr{n} (of length $k$)} \\
\Leftrightarrow &
\big| \Sub_w (n) \big| =k \\
\Leftrightarrow & \Num_w(u)\in\{ 0,1 \} \quad \forall\, u\in\A^{n}
\end{array} \]

\item We identify $\BGr[\A ]{n+m} $ with $L_m\BGr[\A ]{n}$.
Let $[w] $ corresponds to the closed walk $P$ in \BGr[\A ]{n}.
Then for every $m\in\NN_0 $
$[w] $ corresponds also to the closed walk $\hat{L}_mP$ in \BGr[\A ]{n+m}.

\item We identify $\BGr[\A ]{n+1} $ with $L\BGr[\A ]{n}$.
Let $[w] $ corresponds to the closed walk $P$ in \BGr[\A ]{n} and
to the closed walk $P'$ in \BGr[\A ]{n+1}.
$P'$ is a cycle if and only if $P$ is a (closed) path.

\item If there exists $n\in\NN $ such that  $[w] $ is a closed path in \BGr{n},
then $[w]$ is a primitive cyclic sequence.

\end{enumerate}
\end{prop}

\smallskip\noindent
{\bf Proof:} (i),(ii),(iii) and (iv) follows from Proposition~\ref{cycleprop1} and
the one-to-one corresponding between cyclic sequences of length $k$ and
words of length $n+k$.
Therefore we have only to show (v).
Let $[w]=[w_0\ldots w_{k-1}]$ be a cyclic sequence,
such that $[w]$ is a closed path
for some $n\in\NN $. Let us assume that $w$ is not primitive. Then there
are $u\in\A^+ $ and $p\ge 2 $ with $w=u^p $. It follows:
\[ w_0\ldots w_{n} =w_{|u|}\ldots w_{|u|+n}
=w_{2|u|}\ldots w_{2|u|+n}=\ldots = w_{(p-1)|u|}\ldots
w_{(p-1)|u|+n-1} \, .\]
Since $(p-1)|u| < |w|=k $, we obtain
that $\Num_w (w_0\ldots w_{n} )\ge p \ge 2 $.
This is a contradiction, because from (i) follows that
$\Num_w (v)\le 1 $ for all $v\in\A^{n+1} $.\qed\\

\newpage

\section{Cycles and 1-factors of \BGr{n}}
In the rest of this chapter we suppose that $|\A | =q<\infty $. Then
$\BGr[\A ]{n} $ is a finite connected $q$-regular graph with $q^n$
vertices and $q^{n+1}$ edges. Since $\BGr[\A ]{n}\cong\BGr{n} $,
it is sufficient to restrict our considerations to the graphs \BGr{n}.
Furthermore we identify $\BGr{n+m} $ with $L_m\BGr{n} $ without writing
this explicitly.

\begin{prop} \label{propdebrfinit}
For every $n\in\NN_0, 1\le k\le q $, \BGr{n} contain an Euler circuit, a Hamilton
circuit and a $k$-factor.
\end{prop}

\smallskip\noindent
{\bf Proof:} Obviously \BGr{n} contain an Euler circuit, because \BGr{n} is
$q$-regular, i.e. an Euler graph.
Since $\BGr{n}$ is finite and $q$-regular, it follows from
Proposition~\ref{factor} that there exists a $k$-factor in \BGr{n}.
Any loop in \BGr{0} is an Euler and a Hamilton circuit.
While $\BGr{n}\cong L\BGr{n-1} $, it follows from the remarks at the end of
Proposition~\ref{linegraph2}, that \BGr{n} has a Hamilton circuit for
$n\ge 1$.Especially if $E$ is an Euler circuit in \BGr{n-1}, then
$\hat{L}E $ is a Hamilton circuit in \BGr{n}.\qed\\

\smallskip
Let $\A =\{ 0,1,2,\ldots ,q-1 \} $.
>From Proposition~\ref{cycleprop2} (ii) follows, that a cyclic
sequence $[w]$ of length $q^n $ is a Hamilton circuit in \BGr{n} if and
only if $\Sub_w (n)=q^n $. This means every word of $\A^n $ occurs
exactly one time as a subword in $[w]$. Such sequences are known
as $q$-ary de Bruijn sequences. From the theorem above follows, that
there are $q$-ary de Bruijn sequences of length $q^n $ for every
$q\ge 2 $ and $n\in\NN $.
The next theorem answer the question
of their number.

\smallskip
\begin{theorem}\label{fullcyc}
There exist $((q-1)!)^{q^{n-1}}\cdot q^{q^{n-1}-n} $ $q$-ary de Bruijn
sequences of length $q^n $ and this is also the number of Hamilton
circuits in \BGr{n}(, where we do not distinguish between cycles and
sequences which differs only
in the starting vertex.
\end{theorem}

\smallskip
The theorem was first shown
by de Bruijn \cite{bruijn}, Good \cite{good} and
Flye-Saint Marie \cite{flye} independently. A proof for the binary case
can also be found in \cite{lint1}. An overview of de Bruijn sequences, their history
and their constructions ca be found in \cite{bruijn},\cite{fred}, \cite{stein},
and \cite{gol1}. However, in this section we focus on cycles of arbitrary
length in \BGr{n}. We show, that there exists cycles in \BGr{n} of length $L$, for
every $1\le L\le q^n $. We give two different proofs. First we construct
cycles of arbitrary lengths in \BGr[2]{n} with the help of a maximal linear cycle.
This construction was given by Golomb in \cite{gol1}. Then we show
in the general case, that there exist cycles of arbitrary length in \BGr{n}.
This was first shown by Lempel in \cite{lempel}. Indeed Lempel's proof
of the $q$-ary case does not give a construction of the cycles.

\pagebreak

\subsection*{Successor maps of 1-factors of \BGr{n} }
Let $\Lambda $  be a 1-factor of \BGr{n}, then $\Lambda $ is the
union of vertex disjoint cycles $ P_1,\ldots ,P_k $. We denote with
$L_1,\ldots ,L_k $ the lengths of these cycles.
We can understand the cycles as (primitive) cyclic
sequences, $[w_1],\ldots ,[w_k] $ with $w_i \in\A^{L_i } $ for
$1\le i \le k $. Let $w_i =w_{0,i}\ldots w_{L_i-1,i} $ for all $1\le i\le k $, where
$w_{0,i},\ldots ,w_{L_i-1,i} \in\A $.
Every $t$-shift $[w_i]_t $ of the cycle $[w_i] $, differs from $[w_i] $
only in its starting and end vertex.\\

Since every vertex of \BGr{n} lays on a unique cycle of the $1$-factor, it follows,
that  $\Sub_{w_1} (n),\ldots ,\Sub_{w_k} (n) $ is a partition of
$\A^n $. This means, that every word in $\A^n $ is a subword of a
unique sequence $[w_1],\ldots ,[w_k ] $.
Let us suppose, that $v\in\A^n $  is a subword of $[w_i]$ for some
unique $i\in \{ 1,\ldots ,k \} $. While $[w_i] $ denotes a cycle
in \BGr{n}, it follows that $\Num_w(v)=1 $. Therefore we obtain an
unique $0\le j < L_i $ with $v=w_{j,i}\ldots w_{j+n-1,i} $. Thus
we can define a map $F:\A^n\rightarrow \A $ by:
\begin{equation}\label{shuffel1}
F(w_{j,i}\ldots w_{n+j-1,i}):=w_{n+j,i} \mbox{ for all } 1\le i
\le k \, ,\, 0\le j < L_i
\end{equation}
Then for every $v:=\A^n $ the vertex
$v_2\ldots v_nF(v) $ is the unique successor vertex of $v$ in $\Lambda $,
where $v=v_1\ldots v_n $ and $v_1,\ldots ,v_n\in\A $.

\smallskip\noindent
In the same way we can define a map $\tilde{F}:\A^n\rightarrow \A $ such
that $\tilde{F}(v)v_1\ldots v_{n-1} $ is the unique antecessor vertex of
$v$ in $\Lambda $. This map is given by:
\begin{equation}\label{shuffel2}
\tilde{F}(w_{j,i}\ldots w_{n+j-1,i}):=w_{j-1,i} \mbox{ for all }
1\le i \le k \, ,\, 0\le j < L_i
\end{equation}

\smallskip\noindent
We call $F $ the {\em successor map} and $\tilde{F} $ the {\em antecessor map}
of $\Lambda $. These maps have the properties:
\begin{equation}\label{shuffel3}
\begin{array}{l}
\mbox{For every }  u\in\A^{n-1} \mbox{ the map } F_u:\A \rightarrow
\A
\mbox{ which is given by }\\
F_u(a):=F(au) \mbox{ is a permutation of } \A \, .
\end{array}
\end{equation}
\begin{equation} \label{shuffel3b}
\begin{array}{l}
\mbox{For every } u\in\A^{n-1} \mbox{ the map } \tilde{F}_u:\A
\rightarrow \A
\mbox{ which is given by }\\
\tilde{F}_u(a):=F(ub) \mbox{ is a permutation of } \A \, .
\end{array}
\end{equation}

\smallskip\noindent
We show only (\ref{shuffel3}), because (\ref{shuffel3b}) follows the same
way.
If $u\in\A^{n-1} \, ,\, a,b\in\A $, then
the vertex $au\in\A^n $ is the antecessor vertex of $uF_u(a)\in\A^n $ in
$\Lambda $. Since the antecessor vertex is unique, it follows that
$uF_u(a)\neq uF_u(b)$ for $a\neq b $. Therefore $F_u $ is a one-to-one
map. While $F_u $ is defined for all $a\in\A $, it follows that $\F_u $ is
a permutation of $\A $.\\

\pagebreak

\begin{prop}\label{succmap}
$\F:\A^n \rightarrow \A $ is a successor map of a (unique) $1$-factor of \BGr{n}
if and only if $F $ fulfill (\ref{shuffel3}).
\end{prop}

\smallskip\noindent
{\bf Proof:}
We have shown already, that a successor map of a $1$-factor
has property (\ref{shuffel3}).
Let  $F:\A^n \rightarrow \A $ be a map which fulfill
property (\ref{shuffel3}). Let $v\in\A^n $ we define the
sequence $(w_l )_{l\in\NN_0 } $ by:
\begin{equation}\label{shuffel4}
w_0\ldots w_{n-1}:= v \mbox{ and by induction } w_l:=
F(w_{l-n}\ldots w_{l-1}) \mbox{ for } l \ge n \, .
\end{equation}

\noindent
We claim:

\begin{equation}\label{successorgl1}
(w_l)_{l\in\NN_0 } \mbox{ is perodical } \, .
\end{equation}

\noindent
While $\left|\A\right|<\infty $, there are $i<j $ such that
$w_{i}\ldots w_{i+n-1} = w_{j}\ldots w_{j+n-1} $ . For $i>0 $ and
$u:=w_i\ldots w_{i+n-2} $ we get $F_u (w_{i-1})=w_{i+n-1}=w_{j+n-1}= F_u(w_{j-1}) $.
Since $F_u $ is a permutation of $\A $, we obtain $w_{i-1} = w_{j-1} $.
It follows that $w_{i-1}\ldots w_{i+n-2} = w_{j-1}\ldots w_{j+n-2} $ and
by induction we find an $L_v\in\NN $ with
\[v=w_0\ldots w_{n-1} =w_{L_v }\ldots w_{L_v+n-1} \, .\]

\noindent
Let us choose $L_v$ minimal, then by definition of
$(w_l)_{l\in\NN_0 } $ we have that the sequence is periodical with
period $L_v $. This shows (\ref{successorgl1}).\\

\smallskip\noindent
Let $w^v:=w_0\ldots w_{L_v-1}$. By (\ref{successorgl1}) we obtain, that
$[w^v] $ is a cyclic sequence of length $L_v $. We claim:

\begin{equation}\label{successorgl2}
[w^v] \;\mbox{ is a cycle.}
\end{equation}

\noindent
It is sufficient to show, that
\[ |\Sub_{w^v} (n) | = L_v \, .\]
Let us assume, that $|\Sub_{w^v} (n) | < L_v $. Then there are
$0\le i< j < L_v$ with $w_{i}\ldots w_{i+n-1} = w_{j}\ldots w_{j+n-1} $
with the same argumentation as above we obtain that
\[ w_{0}\ldots w_n = w_{j-i}\ldots w_{j-i+n-1} \, .\]
This is a contradiction, because we have chosen $L_v$ minimal and $(j-i)< L_v $.
It follows that $|\Sub_{w^v} (n) | = L_v $. By Proposition~\ref{cycleprop2}
follows that $[w^v] $ is a cycle in \BGr{n} of length $L_v$.\\

\pagebreak\noindent
We call $[w^v] $ the cycle which {\em is generated by $F $ and $v$}.
For $w\in\A^L $ with $w=w_0\ldots w_{L-1} $, $w_0,\ldots ,w_{L-1}\in\A $ we obtain:

\begin{equation}\label{Fgeneratedseq2} \begin{array}{ll}
\quad & [w] \mbox{ is the cycle generated by $F$ and $v$ } \\
\Leftrightarrow & w_0\ldots w_{n-1} =v \; \mbox{ and } \; F(w_l\ldots
w_{l+n-1} )=w_{l+n} \quad\forall\, l\in\ZZ
\end{array}
\end{equation}

\smallskip
Let $v,v'\in\A^n $.
Let $[w] $ be the cycle which is generated by $F$ and $v$ and let $[w'] $ be the
cycle which is generated by $F$ and $v'$.
By (\ref{Fgeneratedseq2}) follows, that
$[w]$ and $[w']$ have a vertex (a subword of length $n$) in common
if and only if $[w']=[w]_t $ for some $t\in\ZZ $.
In this case both cycles corresponds to the same subgraph in \BGr{n}.
Let us denote with $\C_v $ the
$1$-regular subgraph which corresponds to the cycle generated by $F$ and
$v$. It follows, that $\Lambda :=\bigcup\limits_{v\in\A^n }\C_v $
is the union of vertex disjoint cycles and every  $v\in\A^n $
is a vertex of $\Lambda $.
Therefore $\Lambda $ is a 1-factor of \BGr{n}.
Obviously $\Lambda $ has $F$ as its successor map.\qed\\

\medskip
Proposition~\ref{succmap} gives us a one-to-one relation between 1-factors of
$\Lambda $ and maps $F:\A^n \rightarrow \A $ with property
(\ref{shuffel3}). Therefore we call maps with
property (\ref{shuffel3}) successor maps. For a successor map $F$ we
obtain the corresponding $1$-factor by taking the union of all cycles
generated by $F$ and vertices $v$ of \BGr{n}. With the same
arguments, it can be shown, that there is also a one-to-one
correspondence of 1-factors and maps $\tilde{F}:\A^n\rightarrow \A $
with property (\ref{shuffel3b}). Therefore we call maps
with property (\ref{shuffel3b}) antecessor maps.\\

Every $1$-factor of $\BGr{n} $ can be constructed by choosing
for every $u\in\A^{n-1} $ a permutation $F_u $ of $\A $.
Then the map given by $F(au):=F_u (a) $ for all $u\in\A^{n-1} ,a\in\A $, is
a successor map of a (unique) $1$-factor in \BGr{n}.
Since the number of
permutations of $\A $ is $q! $ and $\left| \A^{n-1} \right|
=q^{n-1} $ it follows:

\begin{prop}\label{1facnumber}  There are $q^{q!\cdot (n-1)} $ different 1-factors of
\BGr{n}.
\end{prop}

\smallskip
One might ask, wether a given cycle in $\BGr{n} $ can be extended to a
1-factor. The next lemma shows, that this is possible for
every cycle in \BGr{n}.
\begin{lemma}\label{shuffel4}
Let $\Gamma $ be a 1-regular subgraph of $\BGr{n} $. (i.e. $\Gamma $ is the union
of vertex disjoint cycles), then there exists  a 1-factor $\Lambda $ of
$\BGr{n} $ which is an extension of $\Gamma $.
This means $\Gamma\subseteq\Lambda $.
\end{lemma}

\pagebreak
\noindent
{\bf Proof:} Let $\Gamma $ be a $1$-regular subgraph of \BGr{n} and let
$\V $ be the vertex set of $\Gamma $.
Then $\Gamma $ is the vertex disjoint union of some cycles $P_1,\ldots ,P_k $.
Let $[w_1],\ldots , [w_k] $ be cyclic sequences which corresponds to these cycles.
Let $L_i $ be the length of the sequence $[w_i]$ and
$w_i=w_{0,i}\ldots w_{{L_i-1},i } $ with $w_{0,i},\ldots ,w_{L_i-1,i }\in\A $
for all $1\le i \le k $. \\

\noindent
Since the cyclic sequences are vertex disjoint cycles, there are
unique $1\le j \le k $ and $0\le i \le L_j $ for
every $v\in\V $, such that $v = w_{i,j} \ldots w_{i+n-1,j} $.
Thus we can define the map $F: \V \rightarrow \A $ as:
\[ F(v):=w_{i+n,j} \;\mbox{ for all}\; v\in\V \, .\]

\noindent
Let $v=v_1\ldots v_n \in\A^n $ with $v_1,\ldots , v_n \in\A $. We obtain that
$v_2\ldots v_nF(v)\in\V $ is the unique successor vertex of $v$ in
$\Gamma $. For $u\in\A^{n-1} $ we define the set $\A_u $ as:
\[ \A_u :=\{ a | v=au ,a\in\A ,au\in\V \} \, .\]

\noindent
If $F(au)=F(bu) $ for some $a,b\in\A_u $ with
$a\neq b$, it follows that $au ,bu\in\V $ are two different antecessor vertices of
the vertex $uF(au)=uF(bu) $ in $\Gamma $. This is a contradiction,
because $\Gamma $ is $1$-regular. We conclude that
the map $F_u:\A_u \rightarrow \A$  is a one-to-one map, where $F_u $ is defined
for $u\in\A^{n-1} $ as $F_u(a):=F(au) $ for all $a\in\A_u$.
(If $au\not\in\V $ for all $a\in\A $, then $F_u $ is the empty map.)
Since $F_u $ is one-to-one, we can extend $F_u $ to a
permutation of $\A $. This gives us an extension of the map
$F :\V\rightarrow \A $ to a map $F':\A^n\rightarrow \A $. Obviously $F' $
fulfill the property (\ref{shuffel3}). Therefore $F' $ is a successor map
of some $1$-factor $\Lambda $. Since $F' $ is an extension of $F $, it follows
by the definition of $F$, that $\Gamma \subseteq\Lambda $.\qed\\

\subsection*{Maximal linear cycles in \BGr[2]{n} }
A map $F:\A^n \rightarrow \A $ is called a {\em linear map},
if there exists $c_1,\ldots ,c_n \in\A $ such that
\[ F(w):= c_1w_1+\ldots +c_nw_n \mod q \;\mbox{ for }\; w=w_1\ldots w_n\in\A^n \, ,\]

\noindent
Furthermore the map $F $ is called a {\em linear successor map}, if
$c_1 $ is not a divisor of zero in $\ZZ_q $. This means $c_1a\mod q\neq 0 $
for all $a\in\A $. For example $c_1 = 1 $ is not a divisor of zero for all $q\ge 2 $,
but $2$ is a divisor of zero in $\ZZ_4 $, since $2\cdot 2\mod 4=0 $.\\

\pagebreak
Let us show, that every linear successor map $F$ is really a successor map.\\
Let $u:=u_1\ldots u_{n-1} $ and $a,b,u_1,\ldots u_{n-1} \in\A $,
then:
\[
\begin{array}{ll}
\quad & F_u(a) =F_u(b) \Leftrightarrow
c_1a+c_2u_1+\ldots +c_nu_n\mod q =c_1b+c_2u_1+\ldots +c_nu_n \mod q \\
\Leftrightarrow & c_1a\mod q =c_1b\mod q \\
\Leftrightarrow & c_1(a-b)\mod q =0 \, . \\
\end{array}
\]
Since $c_1 $ is not a divisor of zero, by the last equation
follows:
\[ (a-b) \mod q= 0\Leftrightarrow a=b \, .\]
We conclude that $F$ has property  (\ref{shuffel3}), i.e. $F$ is a successor map
of some $1$-factor of \BGr{n}.\\

If $F$ is a linear successor map, then the cycle which is generated by $F$ and
$0^n$ is the loop at the vertex $0^n $.
Therefore $q^n-1 $ is the maximal possible length
of a cycle $[w] $ which is generated by a linear successor map $F$ and a vertex
$v\in\A^n$. If $[w] $ is of maximum length $q^n-1$, then
every word in $\A^n -\{0^n\} $ is a subword of $[w] $
and for every word
$v'\in\A^n -\{ 0^n\} $ the cycle $[w'] $ which is generated by $F$ and $v'$
is a $t$-shift of $[w] $ for some $t\in\ZZ $.

\begin{Def}\label{maxlincyc}{\upshape (and proposition)}\\
\begin{enumerate}
\renewcommand{\labelenumi}{(\roman{enumi})}
\item The cyclic sequence $[w] $ is called
a {\em maximal linear cycle }in \BGr{n}, if $[w] $ has length $q^n-1$ and
is generated by a linear successor map $F:\A^n\rightarrow \A $ and some
$v\in\A^n$.
\item A linear successor map $F:\A^n\rightarrow \A $ is called
{\em maximal linear (successor) map }, if $F $ generates for some $v\in\A^n $
a maximal linear cycle.

\item Let $[w] $ be a maximal linear cycle which is generated by $F$ and
$v\in\A^n $. Then $\Sub_w (n) =q^n-1 $, $\Num_w (0^n )=0 $ and
$\Num_w (u) =1 $ for all $u\in\A^n-\{ 0^n \} $. If $[w'] $ is another
cyclic sequence, which is generated by $F $ and some $v'\in\A^n-\{ 0^n \} $,
then $[w'] = [w]_t $ for some $0\le t < q^n-1 $.
\end{enumerate}
\end{Def}

\pagebreak
The question rises, wether maximal linear cycles exist in \BGr{n}?
We will give an answer only for the binary case.

\noindent
The Euler $\varphi $-function is defined as follows: \\
For $n>1 $, let
\begin{equation}
n=\prod\limits_{i=1}^m p_i^{k_i} \mbox{ with } p_1<\ldots <p_m
\mbox{ primes, } k_1,\ldots ,k_m\in\NN
\end{equation}
be the unique factorization of n into primes. Then $\varphi $ is defined
as:
\begin{equation}
\phi (n) :=\left\{ \begin{array}{cl}
1 & \quad\mbox{ if } n=1 \\
\prod\limits_{i=1}^m p_i^{k_i-1}(p_i-1) & \quad\mbox{ if } n>1
\end{array}\right.
\end{equation}
For $q\ge 2 $ the function  $\lambda_q :\NN \rightarrow
\NN $ is given by:
\begin{equation}
\lambda_q (n) := \frac{\phi (q^n-1)}{n}
\end{equation}

\smallskip
\begin{theorem}[Golomb \cite{gol1}] \label{cycnumber}
Let  $\A :=\{ 0, 1\}$. Then
there exists $\lambda_2(n) $  maximal linear maps
$F:\A^n\rightarrow \A $.
\end{theorem}
A proof of the theorem  can be found in \cite{gol1}.\\

\smallskip
Since $\lambda_2(n) \ge 1$ for all $n\in\NN $, it follows that there
exist maximal linear cycles in \BGr[2]{n} for all $n\in\NN $. Furthermore
we obtain the following proposition:

\begin{prop}\label{maxcycprop}
Let $n\in\NN $. There are $\lambda_2(n)\cdot (2^n-1) $ maximal linear binary cycles
and \BGr[2]{n} contains $\lambda_2 (n) $ different subgraphs
of maximal linear cycles.
\end{prop}

\pagebreak

\subsection*{Cycles of arbitrary lengths in \BGr[2]{n} obtained from maximal
linear cycles}

\smallskip
In this section we give a construction of cycles in \BGr[q]{n}
of arbitrary length, by splitting a maximal cycle into two cycles of length
$L $ and length $q^n-1-L $. The construction was proposed by Golomb in \cite{gol1}.
It works for every maximal cycle
of \BGr{n}. Therefore every maximal cycle in \BGr{n} gives us cycles
of length $L$ in \BGr{n} for every $1\le L \le q^n-1 $. Since there exists
maximal linear cycles in \BGr[2]{n} for every $n\in\NN $, it follows
that there exist cycles of length $L$ in \BGr[2]{n} for every $1\le L \le 2^n-1 $.
While $10^{n-1}\rightarrow 0^{n-1}1 $ is an edge of every
maximal linear cycle in \BGr{n},
we obtain an Hamilton circuit for \BGr{n} by replacing the edge
$10^n\rightarrow 0^n1 $ of a maximal linear cycle
with the path $10^{n-1}\rightarrow 0^n\rightarrow 0^{n-1}1 $.
Therefore we can obtain cycles of arbitrary lengths in \BGr{n} from only one
maximal linear cycle of \BGr{n}. Constructions of maximal linear cycles in \BGr{n}
with primitive polynomials ca be found in \cite{gol1}.\\

\smallskip\noindent
Let $F:\A^n \rightarrow \A $ be a maximal linear successor map.
\begin{equation}\label{group0}\begin{array}{l}
F(v_1\ldots v_n):= c_1v_1+\ldots c_nv_n \quad\mbox{ for all }\;
v_1,\ldots, v_n\in\A \;\mbox{ with }\\
c_1,\ldots ,c_n\in\A \;
\mbox{ such that } c_1 \;\mbox{ is not a divisor of zero in } \ZZ_q \, .
\end{array}
\end{equation}
The cyclic sequence $[w] $ which is generated by $F$ and $10^{n-1}\in\A^n $
is a maximal linear cycle, i.e. $|w|=|\Sub_w (n)|=
q^n-1 \, ,\; \Sub_w (n)=\A^n -\{ 0^n \} $ and every word in $\A^n
-\{ 0^n \} $ occurs exactly one time as a subword in $[w] $.\\

\noindent
$[w]_t $ is for all $t\in\ZZ $ a maximal linear
cycle, too. It differs from $[w] $ only in the start- and end vertex.
Let $v\in\A^n-\{ 0^n \} $, then there exists a unique $t$ with $0\le t < q^n-1 $
such that the cycle which is generated by $F$ and $v$
is the maximal linear cycle $[w]_t $.\\

\noindent
Let
\begin{equation}\label{group1}
[{\bf 0}]:=[0^{q^n-1}] \;\mbox{ and }\; \G:=\big\{ [{\bf 0}], [w]
,[w]_1 ,\ldots ,[w]_{q^n-2} \big\} \, .
\end{equation}
Then $|\G |=q^n $. We define the binary operations $\oplus $ and
$\ominus $ on $\G $ by:
\noindent
{\small
\begin{equation}\label{group2}
\begin{array}{l}
\mbox{For } [w'],[w'']\in\G  \;\mbox{ with } w'=w'_0\ldots
w'_{q^n-2 } \, ,\; w''=w''_0\ldots w''_{q^n-2 }, \\
w'_l,w''_l\in\A \;\mbox{ for all }\; 0\le l< q^n-1 :\\[1mm]

[w'] \oplus [w''] := [u_0\ldots u_{q^n-2}] \mbox{ with }
u_l:= (w'_l+w''_l)\mod q \; \mbox{ for } 0\le l < q^n-1\\[1mm]

[w'] \ominus [w''] := [u_0\ldots u_{q^n-2}] \mbox{ with } u_l:=
(w'_l-w''_l)\mod q \; \mbox{ for } 0\le l < q^n-1
\end{array}
\end{equation} }
\noindent
This means, $\oplus $ is the componentwise addition mod $q$ of cyclic sequences.

\pagebreak\noindent
\begin{lemma}\label{goloumb1}
$\big( \G ,\oplus , [{\bf 0}] \big) $ is an Abelian group
with identity $[{\bf 0}] $ and $[{\bf 0}]\ominus [w'] $ is the
negative element of $[w'] $ for all $[w']\in\G $, i.e.
$([w']\oplus [w''])\, ,\, ([w']\ominus [w''])\in\G $ for all
$[w'],[w'']\in\G $
\end{lemma}

\smallskip\noindent
{\bf Proof:} Let $w=w_0\ldots w_{q^n-2} $ with $w_0,\ldots
,w_{q^n-2}\in\A $.
We have to show:\\

\begin{enumerate}
\renewcommand{\labelenumi}{(\roman{enumi})}

\item $([w'] \oplus [w''])\oplus [w'''] =[w'] \oplus ([w'']\oplus
[w'''])$ for all $[w'],[w''],[w''']\in\G $,

\item $[w'] \oplus
[w''] =[w''] \oplus [w'] $ for all $ [w'],[w'']\in\G$,

\item $[w']
\oplus [{\bf 0}] =[w'] $ for all $ [w']\in\G$,

\item For all
$[w']\in\G  $ there is $[w'']\in\G $ with $[w'']= [{\bf 0}]\ominus
[w'] $ and \\
$[w']\oplus [w''] = [w']\ominus [w'] =[{\bf 0}] $,

\item
$[w']\oplus [w''] \in G $ for all $[w'],[w'']\in\G  $.

\end{enumerate}
Obviously (i)-(iii) hold by the definition of $\oplus $ and (iv)
holds for  $[w']=[{\bf 0}] $.Let $[w']=[w]_t $ for
some $0\le t< q^n-1 $. Let $w'=w'_0\ldots w'_{q^n-2}=
w_{t}\ldots w_{t+q^n-1} $ , $w''_l := (-w'_{l})\mod q $ for all $l$
and
\[ [w'']:=[{\bf 0}]\ominus [w']=[ w''_0\ldots w''_{q^n-2}] \, . \]
We obtain $(w''_0\ldots w''_{n-1})\, ,\, (w'_0\ldots
w'_{n-1}) \neq 0^n $, because $\Sub_w(n) =\A^n-\{ 0^n \} $.

\smallskip\noindent
Since $F:\A^n \rightarrow \A $ is linear and $[w'] $ is
generated by $F$ and $w'_0\ldots w'_{n-1} $ we obtain for all
$l\in\ZZ $ :
\begin{eqnarray*}
0 & = & F \bigg( (w'_l+w''_l \mod q)\ldots
(w'_{l+n-1}+w''_{l+n-1}\mod q) \bigg) \\
\quad & = & F \big( w'_l\ldots w'_{l+n-1} \big) +
F \big( w''_l\ldots w''_{l+n-1}\big) \mod q \\
\quad & = & w'_{l+n}+F \big( w''_l\ldots w''_{l+n-1}\big) \mod q \, .
\end{eqnarray*}
It follows that:
\[ F \big( w''_l\ldots w''_{l+n-1}\big)=(-w'_{l+n})\mod q =w''_{l+n}
\quad \forall\, l\in\ZZ \, . \]
The cyclic sequence generated by
$F$ and $w''_0\ldots w''_{n-1} $ satisfies the above equation as well.
Since $F$ is maximal and $w''_0\ldots w''_{n-1}\neq 0 $, the
sequence has length $q^n-1 $. Therefore $[w''] $ is the
maximal linear cycle generated by $F$ and $w''_0\ldots w''_{n-1}$.
While $\Sub_w (n) =\A^n-\{ 0^n \} $, there exists a
(unique) $s\in\{ 0\ldots ,q^n-2 \} $ such that
$w''_0\ldots w''_{n-1}=w_{s}\ldots w_{s+n-1} $.
Thus we obtain $[w'']=[w]_s\in\G $.
By the definition of $[w''] $ it follows that
$[w']\oplus [w''] = [w']\ominus [w'] = [{\bf 0 }] $. This shows (iv).\\

\pagebreak\noindent
Let $[w'],[w'']\in G $. If one of
the cyclic sequences is equal to $[{\bf 0 }] $, then (v) is obviously true.
Thus let $[w'],[w'']\neq [{\bf 0 }]$.
Then both sequences are generated by F and their first n terms.
Let $[v_0\ldots v_{q^n-2}] :=[w']\oplus [w''] $
Since $F$ is linear, it follows:
\begin{equation}\label{groupgl1}
 F(v_{l}\ldots v_{l+n-1} )=v_{l+n} \quad \forall\, l\in\ZZ \, .
\end{equation}
Thus $[v_0\ldots v_{q^n-2}] =[{\bf 0 }] $ if and only if $v_0\ldots v_{n-1} = 0^n $.
In this case we have $[w']\oplus [w'']\in\G $.
Let us suppose that $v_0\ldots v_{n-1} \neq 0^n $.
While $F$ is maximal the cyclic sequence which is
generated by $F$ and $v_0\ldots v_{n-1} $ also satisfy
(\ref{groupgl1}), i.e. this sequence has length $q^n-1 $.
Therefore $[w']\oplus [w''] $ is the cyclic sequence which is generated by $F$ and
$v_0\ldots v_{n-1} $. Since $\Sub_w (n) =\A^n-\{ 0^n \}$,
we obtain that $v_0\ldots v_{n-1}=w_{t}\ldots w_{t+n-1} $ for
some (unique) $0\le t < q^n-1 $. It follows that
$[w']\oplus [w''] =[v_0\ldots v_{q^n-2}]=[w]_t\in\G $. This shows (v).\qed\\

\medskip
Let $\Gamma:=(\V ,\E ) $ be a digraph without multiple edges.
Let $P$ be a cycle in $\Gamma $. Since $\Lambda $ has no multiple edges, we
can interpret $P$ as a sequence $v_1\ldots v_n $ of vertices with
$v_i\in\V $ for all $1\le i\le n $.
$P$ can be {\em split into two cycles} $P_1 ,P_2 $,
if there exists i,i $1\le i<j \le n $ such that
\[ P_1=v_1\ldots v_iv_{j+1}\ldots v_{n+1} \, ,\quad P_2=v_{i+1}\ldots v_jv_{i+1} \, .\]
This is possible if and only if
$v_i\rightarrow v_{j+1}  $ and $v_j\rightarrow v_{i+1}  $ are edges of $\Gamma $.
Obviously these edges are not contained in the cycle $P$.\\

\smallskip

\begin{theorem}[Goloumb \cite{gol1}]\label{goloumb2}
Let $\, 2\le L \le q^n-2 $. Every maximal linear cycle in
\BGr{n} can be split into two cycles of length $L$ and $q^n-L-1 $.
\end{theorem}

\smallskip\noindent
{\bf Proof  :} Let $[w] $ be a maximal linear cycle in \BGr{n}, where
$w=w_0\ldots w_{q^n-2} $ with
$w_0,\ldots ,w_{q^n-2}\in\A $.
Let $v\in\A^n $ and $F:\A^n \rightarrow \A $ be a maximal linear map, such
that $[w] $ is generated by $F $ and $v$.
\[ \begin{array}{l}
F(u_1\ldots u_n):= c_1u_1+\ldots c_nu_n \mod q \quad \mbox{ for
all }
u_1,\ldots ,u_n\in\A \\
\mbox{with } c_1,\ldots , c_n\in\A \mbox{ and } c_1\neq 0 \mbox{
is not a divisor of zero in } \ZZ_q  \end{array} \, .\]
Since $F$ is maximal, the cycle
which is generated by $F$ and $10^{n-1}\in\A^n $,
is a $t$-shift of $[w]$ for some $0\le t < q^n-1 $, i.e.
$[w]$ and the cycle  which is generated by $F$ and $10^{n-1} $ have the same
subgraph in \BGr{n}.
Therefore it is sufficient to show the theorem for
$w_0\ldots w_{n-1}=10^{n-1} $.\\

\smallskip\noindent
Thus let $w_0\ldots w_{n-1}=10^{n-1} $.
Let $\oplus , \ominus $ be as in Lemma~\ref{goloumb1} the componentwise addition and
substraction modulo $q $ of cyclic sequences.
We have $[w] \neq [w]_L $ for all L with $1\le L  < q^n -1 $, because $[w] $
is a cycle in \BGr{n}.
Therefore
\[ w_0\ldots w_{n-1} \neq w_{L}\ldots w_{L+n-1} \, .\]

\pagebreak
By Lemma~(\ref{goloumb1}) follows, that for every
$1\le L < q^n-1 $ there is a unique $0\le M < q^n-1$ such that:
\begin{equation} \label{golnr1}
[w]\ominus  [w]_L=[w]_M \Rightarrow [w] =[w]_L\oplus [w]_M \, .
\end{equation}
Since $[w]_M =[w_{M}\ldots w_{M+q^n-2}] $ is also maximal linear,
it follows that every word in $\A^n -\{ 0^n \}$ occurs as a subword
in $[w]_M$.
Thus, for every $1\le a\le (q-1) $ there exists an $m$ with $0\le
m \le (q^n-2) $, such that $w_{m+M}\ldots w_{m+M+n-1} =0\ldots 0a $.
With (\ref{golnr1}) follow:
\begin{eqnarray*}
w_m & = & w_{m+L} \, ,\\
w_{m+1} & = & w_{m+L+1} \, , \\
\quad & \vdots & \quad \, ,\\
w_{m+n-2} & = & w_{m+L+n-2}  \, ,\\
w_{m+n-1} & = &  w_{m+L+n-1} + a \mod q \, .
\end{eqnarray*}
It follows that we can split the maximal linear cycle $[w] $ into
two cycles of length $L$ and $(q^n-1-L)$. These cycles are given by:\\

\smallskip
\begin{quote}
{\bf Cycle of length $L$:}
\begin{eqnarray*}
\qquad &\qquad & w_m\ldots w_{m+n-1} \\
\qquad & \stackrel{w_{m+n}}{\longrightarrow} & w_{m+1}\ldots w_{m+n} \\
\qquad & \longrightarrow & \ldots \\
\qquad & \stackrel{w_{m+L+n-3}}{\longrightarrow} & w_{m+L-2}\ldots w_{m+L+n-3}\\
\qquad & \stackrel{w_{m+L+n-2}}{\longrightarrow} &
w_{m+L-1}\ldots w_{m+L+n-2} =w_{m+L-1}w_{m}\ldots w_{m+n-2}\\
\qquad & \stackrel{w_{m+n-1}}{\longrightarrow} &  w_m\ldots w_{m+n-1}\\
\qquad &\qquad & \qquad \\
\end{eqnarray*}

{\bf Cycle of length $(q^n-1-L)$:}
\begin{eqnarray*}
\qquad &\qquad & \qquad \\
\qquad & \qquad & w_{m+L}\ldots w_{m+L+n-1} \\
\qquad & \stackrel{w_{m+L+n}}{\longrightarrow} & w_{m+L+1}\ldots w_{m+L+n} \\
\qquad & \longrightarrow & \ldots \\
\qquad & \stackrel{w_{m+q^n+n-3}}{\longrightarrow} &
w_{m+q^n-3}\ldots
w_{m+q^n+n-3}\\
\qquad & \stackrel{w_{m+q^n+n-2}}{\longrightarrow} &
w_{m+q^n-2}\ldots w_{m+q^n+n-2} =w_{m+q^n-2}w_{m}\ldots w_{m+n-2}\\
\qquad &\quad &= w_{m+q^n-2}w_{m+L}\ldots w_{m+L+n-2}\\
\qquad & \stackrel{w_{m+L+n-1}}{\longrightarrow} & w_{m+L}\ldots
w_{m+L+n-1} \hspace{3cm}\mbox{\qed }
\end{eqnarray*}
\end{quote}
\pagebreak

\begin{prop}\label{goloumb3}
Let $\A =\{ 0,1 \} $. There exists a cycle of length $L$ in \BGr[2]{n}
for every $1\le L \le 2^n $.
\end{prop}

\smallskip\noindent
{\bf Proof:} By Proposition~\ref{maxcycprop} there exists a maximal linear
cycle in \BGr[2]{n}. If \\
$2\le L \le 2^n-1 $, it follows from Theorem~\ref{goloumb2},
that there exists a cycle of length $L$ in \BGr[2]{n}. If $L=2^n$, we obtain
a cycle of length $2^n $ by replacing the edge
$10^{n-1}\rightarrow 0^{n-1}1 $ of a maximal linear cycle with the path
$10^{n-1}\rightarrow 0^n \rightarrow 0^{n-1}1 $. If $L=1 $, then the loop
$0^n\rightarrow 0^n $ is a cycle of length $1$ in \BGr[2]{n}. \qed.\\

\bigskip
\subsection*{Cycles of arbitrary lengths in \BGr{n}}

\begin{theorem}[Lempel\cite{lempel}]\quad\\ Let $1\le L\le q^n$. Then
\BGr{n} contains a cycle of length~$L$.
\end{theorem}
Obviously \BGr{n} contain a cycle of length $1$ for every $n\in\NN_0 $.
For example the loop $0^n\stackrel{0}{\rightarrow } 0^n $ is such a cycle.
Let $n\in\NN $ and $\Lambda $ be a connected Euler subgraph in $\BGr{n-1}$ with $L$
edges. If $E$ is an Euler circuit for $\Lambda $, then from
Proposition~\ref{linegraph2} (iii) follows that $\hat{L}E $ is a cycle of length $L$
in \BGr{n}. Therefore it is sufficient to show, that \BGr{n} contain
a connected Euler subgraph with $L$ edges for every $1\le L\le q^{n+1} $ and
$n\in\NN_0 $.

\begin{lemma}[Lempel]\label{lempel} Let $n\in\NN_0 $.
For every  $1\le L \le q^{n+1} $ there exists a connected Euler
subgraph in $\BGr{n} $ with $L$ edges.
\end{lemma}
\smallskip\noindent
{\bf Proof: } We proof the lemma by induction on $n$.\\
\noindent
{\bf n=0}\\
For $L$ with $1\le L \le q $ we can choose $L$ loops in $\BGr{0} $.
This gives us a connected Euler subgraph of $\BGr{0} $ with $L$
edges. Therefore the lemma holds for $n=0 $.\\

\smallskip\noindent
{\bf n-1$\rightarrow $ n}\\
Let us assume that there exists a connected Euler subgraph in
\BGr{n-1} with $L$ edges for every $1\le L \le q^n $.
Let $1\le L \le q^{n+1} $. We show that there exists a connected Euler subgraph in
\BGr{n} with $L$ edges. We distinguish two cases.\\

\pagebreak\noindent
{\em Case 1:} $L\le q^n$ \\
By induction hypothesis \BGr{n-1} contains a connected
Euler graph with $L$ edges. If  $P$ is an Euler circuit for this subgraph,
then $\hat{L}P $ is a cycle of length $L$ in \BGr{n}, where we identify
$\BGr{n}$ with $L\BGr{n-1} $.
This follows from Proposition~\ref{linegraph2} (iii). Since a cycle of
length $L$ is a connected $1$-regular graph with $L$ edges, we have found a
connected Euler subgraph in $\BGr{n} $ with $L$ edges.\\

\medskip\noindent
{\em Case 2:} $q^n< L\le q^{n+1} $\\
Let $1\le m\le (q-1) \,$, $0<k\le q^n $ such that $L=mq^n+k $ and
$L':= q^n-k <q^n $.\\

\smallskip\noindent
If $L'>0 $, then by the induction hypothesis it follows that \BGr{n-1}
contains a connected Euler graph with $L'$ edges.
Let $E$ be an Euler circuit in this graph. Then from
Proposition~\ref{linegraph2} (iii) follows that
$C:=\hat{L}P $ is a cycle of length $L'$ in \BGr{n}.
If $L'=0$, then we choose an arbitrary vertex in \BGr{n}. This gives us
a cycle $C$ of length $0$ in \BGr{n}.
>From Lemma~\ref{shuffel4}
follows that there is a $1$-factor $\Gamma_1 $ in $\BGr{n} $
which contains the cycle $C$.
While $m+1\le q $ and $\BGr{n}$ is $q$-regular, from Proposition~\ref{factor} follows
that there exists a $(m+1)$-factor $\Gamma_2 $ of \BGr{n}
with $\Gamma_1 \subseteq \Gamma_2 $.
Let $\Gamma_3 $  be the
complement of $C$ in $\Gamma_2$. We obtain $\Gamma_3 $ if we delete in $\Gamma_2 $
all edges of $C$.
Since $\Gamma_2 $ is $m+1$-regular and $m+1\ge 2$, it follows, that
the graph $\Gamma_3 $ is a spanning Euler subgraph in $\BGr{n} $ without isolated
vertices. Since $C$ is contained in $\Gamma_1\subseteq\Gamma_2 $
and has $L'$ edges, we obtain for the number of edges in $\Gamma_3 $:

\[ (q^n-L')+mq^n=(q^n-(q^n-k))+mq^n = mq^n+k =L \, .\]

\smallskip\noindent
Let $\Lambda_1,\ldots ,\Lambda_p $ be the connected components of
$\Gamma_3 $. If $p=1 $, then the proof is finished, because in this case
$\Lambda_3 $ is a connected Euler subgraph of \BGr{n} with $L$
edges.

\smallskip\noindent
Thus assume, that $p\ge 2 $.
While \BGr{n} is connected and $\Gamma_3 $ is a spanning subgraph of $\BGr{n} $,
we conclude, that there is an edge $e$ in $\BGr{n} $
whose initial and terminal vertices are in two different connected components of
$\Gamma_3 $. Let $w_1\ldots w_{n+1}\in\A^{n+1} $ be the label of $e$,
where $w_1,\ldots ,w_{n+1}\in\A $.
Then $w:=w_1\ldots w_n $ is the initial vertex of $e$ and $w':=w_2\ldots w_{n+1} $
is the terminal vertex. Let $w$ contained in $\Lambda_i $ and
$w' $ contained in $\Lambda_j $ for some $i,j $ with $i < j $.
While $\Gamma_3 $ is an Euler graph without isolated vertices,
it follows that
there is an edge in $\Lambda_i $ with initial vertex $w$ and an
edge in $\Lambda_j $ with terminal vertex $w'$.
This means, there are $a,b\in\A $ with $a\neq w_1 $ and $b\neq w_{n+1} $, such
that $w_1\ldots w_nb $ is an edge in $\Lambda_i $ and $aw_2\ldots w_{n+1} $
is an edge in $\Lambda_j $.
It follows that $aw_2\ldots w_nb $ is an edge, different to $w_1\ldots w_{n+1} $,
which also connect  $\Lambda_i $ with $\Lambda_j $.
Especially the edge $aw_2\ldots w_nb$ is not an edge in $\Gamma_3 $.
We delete the edges $w_1\ldots w_nb $ and
$aw_2\ldots w_{n+1} $ in $\Gamma_3 $ and replace them by the edges
$w_1\ldots w_{n+1} $ and $aw_2\ldots w_nb $.

\smallskip\noindent
Thus we obtain a new spanning subgraph $\Gamma_4 $ of $\BGr{n} $
with $L$ edges which is also an Euler graph without isolated vertices,
but has only $(p-1) $ connected components. Especially
the connectivity components of $\Gamma_4 $ are

\[ \Lambda_i\cup\Lambda_j ,\Lambda_1,\ldots ,\Lambda_{i-1} ,\Lambda_{i+1},
\ldots \Lambda_{j-1} ,\Lambda_{j+1} ,\ldots ,\Lambda_p \, . \]

\smallskip\noindent
If we proceed to connect the connectivity components of $\Gamma_3 $
in this way, we obtain after $(p-1) $ steps a connected spanning
Euler subgraph of $\BGr{n} $ with $L$ edges. \qed \\

\section{Regular subgraphs of \BGr{n} }

In this section we study  $k$-regular subgraphs of \BGr{n}, i.e.
we deal with the question, wether there exists a $k$-regular subgraph
of \BGr{n} for a given number of vertices. Throughout this section
we denote with $\A $ the alphabet $\A =\{ 0,\ldots q-1 \} $ for
some $q\ge 2$.\\

\smallskip
Let $\Lambda \subseteq (\V ,\E ) $ be a subgraph of \BGr{n}.
If we understand the edges in $\Lambda $ as words of length $n+1$, then
we obtain that $\Lambda $ is $k$-regular if and only if there exist
for every $v\in\V $ unique sets $\B_a, \B_s\subseteq\A $ such that
$\B_av , v\B_s\subseteq \E $ and $|\B_a |=|\B_v |= k $.
In this case $\B_av $ are the edges of $\Lambda $ incident to $v$ and
$v\B_s $ are the edges of $\Lambda $ incident from $v$. Furthermore we obtain
that $k|\V | =|\E | $.\\

\smallskip
Since a cycle of length $L $ is also a connected $1$-regular graph with $L$
vertices, we obtain by Lempel's Theorem~\ref{lempel}:

\begin{prop}
For every $n\in\NN_0 $ and $L\in\NN $ with $1\le L\le q^n $ there
exists a connected $1$-regular subgraph in \BGr{n} with L vertices.
\end{prop}

\noindent
Let $2\le k \le q $. Since $\BGr[k]{n}$ is a $k$-regular subgraph of
\BGr{n} we obtain:

\begin{prop}
Let $1\le k \le q $. For every $n\in\NN_0 $ there exists  a
connected $k$-regular subgraph in \BGr{n} with $k^n$ vertices and
$k$ loops.
\end{prop}

\begin{prop}\label{linereg}
If there exists $k$-regular subgraph of \BGr{n} with $L$
vertices and $l$ loops then there exists also a
$k$-regular subgraph in \BGr{n+m} with $L\cdot k^m$ vertices and $l$ loops.
This holds also
for "connected $k$-regular subgraph" in place of "$k$-regular subgraph".
\end{prop}

\pagebreak\noindent
{\bf Proof:} Let $\Lambda =(\V,\E ) $ be a $k$-regular subgraph of
\BGr{n} with $L$ vertices and $l$ loops. By Proposition~\ref{linegraph1}  follows
that $L_m\Lambda $ is a $k$-regular subgraph with $L\cdot k^m$
vertices and $l$ loops. Furthermore we obtain that $L_m\Lambda $ is connected, if
$\Lambda $ is connected. By identifying $L_m\BGr{n} $ with $\BGr{n+m} $
we obtain the proposition.\qed\\

\medskip
Since any connectivity component of a $k$-factor of \BGr{n} is by
itself a connected $k$-regular subgraph of $\BGr{n} $, we can find
$k$-regular subgraphs of \BGr{n+m} by taking the $m$-iterated lingraph
of connected components of $k$-factors of \BGr{n}.
However for
$k\ge \left\lceil \frac{q}{2}\right\rceil $ every $k$-factor of
\BGr{n} is a connected graph.\footnote{We omit a prove of this statement.}\\

\smallskip\noindent
We continue with an example of Proposition~\ref{linereg}.

\
\begin{example}\upshape
Let $2\le k \le q $. Let $A \in\A $ be set of letters with $|A|\ge  k $
and let $\varphi :\{ 0,\ldots |A|-1\} \leftrightarrow A $ be a bijection.
We define the subgraph\\
$\No (A,\varphi ,k)=(\V , \E) \subseteq (\A , \A^2 ) $ in $\BGr{1} $
as follows:\\

\noindent
Let $[w] $ be the cyclic sequence
\[ [w]:=[w_0\ldots w_{|A|-1}] \;\mbox{ with }\;
w_l:=\varphi (l) \quad \forall\; 0\le l < |A| \, .\]
Since $\varphi $ is a bijection, it follows that $[w] $ is a cycle in
\BGr{1}  of length bigger than or equal to $k$.
We define the vertex set $\V $ and the edge set $\E $ of $\No (A,\varphi ,k) $
as:
\[\begin{array}{lcl}
\V & := & A =\Sub_w (1) \subseteq \A^1 \, ,\\
\E & := & \{ w_lw_{l+i} \,|\, l\in\ZZ \, ,\, 0\le i <k \} \, .
\subseteq \A^2
\end{array}\]
Every letter in $A=\Sub_w(1) $ occurs exactly one time in
$w$ and $|w|=|A|\ge k $. Hence we obtain, that for every
$a\in \V= \Sub_w (1)  $
there exist $k$ unique letters $b_1,\ldots b_k \in A $ with
$ab_j\in\E $ and  further on $k$ unique letters $c_1\ldots c_k \in A $
with $c_ja \in\E $ for  $j\in\{ 1,\ldots ,k\} $.
Therefore $\No (A,\varphi ,k)$ is a $k$-regular subgraph in \BGr{1} with
$|A| $ vertices. Furthermore $aa$ is a loop in $\No (A,\varphi ,k)$ for
every $a\in A $. Obviously $\No (A,\varphi ,k)$ contains no other loops.
Finally we obtain that $\No (A,\varphi ,k)$ is connected, because
$[w] $ is a Hamilton circuit for $\No (A,\varphi ,k)$.\\

\noindent
Let $0\le p\le q-k $ and $A :=\{ 0,\ldots , k+p-1 \} $. It follows
that $\No (A, id_A ,k)$ is a connected $k$-regular subgraph of \BGr{1}
with $k+p$ vertices and a loop at each vertex.
Thus we obtain by Proposition~\ref{linereg} the following result:
\end{example}
\begin{prop}\label{normalgraph}
Let $2\le k \le q $ and $0\le p \le q-k $.
There exists a connected $k$-regular subgraph in \BGr{n} with
$k^n+p\cdot k^{n-1} $ vertices and $k+p $ loops for every $n\in\NN $.
\end{prop}

\pagebreak
\subsection*{Cyclic sequences of regular subgraphs of \BGr{n}}

Let $\Lambda $ be a $k$-regular subgraph of \BGr{n} and $E$ be an
Euler circuit for $\Lambda $. Since $\Lambda $ has $kL $ edges, the closed
path $E $ has length $kL$. Let $[w] $ be the corresponding cyclic sequence of length
$kL $. Since $E$ is a closed path with $L$ vertices and runs through every of its
vertices exactly $k$ times, it follows that $[w] $ has the following properties:

\begin{equation}\label{regcyc1}
\begin{array}{ll}
\mbox{(a)} & \Num_w (v) =k \;\mbox{ for all }\; v\in\Sub_w (n) \;\mbox{ and }\;
|\Sub_w (n) |=L \, ,\\[2mm]
\mbox{(b)} & \Num_w (v) =1 \;\mbox{ for all }\; v\in\Sub_w (n+1) \, ,\\[2mm]
\mbox{(c)} & |\Sub_w (n+1) |= |w| =kL\, .\\[2mm]
\end{array}
\end{equation}
Obviously we have:
\[ \mbox{(a) and (b) holds if and only if  (a) and (c) holds.} \]

\noindent
Let $[w] $ be a cyclic sequence which fulfill the above properties.
By (b) and (c) follows, that $[w] $ is a closed path in \BGr{n}of length $kL$.
By (a) follows, that the path runs through every of its vertices exactly $k$ times.
Therefore $[w] $ is an Euler circuit for some $k$-regular subgraph of \BGr{n}.\\

We call a cyclic sequence which fulfill the properties (a), (b) and (c)
a \\
{\em $(k,L,n)$-regular sequence}. Obviously a (k,L,n)-regular sequence has
length $kL$. From the above remarks follows:
\begin{prop}\label{regcycprop1}\quad\\
\begin{enumerate}
\renewcommand{\labelenumi}{(\roman{enumi})}

\item There exists a $(k,L,n)$-regular sequence if and only if there
exists a $k$-regular subgraph of \BGr{n} with $L$ vertices.

\item If $\Lambda $ is a $k$-regular subgraph of \BGr{n} with $L$ vertices,
then every Euler circuit $[w] $ of $\Lambda $ is a $(k,L,n)$-regular sequence.

\item If $[w]$ is a $(k,L, n)$-regular sequence, then $[w] $ is an Euler circuit
of some $k$-regular subgraph of \BGr{n} with $L$ vertices, where
$\Sub_w (n) $ is the vertex set and $\Sub_w (n+1) $ the edge set of
the subgraph.

\end{enumerate}
\end{prop}

\smallskip

\begin{lemma}\label{klnseq1}
Let $w\in\A^+ $ and let $[w] $ be a $(k,L,n)$-regular cyclic sequence.
Then
\[ \Num_w (u)\ge k^{n+1-l} \quad \forall\, u\in\Sub_w(l) \, ,\; 1\le l\le n+1 \, ,\]
i.e. every letter of $w$ occurs at least $k^n$ times in $w$.
\end{lemma}

\pagebreak\noindent
{\bf Proof:} Let $w=w_0\ldots w_{L\cdot k} $ with $w_0,\ldots
,w_{L\cdot k-1}\in\A $.
We show the lemma by induction on $l$:\\

\smallskip\noindent
By definition (\ref{regcyc1}) (a) and (b) of $(k,L,n)$-regular sequences, we
obtain that the lemma holds for $l\in\{ n,n+1 \} $.\\

\smallskip\noindent
Let $0<l<n $ and let us assume that the lemma holds for $l+1$:
\begin{equation}\label{indhypcycseq}
\Num_w (v)\ge k^{n+1-(l+1)} =k^{n-l} \;\mbox{ for all } v\in\Sub_w (l+1)\, .
\end{equation}

\smallskip\noindent
We have to show that the lemma holds for $l$ as well.
If $u\in\Sub_w (l) $, then there is $0\le i <k\cdot L $ such
that $u=w_i\ldots w_{i+l-1} $. It follows:
\[ uw_{i+l}\ldots w_{i+n-1}\in\Sub_w (n) \, .\]
By the property (\ref{regcyc1}) (a) of $(k,L,n)$-regular sequences follows:
\[ \Num_w (uw_{i+l}\ldots w_{i+n-1}) = k \, .\]
This shows, that there exists a (unique) set $A\subseteq \A $ with
\[ |A|=k \;\mbox{ and } \; Auw_{i+l}\ldots w_{i+n-1}\subseteq \Sub_w (n+1) \, ,\]
i.e. we obtain
\[ Au\subseteq \Sub_w (l+1) \, .\]
By induction hypothesis (\ref{indhypcycseq})) follows
$\Num_w (au)\ge k^{n-l} $ for all $a\in A $.

\smallskip\noindent
Thus we obtain:
\[ \Num_w (u) \ge \sum\limits_{a\in A}\Num_w (au)\ge |A|\cdot k^{n-l}
= k^{n+1-l} \, .\]
The first inequality holds, because
$\Num_w (u) $ is the number of occurrence of $u$ as a subword in $[w] $
and therefore $\Num_w (u)=\sum\limits_{a\in\A } \Num_w (au) $.\\

\noindent
This shows that the lemma holds for all $l$ with $0<l \le n+1 $.
The set of letters in $w$ is given by $\Sub_w (1) \subseteq \A $.
Since $\Num_w (a)\ge k^{n+1-1} =k^n $, every letter in $\Sub_w (1) $
occurs in $w$ at least $k^n $ times. \qed\\

\medskip
We obtain with the one-to-one correspondence in Proposition~\ref{regcycprop1}
between \\
$(k,L,n)$-regular sequences and $k$-regular subgraphs of \BGr{n} with
$L$ vertices:

\begin{theorem}\label{klnseq2}
Let $n\in\NN $ and $1\le k < q^n $. For every $L\in\NN $ with
$L<k^n $ or $k^n< L < k^n+k^{n-1} $ there does not exist  a $k$-regular
subgraph in $\BGr{n} $ with L vertices.
Indeed there exists a connected
$k$-regular subgraph in $\BGr{n} $ with $k^n $ vertices as well as there
exist one with $k^n-k^{n-1}$ vertices.
\end{theorem}

\smallskip\noindent
{\bf Proof:}
We show first the second part of the lemma.
Let $k\ge 2$. Obviously \BGr[k]{n} is
a connected $k$-regular subgraph of \BGr{n} with $k^n $ vertices.
>From Proposition~\ref{normalgraph} follows, that there
exists a connected $k$-regular subgraph in \BGr{n} with
$k^n+k^{n-1} $ vertices if $k\ge 2 $. If $k=1 $, then from
Lempel's Theorem~\ref{lempel} follows that there exists a cycle
of length $1=1^n $ and a cycle of length $2=1^n+1^{n-1} $.This shows
the second part of the lemma.\\

\smallskip
We show that the first part of the lemma holds for connected $k$-regular graphs.
Let $L< k^n+k^{n-1} $ and $\Lambda $ be a connected $k$-regular
subgraph in \BGr{n} with $L$ vertices. By Proposition~\ref{regcycprop1}
follows that there exists a $(k,L,n)$-regular sequence $[w] $
for some $w\in\A^+ $ with $|w|=k\cdot L <k^{n+1}+k^{n-1}$,
such that $[w] $ is an Euler circuit of $\Lambda$.
Let $\A' :=\Sub_w (1) $ be the set of letters which occurs in $w$.
It follows that $\Lambda $ is also a connected $k$-regular subgraph of
\BGr[\A']{n} with L vertices.
Since $\BGr[|\A' |]{n} \cong \BGr[\A']{n}$,
there exists a connected $k$-regular subgraph $\tilde{\Lambda } $
with L vertices in $\BGr[|\A' |]{n} $, where $\tilde{\Lambda } \cong \Lambda$.
With Lemma~\ref{klnseq1} follows, that every letter in $w$ occurs at
least $k^n $ times in $w$. Thus:
\begin{equation}\label{cycseqnum1}
\begin{array}{ll}
\quad & k\cdot L= |w| = \sum\limits_{a\in\A'} \Num_w (a) \ge |\A'|\cdot k^n\\
\Rightarrow & \frac{L}{k^{n-1}}=\frac{|w|}{k^n} \ge |\A'| \, .
\end{array}
\end{equation}
If $|\A' | < k $, then there does not exist a $k$-regular subgraph in
$\BGr[|\A' |]{n} $. Therefore from (\ref{cycseqnum1}) follows, that
$L\ge k^n $.
Thus let $|\A' |\ge k $. Since $L < k^n+k^{n-1} $ it follows by (\ref{cycseqnum1}),
that $|\A' | = k $.
Further on \BGr[k]{n} is the only $k$-regular subgraph in \BGr[k]{n}.
We conclude $\Lambda \cong \tilde{Lambda }=\BGr[k]{n} $ and  $L=k^n $.
This shows that there does not exist a connected $k$-regular subgraph of \BGr{n} with
$L$ vertices, if $1\le L < k^n $ or $ k^n < L < k^n+k^{n-1} $.
Therefore the lemma holds for connected $k$-regular subgraphs of \BGr{n}.
Furthermore we obtain that every connected $k$-regular subgraph of \BGr{n} with
$k^n $ vertices is isomorphically to $\BGr[k]{n} $.\\

\smallskip
Let us assume that there exist a unconnected $k$-regular subgraph $\Lambda $
of \BGr{n} with $L<k^n +k^{n-1}$ vertices.
Let $\Lambda_1 ,\Lambda_2 $ be two connectivity components of $\Lambda $.
Then $\Lambda_1 ,\Lambda_2 $ are vertex disjoint connected $k$-regular subgraphs in
\BGr{n} with less than $k^n +k^{n-1}$ vertices. It follows that
$\Lambda_1 ,\Lambda_2 $ are isomorphically to $\BGr[k]{n} $. Therefore each of them
has $k^n $ vertices. It follows, that $2k^n \le L $. This is a contradiction, because
$k^n +k^{n-1} \le 2k^n $. Therefore the lemma holds also for unconnected
graphs.\qed\\

\pagebreak

\chapter{Fix-free codes obtained from $\pi $-systems}

\setlength{\parindent}{0pt}

\setlength{\parskip}{0.5ex}

In this chapter we will proof a generalization of a theorem of Yekhanin
\cite{yek1}(2001), which shows that the $\frac{3}{4}$-conjecture holds for binary
codes if the Kraftsum of the first level which occurs in the code together
with it neighboring level is bigger than $\frac{1}{2} $.
To show this, Yekhanin claimed two lemmas which
imply the theorem. However in \cite{yek1} no proof was given for the lemmas and
due to my knowledge, no proof was published in other papers.
The theorem and the sketch of
the proof given in \cite{yek1} is the following:

\smallskip
\begin{theorem}[Yekhanin] \label{yek}
Let $|\A | $ =2 and $(\alpha_l )_{l\in\NN} $ be a sequence of nonnegative integers with
$\sum\limits_{l=1}^{\infty}\alpha_l \left( \frac{1}{2} \right)^l \le \frac{3}{4} $.
If there exists an $n\in\NN $ such that $\alpha_1=\ldots =\alpha_{n-1} =0 $ and
$\frac{\alpha_n}{2^n}+\frac{\alpha_{n+1}}{2^{n+1}}\ge \frac{1}{2}$, then there exists
a fix-free code $ \C \subseteq \A^* $ which fits to $(\alpha_l )_{l\in\NN} $.
\end{theorem}

For proving the theorem, Yekhanin introduced in \cite{yek1} a special kind of
fix-free codes, which he called $\pi $-systems:

\begin{Def} \label{pisyst}
Let $|\A |=2 $, we say $\D \subseteq \bigcup\limits_{l=1}^n \A^l $ is a
$\pi_2$-system if $\D $ is fix-free with Kraftsum $\frac{1}{2} $
and
\begin{equation}\label{defprop0}
|\sfs[n](\D )|=|\pfs[n](\D )|=|\A^{-1}\pfs[n](\D )|=|\sfs[n](\D )\A^{-1}|
\end{equation}
\end{Def}
Instead of (\ref{defprop0}) Yekanin defined in \cite{yek1} $\pi $-systems
with the following property:
\begin{equation}\label{defprop1}
|\A^n-\sfs[n](\D )|=|\A^n- \pfs[n](\D )|=
|\A^{-1}\big( \A^n- \pfs[n](\D ) \,\big)|=|\big(\A^n-\sfs[n](\D )\, \big) \A^{-1}|
\end{equation}

If $\D $ is fix-free, from $\Kraft (\D ) =\frac{1}{2} $ follows, that
$|\pfs[n](\D ) | =|\sfs[n](\D ) | =2^{n-1} $. Therefore the next proposition
shows that for $\D $ fix-free with $\Kraft (\D ) =\frac{1}{2} $, the properties
(\ref{defprop0}) and (\ref{defprop1}) are equivalent.

\begin{prop} \label{yekprop0}
Let $|\A |=q\ge 2 $, $\X\subseteq \A^n $ and $\X^c:=\A^n-\X $ then we
have:\\
\[\begin{array}{ll}
\quad & |\X |=|\A^{-1}\X | \ge q^{n-1} \\
\Leftrightarrow  & |\X |=|\A^{-1}\X |=q^{n-1}\\
\Rightarrow & |\A^{-1}\X^c|=q^{n-1} \mbox{ and } |\X^c |=(q-1)q^{n-1}
\end{array} \]

\[\begin{array}{ll}
\quad & |\X |=|\X\A^{-1} | \ge q^{n-1} \\
\Leftrightarrow & |\X |=|\X\A^{-1} |=q^{n-1}\\
\Rightarrow & |\X^c\A^{-1}|=q^{n-1} \mbox{ and } |\X^c |=(q-1)q^{n-1}
\end{array} \]

\end{prop}

{\bf Proof: } From $\X\subseteq\A^n $ follows
$|\A^{-1}\X |\le |\A^{n-1} | =q^{n-1} $ and $|\X\A^{-1} |\le q^{n-1} $. Obviously
the equivalents in the proposition holds.
Let $|\X |=|\A^{-1}\X |=q^{n-1} $. Then there exists
for every $w\in \A^{n-1} $ a letter $a_w\in \A $ with $ a_ww\in \X $ and $a_w $ is
unique because of the first equality. Therefore $\X^c $ is the
(disjoint) union of the sets $(\A-\{ a_w \})w$ with $w\in\A^{n-1} $.
This shows $|\A^{-1}\X^c|=q^{n-1}$ and $|\X^c |=(q-1)q^{n-1}$ follows directly from
$|\X |=q^{n-1} $. The second part of the proof of the proposition follows
same steps.\qed\\

\smallskip
Theorem \ref{yek} follows from the two lemmas below:
\begin{lemma} \label{yek1}
Let $|\A | $ =2 and $(\alpha_l )_{l\in\NN} $ be a sequence of nonnegative integers with\\
$\sum\limits_{l=1}^{\infty}\alpha_l \left( \frac{1}{2} \right)^l \le \frac{3}{4} $.
If there exists an $n\in\NN $ and a $\pi_2 $-system $\D $ such that
$|\A^l\cap \D|= \alpha_l $ for all $1\le l\le n $ and
$|\A^{n+1}\cap \D|\le  \alpha_{n+1} $, then there exists fix-free extension
$\C $of $\D $ which fits $(\alpha_l )_{l\in\NN} $
\end{lemma}

Furthermore in the lemma above the codewords in $(\C -\D ) $ can chosen arbitrary
by induction on the codeword lengths.

\smallskip
\begin{lemma} \label{yek2}
Let $n\in\NN $,
$\beta_1 =\ldots =\beta_{n-1}=0 $ and $\beta_n ,\, \beta_{n+1}\in\NN $
such that $\frac{\beta_n}{2^n}+\frac{\beta_{n+1}}{2^{n+1}}= \frac{1}{2}$, then there
exists a $\pi_2 $-system $\D\subseteq \A^{n+1} $ with
$|\A^l\cap \D| = \beta_l $ for $1\le l\le n+1$.
\end{lemma}

\smallskip
In the next two sections we prove a generalization of the theorem for arbitrary
finite alphabets. Therefore we give in the next section a more general
definition of $\pi $-systems and a generalization of Lemma \ref{yek1}.
In the second section of this chapter we show that there is a one-to-one correspondence
between two level $\pi $-systems
$\D \subseteq \A^n\cup\A^{n+1} $ and regular
subgraphs in $\BGr{n-1} $, whereas the edges\footnote{
Like in Chapter 3 we label the edges of $\BGr{n-1} $ with words of length $n$.}
of the corresponding regular subgraph are the codewords in $\D $ of length $n$.
Especially for $|\A | = 2 $ every cycle in \BGr[2]{n-1} is a 1-regular subgraph
and as it was shown in the previous chapter,
for every $1\le \beta_n \le 2^{n-1} $ there exists a $\beta_n $ length cycle
in \BGr[2]{n-1}.
With the one-to-one correspondence between
regular subgraphs and two level $\pi $-systems of the form $\D \subseteq\A^n\cup\A^{n+1}$
we get Lemma~\ref{yek2}. Finally we show in the second section of this chapter
a generalization of Theorem \ref{yek} for arbitrary finite alphabets.
However, because of the one-to-one
correspondence between regular subgraphs and $\pi $-systems, in the general form of the
theorem occurs a additional condition. This condition is the existence
of regular subgraphs in \BGr{n-1} with certain numbers of vertices.

\section{Extensions of $\pi $-systems}
In this section we give a generalization of Lemma \ref{yek1}
and introduce $\pi $-systems
for arbitrary finite alphabets.
For this we need some remarks
about sets $\X\subseteq \A^n $ with the property $|\A^{-1}\X |=|\X |$ or
$|\X\A^{-1} |=|\X |$.
\begin{prop} \label{yekprop1}
Let $|\A |=q\ge 2 $.\\
\begin{enumerate}
\renewcommand{\labelenumi}{(\roman{enumi})}

\item Let $\X\subseteq\A^n $ then\\
$|\X |=|\A^{-1}\X | \Leftrightarrow $ if $w_1\ldots w_n\in\X ,a\in\A-\{ w_1\}$ then $aw_2\ldots w_n\not\in\X $\\
$|\X |=|\X \A^{-1}| \Leftrightarrow $ if $w_1\ldots w_n\in\X ,a\in\A-\{ w_n\}$ then
$w_1\ldots w_{n-1}a\not\in\X $

\item Let $\X\subseteq\A^n $ then\\
$|\X |=|\A^{-1}\X | \Leftrightarrow |\A^{-1}\X\A^l |=|\X\A^l | \;\forall\, l\in\NN $\\
$|\X |=|\X \A^{-1}| \Leftrightarrow |\A^l\X\A^{-1} |=|\A^l\X | \;\forall\, l\in\NN $

\item Let $\X\subseteq\bigcup\limits_{l=1}^n\A^l $ and $ \X_l:=\X\cap\A^l $ then\\
$ |\A^{-1}\X |=|\X | \Leftrightarrow | \A^{-1}\X_l | =|\X_l  | \;\forall\, l\in\NN $\\
$ |\X \A^{-1}|=|\X | \Leftrightarrow | \X_l\A^{-1} | =|\X_l  | \;\forall\, l\in\NN $

\item Let $\X\subseteq\bigcup\limits_{l=1}^n\A^l $ then we have for every $N\ge n$\\
$ | \pfs[n] (\X )| =|\A^{-1}\pfs[n] (\X ) | \Leftrightarrow | \pfs[N] (\X )| =|\A^{-1}\pfs[N] (\X ) | $\\
$ | \sfs[n] (\X )| =|\sfs[n] (\X )\A^{-1} | \Leftrightarrow | \sfs[N] (\X )|
=|\sfs[N] (\X )\A^{-1} | $

\end{enumerate}

\end{prop}
{\bf Proof:} (i) is obvious. For (ii) we have
\[\begin{array}{lrcl}
\quad & \left|\X \right| & = &\left|\A^{-1}\X \right| \\[1mm]
\Leftrightarrow & \left|\X \right| \cdot |\A^l |
& = &\left|\A^{-1}\X \right|\cdot |\A^l |\\[1mm]
\Leftrightarrow &
\left|\X\A^l \right| & = & \left|\A^{-1}\X\A^l \right|
\end{array} \]
This shows the first part of (ii), the second part (ii) follows the same steps.\\
For the first part of (iii) take in account that:
\[ |\X |  =  |\A^{-1}\X | \Leftrightarrow
 \sum\limits_{l=1}^n | \X_l |  = \sum\limits_{l=1}^n
|\A^{-1}\X_l | \, . \]
Since the terms in the sums are nonnegative and
$|\X_l | \ge | \A^{-1}\X_l | $ for all $l$ with $1\le l\le n $,
the second equation holds only if $|\X_l |  =  |\A^{-1}\X_l | $ for all
$l$ with $1\le l\le n $. In the same way follows the second part of (iii).\\
(iv) follows now from (ii), because
\[ \pfs[N] (\X ) = \pfs[n](\X )\A^{N-n} \, ,\quad
\A^{-1}\pfs[N] (\X ) = \A^{-1}\pfs[n](\X )\A^{N-n} \]
\[ \sfs[N] (\X ) =
\A^{N-n}\sfs[n](\X )  \, ,\quad
\sfs[N] (\X )\A^{-1} = \A^{N-n}\sfs[n](\X )\A^{-1} \] \qed\\

\begin{lemma}\label{yeklemma0}
Let $\X  \subseteq\bigcup\limits_{l=1}^n\A^l \, ,\, N\ge n$
and $\X_l:=\X\cap \A^l $ then:
\begin{enumerate}
\renewcommand{\labelenumi}{(\roman{enumi})}

\item
\[
\begin{array}{ll}
\quad & \left|\pfs[N](\X ) \right| =\left|\A^{-1}\pfs[N](\X ) \right|
\quad \mbox{and $\X $ is prefix-free} \\
\Leftrightarrow & \A^{-1}\X \quad\mbox{is prefix-free and }
\left|\A^{-1}\X_l  \right| = \left|\X_l \right| \quad\mbox{ for all  }
1\le l\le n ,\\
\Leftrightarrow & \A^{-1}\X \quad\mbox{is prefix-free and }
\left|\A^{-1}\X  \right| = \left|\X \right|.
\end{array}
\]
\item
\[
\begin{array}{ll}
\quad & \left|\sfs[N](\X ) \right| =\left|\A^{-1}\sfs[N](\X ) \right|
\quad \mbox{and $\X $ is suffix-free} \\
\Leftrightarrow & \X\A^{-1} \quad\mbox{is suffix-free and }
\left|\X_l \A^{-1} \right| = \left|\X_l \right| \quad\mbox{ for all  }
1\le l\le n, \\
\Leftrightarrow & \X\A^{-1} \quad\mbox{is suffix-free and }
\left|\X \A^{-1} \right| = \left|\X \right|.
\end{array}
\]
\end{enumerate}
\end{lemma}
{\bf Proof:} Let
$ \left|\pfs[N](\X ) \right| =\left|\A^{-1}\pfs[N](\X ) \right|
\quad \mbox{and $\X $ be prefix-free} $. If we assume that $\A^{-1}\X $ is not
prefix-free, then there exists $u,v\in\A^{-1}\X $ and $a,b\in\A $ such that $u=vu' $
for some $u'\in\A^+ $ and $au,bv\in\X $. Then $a\neq b $, because $\X $ is
prefix-free. It follows:
\[ au\A^{N-|u|-1} \subseteq \pfs[N](\X ) \;\mbox{ and }\;
bvu'\A^{N-|vu'|-1}=bu\A^{N-|u|-1} \subseteq \pfs[N](\X )\]
By Proposition~\ref{yekprop1} (i) and $a\neq b $ follows
$\left|\pfs[N](\X ) \right| >\left|\A^{-1}\pfs[N](\X ) \right| $. This is a
contradiction. Therefore $\A^{-1}\X $ is prefix-free.\\
If we take in account that also $\X $ is prefix-free, we obtain:
\[
\left|\pfs[N](\X ) \right| =\left|\A^{-1}\pfs[N](\X ) \right|
\Leftrightarrow \sum\limits_{l=1}^n |\X_l|\cdot |\A^{N-l} | =
\sum\limits_{l=1}^n |\A^{-1}\X_l|\cdot |\A^{N-l} |
\]
While $0\ge |\A^{-1}\X_l|\cdot |\A^{N-l} | <|\X_l|\cdot |\A^{N-l} | $
for all $l$ with $1\le l\le n $, the second equation holds only if
$ |\X_l| = |\A^{-1}\X_l| $ holds for all $l$ with $1\le l\le n $.
This shows that for the prefix-free sets $\X $ and $\A^{-1}\X $ the following
equivalence is true:
\begin{equation}\label{yeklemgl1}
\left|\pfs[N](\X ) \right| =\left|\A^{-1}\pfs[N](\X ) \right|
\Leftrightarrow
|\X_l| = |\A^{-1}\X_l| \quad\forall\, 1\le l\le n
\end{equation}
Therefore we obtain $\A^{-1}\X $ is prefix-free and
$|\X_l| = |\A^{-1}\X_l| \quad\forall\, 1\le l\le n $.\\

\smallskip
Let us assume that $\A^{-1}\X $ is prefix-free and
$|\X_l| = |\A^{-1}\X_l| \quad\forall\, 1\le l\le n $.
Then from the assumption that $\A^{-1}\X $ is prefix-free follows that
$\X $ is prefix-free as well and by (\ref{yeklemgl1}) we obtain,
that also $ \left|\pfs[N](\X ) \right| =\left|\A^{-1}\pfs[N](\X )\right| $ holds.
This shows the first equivalence of (i).\\

The second equivalence of (i) follows by Proposition \ref{yekprop1} (iii).
The proof for (ii) follows the same step as the proof of (i). \qed\\

\smallskip
\begin{Def} \label{yekdef2}\quad\\

Let $|\A |=q \ge 2$, $1\le k\le q $ and $n\in\NN $. We call a set
$\D \subseteq \bigcup\limits_{l=1}^n \A^l $ a
{\em $\pi_q(n;k)$-system } if $\D $ is fix-free, and there exists a partition
of $\D $ into $k$ sets $\D_1 ,\ldots ,\D_k $ for which the following three
equivalent properties holds.
\begin{enumerate}
\renewcommand{\labelenumi}{(\arabic{enumi}):}

\item For all $1\le i \le k  $ holds:
\begin{eqnarray*}
q^{n-1} & = & |\pfs[n](\D_i)| = |\A^{-1}\pfs[n](\D_i)| \\
\quad   & = & |\sfs[n](\D_i)| = |\sfs[n](\D_i)\A^{-1}|
\end{eqnarray*}

\item $\Kraft (\D ) =\frac{k}{q} $ and for all i with $1\le i \le k $ holds:
\[ |\pfs[n](\D_i)| = |\A^{-1}\pfs[n](\D_i)| \;\mbox{ and }\;
|\sfs[n](\D_i)| = |\sfs[n](\D_i)\A^{-1}| \]

\item For all $1\le i\le k $ the set
$\;\A^{-1}\D_i \,$ is maximal prefix-free, $\;\D_i\A^{-1} $ is maximal suffix-free
and $|\A^{-1}\D_i |= |\D_i\A^{-1} | =|\D_i | $.

\end{enumerate}

The sets $\D_1,\ldots ,\D_{k} $ are called
a {\em $\pi $-partition of $\D $ }\\

For $\alpha_1,\ldots ,\alpha_n \in\NN $ we call a $\pi_q(n;k) $-system $\D $ a
$\pi_q(\alpha_1,\ldots ,\alpha_n ;k) $-system if
$\quad \left| \D\cap\A^l\right| =\alpha_l $ for all $1\le l\le n $.

\end{Def}

\smallskip
We show that (1)-(3) in the definition are all equivalent.
Therefore let $\D \subseteq \bigcup\limits_{l=1}^n \A^l $ be a fix-free code and
$\D_1 \ldots ,\D_k $ be a partition of $\D $.\\

\pagebreak
{\bf $(1)\Rightarrow (3)$:}\\
Let $\D_1 \ldots ,\D_k $ such that (1)
holds. Since $\D $ is fix-free, all $\D_i $ are fix-free.
With Lemma~\ref{yeklemma0} and property (1) follows that
$\A^{-1}\D_i $ is prefix-free and $ \D_i\A^{-1} $ is suffix-free.\\

We obtain for all $1\le i\le k$:
\[
\begin{array}{lcl}
q^{n-1} & = & \big| \A^{-1} \pfs[n] (\D_i ) \big| =
\big| \A^{-1}\bigcup\limits_{l=1}^n (\D_i\cap\A^l)\A^{n-l} \big| =
\big| \bigcup\limits_{l=1}^n \A^{-1}(\D_i\cap\A^l)\A^{n-l} \big|\\
\begin{array}{l} \mbox{\tiny $\A^{-1}\D $ is }\\
\mbox{\tiny prefix-free } \end{array} & = &
\sum\limits_{l=1}^n \big|\A^{-1}(\D_i\cap\A^l )\A^{n-l} \big|=
\sum\limits_{l=1}^n \big| \A^{-1} D_i\cap \A^{l-1} \big|\cdot q^{n-l} \\
\quad & = &
\sum\limits_{l=0}^{n-1} \big| \A^{-1}\D_i \cap \A^l \big|\cdot q^{n-l-1}\, .

\end{array}
\]
It follows:
\[ \Kraft (\A^{-1}\D_i )= \sum\limits_{l=0}^{n-1}
\big| \A^{-1}\D_i \cap \A^l \big|\cdot q^{-l} =1 \, .\]
This shows that $\A^{-1}\D_i $ is maximal prefix-free.
In the same way follows that
$\D_i\A^{-1} $ is maximal suffix-free. Furthermore we obtain
$|\A^{-1}\D_i |=|\D_i\A^{-1}| =|\D_i | $ from Lemma \ref{yeklemma0}. Therefore (3)
holds for the sets $\D_1,\ldots ,\D_k $.\\

\smallskip
{\bf $(3)\Rightarrow (2)$:}\\
Let $\D_1 \ldots ,\D_k $ be such that (3)
holds. Then from Lemma~\ref{yeklemma0} follows
\[ |\pfs[n](\D_i)| = |\A^{-1}\pfs[n](\D_i)| \;\mbox{ and }\;
|\sfs[n](\D_i)| = |\sfs[n](\D_i)\A^{-1}| \;\mbox{ for all }\; 1\le i\le k  \, .\]
Therefore we have to show: $\Kraft (\D ) =\frac{k}{q} $. Because
of Lemma \ref{yeklemma0} we have\\
$\big|\A^{-1}(\D_i\cap\A^l)\big| =\big|\D_i\cap\A^l \big| $ for all $1\le i\le k $
and $1\le l\le n $.
Since $\A^{-1}\D_i $ is maximal prefix-free ($\Kraft (\A^{-1}\D_i )=1 $),
we obtain :
\[
\begin{array}{lcl}
1 & = & \Kraft (\A^{-1} \D_i ) =
\sum\limits_{l=0}^{n-1} \big| \A^{-1}\D_i \cap \A^l \big| \cdot q^{-l} =
\sum\limits_{l=0}^{n-1} \big| \A^{-1}(\D_i\cap\A^{l+1})\big|\cdot q^{-l} \\
\quad & = &
\sum\limits_{l=1}^n \big|\A^{-1}(\D_i\cap \A^l )\big| \cdot q^{-l+1} =
q\cdot \sum\limits_{l=1}^n \big|\D_i \cap\A^l \big| \cdot q^{-l} =
q\cdot\Kraft (\D_i )
\end{array}
\]
Therefore $\Kraft (\D_i ) =\frac{1}{q} $ for all $1\le i\le k $.
Because $\D_1,\ldots ,\D_k $ is a partition of $\D $ it follows:
\[ \Kraft (\D ) =\Kraft (\D_1 )+\ldots +\Kraft (\D_k ) = \frac{k}{q}  \]
This shows that (2) holds for $\D_1,\ldots ,\D_k $.\\

\pagebreak
{\bf $(2)\Rightarrow (1)$:}\\
Let $\D_1 \ldots ,\D_k $ be such that (2)
holds. We have to show that\\
$\big|\pfs[n] (\D_i ) \big| = \big| \sfs[n] (\D_i ) \big| = q^{n-1} $ holds for all
$1\le i\le k $. Since $\A^{-1}\pfs[n] (\D_i ) \subseteq \A^{n-1} $ from
(2) follows :

\begin{equation}\label{yekeq3}
\left|\pfs[n] (\D_i)\right| =\left|\pfs[n] \A^{-1}(\D_i)\right| \le q^{n-1}
\;\forall\, 1\le i \le k\, .
\end{equation}

While the Kraftsum of $\D $ is equal to $\frac{k}{q}$, we get:

\begin{eqnarray*}
k\cdot q^{n-1} & = &
\sum\limits_{l=1}^n \left|\D\cap \A^l\right|\cdot q^{n-l} \\
\mbox{\footnotesize ($\D $ is prefix-free) } & = & \left| \pfs[n] (\D )\right| \\
\mbox{\footnotesize ( the $\D_i$'s are a partition of $\D$ )} & = &
\sum\limits_{i=1}^{k} \left| \pfs[n] (\D_i )\right| \, .
\end{eqnarray*}

>From the last equality and (\ref{yekeq3}) follows:
\[ \left| \pfs[n] (\D_i )\right| = q^{n-1} \;mbox{ for all }\;  1\le i \le k \, . \]
Similar arguments show that  also $\left| \sfs[n] (\D_i )\right| = q^{n-1} $ holds
for all $1\le i\le k $. Thus $\D_1, \ldots ,\D_k $ have property (1) as well.\\

\smallskip
It follows, that for a fix-free code $\D $ with partition $\D_1,\ldots ,D_k $
(1), (2) and (3) in the Definition \ref{yekdef2} are all
equivalent.
Furthermore we get from (2), that
the Definition \ref{yekdef2} of $\pi_q(n;1) $-systems coincides for $q=2 $
with the first definition of $\pi_2 $-systems.\\

\smallskip
\begin{lemma} \label{yeklemma1} Let $| \A |=q<\infty $ \\

\begin{enumerate}
\renewcommand{\labelenumi}{(\roman{enumi})}
\renewcommand{\labelenumii}{\alph{enumii})}
\item Let $ \Y\subseteq\A^n , \X\subseteq\A^{n-1} $ then we have:\\
\begin{enumerate}
\item If $ |\Y\A^{-1} |=|\Y |=q^{n-1} \mbox{ then } |\X\A\cap\Y |=|\X | $.
\item If $ |\A^{-1}\Y |=|\Y |=q^{n-1} \mbox{ then } |\A\X\cap\Y |=|\X | $.
\end{enumerate}

\item Let $\X ,\Y \subseteq\A^n $ for some $n\ge 1 $ then we have:\\
$ |\A^{-1}\X |=|\X | \mbox{ and } |\Y\A^{-1} |=|\Y | \Rightarrow | \X\A\cap \A\Y |
\ge |\X |+|\Y | - q^{n-1} $

\end{enumerate}
\end{lemma}

\pagebreak
{\bf Proof:}\\
\renewcommand{\labelenumi}{(\roman{enumi}):}
\begin{enumerate}
\item Let $\A := \{ a_1,\ldots ,a_q \}$.
We prove only part a), because the proof of part b) is analogously.
Therefore let $ \Y\subseteq\A^n , \X\subseteq\A^{n-1} $ and
$|\Y\A^{-1} |=|\Y |=q^{n-1}$. Let $\Y_l:=\Y a_l^{-1} $, then $\Y $
is the disjoint union of $\Y_1a_1,\ldots ,\Y_qa_q $. We claim, that
$\Y_1,\ldots \Y_q $ are pairwise disjoint. Assume that
$\Y_l\cap\Y_k \neq \emptyset $ for some $l\neq k $, then there exists some
$w\in\A^{n-1} $, such that $wa_l ,wa_k\in\Y $. This is a contradiction,
because $|\Y\A^{-1}|=|\Y | $.
Therefore $\Y\A^{-1} $ is the disjoint union of $\Y_1,\ldots ,\Y_q $. Since
$|\Y\A^{-1} | =q^{n-1} =|\A^{n-1} | $, the sets
$\Y_1,\ldots ,\Y_q $ are a partition of $\A^{n-1} $.\\
Thus we get:
\[ |\X\A\cap\Y | =\sum\limits_{l=1}^q|\X\A\cap\Y_la_l |
=\sum\limits_{l=1}^q|\X\cap\Y_l |=|\X\cap\A^{n-1} | =|\X | \, .\]

\item By $|\A^{-1}\X |=|\X | $ and $|\Y\A^{-1} | =|\Y | $ we have:
\begin{eqnarray*}
q^{n-1}=|\A^{n-1}| & \ge & |\A^{-1}\X \cup \Y\A^{-1} | \\
\quad & = &|\A^{-1} \X | + |\Y\A^{-1}| -|\A^{-1}\X\cap \Y\A^{-1} |\\
\quad & = & |\X |+|\Y | -|\A^{-1}\X\cap\Y\A^{-1} |
\end{eqnarray*}
and therefore we obtain:
\begin{equation}\label{yeklem2gl1}
|\A^{-1}\X\cap \Y\A^{-1} |\ge |\X |+|\Y | - q^{n-1} \, .
\end{equation}

For every $w\in\A^{-1}\X\cap\Y\A^{-1} $ there exist $a,b\in\A $ with
$aw\in\X $ and $wb\in\Y $. It follows that $awb\in\X\A\cap\A\Y $.
Since $|\A^{-1}\X |=|\X | $ and $|\Y\A^{-1}| =|\Y |$,
the letters $a,b $ are unique. Vice versa, for $v\in\X\A\cap\A\Y $ there
are $a,b\in\A $ and $w\in\A^{n-1} $ such that $aw\in\X $ and $wb\in\Y $.
It follows that $w\in\A^{-1}\X\cap\Y\A^{-1} $. This gives
us a one-to-one map from $\A^{-1}\X\cap\Y\A^{-1} $ onto $\X\A\cap\A\Y $,
and therefore
\[|\A^{-1}\X\cap\Y\A^{-1}| = |\X\A\cap\A\Y | \, .\]
Together with (\ref{yeklem2gl1}) follows:
$ | \X\A\cap \A\Y | \ge |\X |+|\Y | - q^{n-1} $. \qed
\end{enumerate}

\pagebreak
The following theorem shows that $\pi_q(n) $-systems can always be extended
to a fix-free
code, if the Kraftsum is smaller than or equal to $\frac{3}{4} $.
It is a generalization of Lemma \ref{yek2} for arbitrary finite alphabets .
\begin{theorem}[$\pi $-system extension theorem] \label{piext}
For $ |\A |=q\ge 2 $  and $1\le k < q $ let

\[ \gamma_k :=\left\{\begin{array}{l@{\mbox{ for }}l}
\frac{1}{2}+\frac{k}{2q} & 1\le k\le\left\lfloor\frac{q}{2}\right\rfloor \\
\left(\frac{q-k}{q}\right)^2 +\frac{k}{q} &
\left\lfloor\frac{q}{2}\right\rfloor < k < q \end{array}\right.\]

Let $(\alpha_l)_{l\in\NN } $ be a sequence of nonnegative integers
and let $n\in\NN $, $1\le \beta \le \alpha_n $ be such that:

\[ \sum\limits_{l\in\NN }^{\infty } \alpha_lq^{-l } > \frac{k}{q}
\;\mbox{ and }\; \beta q^{-n} +\sum\limits_{l=1}^{n-1} \alpha_l q^{-l} = \frac{k}{q}. \]

If $\;\sum\limits_{l=1}^{\infty }\alpha_l q^{-l}\le \gamma_k $, then for
every $\pi_q (\alpha_1,\ldots ,\alpha_{n-1} ,\beta ;k) $-system
there exists a fix-free extension which fits to $(\alpha_l)_{l\in\NN } $.

\end{theorem}

Note that $\gamma_k > \frac{k}{q} $ for $1\le k\le q $
and that there exist unique $\beta ,n\in\NN $ with the properties in the theorem.\\

Furthermore the proof of the theorem will show, that an extension $\C $ of a\\
$\pi_q(\alpha_1,\ldots ,\alpha_{n-1}\beta ;k) $-system $\D $ can be
constructed as follows:
\begin{enumerate}
\renewcommand{\labelenumi}{\arabic{enumi}.}
\item add to $\D $  $(\alpha_n-\beta) $ arbitrary codewords of length $n$ which
are not in $\bfs[n] (\D ) $ to obtain a fix-free $\C_0 $.
\item For $m\in\NN $ add to $\C_{m-1} $ $\alpha_{n+m} $ arbitrary codewords of
length $(n+m)$ which are not in $\bfs[n+m] (\D_{m-1} ) $ to obtain a fix-free $\C_m $.
\item Take the union of all $\C_m $'s to obtain the fix-free extension $\C $.

\end{enumerate}

{\bf Proof:} Let $q$ and $k$ as in the theorem. We claim:
\begin{equation}\label{gamma}
\gamma_k =\min \bigg\{ \frac{1}{2}+\frac{k}{2q}  \quad ,\: \bigg(\frac{q-k}{q}\bigg)^2+\frac{k}{q}
\bigg\}\quad \forall\, 1\le k < q \, .
\end{equation}
Let $f(x) := \frac{q+x}{2q} $ and
$g(x) := \left(\frac{q-x}{q}\right)^2+\frac{x}{q} $, then:
\begin{equation}\label{gamma1}
f(0)=\frac{1}{2}<1 =g(0) \;\mbox{ and }\; f(x)=g(x)
\Leftrightarrow x\in \{ \frac{q}{2} ,q \}
\end{equation}
Since $f $ and $g$ are continuous functions, equation (\ref{gamma}) follows
from (\ref{gamma1}).\\

\smallskip
Let $(\alpha_l )_{l\in\NN}$ be a sequence of nonnegative integers with
$\sum\limits_{l=1}^{\infty } \alpha_l q^{-l} \le \frac{3}{4} $ and choose
$\beta, n\in\NN $ such that
$\beta q^{-n}+\sum\limits_{l=1}^{n-1}\alpha_lq^{-l} = \frac{k}{q} $
Let $\D $ be a $\pi_q(\alpha_1 ,\ldots , \alpha_{n-1},\beta ;k )$-system, with
$\pi $-partition $\D_1 , \ldots ,\D_{k} $. We will show by a simple
induction on the codeword length, that the construction of a fix-free code which fits
$(\alpha_l )_{l\in\NN}$ is possible in the way described above.
Because of (1) in the definition of $\pi $-systems and Proposition~\ref{yekprop1}~(iv)
we get for all $ m\in\NN $ and $1\le i\le k $:
\begin{equation} \label{yekineq1}
\left|\A^{-1}\pfs[n+m] (\D_i)\right| =\left|\pfs[n+m] (\D_i)\right| =
q^m \left|\pfs[n] (\D_i)\right| = q^{n+m-1}\, ,
\end{equation}

\begin{equation} \label{yekineq2}
\left|\sfs[n+m] (\D_i)\A^{-1}\right| =\left|\sfs[n+m] (\D_i)\right| =
q^m \left|\sfs[n] (\D_i)\right| = q^{n+m-1} \, .
\end{equation}

Since $\D $ is fix-free and the disjoint union of the $\D_i $'s, it follows:

\begin{equation} \label{yekineq3}
\left| \pfs[n+m] (\D ) \right| = \left| \sfs[n+m] (\D ) \right| = q^{n+m-1}k
  \;\mbox{ for all }\; m\in\NN \, .
\end{equation}

\smallskip
{\bf Case m=0:}\\
We show, that the cardinality of the bifix-shadow of $\D $ on the $n$-th level
is smaller than $\left| \A^n \right|- (\alpha_n -\beta )  =q^n+\beta -\alpha_n  $.
Then we can add $(\alpha_n -\beta ) $ codewords of length $n$ to $\D $ and obtain a
fix-free code $\C_0 \supseteq \D $ which fits to $(\alpha_1 ,\ldots ,\alpha_n) $.\\

Let for $1\le i\le k $
\[ \begin{array}{l@{\quad ,\quad }l}
\F_i := \D_i\cap \A^n & \F:=\bigcup\limits_{i=1 }^{k}\F_i
= \D \cap \A^n  \, ,\\
\E_i := \D_i - \F_i &
\E:=\bigcup\limits_{i=1 }^{k} \E_i = \D - \F  \end{array} \, ,\]

because the $\D_i $'s are pairwise disjoint and $\D $ is fix-free it follows that also
$\F_1 ,\ldots \F_k $, \\
$\E_1 ,\ldots \E_k $ are pairwise disjoint and
fix-free. Furthermore we obtain \\
for all $1\le i\le k  $:

\[ \pfs[n] (\F_i ) =\sfs[n] (\F_i )= \F_i \, ,\quad
\pfs[n] (\F ) =\sfs[n] (\F ) =\F  \, ,\]

\[ \pfs[n] (\E_i ) = \pfs[n-1] (\E_i )\A \mbox{ and }
\sfs[n] (\E_i ) = \A\sfs[n-1] (\E_i ) \, .\]

\pagebreak
With Lemma \ref{yeklemma1} (i) and property (1) of the $\D_i $'s
in the definition of $\pi $-systems follows
for all $1\le i,j \le k $ :

\begin{equation} \label{yekineq4}
\left| \pfs[n] (\E_i ) \cap \sfs[n] (\D_j ) \right|  =
\left| \pfs[n-1] (\E_i )\A  \cap \sfs[n] (\D_j ) \right| =
\left| \pfs[n-1] (\E_i) \right|\, ,
\end{equation}

\begin{equation} \label{yekineq5}
\left| \sfs[n] (\E_i ) \cap \pfs[n] (\D_j ) \right|  =
\left| \A\sfs[n-1] (\E_i )  \cap \pfs[n] (\D_j ) \right| =
\left| \sfs[n-1] (\E_i) \right|\, .
\end{equation}

The sets $\pfs[n] (\E_1),\ldots , \pfs[n] (\E_k)$ as well as the sets
$\pfs[n] (\D_1),\ldots , \pfs[n] (\D_k)$ are pairwise disjoint, because
$\E ,\D $ are fix-free and both the $\E_i $'s and $\D_i $'s are pairwise disjoint.
It follows:
\[
\begin{array}{ccl}
\left| \pfs[n] (\E ) \cap \sfs[n] (\D ) \right| & = &
\bigg| \bigcup\limits_{i=1}^{k} \pfs[n] (\E_i ) \cap
\bigcup\limits_{j=1}^{k} \pfs[n] (\D_j ) \bigg| \\
\qquad & = & \bigg|\bigcup\limits_{i,j=1}^{k}
\left( \pfs[n] (\E_i )\cap \sfs[n] (\D_j ) \right) \bigg| \\
\qquad & = & \sum\limits_{i,j=1}^{k}
\left| \pfs[n] (\E_i) \cap \sfs[n] (\D_j) \right| \\
\mbox{\footnotesize with (\ref{yekineq4})} & = &
\sum\limits_{i,j=1}^{k}\left|\pfs[n-1] (\E_i ) \right|
 =   k\cdot \sum\limits_{i=1}^{k }\left| \pfs[n-1] (\E_i) \right| \\
\qquad & = & k\cdot \sum\limits_{i=1}^{k }
\frac{\left| \pfs[n] (\E_i) \right| }{q}
 =  \frac{k}{q} \cdot \left|\pfs[n] (\E )\right| \, .\\
\end{array}
 \]

Therefore we have:
\begin{equation} \label{yekineq6}
\left| \pfs[n] (\E ) \cap \sfs[n] (\D ) \right|  =
\frac{k}{q} \cdot \left|\pfs[n] (\E )\right| \, .
\end{equation}

For the codewords of $\D $ which are contained in the prefix-shadow and
in the suffix-shadow we obtain:
\[
\begin{array}{ccl}
\left| \pfs[n] (\D )\cap \sfs[n] (\D ) \right| & = &
\left|\left( \pfs[n](\E ) \cap \sfs[n] (\D )  \right) \cup
\left(\F  \cap \sfs[n] (\D )  \right)  \right| \\[0.2cm]
\mbox{\parbox{3.5cm}{ \scriptsize  $\pfs[n] (\E )\cap \F  =\emptyset $
because $\D=\E\stackrel{\cdot}{\cup}\F $ is prefix-free }} & = &
\left|\pfs[n] (\E )\cap\sfs[n] (\D )\right| +
\left| \F\cap \sfs[n] (\D ) \right|  \\[0.4cm]
\mbox{\scriptsize $ \F\subseteq \D $ and $\D $ is suffix-free } & = &
\left| \pfs[n] (\E )\cap \sfs[n] (\D ) \right| + \left|\F \right| \\[0.2cm]
\mbox{\scriptsize equation (\ref{yekineq6}) } & = &
\frac{k}{q} \left| \pfs[n] (\E )\right| +
\left| \F \right| \\[0.2cm]
\qquad & \ge & \frac{k}{q}\cdot \left( \left| \pfs[n] (\E )\right| +
\left| \F \right| \right)  = \frac{k}{q}\cdot \left| \pfs[n] (\D ) \right| \\[0.2cm]
\mbox{\scriptsize with (\ref{yekineq3}) for $m=0 $ }
& = & \frac{k}{q}\cdot  q^{n-1} \cdot k =k^2\cdot q^{n-2} \, .
\end{array}
\]

With the above equation and (\ref{yekineq3}) for $m=0 $ follows:

\[
\begin{array}{rcccl}
\left| \bfs[n] (\D ) \right| & = & \left|\pfs[n] (\D ) \right| +
\left|\sfs[n] (\D ) \right| - \left| \pfs[n] (\D )\cap \sfs[n] (\D ) \right|
& \le & 2 \left|\pfs[n] (\D ) \right| - q^{n-2} \cdot k^2 \\
\qquad & = & 2q^{n-1}k - q^{n-2} k^2 & = & q^{n-2} k( 2q-k )
\end{array}
\]

\pagebreak
Thus:

\begin{equation}\label{yekineq7}
\left| \bfs[n] (\D ) \right| = q^{n-2}k( 2q-k) \, .
\end{equation}

Since the Kraftsum of $(\alpha_n )_{n\in\NN } $ is smaller than or equal to
$\gamma_k $ and $\D $ is a \\
$\pi_q (\alpha_1,\ldots ,\alpha_{n-1},\beta ;k)$-system,
we obtain by (\ref{yekineq3}):

\[ \gamma_k \cdot q^n \ge \sum\limits_{l=1}^n \alpha_lq^{n-l} =
\left| \pfs[n] (\D) \right| +(\alpha_n-\beta )
=k\cdot q^{n-1} +(\alpha_n-\beta ) \, .
\]

It follows:

\begin{equation}\label{yekineq8}
\alpha_n-\beta \le q^n\bigg( \gamma_k -\frac{k}{q} \bigg) \, .
\end{equation}

By (\ref{yekineq7})  and (\ref{yekineq8})we obtain:

\begin{equation}\label{yekineq9}
\left| \bfs[n] (\D ) \right| + (\alpha_n - \beta ) \le
q^n \bigg( \frac{k (2q-k ) }{q^2} + \gamma_k -\frac{k}{q}  \bigg) \, .
\end{equation}

>From $(\ref{gamma}) $ follows, that the term inside
paranthesis of the right-hand side of equation
(\ref{yekineq9}) is smaller than or equal to one.

\[\begin{array}{rl}
(\ref{gamma})\Rightarrow &\gamma_k \le \left(\frac{q-k}{q}\right)^2+\frac{k}{q} \\
\Leftrightarrow & q^2\gamma_k \le (q-k)^2+kq= q^2-2kq+k^2 +kq \\
\Leftrightarrow & k(2q-k)+q^2\gamma_k -kq \le q^2 \\
\Leftrightarrow & \frac{k(2q-k)}{q^2} +\gamma_k -\frac{k}{q} \le 1 \, .
\end{array}\]

Therefore we conclude:
\[  \left| \bfs[n] (\D ) \right| + (\alpha_n - \beta ) \le q^n =|\A^n | \, . \]

This shows, that we can choose  $(\alpha_n - \beta ) $ codewords
$c_1,\ldots c_{\alpha_n- \beta} \in\A^n-\bfs[n] (\D ) $ of length $n$
which are not in the bifix-shadow of $\D $. Then
$\C_0 :=\D \cup \{ c_1,\ldots ,c_{\alpha_n -\beta} \} $ is a fix-free code
which extend $\D $, whereas $\left|\C_0\cap \A^l \right| =\alpha_l $ for
all $1\le l \le n $.\\

\pagebreak
{\bf m $\rightarrow $ m+1 :} \\
Let $\C_m $ be a fix-free extension of $\D $ which
fits to $\alpha_1, \ldots ,\alpha_{n+m} $.
More precisely :
\begin{quote}
$\C_m $ with  $\D \subseteq\C_m\subseteq \bigcup\limits_{l=1}^{n+m}\A^l $ is fix-free and
$\left| \C_m \cap \A^l \right| = \alpha_l $\\
for all $1\le l \le n+m $.
\end{quote}

We will show that there exists $\alpha_{n+m+1} $ codewords of length $(n+m+1)$
which are not in the bifix-shadow of $\C_m $. If we add this codewords
to $\C_m $  we obtain a
fix-free extension $\C_{m+1} $ of $\D\subseteq \C_m $ which fits to
$(\alpha_1,\ldots ,\alpha_{n+m+1}) $.\\

We define $\X $ and $M$ as:

\[ \X := \C_m -\D \quad ;\quad M:=n+m+1. \]

Because  $\C_m $ is fix-free, $\C_m = \X \cup \D $ and
$\X\cap \D =\emptyset $ we obtain:
{\small
\begin{eqnarray}
\left| \bfs[M] (\C_m ) \right| & = &
\left| \bfs[M] (\X ) \cup  \bfs[M] (\D )\right|  \nonumber \\
\qquad & = & \left|  \bfs[M] (\D ) \right| + \left|  \bfs[M] (\X ) \right|
- \left| \bfs[M] (\X )\cap \bfs[M] (\D ) \right| \nonumber\\
\qquad & = & 2 \left|  \pfs[M] (\D ) \right| + 2\left|  \pfs[M] (\X ) \right|
- \left| \pfs[M] (\D )\cap \sfs[M] (\D ) \right|
- \left| \pfs[M] (\X )\cap \sfs[M] (\X ) \right|\nonumber \\
 \qquad & \quad &
- \left| \bfs[M] (\X )\cap \bfs[M] (\D ) \right| \nonumber \\
 \qquad & \le & 2\left| \pfs[M] (\D ) \right|
+ 2\left|  \pfs[M] (\X ) \right|
- \left| \pfs[M] (\D )\cap \sfs[M] (\D ) \right|
- \left| \bfs[M] (\X )\cap \bfs[M] (\D ) \right| \, .\nonumber \\ \label{yekineq12}
\end{eqnarray} }
For the terms in the sum on the right-hand side of inequality (\ref{yekineq12}) we
get:\\

\begin{equation}\label{yekineq13}
 \left| \pfs[M] (\D ) \right| = q^{M-1}k\, .
\end{equation}
This follows from (\ref{yekineq3}).

\begin{equation} \label{yekineq14}
\left| \pfs[M] (\D ) \cap \sfs[M] (\D )\right| =
q^{M-2} k^2
\end{equation}

Whereas the above equation follows from:

\[
\begin{array}{lcl}
\left| \pfs[M] (\D )\cap \sfs[M] (\D )  \right| & = &
\left| \bigcup\limits_{i,j=1}^{k }
\left(\pfs[M] (\D_i ) \cap \sfs[M] (\D_j ) \right)   \right| \\[0.4cm]
\mbox{\parbox{3.5cm}{\scriptsize  $\D $ is fix-free and the
disjoint union of the $D_i $  }} & = &
\sum\limits_{i,j = 1}^{k } \left|\pfs[M] (\D_i )\cap
\sfs[M] (\D_j )  \right| \\
\quad & = & \sum\limits_{i,j = 1}^{k}
\left|\pfs[M-1] (\D_i )\A \cap \sfs[M] (\D_j )  \right| \\
\mbox{\scriptsize (\ref{yekineq2}) and Lemma \ref{yeklemma1} (i) } & = &
\sum\limits_{i,j = 1}^{k }
\left| \pfs[M-1] (\D_i ) \right| \\
\mbox{\scriptsize with (\ref{yekineq1}) } & = &
q^{M-2} k^2 \, .
\end{array} \]

Let us determine the value of $\left| \bfs[M] (\X ) \cap \bfs[M] (\D )\right| $.
We have :

\[ \bfs[M] (\X )\cap \bfs[M] (\D ) =
\big( \pfs[M] (\X ) \cap \sfs[M] (\D ) \big) \stackrel{\cdot }{\cup }
\big( \sfs[M] (\X ) \cap \pfs[M] (\D ) \big) \, .\]

This holds, because
$\C_m $ is fix-free and the union of $\D $ and $\X $ is disjoint. Therefore we get
$ \sfs[M] (\X )\cap \sfs[M] (\D ) =  \pfs[M] (\X )\cap \pfs[M] (\D ) = \emptyset $.
Furthermore we have \\
$\pfs[M] (\D ) \cap \pfs[M] (\X ) = \emptyset $. It follows:

\begin{eqnarray}
\left| \bfs[M] (\X )\cap \bfs[M] (\D ) \right| & = &
\left| \pfs[M] (\X ) \cap \sfs[M] (\D ) \right| +
\left| \sfs[M] (\X ) \cap \pfs[M] (\D ) \right| \nonumber \\
\qquad & = & \left| \pfs[M-1] (\X )\A \cap \sfs[M] (\D ) \right| +
\left| \A\sfs[M-1] (\X ) \cap \pfs[M] (\D ) \right| \, . \nonumber
\end{eqnarray}

$\D =\D_1\cup\ldots \cup\D_k$ is fix-free and the
union of the $\D_i $ is disjoint. Therefore from the above equation follows:
\begin{equation} \label{yekineq16}
\left| \bfs[M] (\X )\cap \bfs[M] (\D ) \right| =
 \sum\limits_{i=1}^{k }\big(
\left| \pfs[M-1] (\X )\A \cap \sfs[M] (\D_i ) \right| +
\left| \A\sfs[M-1] (\X ) \cap \pfs[M] (\D_i ) \right| \big) \, .
\end{equation}

By (\ref{yekineq1}), (\ref{yekineq2}) and Lemma \ref{yeklemma1} (i)
for all $ 1\le i \le k $ follows:

\begin{eqnarray*}
\left| \pfs[M-1] (\X )\A \cap \sfs[M] (\D_i ) \right| & = &
\left| \pfs[M-1] (\X ) \right| = \frac{\left| \pfs[M] (\X ) \right| }{q} \quad
\mbox{ and } \\
\left| \A\sfs[M-1] (\X )\cap \pfs[M] (\D_i ) \right| & = &
\left| \sfs[M-1] (\X ) \right| = \left| \pfs[M-1] (\X ) \right| =
\frac{\left| \pfs[M] (\X ) \right| }{q} \, .
\end{eqnarray*}

>From the above equations and (\ref{yekineq16}) we obtain:

\begin{equation} \label{yekineq17}
\left| \bfs[M] (\X ) \cap \bfs[M] (\D ) \right| = \frac{2k}{q}
\left| \pfs[M] (\X ) \right| \, .
\end{equation}

For the cardinality of $\bfs[M] (\C_m ) $ we obtain from
(\ref{yekineq12}), (\ref{yekineq13}) ,(\ref{yekineq14}) and
(\ref{yekineq17}):

\begin{equation} \label{yekineq18}
\left| \bfs[M] (\C_m )  \right| \le k\cdot\frac{2q-k }{q^2} q^M
- \frac{2k}{q} \left| \pfs[M] (\X ) \right| + 2\left| \pfs[M] (\X ) \right| \, .
\end{equation}

We have $\pfs[M]( \X )  = \pfs[M] (\C_m -\D ) = \pfs[M] (\C_m ) - \pfs[M] (\D ) $,
because $\C_m $ is fix-free  and $\D \subseteq \C_m $.
If we take into account that
$\C_m $ fits to $\alpha_1,\ldots ,\alpha_{m+n} $ (whereas $m+n=M-1 $)
and the Kraftsum of $(\alpha_l )_{l\in\NN }$ is smaller than or equal to $\frac{k}{q} $,
we obtain that the following equalities and inequalities are true.

\pagebreak
\begin{eqnarray*}
\left| \pfs[M] (\X ) \right| +\alpha_M & = &
\left| \pfs[M] (\C_m ) \right| +\alpha_M  - \left| \pfs[M] (\D ) \right|
\stackrel{\tiny (\ref{yekineq3})}{=}
\left| \pfs[M] (\C_m ) \right| +\alpha_M - q^{M-1} k \\
\qquad & = &  \sum\limits_{l=1}^{M-1} \alpha_lq^{M-l} +\alpha_M - q^{M-1}k =
\bigg( \sum\limits_{l=1}^M \alpha_lq^{-l}
-\frac{k}{q} \bigg)\cdot q^M \\
\qquad & \le &
\bigg( \gamma_k -\frac{k}{q} \bigg)\cdot q^M \, .
\end{eqnarray*}

\[\Rightarrow
\left| \pfs[M] (\X ) \right| \le \left| \pfs[M] (\X ) \right| +\alpha_M
\le \bigg( \gamma_k -\frac{k}{q} \bigg)\cdot q^M
\]

By the above equation and (\ref{yekineq18}) follows:

\begin{eqnarray*}
\left| \bfs[M] (\C_m )  \right| & \le & k\cdot \frac{2q-k }{q^2} q^M
- \frac{2k}{q} \bigg( \gamma_k -\frac{k}{q} \bigg)\cdot q^M
+ 2\bigg( \gamma_k -\frac{k}{q} \bigg)\cdot q^M \\
\qquad & = & \bigg(\, \frac{k (2q-k)}{q^2} +\bigg( 2- \frac{2k}{q}  \bigg)
\bigg( \gamma_k -\frac{k}{q}  \bigg)\, \bigg) \cdot q^M\, .
\end{eqnarray*}

>From (\ref{gamma}) follows that
the term inside the big paranthesis on the right hand side of the above equation
is smaller than or equal to one.

\[\begin{array}{rl}
(\ref{gamma})\Rightarrow & \gamma_k \le \frac{1}{2}+\frac{k}{2q} \\
\Leftrightarrow & 2q\gamma_k -2k \le q-k \\
\mbox{\scriptsize ($0\le k < q$)} \Leftrightarrow  &
(2q\gamma_k -2k)(q-k) \le (q-k)^2 \\
\Leftrightarrow & 2qk-k^2 +(2q-2k)(q\gamma_k -k )\le q^2 \\
\Leftrightarrow & \frac{k (2q-k)}{q^2} +\bigg( 2- \frac{2k}{q}  \bigg)
\bigg( \gamma_k -\frac{k}{q} \bigg) \le 1
\end{array}\]

Therefore we conclude:
\[
\left| \bfs[M] (\C_m )  \right|+\alpha_{n+m+1} =
\left| \bfs[M] (\C_m )  \right|+\alpha_M \le q^M = \left| \A^{n+m+1} \right| \, .
\]

This shows that we can add $\alpha_{n+m+1} $ codewords of length
$(n+m+1) $ to $\C_m $, which are not in the bifix-shadow of $\C_m $.
In this way we obtain
a fix-free code $\C_{m+1} $ with $\C_{m+1}\supseteq\C_m\supseteq \D $
and $\left| \C_{m+1}\cap\A^l \right| =\alpha_l $ for all $ 1\le l \le n+m+1 $.\\

\smallskip
Let $\C := \bigcup\limits_{l=0}^{\infty } \C_l $.
Because of $\D\subseteq\C_0 \subseteq \C_1 \subseteq\C_2 \ldots $, the set $\C $ is
fix-free and $\left| \C_{m+1}\cap\A^l \right| =\alpha_l $ for all $ l\ge 1 $.
Therefore $\C $ is a fix-free extension of $\D $ which fits to
$(\alpha_l )_{l\in\NN }$. \qed\\

\pagebreak
\begin{cor}\label{piext2}
Let $(\alpha_l )_{l\in\NN } $ a sequence of nonnegative integers
with
\[\big\lceil\frac{q}{2}\big\rceil\frac{1}{q} <
\sum\limits_{l=1}^{\infty }\alpha_lq^{-l} \le \frac{3}{4} \]
and $n\in\NN $, $1\le \beta \le\alpha_n $ such that:
\[ \beta q^{-n} +\sum\limits_{l=1}^{n-1}\alpha_lq^{-l} =
\bigg\lceil\frac{q}{2}\bigg\rceil\frac{1}{q} \]
Then for every
$\pi_q\big(\alpha_1,\ldots ,\alpha_{n-1} ,\beta ,\big\lceil\frac{q}{2}\big\rceil\big)$-system
there exists a fix-free extension which fits
$(\alpha_l )_{l\in\NN } $.\\
\end{cor}

\smallskip
{\bf Proof :} We have to show that
$\gamma_{\left\lceil\frac{q}{2}\right\rceil } \ge\frac{3}{4} $ for all $q\ge 2$. \\
For even $q$ we obtain:

\[ \gamma_{\left\lceil\frac{q}{2}\right\rceil } =
\frac{1}{2}+\frac{\frac{q}{2}}{2q}=\frac{3}{4} \, .\]

For odd $q$ let $q=2t+1 $:

\[\begin{array}{ll}
\quad & \gamma_{\left\lceil\frac{q}{2}\right\rceil } =
\left(\frac{q-\left\lceil\frac{q}{2}\right\rceil}{q}\right)^2 +
\frac{\left\lceil\frac{q}{2}\right\rceil}{q}
=\frac{\left\lfloor\frac{q}{2}\right\rfloor^2 +q\left\lceil\frac{q}{2}\right\rceil}{q^2}
\ge \frac{3}{4} \\
\Leftrightarrow &
4\left\lfloor\frac{q}{2}\right\rfloor^2 + 4q\left\lceil\frac{q}{2}\right\rceil
\ge 3q^2 \\
\Leftrightarrow & 4t^2+4(2t+1)(t+1)\ge 3(2t+1)^2 \\
\Leftrightarrow & 12t^2+12t+4\ge 12t^2+12t+3 \\
\Leftrightarrow & 4\ge 3
\end{array} \]

Therefore
$\gamma_{\left\lceil\frac{q}{2}\right\rceil } \ge\frac{3}{4} $ for all $q\ge 2$.
The corollary follows from Theorem \ref{piext}.\qed\\

\smallskip
For the binary case $|\A |=2 $ we obtain from the corollary  Lemma \ref{yek1}.\\

The table below shows the values from $\gamma_k $  for $q\in\{ 2,3,4,5,6 \} $

\[\begin{array}{|l|r|r|r|r|r|r|} \hline
q\backslash k & 1 & 2 & 3 & 4 & 5 \\ \hline
2 &\frac{3}{4} &&&&\\ \hline
3 & \frac{2}{3} & \frac{7}{9} &&&\\ \hline
4 & \frac{5}{8} & \frac{3}{4} & \frac{13}{16} &&\\ \hline
5 & \frac{3}{5} & \frac{7}{10} & \frac{19}{25} & \frac{21}{25} & \\ \hline
6 & \frac{7}{12} & \frac{2}{3} & \frac{3}{4} & \frac{7}{9} & \frac{31}{36} \\
\hline
\end{array}\]

\pagebreak
Next we give some easy example fors $\pi $-systems:
\begin{example}\label{expiext1}\upshape
Let $|\A | = q\ge 2  $
and $k,d \in\NN $ such that $1\le  d < q $ and\\
$k\le \min \{ d,q-d \} $.
Furthermore let
$\X ,\Y $ be a partition of $\A $ with

\[ |\X |= d \qquad ;\qquad
|\Y |= q-d  \, .\]

Since $k\le \min \{ d,q-d \} $ we can choose permutations of $\X \;$
$\varphi_1,\ldots ,\varphi_k :\X \leftrightarrow \X $ and permutations of $\Y\; $
$\phi_1,\ldots ,\phi_k :\Y \leftrightarrow \Y $ with the property:

\begin{equation}\label{yekexgl1}
\varphi_i (x) \neq\varphi_j (x) \;\mbox{ and }\;
\phi_i(y) \neq \phi_j (y) \quad\mbox{ for all }\; i\neq j \, ,\; x\in\X
\, ,\; y\in\Y
\end{equation}
For example, let $\X =\{ x_0,\ldots x_{d-1} \} $ and
$\Y =\{ y_0,\ldots ,y_{q-d-1} \}$
then it is possible to choose the $\varphi_i $ and the $\phi_i $ as
\[ \varphi_i (x_l ) := x_{l+i-1 \,\mbox{\footnotesize $\mod $} d} \;\mbox{ and }\;
\phi_i (y_m ) := y_{m+i-1 \,\mbox{\footnotesize $\mod $} q-d} \]
for all $1\le i\le k \,$, $0\le l \le d-1 \, $, $ 0\le m\le q-d-1 $.\\

For $1\le i \le k $ and $n\ge 2 $ we define:
\[ \D_i  :=
\bigcup\limits_{y\in\Y }
y\Y^{n-2}\phi_i (y) \;\cup\; \bigcup\limits_{m=0}^{n-2}\bigcup\limits_{x\in\X }
x\Y^m\varphi_i(x) \, ,\]

\[ \D := \D_1\cup\ldots \cup\D_k  \, .\]

By (\ref{yekexgl1} ) the sets $\D_1,\ldots ,\D_k $ are pairwise disjoint
and $\D $ is fix-free, because $\X, \Y $ is a partition of $\D $.
While the $\varphi_i $'s are permutations of $\X $ and the
$\phi_i $'s are permutations of $\Y $, we obtain for $1\le i\le k $:

\[\begin{array}{lcl@{\quad ; \quad }lcl}
\A^{-1}\D_i & = & \Y^{n-1} \cup \bigcup\limits_{m=0}^{n-2}\Y^m\X &
\D_i\A^{-1} & = & \Y^{n-1} \cup \bigcup\limits_{m=0}^{n-2}\X\Y^m \, .\\
\end{array}\]

It follows:
\[ |\A^{-1}\D_i |= |\D_i\A^{-1}| =|\D_i | =
(q-d)^{n-1} + \sum\limits_{m=0}^{n-2} d\cdot (q-d)^m \, . \]

\pagebreak
Obviously $\A^{-1}\D_i $ is prefix-free and
$\D_i\A^{-1} $ is suffix-free. The equation below shows, that they are
maximal, too.

\[\begin{array}{lcl}
\Kraft (\A^{-1}\D_i )  =  \Kraft (\D_i\A^{-1} ) & = &
(q-d)^{n-1} \cdot q^{-n+1}+
\sum\limits_{l=1}^{n-1} d\cdot (q-d)^{l-1} \cdot q^{-l} \\
\quad & = & \big( \frac{q-d}{q}\big)^{n-1} +
\frac{d}{q} \cdot \sum\limits_{l=0}^{n-2} \big(\frac{q-d}{q}\big)^{l} \\
\quad & = & \big( \frac{q-d}{q}\big)^{n-1} +
\frac{d}{q} \cdot
\frac{1-\left(\frac{q-d}{q}\right)^{n-1}}{\hspace*{-8.5mm}1-\frac{q-d}{q}}\\[3mm]
\quad & = & \big( \frac{q-d}{q}\big)^{n-1} +
\frac{d}{q} \cdot
\frac{1-\left(\frac{q-d}{q}\right)^{n-1}}{\frac{d}{q}}\\[3mm]
\quad & = & \big( \frac{q-d}{q}\big)^{n-1} +1 -\big( \frac{q-d}{q}\big)^{n-1} =1
\end{array}\]

Therefore $\A^{-1}\D $ is  maximal prefix-free and
$\D\A^{-1} $ is maximal suffix-free. This shows that(3) in
Definition \ref{yekdef2} holds for $\D_1,\ldots ,\D_k $. Therefore
$\D $ is a $\pi_q (n;k) $-system.\\

For the numbers of codewords of length $l$ we obtain:

\[\begin{array}{lcll}
|\D\cap \A | & = & |\D\cap \A^l |=0 &
\mbox{\hspace*{-3cm}for }\;l> n\ge 2 \, ,\\
|\D\cap \A^l| & = & k\cdot d\cdot (q-d)^{l-2}& \mbox{\hspace*{-3cm}for }\;2\le l <n  \, , \\
|\D\cap \A^n| & = & k\cdot d\cdot (q-d)^{n-2} + k\cdot (q-d)^{n-1}
 =k\cdot q\cdot (q-d)^{n-2} & \, .\\
\end{array}\]

Therefore by Theorem \ref{piext} we obtain the following proposition:

\end{example}

\begin{prop}\label{ex1prop1}
Let $|\A |= q\ge 2 $, $n\ge 2 $, $1\le d < q $, $k\le\min \{ d,q-d \} $
and $(\alpha_l )_{l\in\NN } $ be a sequence of
nonnegative integers with
$\sum\limits_{l=1}^{\infty } \alpha_l\cdot q^{-l} \le \gamma_k $
whereat $\gamma_k $ is chosen as in Theorem \ref{piext}.
If $\alpha_1 =0 $,
$\alpha_l =k\cdot d\cdot (q-d)^{l-2}  $ for $2\le l < n $ and
$\alpha_n\ge k\cdot q\cdot (q-d)^{n-2} $
then there exists a fix-free code which fits
$(\alpha_l )_{l\in\NN } $.
\end{prop}

\smallskip
For even $q$ we can choose $k=d=q-d =\frac{q}{2} $. Because of
$\gamma_{\frac{q}{2}}=\frac{3}{4} $, we obtain in this case:
\begin{prop}\label{ex1prop2}
Let $|\A |= q $ with q even, $n\ge 2 $  and $(\alpha_l )_{l\in\NN } $ be a sequence of
nonnegative integers with
$\sum\limits_{l=1}^{\infty } \alpha_l\cdot q^{-l} \le \frac{3}{4} $
If $\alpha_1 =0 $, $\alpha_l =\big(\frac{q}{2}\big)^l $ for $2\le l < n $ and
$\alpha_n \ge q\cdot \big(\frac{q}{2}\big)^{n-1} $
then there exists a fix-free code which fits to
$(\alpha_l )_{l\in\NN } $.
\end{prop}

For the binary case  we conclude:
\begin{prop}\label{ex1prop3}
Let $|\A |=2 $ and $(\alpha_l )_{l\in\NN } $ be a sequence of
nonnegative integers with
$\sum\limits_{l=1}^{\infty } \alpha_l\cdot q^{-l} \le \frac{3}{4} $. If there
exists an $n\ge 2 $ such that
$\alpha_0 =0 $, $\alpha_l =1 $ for $2\le l< n $ and $\alpha_n\ge 2 $ then there
exists a fix-free code $\C\subseteq \A^+ $ which fits to
$(\alpha_l )_{l\in\NN } $.
\end{prop}

\pagebreak
\begin{example}\label{expiext2} \upshape
Let $\A :=\{ 0,\ldots ,q-1 \} $ for some $q\ge 2 $.
We will show that for any $n\in\NN $ and
$1\le k < q $ there exist one level $\pi_q(n;k) $-systems. If $n=1 $ then
we can choose $\D := \{ 0,\ldots ,k-1 \} $ and $\D_i :=\{ i \} $ for
$0 \le i \le k-1 $. Then $\D $ is a $\pi_q (1;k )$-system with $\pi $-partition
$\D_0 ,\ldots , \D_{k-1} $. Thus let us assume that $n \ge 2 $. We choose
permutations $\varphi_0 ,\ldots ,\varphi_{k-1} :\A \leftrightarrow \A $ with the
property:
\begin{equation}\label{1levelpi1}
\varphi_i (a) \neq \varphi_j (a) \;\mbox{ for all }\; i\neq j\, ,\quad a\in\A
\end{equation}
For example, if we choose $\varphi_i (a) := a+i\mod q $ for all $a\in\A $ and
$0\le i\le k-1 $, then $\varphi_0 ,\ldots ,\varphi_{k-1} $ are permutations
of $\A $ for which (\ref{1levelpi1}) holds.\\

We define the sets $\D_0 ,\ldots ,\D_{k-1}\subseteq\A^n $ as:
\[ \D_i :=\bigcup\limits_{a=0 }^{q-1} a\A^{n-2}\varphi_i (a)\subseteq\A^n
\;\mbox{ for }\; 0 \le i \le k-1 \]
Because of (\ref{1levelpi1}) the sets $\D_0,\ldots ,\D_{k-1} $
are pairwise disjoint. They are a partition
of $\D $, where
\[ \D :=\bigcup\limits_{i=0}^{k-1} \D_i \subseteq \A^n .\]
Furthermore $\D $ is fix-free, because it is a subset of $\A^n $.\\

Because $\varphi_0 ,\ldots ,\varphi_{k-1} $ are permutations of $\A $ we obtain
that for any $0\le i\le k-1 $ the sets $A^{n-2} \varphi_i(0), \ldots ,
\A^{n-2}\varphi_i(q-1) $ are a partition of $\A^{n-1} $
For all
$0\le i\le k-1 $ follows:
\[\begin{array}{lcl}
|\D_i | & = & \sum\limits_{a=0}^{q-1} \big| a\A^{n-2}\varphi_i (a) \big| =q\cdot q^{n-2}
= q^{n-1}, \\
|\A^{-1}\pfs[n] (\D_i ) |  & = & |\A^{-1}\D_i | =
\big| \bigcup\limits_{a=0}^{q-1} \A^{n-2}\varphi_i (a) \big| =|\A^{n-1} |=q^{n-1},\\
|\sfs[n] (\D_i )\A^{-1} |  & = & |\D_i\A^{-1} | =
\big| \bigcup\limits_{a=0}^{q-1} a\A^{n-2} \big| =|\A^{n-1} |=q^{n-1}.
\end{array} \]
Therefore (1) in the Definition of $\pi $-systems holds.
This means $\D $ is a one level $\pi_q (n;k) $-system with $\pi $-partition
$\D_0,\ldots ,\D_{k-1} $. This shows that for every $n\in\NN $ and
$1\le k < q $, there exists a $\pi_q (n;k) $-system $\D $ with $\D\subseteq\A^n $.
\end{example}

\pagebreak
By Theorem \ref{piext} and Corollary \ref{piext2} we conclude that the following
proposition holds:
\begin{prop}\label{1levelpiprop1}
Let $|\A | =q\ge 2 $, $1\le k < q $, $\gamma_k $ as in Theorem \ref{piext}
and $(\alpha_l)_{l\in\NN } $ be a sequence of nonnegative integers.
\begin{enumerate}
\item If $\sum\limits_{l=1}^{\infty } \alpha_l \cdot q^{-l} \le \gamma_k $,
$\alpha_1 =\ldots =\alpha_{n-1}= 0 $ and $\alpha_n \ge \frac{k}{q} $ for some
$n\in\NN $, then there exists a fix-free code $\C\subseteq\A^+ $ which fits
$(\alpha_l)_{l\in\NN } $.

\item If $\sum\limits_{l=1}^{\infty } \alpha_l \cdot q^{-l} \le\frac{3}{4} $,
$\alpha_1 =\ldots =\alpha_{n-1}= 0 $ and
$\alpha_n \ge \left\lceil\frac{q}{2}\right\rceil \frac{1}{q} $ for some
$n\in\NN $, then there exists a fix-free code $\C\subseteq\A^+ $ which fits
$(\alpha_l)_{l\in\NN } $.

\end{enumerate}
\end{prop}

\smallskip
For the binary case we obtain:
\begin{prop}\label{1levelpiprop2}
Let $|\A |=2 $ and $(\alpha_l )_{l\in\NN } $ be a sequence of
nonnegative integers with
If $\sum\limits_{l=1}^{\infty } \alpha_l \cdot \left(\frac{1}{2}\right)^2 \le\frac{3}{4} $,
$\alpha_1 =\ldots =\alpha_{n-1}= 0 $ and
$\alpha_n \ge  \frac{1}{2} $ for some
$n\in\NN $, then there exists a fix-free code $\C\subseteq\A^+ $ which fits
$(\alpha_l)_{l\in\NN } $.
\end{prop}

\begin{example}\label{expiext3}\upshape
Let $|\A | =q\ge 2  $
and $k,d \in\NN $ such that $1\le  d < q $ and\\
$k\le \min \{ d,q-d \} $.
Furthermore let
$\X ,\Y $ a partition of $\A $ with

\[ |\X |= d \qquad ;\qquad
|\Y |= q-d \]
As in Example \ref{expiext1} we can choose permutations
$\varphi_1,\ldots ,\varphi_k :\X \leftrightarrow \X $ and permutations
$\phi_1,\ldots ,\phi_k :\Y \leftrightarrow \Y $ with (\ref{yekexgl1}).\\

For $n\ge 3 $ and $1\le i\le k $ we define the sets
$\D_i ,\E_i ,\F_i ,\G_i $ and $\D $ as:
\[\begin{array}{lcl}
\E_i & := & \bigcup\limits_{l=1}^{n-2}\bigcup\limits_{x\in\X }x\Y^l\varphi_i(x) \\
\F_i & := & \bigcup\limits_{l=1}^{n-2}\bigcup\limits_{y\in\Y} y\X^l\phi_i(y) \\
\G_i & := & \bigcup\limits_{x\in\X }x\X^{n-2}\varphi_i(x)\;\cup\;
\bigcup\limits_{y\in\Y }y\Y^{n-2}\phi_i (y) \\
\D_i   & := & \E_i\cup\F_i \cup \G_i \\
\D & = & \D_1\cup\ldots\cup \D_k
\end{array}\]
Obviously $\D $ is fix-free.
The permutations $\varphi_1,\ldots \varphi_k $ and $\phi_1 ,\ldots ,\phi_k $
fulfill (\ref{yekexgl1}). Therefore $\D_1,\ldots ,\D_k $ are pairwise
disjoint.
For $1\le i\le k $ the sets $\E_i ,\F_i ,\G_i $ are pairwise disjoint and
$\varphi_i $ and $\phi_i $ are permutations of $\X $ and $\Y $. It follows:
\begin{equation}\label{yekexgl2}
|\D_i | =|\E_i | +|\G_i | +|\F_i | =d\cdot \sum\limits_{l=1}^{n-2} (q-d)^l
+(q-d )\cdot \sum\limits_{l=1}^{n-2} d^l + d^{n-1} +(q-d)^{n-1}.
\end{equation}

Because $\varphi_i $ and $\phi_i $ are permutations of $\X $ and $\Y $ we obtain:

\[\begin{array}{lcllcl}
\A^{-1}\D_i & = & \A^{-1}\E_i\cup \A^{-1}\F_i\cup\A^{-1}\G_i &\quad , \quad
\D_i\A^{-1} & = & \E_i\A^{-1}\cup \F_i\A^{-1}\cup\G_i\A^{-1} ,\\
\A^{-1}\E_i & = & \bigcup\limits_{l=1}^{n-2} \Y^l\X &\quad , \quad
\E_i\A^{-1} & = & \bigcup\limits_{l=1}^{n-2} \X\Y^l ,\\
\A^{-1}\F_i & = & \bigcup\limits_{l=1}^{n-2} \X^l\Y &\quad , \quad
\F_i\A^{-1} & = & \bigcup\limits_{l=1}^{n-2} \Y\X^l .
\end{array}\]
\[ \A^{-1}\G_i  =  \G_i\A^{-1} =\X^{n-1}\cup \Y^{n-1} \]

Obviously $\A^{-1}\D_i $ is prefix-free and
$\D_i\A^{-1} $ is suffix-free for $n\ge 3 $. Furthermore
$\A^{-1}\E_i ,\A^{-1}\G_i ,\A^{-1}\F_i $ are pairwise disjoint and also
$\E_i\A^{-1} ,\F_i\A^{-1} ,\G_i\A^{-1} $ are pairwise disjoint. Therefore

\[ |\A^{-1}\D_i | = |\A^{-1}\E_i | +|\A^{-1}\F_i | +|\A^{-1}\G_i |=
d\cdot \sum\limits_{l=1}^{n-2} (q-d)^l
+(q-d )\cdot \sum\limits_{l=1}^{n-2} d^l + d^{n-1} +(q-d)^{n-1} \, . \]

The same way follows $ |\D_i\A^{-1} | = d\cdot \sum\limits_{l=1}^{n-2} (q-d)^l
+(q-d )\cdot \sum\limits_{l=1}^{n-2} d^l + d^{n-1} +(q-d)^{n-1} $.

By (\ref{yekexgl2}) follows:
\[ |\A^{-1}\D_i | =|\D_i\A^{-1}| =|\D_i | \, .\]

Let us show that $\A^{-1}\D_i $ and $\D_\A^{-1} $ are maximal as well.

\[ \Kraft (\A^{-1}\G_i ) =\Kraft (\G_i \A^{-1} ) =
\left(\frac{d}{q}\right)^{n-1} + \left(\frac{q-d}{q}\right)^{n-1} \]

\[\begin{array}{lcl}
\Kraft (\A^{-1}\E_i ) & = & \Kraft (\E_i\A^{-1} ) =
\sum\limits_{l=2}^{n-1} d\cdot (q-d)^{l-1} \cdot q^{-l}
= d\cdot \sum\limits_{l=0}^{n-3} (q-d)^{l+1} \cdot q^{-l-2} \\[2mm]
\quad & = &  \frac{d\cdot (q-d)}{q^2}\cdot \sum\limits_{l=0}^{n-3}
\bigg(\frac{q-d}{q}\bigg)^l =
\frac{d\cdot (q-d)}{q^2 } \cdot \frac{1-(\frac{q-d}{q})^{n-2}}{\hspace*{-7mm}1-\frac{q-d}{q}}
=\frac{q-d}{q} -\big(\frac{q-d}{q}\big)^{n-1}

\end{array}\]

Same way:

\[ \Kraft (\A^{-1}\F_i )  =  \Kraft (\F_i\A^{-1} ) =
\frac{d}{q}-\left( \frac{d}{q} \right)^{n-1} \]
$\A^{-1}\E_i\, \;\A^{-1}\F_i\, \;\A^{-1}\G_i $ and
$\E_i\A^{-1}\, ,\;\F_i\A^{-1}\, ,\;\G_i\A^{-1} $ are pairwise disjoint, therefore:
\[\begin{array}{lcl} \Kraft(\A^{-1}\D_i )& = &\Kraft (\A^{-1}\E_i ) +
\Kraft (\A^{-1}\F_i ) +\Kraft (\A^{-1}\G_i )\\
\quad  & = &\frac{q-d}{q} -\left(\frac{q-d}{q}\right)^{n-1} +
\frac{d}{q}-\left( \frac{d}{q} \right)^{n-1}+
\left(\frac{d}{q}\right)^{n-1} + \left(\frac{q-d}{q}\right)^{n-1}=1
\end{array}\]

In the same way follows:
\[ \Kraft(\D\A^{-1} ) = \Kraft (\E_i\A^{-1} ) +
\Kraft (\F_i\A^{-1} ) +\Kraft (\G_i\A^{-1} ) =1 \, .
\]

This shows, that $\A^{-1}\D_i $ is maximal prefix-free and
$\D\A^{-1} $ maximal suffix-free. Therefore $\D $ is a
$\pi_q (n;k) $-system for all $n\ge 3 $.

The numbers of codewords of length $l$ is given by:

\[\begin{array}{lcll}
|\D\cap \A^l | & = & 0 &
\mbox{ for }\;l> n\ge 3\; \mbox{ or }\; l\in\{ 1,2 \} \\
|\D\cap \A^l| & = & k\cdot \big( d\cdot (q-d)^{l-2}+(q-d)\cdot d^{l-2}\big)&
\mbox{ for }\;3\le l <n \\
|\D\cap \A^n| & = & k\cdot q\cdot\big( (q-d)^{n-2} +d^{n-2} \big) & \\

\end{array}\]
Similar like in example 1 we obtain with Theorem \ref{piext} the following propositions:
\end{example}

\begin{prop}\label{ex3prop1}
Let $|\A |= q\ge 2 $, $n\ge 3 $, $1\le d < q $, $k\le\min \{ d,q-d \} $
and $(\alpha_l )_{l\in\NN } $ be a sequence of
nonnegative integers with
$\sum\limits_{l=1}^{\infty } \alpha_l\cdot q^{-l} \le \gamma_k $
where $\gamma_k $ is chosen as in Theorem \ref{piext}.
If $\alpha_1 =\alpha_2 = 0 $,
$\alpha_l =k\cdot \big( d\cdot (q-d)^{l-2}+(q-d)\cdot d^{l-2}\big)  $ for $3\le l < n $ and
$\alpha_n\ge k\cdot q\cdot\big( (q-d)^{n-2} +d^{n-2} \big)  $
then there exists a fix-free code which fits to
$(\alpha_l )_{l\in\NN } $.

\end{prop}

\smallskip
For $q$ even and $k=d=q-d =\frac{q}{2} $  we obtain:
\begin{prop}\label{ex3prop2}
Let $|\A |= q $ with q even, $n\ge 3 $  and $(\alpha_l )_{l\in\NN } $ be a sequence of
nonnegative integers with
$\sum\limits_{l=1}^{\infty } \alpha_l\cdot q^{-l} \le \frac{3}{4} $
If $\alpha_1 =\alpha_2 = 0 $,
$\alpha_l =2\cdot \big(\frac{q}{2}\big)^l $ for $3\le l < n $ and
$\alpha_n \ge 2q \big(\frac{q}{2}\big)^{n-1} $
then there exists a fix-free code which fits to
$(\alpha_l )_{l\in\NN } $.
\end{prop}

Finally we obtain for the binary case:
\begin{prop}\label{ex3prop3}
Let $|\A |=2 $ and $(\alpha_l )_{l\in\NN } $ be a sequence of
nonnegative integers with
$\sum\limits_{l=1}^{\infty } \alpha_l\cdot q^{-l} \le \frac{3}{4} $. If there
exists an $n\ge 2 $ such that
$\alpha_0 =0 $, $\alpha_l =2 $ for $3\le l< n $ and $\alpha_n\ge 4 $ then there
exists a fix-free code $\C\subseteq \A^+ $ which fits to
$(\alpha_l )_{l\in\NN } $.
\end{prop}

\newpage
\section{Generation of $\pi$ -systems by regular subgraphs of \BGr{n} }

\begin{lemma}\label{piexistlemma} Let $| \A |=q\ge 2$ and $\X \subseteq\A^n $
for some $n\ge 1 $. Then:\\
$ |\A^{-1}\X | = |\X\A^{-1} |=|\A\X\cap\X\A |=|\X | $
if and only if $\X $ is the edge-set of a $1$-regular subgraph
in $\BGr{n-1} $
\end{lemma}

{\bf Proof:} Let $\X\subseteq \A^n $ be the edge set of a $1$-regular
subgraph \\
$\Lambda :=(\V ,\X )\subseteq(\A^{n-1},\A^{n} ) $ in
\BGr{n-1}, where $\V $ should denote the vertex set of $\Lambda $.
The set $\X\A^{-1} $ is the set of vertices
which are initial vertices of some edge in $\X $, and the set
$\A^{-1}\X $ is the set of vertices, which are terminal vertices of some
edge in $\X $. Since $\X $ is the edge set of a regular subgraph in \BGr{n-1},
it follows that:
\[ \A^{-1}\X =\V =\X\A^{-1} \, . \]

For a 1-regular graph the number of vertices is equal to
the number of edges, therefore we obtain:

\[ |\X |=|\V | =|\A^{-1}\X |=|\X\A^{-1}| \, .\]

\smallskip
$\Lambda $ is 1-regular and therefore for every $v\in\V $ there exist a
unique edge in $\X $ which is incident to $v$ and a unique edge in $\X $
which is incident from $v$. It follows, that for every $v\in\V $
there exist unique $a,b\in\A $ with
$av,vb\in\X $ and then also $avb\in\A\X\cap\X\A $ holds.
This shows, that there exists a
one-to-one map $G:\V \rightarrow\A\X\cap\X\A $.\\

\smallskip
Let $w\in\A\X\cap \X\A  $, then there exist $a,b\in\A $
and $v\in\A^{n-1} $ with $av,vb\in\X $ and
$w=avb$. While $av$ is an edge in $\Lambda $, we obtain $v\in\V $
and it follows, that $G(v)=avb=w $. This shows, that $G $ is a bijection
and therefore we obtain:

\[ |\X |=|\V |=|\A\X\cap\X\A| \, .\]

This shows the first part of the lemma.\\

\smallskip
Thus let us show the other direction of the lemma.
Let $\X \subseteq\A^n $ be a set with
$|\X |= |\A^{-1}\X | =|\X\A^{-1} |=|\A\X\cap\X\A| $.
Moreover let $\Lambda = (\V ,\X ) $ be the
subgraph (without isolated vertices ) of \BGr{n-1} with edge-set
$\X $. We have to show that $\Lambda $ is one-regular.\\

First we have:
\begin{equation}\label{piexistgl2}
|(\A\X\cap\X\A )\A^{-1}| =|\A\X\cap\X\A | \, .
\end{equation}

To show (\ref{piexistgl2}), let us assume that $|(\A\X\cap\X\A )\A^{-1}| <
|\A\X\cap\X\A | $.
Then there exist $a,b,c\in\A $ and $w\in\A^{n-1} $ with $a\neq b $
and $cwa,cwb\in\A\X\cap\X\A $. It follows, that $wa,wb\in\X $.
This is a contradiction, because $|\X | =|\X\A^{-1} | $ \, .\\

In the same way we obtain:
\begin{equation}\label{piexistgl3}
|\A^{-1}(\A\X\cap\X\A )| =|\A\X\cap\X\A | \, .
\end{equation}

>From $\A^{-1}(\A\X\cap\X\A ),\,(\A\X\cap\X\A )\A^{-1}\subseteq \X
$, $|\A\X\cap\X\A |=|\X | $,
(\ref{piexistgl2}) and (\ref{piexistgl3}) follows:

\begin{eqnarray}\label{piexistgl4}
\X & = & \A^{-1}( \A \X \cap \X \A ) = \A^{-1} \X \A \cap \X \\
\label{piexistgl5}\quad & = & ( \A \X \cap \X \A ) \A^{-1} = \A \X
\A^{-1} \cap \X \, .
\end{eqnarray}

Let $bv\in\X $ with  $b\in\A $ and $v\in\A^{n-1} $.
>From (\ref{piexistgl4}) follows that, there exists a letter
$a\in\A $ with $va\in\X $
and from  $|\X\A^{-1}|=|\X | $ follows, that the letter $a$ is unique.
Furthermore $v\in\V $ because $bv $ is an edge in $\Lambda $.
Thus we have:

\begin{equation}\label{eigen1}
\begin{array}{l}
\mbox{Let $v$ be a vertex of $\Lambda $,
such that there is at least one
edge of $\Lambda $ with}\\
\mbox{terminal vertex $v$. Then there exists an unique edge of
$\Lambda $ with initial}\\
\mbox{vertex $v$.}\\
\end{array}
\end{equation}

In the same way we obtain from
(\ref{piexistgl5}) and $|\A^{-1}\X|=|\X | $:

\begin{equation}\label{eigen2}
\begin{array}{l}
\mbox{Let $v$ be a vertex of $\Lambda $,
such that there is at least one
edge of $\Lambda $ with}\\
\mbox{initial vertex $v$. Then there exists an unique edge of
$\Lambda $ with terminal}\\
\mbox{vertex $v$.}\\
\end{array}
\end{equation}

>From (\ref{eigen1}) and (\ref{eigen2}) follows, that $\Lambda $ is
$1$-regular.\qed\\

\pagebreak
\begin{theorem}\label{piexisttheorem}
Let $|\A | =q\ge 2 $, $n\ge 1 $ and $1\le k< q$.
\begin{enumerate}
\renewcommand{\labelenumi}{(\roman{enumi})}
\item Let $\D $ be a two level $\pi_q(n+1;k)$-system with
$\D\subseteq\A^n\cup\A^{n+1} $ or a one level $\pi_q(n;k) $-system
with $\D\subseteq\A^n $.Then there exists $1\le L \le q^{n-1}$ such that
for any $\pi $-partition  $\D_1,\ldots ,\D_k $ of $\D $:

\[\begin{array}{lcccccccc}
L & = & |\D_1\cap\A^n | & = & |\D_2\cap\A^n | & = & \ldots & = & |\D_k\cap\A^n|\\
q^n-Lq & = & |\D_1\cap\A^{n+1} | & = & |\D_2\cap\A^{n+1} | & = & \ldots & = &
|\D_k\cap\A^{n+1}|
\end{array}\]
i.e., $|\D\cap\A^n | =kL \, $, $|\D\cap\A^{n+1}| =kq(q^{n-1}-L) $ and $\D $ is
a one level $\pi $-system iff $L=q^{n-1} $

\item Let $1\le L < q^{n-1} $, then there exists a two level $\pi_q(n;k) $-system\\
$\D\subseteq\A^n\cup\A^{n-1} $ with $kL=|\D\cap\A^n | $ if and only if there
exists a $k$-regular subgraph in \BGr{n-1} with $L$ vertices.

\item $\D\subseteq\A^n $ is a (one level) $\pi_q(n;k)$-system with
$\pi $-partition $\D_1,\ldots ,\D_k $
if and only if
$\D $ is the edge set of a $k$-factor $\Lambda $ in \BGr{n-1} and
$\D_1,\ldots ,\D_k $ are the edge sets of an edge disjoint decomposition
of $\Lambda $ into $1$-factors.

\end{enumerate}

\end{theorem}

\smallskip
>From (i) follows, that there exists two level $\pi_q
(n+1;k) $-systems $\D\subseteq \A^n\cup\A^{n+1} $ only of the form
$|\D\cap\A^n |=kL $ for some $1\le L <q^{n-1} $.
As the proof of the theorem will show any such $\pi $-system and any
$\pi $-partition for two level $\pi $-systems can be constructed as
described below.

\begin{quote}
\begin{constr}\label{piconstr1} \upshape\quad\\
\begin{enumerate}
\renewcommand{\labelenumi}{\arabic{enumi}.}

\item Let $\Lambda :=(\V ,\X )\subseteq (\A^n,\A^{n+1} ) $ be a
$k$-regular proper subgraph of \BGr{n-1}
with $L=|\V |$.

\item
Choose a decomposition of $\Lambda $ into
$k$ edge disjoint 1-factors
$\Lambda_1,\ldots ,\Lambda_k $ of
$\Lambda $. Let $\X_i $
denote the edge set of the $\Lambda_i $ for all $1\le i\le k$.

\item Choose permutations $\varphi_1,\ldots
,\varphi_k :\A \longleftrightarrow \A $ with
the property:
\[ \varphi_i (a) \neq \varphi_j(a) \quad \forall\, a\in\A\, ,\;
i\neq j \] and define
\[\begin{array}{lcl}
\V^c & := &\A^{n-1}-\V  \, , \\
\Y_i & := &\bigcup\limits_{a\in\A }a\V^c\varphi_i(a)\quad\forall\, 1\le i\le k \, ,\\
\Y & := & \Y_1\cup\ldots\cup\Y_k \, .
\end{array}\]

\pagebreak
\item Let $\D :=\X\cup\Y\subseteq \A^n\cup\A^{n+1} $ and $\D_i
:=\X_i\cup \Y_i $ for all $1\le i\le k $
\end{enumerate}
$\D\subseteq\A^n\cup\A^{n-1}  $ is a two level $\pi_q(n+1;k) $-system
with\\
$|\D\cap \A^n |=kL $ and $\pi $-partition $\D_1,\ldots ,\D_k $. Furthermore
any such $\D $ and $\pi $-partition $\D_1,\ldots ,\D_k $ of $\D $ can constructed
in such a way.
\end{constr}
\end{quote}

\smallskip
If $\A =\{ 0,\ldots ,q-1\} $, we can choose the permutations
$\varphi_1,\ldots ,\varphi_k $ in step 3 as:
\begin{equation}\label{pipermex}
\varphi_i (a) := a+i-1\mod q \quad \mbox{ for all }\; a\in\A\, ,\quad 1\le i\le k \, .
\end{equation}

\smallskip
If one needs only the $\pi $-system $\D $ without a certain $\pi $-partition
the following construction is possible. Let $\X \subseteq \A^n $ be the edge set
of a $k$-regular subgraph $\Lambda \subseteq \BGr{n-1} $ with $L$ vertices and
$ \A_a := \{\varphi_1 (a), \ldots \varphi_k (a) \} $, where
$\varphi_1,\ldots ,\varphi_k $ are permutations with the property in step 3.
For example, let $\A :=\{ 0,\ldots ,q-1 \} $ and the $\varphi_i $ as in
(\ref{pipermex}), then $\A_a = \{ a \mod q , (a+1)\mod q ,\ldots , (a+k-1)\mod q \} $.\\
We define $\D \subseteq\A^n\cup\A^{n-1} $ as:

\[ \D\cap \A^n := \X  \;\mbox{ and }\;
\D\cap\A^{n+1} := \bigcup\limits_{a\in\A }a(\A^{n-1}-\A^{-1}\X)\A_a  \, .\]

Then $\D $ is a two level $\pi_q(n+1;k)$-system,
because $\A^{-1}\X =\X\A^{-1} $ is the vertex set of $\Lambda $ and therefore
$\D\cap\A^{n+1} $ is the same as the set $\Y $ in step 3.
(i.e., $\A^{-1}\X $ are the vertices
of $\Lambda $ which has at least one antecessor vertex in $\Lambda $ and $\X\A^{-1} $
are vertices which have at least one successor vertex in $\Lambda $.)\\

\smallskip
For a given two level $\pi_q(n+1;k) $-system
$\D\subseteq\A^n\cup\A^{n+1} $
neither the decomposition of $\Lambda $ into 1-factors in step 2, nor the permutations
$\varphi_1,\ldots ,\varphi_k $ in step 3 are unique. The
above construction shows, that $\D $ has in general more than one $\pi $-partition.
In the same way by Theorem \ref{piexisttheorem} (iii) follows, that an
one level $\pi $-systems has more than one $\pi $-partition, because regular subgraphs
in $\BGr{n-1} $ has in general more than one decomposition into edge disjoint
$1$-factors.\\

\smallskip
{\bf Proof of Theorem \ref{piexisttheorem} :}
Let $\D \subseteq\A^n\cup\A^{n+1} $ a be two level $\pi_q(n+1;k)$-system
or a one level $\pi_q(n;k) $-system and $\D_1,\ldots ,\D_k $ a $\pi $-partition
of $\D $. We define:

\[\begin{array}{l@{\quad,\quad }ll}
\X:=\D\cap\A^n & \Y :=\D\cap\A^{n+1} \, , & \\
\X_i :=\D_i\cap\A^n & \Y_i:=\D_i\cap\A^{n+1} &\mbox{for } 1\le i\le k  \, ,\\
\multicolumn{2}{l}{ \mbox{and } L_i :=|\X_i |=|\D_i\cap\A^n | } &
\mbox{for } 1\le i\le k \, .
\end{array} \]

\pagebreak
\begin{claim}\label{piexistclaim1}
$\X \; $ is the edge set of a $k$-regular subgraph in \BGr{n-1}
and $\X_1,\ldots ,\X_k \; $
are the edge sets of edge disjoint $1$-factors of this subgraph.
\end{claim}

>From the properties of $\pi $-partition
follows for all $ 1\le i \le k $:
\[ q^n =| \pfs[n+1] (D_i) | =|\X_i\A | +|\Y_i | =qL_i + |\Y_i | \]
and therefore we have
\begin{equation} \label{pitheoremgl1}
|\Y_i | =q^n -qL_i\;\mbox{ for all }\; 1\le i \le k \, .
\end{equation}

Since the $\D_i $'s are fix-free and\\
$|\pfs[n+1](\D_i )| = |\sfs[n+1](\D_i )| =|\A^{-1}\pfs[n+1](\D_i )|
=|\sfs[n+1](\D_i )\A^{-1}| $, by Lemma~\ref{yeklemma0} follows:

\begin{equation} \label{pitheoremgl2}
L_i=|\X_i |= |\X_i\A^{-1} | =|\A^{-1}\X_i | \mbox{ and }
q^n-qL_i=|\Y_i |= |\Y_i\A^{-1} | =|\A^{-1}\Y_i | \, .
\end{equation}

While $\A^{-1}\D_i $ is prefix-free and $\D_i\A^{-1} $ is suffix-free, it follows
that\\
$\A^{-1}\X_i\A\cap\A^{-1}\Y =\A\X_i\A^{-1}\cap\Y\A^{-1} =\emptyset $. Thus
we obtain:

\begin{equation} \label{pitheoremgl2b}
\A\big(\X_i\A^{-1}\big)\A\cap(\Y_i\A^{-1})\A=
\A\big(\A^{-1}\X_i\big)\A\cap\A(\A^{-1}\Y_i) =\emptyset
\end{equation}

Furthermore we have $ \A\X_i \cap \A (\A^{-1} \Y_i )= \emptyset $,
because $\D_i $ is suffix-free. Therefore we obtain:

\begin{eqnarray*}
 q^{n+1} & \ge & |\A (\A^{-1} \X_i ) \A \cup \A \X_i \cup \A ( \A^{-1} \Y_i ) | \\
  & = & | \A ( \A^{-1} \X_i ) \A | + | \A \X_i |+| \A ( \A^{-1} \Y_i ) |
- | \A ( \A^{-1} \X_i ) \A \cap \A \X_i | \\
 & = & q^2L_i + qL_i +q(q^n-qL_i ) - | \A ( \A^{-1} \X_i ) \A \cap \A \X_i | \\
 & = & qL_i +q^{n+1}- | \A ( \A^{-1} \X_i ) \A \cap \A \X_i | \, .
\end{eqnarray*}

It follows that

\begin{eqnarray*}
& \quad &qL_i=|\A\X_i | \ge |\A (\A^{-1}\X_i )\A \cap \A\X_i | \ge qL_i \\
& \Rightarrow & |\A (\A^{-1}\X_i )\A \cap \A\X_i |=qL_i \\
& \Rightarrow &|\A^{-1}\X_i \A \cap \X_i |=L_i =|\X_i | \\
& \Rightarrow & \A^{-1}\X_i\A\cap\X_i =\X_i \, .
\end{eqnarray*}

>From the last equation follows, that for every $x_1\ldots
x_n\in\X_i $ there exists a letter $a\in\X_i $ such that $x_2\ldots x_na\in\X_i
$. For this letter we have \\
$x_1\ldots x_n a\in\A\X_l\cap\X_i\A\subseteq \A\X_i $.
By $|\X_i\A^{-1}|=|\X_i |$ we obtain,
that the letter $a$ is unique. This shows, that there exists a one-to-one map
from $\X_i $ into $ \A\X_i\cap\X_i\A $. Furthermore this map is
a bijection, because for every \\
$w_1\ldots w_nw_{n+1}\in\A\X_i\cap\X_i\A\subseteq\X_i\A  $ we have $w_1\ldots
w_n\in\X_i $. Thus we conclude:
\begin{equation}
|\X _i|=| \A\X_i\cap\X_i\A |\quad \forall\, 1\le i \le k \, .
\end{equation}

By Lemma \ref{piexistlemma} follows that each $ \X_i $ is
the edge set of a 1-regular subgraph $\Lambda_i :=(\V_i ,\X_i
)\subseteq (\A^n ,\A^{n+1}) $ of $\BGr{n-1} $, where we denote
by $\V_i $ the vertex set of $\Lambda_i $. For $1\le i,j \le k $
with $i\neq j $. We claim:\\

\begin{claim}\label{piexistclaim2}
For every  $x\in\X_j$  with  $x=x_1\ldots x_n  \,$,
$x_1,\ldots x_n\in\A\, $ there exists an unique letter
$x_1'\in\A\, $  with $ x_1\neq x'_1 \, $ and
$x'_1x_2\ldots x_n \in\X_i $.
\end{claim}

Let $x=x_1\ldots x_n\in\X_j $ be as in the claim. From
$\big|\sfs[n+1](\D_i ) \A^{-1} \big| =\big|\sfs[n+1](\D_i ) \big| =q^n $
and Lemma \ref{yeklemma1} (i) follows:

\begin{equation}\label{pitheorem3}
1=\big| \{ x\}\big| =\big| x\A\cap \sfs[n+1] (\D_i ) \big| = \big|
(\A\X_i \cup \Y_i )\cap x\A \big| = \big| \A\X_i \cap x\A \big|
+\big| \Y_i \cap x\A \big| \, .
\end{equation}

Since $\D_i\cup\D_j $ is prefix-free, we obtain
$|\Y_i\cap x\A\big| =0 $ and it follows that \\
$|\A\X_i\cap x\A | =1 $. Thus
we find $a,b\in\A $ and $z_1\ldots z_n\in\X_i \, ,\; z_1,\ldots
,z_n\in\A $ with $ az_1\ldots z_n = x_1\ldots x_nb $. While $\X_i $
is the edge set of a 1-regular subgraph in \BGr{n-1}, the edge
$z_1\ldots z_n $ has a unique antecessor edge in $\X_i $. It means
that there exists an unique letter $x'_1\in\A $ with $x'_1z_1\ldots z_{n-1}
\in\X_i $. It follows that $x'_1x_2\ldots x_n \in\X_i $.
By $\X_i\cap \X_j \subseteq \D_i\cap \D_j =\emptyset $, we obtain that
$x'_1\neq x_1 $. Furthermore from $|\A^{-1}\X_i |=|\X_i | $  it
follows that there is no other $c\neq
x'_1\, ,\;  c\in\A $ with $cx_2\ldots x_n\in\X_i $.
This shows Claim~\ref{piexistclaim2}.\\

\medskip
>From Claim~\ref{piexistclaim2} follows :
\begin{equation}\label{pitheoremgl4}
\A^{-1}\X_i =\A^{-1}\X_j \quad \forall\, 1\le i,j \le k \, .
\end{equation}

By (\ref{pitheoremgl2}) we have:
\begin{equation}\label{Lgleich1}
 L_i=|\X_i |=| \A^{-1}\X_i | =|\A^{-1}\X_j| =|\X_j |=L_j \quad \forall\,
1\le i,j\le k \, .
\end{equation}

Thus let
\begin{equation}\label{Lgleich2}   L:=L_1=\ldots = L_k \, .\end{equation}

Furthermore we have $ \V_i = \A^{-1}\X_i $ for all $1\le i\le k $,
because all $\Lambda_i $'s are
1-regular graphs and therefore every vertex $v\in\V_i $ is the terminal vertex
of a unique edge in $\X_i $.
Thus from (\ref{pitheoremgl4}) follows, that all $\Lambda_i $'s have the same
vertex set. We define:
\[ \V:= \V_1=\ldots =\V_k\; \mbox{ and }\;
\Lambda :=\bigcup\limits_{i=1}^{k }\Lambda_i
=\big( \V ,\X_1\cup \ldots\cup \X_{k }\big) =\big( \V ,\X \big) \, .
  \]

Especially we obtain:
\begin{equation}\label{pitheoremgl4b}
\V =\A^{-1}\X_i =\X_i\A^{-1} =\X\A^{-1}=\A^{-1}\X \quad\forall\, 1\le i\le k \, .
\end{equation}

While the edge sets $\X_1,\ldots ,\X_k $ are pairwise disjoint, it follows
that $\Lambda $ is the union of the $k$ edge disjoint $1$-regular graphs
$\Lambda_1,\ldots ,\Lambda_k $ having
all the same vertex set. Therefore $\Lambda $ is a $k$-regular subgraph of
\BGr{n-1} with $|\V |=L $ vertices and $\Lambda_1,\ldots ,\Lambda_k $
is an edge disjoint decomposition of $\Lambda $ into $1$-factors.
This shows Claim~\ref{piexistclaim1}.\\

\medskip
Furthermore from
(\ref{Lgleich1}), (\ref{Lgleich2}) and (\ref{pitheoremgl1}) follows
part (i) of the theorem.\\

\smallskip
If $\D $ is a one level $\pi $-system
then $\D =\X\, ,\quad \Y=\emptyset $ and $\V =\A^{n-1} $.
In this case $\Lambda $ is a $k$-factor and $\Lambda_1,\ldots ,\Lambda_k $
are $1$-factors of \BGr{n-1} and
from Claim \ref{piexistclaim1}
follows the ``only if'' part of Theorem \ref{piexisttheorem} (iii).
If $\D $ is a two level $\pi $-system, then from Claim \ref{piexistclaim1}
follows the ``only if '' part of Theorem \ref{piexisttheorem} (ii).\\

\smallskip
In the case that $\D $ is a two level $\pi $-system, we show,
that $\D $ and $\D_1,\ldots \D_k $
can be constructed as described in Construction \ref{piconstr1}.

\begin{claim}\label{piexistclaim3}
There exist (unique) permutations $\varphi_1,\ldots ,\varphi_k :
\A \longleftrightarrow \A $ with the property
\[ \varphi_i (a) \neq \varphi_j(a) \quad \forall\, a\in\A\, ,\;
i\neq j \]
such that
$\Y_i =\bigcup\limits_{a\in\A }a(\A^{n-1}-\V )\varphi_i (a) $ for all
$1\le i\le k $.
\end{claim}

Let us assume that there exist $a,b\in\A $ and $v\in\V $ with
$ avb\in\Y_i $ for some $1\le i\le k $.
 Then $ vb\in\A^{-1}\Y_i $ and by (\ref{pitheoremgl4b}) we have
$v\in\A^{-1}\X_i $. This is a contradiction, because $\D $
is a $\pi $-system, i.e., $\A^{-1}\D_i =\A^{-1}\X_i\cup \A^{-1}\Y_i $
is prefix-free. Therefore we obtain:

\begin{equation}\label{piclaim3gl1}
\Y_i\subseteq \A ( \A^{n-1} - \V )\A \;\mbox{ for all }\; 1\le i \le k\, .
\end{equation}

By (\ref{pitheoremgl1}) follows:

\begin{equation}\label{piclaim3gl2}
|\Y_i | =q^n-qL=q(q^{n-1}-|\V |) = |\A |\cdot |\A^{n-1}-\V | \, .
\end{equation}

Let $a,b,c\in\A $ and $w\in\A^{n-1}-\V $ such that $awb\, ,\, awc\in\Y_i $.
Because of \\
$\Y_i =\D_i\cap\A^{n+1} $ and $\D $ is a $\pi $-system follows from
Lemma \ref{yeklemma0}, that $|\Y_i |=|\A^{-1}\Y_i | $ holds. This shows $c=b $.
It follows, that for any $a\in\A $ and $w\in\A^{n-1}-\V $ with $aw\in\Y_i\A^{-1}$
there exists a unique $b\in\A $ with $awb\in\Y_i $. With
(\ref{piclaim3gl1}) and (\ref{piclaim3gl2}) follows, that there exists
a map $\varphi_i :\A\longrightarrow \A $ such that
$\Y_i =\bigcup\limits_{a\in\A }a(\A^{n-1}-\V )\varphi_i (a)  $. Obviously
the map $\varphi_i $ is unique. To show that $\varphi_i $ is a bijection we take
into account that from Lemma \ref{yeklemma0} also follows, that
$|\Y_i |=|\Y_i\A^{-1} | $. This means that for any $w\in\A^{n-1}-\V $ and
$b\in\A $ with $wb\in\Y_i\A^{-1} $ there exists a unique $a\in\A $ such that
$awb\in\Y_i $. This shows that $\varphi_i $ is a one-to-one map, i.e.
a permutation of the alphabet $\A $. Furthermore the sets
$\Y_1,\ldots ,\Y_k $ are pairwise disjoint and therefore
$\varphi_i (a) \neq \varphi_j (a) $ for all $a\in\A $ and $i\neq j $.
This shows that Claim~\ref{piexistclaim3} holds. By
Claim~\ref{piexistclaim1} follows, that any two level $\pi $-system
$\D\subseteq \A^{n}\cup\A^{n-1} $ and any $\pi $-partition of $\D $
is of the form described in Construction~\ref{piconstr1}.\\

\smallskip

We finish the proof, by showing that the set $\D $
in Construction \ref{piconstr1} is a $\pi $-system with $\pi $-partition
$\D_1,\ldots ,\D_k $ and that any edge set of a $k$-factor of \BGr{n-1}
is a one level $\pi_q(n;k) $-system, whereas a $\pi $-partition is given by
the edge sets of an edge disjoint decomposition of the $k$-factor into $1$-factors.
This shows the other direction of Theorem \ref{piexisttheorem} (ii) and (iii).\\

Let $\Lambda :=(\V ,\E ) \subseteq (\A^{n-1} ,\A^n ) $ be a
$k$-regular subgraph of \BGr{n-1} and let $L:=|\V | $.
By Proposition~\ref{factor} we obtain, that there are $k$
edge disjoint 1-factors $ \Lambda_1,\ldots ,\Lambda_k $ of $\Lambda $, i.e. $\Lambda
$ is the edge disjoint union of the $\Lambda_i $'s. Let $ \X_1,\ldots ,\X_k $ be
the edge sets of $\Lambda_1 ,\ldots ,\Lambda_k$. Then
\[ |\X_i |=|\V |=L \mbox{ and } \X_i\cap \X_j =\emptyset \quad
\forall\, 1\le i,j \le k \, .\]

With Lemma \ref{piexistlemma} we obtain:
\begin{equation}\label{piexistgl5}
|\A\X_i \cap\X_i\A | =|\A^{-1}\X_i |=|\X_i\A^{-1} | =|\X_l | =|\V
|=L \quad \forall\, 1\le i\le k \, .
\end{equation}

Let $\V^c:=\A^{n-1} -\V $ and $\varphi_1,\ldots ,\varphi_k :\A\leftrightarrow \A $
be permutations of $\A $ with the property:

\begin{equation}\label{piexistgl6}
\varphi_i (a)\neq \varphi_j (a) \quad\forall\; a\in\A \;\mbox{ and }\; i\neq j \, .
\end{equation}

We define for all $1\le i\le k $:

\[\begin{array}{lcl}
\Y_i & := & \bigcup\limits_{a\in\A }a\V^c \varphi_i (a) \, ,\\
\D_i & := & \X_i \cup\Y_i \, ,\\
\Y   & := & \Y_1\cup \ldots \cup \Y_k \subseteq \A\V^c\A \, ,\\
\D & := & \D_1\cup \ldots \cup \D_k =\X \cup \Y \, .
\end{array} \]

>From (\ref{piexistgl5}) follows:
\begin{equation}\label{piexistgl7}
|\D | =|\X | +|\Y | =L +|\A (\A^{n-1}-\V )| =L+ q^n-qL \, .
\end{equation}

For any subgraph of \BGr{n-1} with vertex set $\V\subseteq\A^{n-1} $
and edge set $\E\subseteq\A^n $, the sets $\V\A $ and $\A\V $ are subsets of $\E $.
Therefore we obtain for
all $1\le i\le k$:

\begin{equation}\label{piexistgl8}
\A\V \subseteq \X \;\mbox{ and }\; \V\A \subseteq \X \, .
\end{equation}

It follows, that $\D =\X \cup \Y $ is fix-free. Furthermore
by property (\ref{piexistgl6}) of the $\varphi_i $'s follows, that
$\Y_1,\ldots ,\Y_k $ are pairwise disjoint and therefore $\D_1,\ldots ,D_k $
is a partition of $\D $. While the $\varphi_i $'s are permutations of $\A $
we have for all $1\le i\le k $

\begin{equation}\label{piexistgl9}
\A^{-1}\Y_i  = (\A^{n-1}-\V )\A  \;\mbox{ and }\;
\Y_i\A^{-1}  = \A(\A^{n-1}-\V ) \, .
\end{equation}

All $\Lambda_i $'s are 1-regular subgraphs with vertex set $\V $, i.e.
for every $v\in\V $ there is an edge in $\X_i $ incident to $v$ and an edge
incident from $v$. It follows, that:

\begin{equation}\label{piexistgl10}
\V =\A^{-1}\X_i =\X_i\A^{-1} \quad\mbox{ for all }\; 1\le i \le k \, .
\end{equation}

By (\ref{piexistgl6}), (\ref{piexistgl9}) and (\ref{piexistgl10}) we obtain
for all $1\le i\le k $:

\[ |\A^{-1}\D_i | = |\A^{-1}\X_i | + |\A^{-1}\Y_i | =
L+q^n -qL =|\D_i | \, .\]

In the same way $|\D_i\A^{-1} | = |\D_i | $ can be shown for all $1\le i\le k $.\\
Furthermore from (\ref{piexistgl9}) and (\ref{piexistgl10}) follows,
that $\A^{-1}\D_i $ is prefix-free and $\D_i\A^{-1} $ is suffix-free.
For the Kraftsum of $\A^{-1}\D_i $ and $\D_i\A^{-1} $ we obtain
\[|\A^{-1}\X_i |\cdot q^{-n+1} + |\A^{-1}\Y_i |\cdot q^{-n}
=Lq^{-n+1} +(q^n-qL)q^{-n} =1 \, .\]
In a similar way we obtain
$|\X_i\A^{-1} |\cdot q^{-n+1} + |\Y_i\A^{-1} |\cdot q^{-n} = 1 $.\\

It follows, that $\A^{-1}\D_i $ is maximal prefix-free and $\D_i\A^{-1} $
is maximal suffix-free. This shows that $\D $ is a $\pi $-system with
$\pi $-partition $\D_1,\ldots \D_k $. Especially, if $1\le L < q^{n-1} $
then $|\Y_i | =q^n-qL > 0 $ and therefore $\D \subseteq\A^{n}\cup\A^{n+1} $
is a two level $\pi_q (n+1;k) $-system with\\
$|\D\cap \A^n |=|\X | =kL $.
This shows part (ii) of Theorem~\ref{piexisttheorem} and
moreover that any two level $\pi $-system $\D\subseteq\A^n\cup \A^{n+1} $ can
be constructed as described in Construction \ref{piconstr1}. If $L=q^{n-1} $
then $\Y =\emptyset $ and $\V = \A^{n-1} $. Therefore $\X $ is the edge set
of a $k$-factor in \BGr{n-1} and $\D \subseteq \A^n $ is a one level
$\pi_q(n;k ) $-system. This shows part (iii) of Theorem \ref{piexisttheorem}.
\qed\\

\smallskip
We give an example for Construction \ref{piconstr1}, by constructing
a two level $\pi_3 (4;2) $-system  for $L= 7 $. Let $\A =\{ 0,1,2 \}$.
We need a $2$-regular
subgraph $\Lambda := (\V ,\X )\subseteq (\A^2 ,\A^3 ) $ in \BGr[3]{2}
with $|\V | = 7 $ and two edge disjoint $1$-factors $\Lambda_1 ,\Lambda_2 $ of
$\Lambda $. With $X_1,\X_2 $ we denote the edge sets of $\Lambda_1 $
and $\Lambda_2 $. The pictures below show such subgraphs in \BGr[3]{2}
and their successor maps $\F ,\F_1 $ and $\F_2 $.\\

\pagebreak
\begin{minipage}{14cm}
\parbox{7cm}{
\begin{center} \BGr[3]{2}\\
\setlength{\unitlength}{0.5mm}
\begin{picture}(140,140)
\put(70,35){\framebox(10,5){\scriptsize 11}}
\put(40,50){\framebox(10,5){\scriptsize10}}
\put(40,85){\framebox(10,5){\scriptsize 00}}
\put(70,100){\framebox(10,5){\scriptsize 02}}
\put(100,85){\framebox(10,5){\scriptsize 22}}
\put(100,50){\framebox(10,5){\scriptsize 21}}
\put(15,30){\framebox(10,5){\scriptsize 01}}
\put(70,130){\framebox(10,5){\scriptsize 20}}
\put(125,30){\framebox(10,5){\scriptsize 12}}
\thicklines
\put(45,59){\vector(0,1){22}}
\put(42,70){\scriptsize 0}

\put(53,92){\vector (2,1){14}}
\put(58,96){\scriptsize 2}

\put(83,99){\vector (2,-1){14}}
\put(92,96){\scriptsize 2}

\put(105,81){\vector (0,-1){22}}
\put(106,70){\scriptsize 1}

\put(97,48){\vector (-2,-1){14}}
\put(90,40){\scriptsize 1}

\put(67,42){\vector (-2,1){14}}
\put(58,40){\scriptsize 0}

\put(96,53){\vector (-1,0){42}}
\put(74,54){\scriptsize 0}

\put(47,58){\vector (2,3){26}}
\put(60,72){\scriptsize 2}

\put(77,97){\vector (2,-3){26}}
\put(89,72){\scriptsize 1}

\put(38,50){\vector (-1,-1){14}}
\put(30,45){\scriptsize 1}

\put(27,34){\vector (1,1){14}}
\put(36,39){\scriptsize 0}

\put(123,34){\vector (-1,1){14}}
\put(113,38){\scriptsize 1}

\put(112,50){\vector (1,-1){14}}
\put(120,44){\scriptsize 2}

\put(77,108){\vector (0,1){19}}
\put(78,116){\scriptsize 0}

\put(73,127){\vector (0,-1){19}}
\put(70,116){\scriptsize 2}

\put(68,127){\vector (-2,-3){22}}
\put(54,110){\scriptsize 0}

\put(103,93){\vector (-2,3){22}}
\put(93,110){\scriptsize 0}

\put(43,81){\vector (-1,-2){22}}
\put(30,60){\scriptsize 1}

\put(129,37){\vector (-1,2){22}}
\put(119,60){\scriptsize 2}

\put(28,30){\vector (4,1){40}}
\put(47,30){\scriptsize 1}

\put(82,40){\vector (4,-1){40}}
\put(98,30){\scriptsize 2}

\qbezier(19,49)(18,100)(66,131)
\put(19,49){\vector (0,-1){11}}
\put(26,92){\scriptsize 1}

\qbezier(131,38)(131,100)(91.5,126)
\put(92,125.5){\vector (-3,2){8}}
\put(123,92){\scriptsize 0}

\qbezier(24,26)(75,0)(120,23)
\put(120,22.5){\vector(2,1){7} }
\put(75,8){\scriptsize 2}

\put(53,82){\circle{8}}
\put(56.5,85.5){\scriptsize 0}
\put(97,82){\circle{8}}
\put(99.5,75.5){\scriptsize 2}
\put(75,44.5){\circle{8}}
\put(80,45){\scriptsize 1}
\end{picture} \end{center} }
\hfill
\parbox{7cm}{
\begin{center} $2$-regular subgraph $\Lambda\subseteq\BGr[3]{2} $ \\
\setlength{\unitlength}{0.5mm}
\begin{picture}(140,140)
\put(40,50){\framebox(10,5){\scriptsize10}}
\put(40,85){\framebox(10,5){\scriptsize 00}}
\put(70,100){\framebox(10,5){\scriptsize 02}}
\put(100,50){\framebox(10,5){\scriptsize 21}}
\put(15,30){\framebox(10,5){\scriptsize 01}}
\put(70,130){\framebox(10,5){\scriptsize 20}}
\put(125,30){\framebox(10,5){\scriptsize 12}}
\thicklines
\put(45,59){\vector(0,1){22}}
\put(42,70){\scriptsize 0}

\put(53,92){\vector (2,1){14}}
\put(58,96){\scriptsize 2}





\put(96,53){\vector (-1,0){42}}
\put(74,54){\scriptsize 0}


\put(77,97){\vector (2,-3){26}}
\put(89,72){\scriptsize 1}

\put(38,50){\vector (-1,-1){14}}
\put(30,45){\scriptsize 1}

\put(27,34){\vector (1,1){14}}
\put(36,39){\scriptsize 0}

\put(123,34){\vector (-1,1){14}}
\put(113,38){\scriptsize 1}

\put(112,50){\vector (1,-1){14}}
\put(120,44){\scriptsize 2}

\put(77,108){\vector (0,1){19}}
\put(78,116){\scriptsize 0}

\put(73,127){\vector (0,-1){19}}
\put(70,116){\scriptsize 2}







\qbezier(19,49)(18,100)(66,131)
\put(19,49){\vector (0,-1){11}}
\put(26,92){\scriptsize 1}

\qbezier(131,38)(131,100)(91.5,126)
\put(92,125.5){\vector (-3,2){8}}
\put(123,92){\scriptsize 0}

\qbezier(24,26)(75,0)(120,23)
\put(120,22.5){\vector(2,1){7} }
\put(75,8){\scriptsize 2}

\put(53,82){\circle{8}}
\put(56.5,85.5){\scriptsize 0}
\end{picture} \end{center} }
\end{minipage} \\[8mm]

\begin{minipage}{14cm}
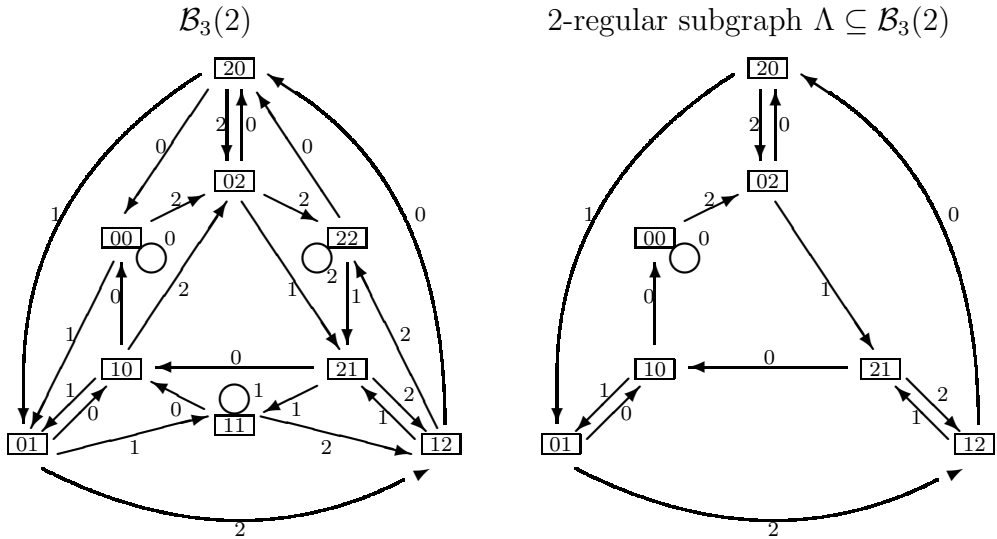
\begin{center} edge disjoint $1$-factors $\Lambda_1 $ and $\Lambda_2 $ of $\Lambda $
\end{center}
\parbox{7cm}{
\begin{center}
\setlength{\unitlength}{0.5mm}
\begin{picture}(140,140)
\put(40,50){\framebox(10,5){\scriptsize10}}
\put(40,85){\framebox(10,5){\scriptsize 00}}
\put(70,100){\framebox(10,5){\scriptsize 02}}
\put(100,50){\framebox(10,5){\scriptsize 21}}
\put(15,30){\framebox(10,5){\scriptsize 01}}
\put(70,130){\framebox(10,5){\scriptsize 20}}
\put(125,30){\framebox(10,5){\scriptsize 12}}
\thicklines
\put(45,59){\vector(0,1){22}}
\put(42,70){\scriptsize 0}

\put(53,92){\vector (2,1){14}}
\put(58,96){\scriptsize 2}





\put(96,53){\vector (-1,0){42}}
\put(74,54){\scriptsize 0}


\put(77,97){\vector (2,-3){26}}
\put(89,72){\scriptsize 1}













\qbezier(19,49)(18,100)(66,131)
\put(19,49){\vector (0,-1){11}}
\put(26,92){\scriptsize 1}

\qbezier(131,38)(131,100)(91.5,126)
\put(92,125.5){\vector (-3,2){8}}
\put(123,92){\scriptsize 0}

\qbezier(24,26)(75,0)(120,23)
\put(120,22.5){\vector(2,1){7} }
\put(75,8){\scriptsize 2}

\end{picture}\end{center} }
\hfill
\parbox{7cm}{
\begin{center}
\setlength{\unitlength}{0.5mm}
\begin{picture}(140,140)
\put(40,50){\framebox(10,5){\scriptsize10}}
\put(40,85){\framebox(10,5){\scriptsize 00}}
\put(70,100){\framebox(10,5){\scriptsize 02}}
\put(100,50){\framebox(10,5){\scriptsize 21}}
\put(15,30){\framebox(10,5){\scriptsize 01}}
\put(70,130){\framebox(10,5){\scriptsize 20}}
\put(125,30){\framebox(10,5){\scriptsize 12}}
\thicklines









\put(38,50){\vector (-1,-1){14}}
\put(30,45){\scriptsize 1}

\put(27,34){\vector (1,1){14}}
\put(36,39){\scriptsize 0}

\put(123,34){\vector (-1,1){14}}
\put(113,38){\scriptsize 1}

\put(112,50){\vector (1,-1){14}}
\put(120,44){\scriptsize 2}

\put(77,108){\vector (0,1){19}}
\put(78,116){\scriptsize 0}

\put(73,127){\vector (0,-1){19}}
\put(70,116){\scriptsize 2}










\put(53,82){\circle{8}}
\put(56.5,85.5){\scriptsize 0}
\end{picture}\end{center} }
\end{minipage}

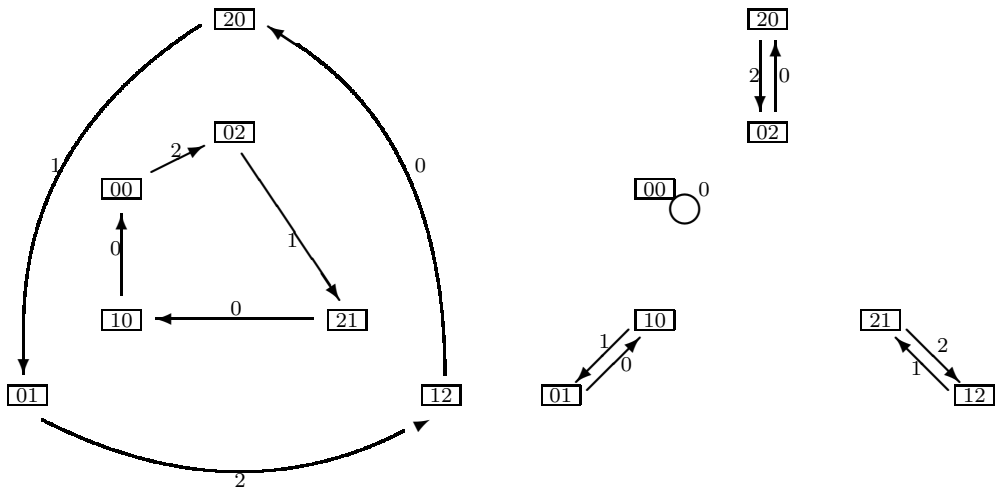
Successor maps $\F ,\F_1 $ and $\F_2 $ of the graphs $\Lambda ,\Lambda_1 $
and $\Lambda_2 $:\\
 {\footnotesize
\[\begin{array}{|l|r|r|r||l|r|r|r||l|r|r|r|}
\hline
&&&&&&&&&&& \\[-3mm]
\mbox{\scriptsize\hspace{-0.5mm}$v\in\A^2 $ }& \mbox{\scriptsize $\F (v)$} & \mbox{\scriptsize $\F_1(v)$} & \mbox{\scriptsize $\F_2(v)$} &
\mbox{\scriptsize\hspace{-0.5mm} $v\in\A^2$} & \mbox{\scriptsize $\F (v)$} & \mbox{\scriptsize $\F_1(v)$} &  \mbox{\scriptsize $\F_2(v)$} &
\mbox{\scriptsize\hspace{-0.5mm} $v\in\A^2$} & \mbox{\scriptsize $\F (v)$} & \mbox{\scriptsize $\F_1(v)$} &  \mbox{\scriptsize $\F_2(v)$} \\

\hline
&&&&&&&&&&& \\[-3mm]
00 & \{ 0,2 \} & 2 & 0 & 01 & \{ 0,2 \} & 2 & 0 & 02 & \{ 0,1 \} & 1 & 0 \\
\hline
&&&&&&&&&&& \\[-3mm]
10 & \{ 0,1 \} & 0 & 1 & 11 & \emptyset & \emptyset & \emptyset & 12 & \{ 0,1 \} & 0 & 1 \\
\hline
&&&&&&&&&&& \\[-3mm]
20 & \{ 1,2 \} & 1 & 2 & 21 & \{ 0,2 \} & 0 & 2 & 22 & \emptyset & \emptyset & \emptyset \\
\hline
\end{array}\] }

For the vertex set $\V $, the set $\V^c $ and the edge sets $\X ,\X_1 ,\X_2 $ of $\Lambda ,\Lambda_1 $
and $\Lambda_2 $ we obtain:

\[ \begin{array}{lcl}
\V & = & \{ 00,01,02,10,12,20,21 \} \quad , \quad
\V^c  =  \A^2-\V =\{ 11, 22 \} \; , \\
\X_1 & = & \{ 000 , 010 ,020 ,101 ,121 ,202 ,212 \} \; , \\
\X_2 & = & \{ 002 ,012 ,021 ,100 ,120 ,201 ,210 \} \; , \\
\X & = & \{ 000,002,010,012,020,021,100,101,120,121,201,202,210,212 \}
\end{array}\]

We define the permutations $\varphi_1,\varphi_2 :\A\leftrightarrow \A $ as:
\[\begin{array}{|c|c|c|}
\hline
a\in\A & \varphi_1 (a) & \varphi_2 (a) \\
\hline
0 & 0 & 1 \\
\hline
1 & 1 & 2 \\
\hline
2 & 2 & 0 \\
\hline
\end{array}\]

Obviously $\varphi_1 (a) \neq \varphi_2 (a) $ holds for all $a\in\A $. If we let
$\Y_i :=\bigcup\limits_{a=0}^2 a \V^c \varphi_i (a)$ for $i\in\{ 1,2 \} $ and
$\Y :=\Y_1\cup \Y_2 $ we obtain:
\[ \begin{array}{lcl}
\Y_1 & = & \{ 0110 ,0220 ,1111,1221,2112,2222 \} \\
\Y_2 & = & \{ 0111 ,0221 ,1112,1222,2110,2220 \} \\
\Y & = & \{ 0110 ,0111,0220,0221,1111,1112,1221,1222,2110,2112,2220,2222 \}
\end{array} \]

Let $\D :=\X \cup \Y $ and $\D_i :=\X_i\cup\Y_i $ for $i\in\{ 1,2 \} $,
then $\D \subseteq\A^3\cup\A^4 $ is a two level $\pi_3 (4;2) $-system with
$\pi $-partition $\D_1 ,\D_2 $ and $|\D\cap\A^3 | =|\X | =14 =2 L $.
For the Kraftsum we obtain:
\[ \Kraft (\D )= |\X |\cdot \frac{1}{3^3} + |\Y |\cdot \frac{1}{3^4}
=\frac{14}{3^3}+\frac{12}{3^4} =\frac{54}{3^4} = \frac{2}{3} \]
\smallskip

Using Theorem \ref{piexisttheorem} we can prove now Lemma \ref{yek2}. This was:

\begin{quote}
Let $n\in\NN $ and $|\A |=2 $. For any $\beta_n ,\beta_{n+1}\in\NN_0 $ with
$\frac{\beta_n}{2^n} +\frac{\beta_{n+1}}{2^{n+1}}=\frac{1}{2} $ there exists
a $\pi_2 (0,\ldots ,0,\beta_n ,\beta_{n+1}\, ;\, 1 ) $-system.
\end{quote}

{\bf Proof of Lemma \ref{yek2}:} Let $n\in\NN $ and $\A =\{ 0,1 \} $. From
$\frac{\beta_n}{2^n} +\frac{\beta_{n+1}}{2^{n+1}}=\frac{1}{2} $ follows,
that $0 \le \beta_n  \le 2^{n-1} $ and $\beta_{n+1}= 2^n-2\beta_n $.
For $\beta_n\neq 0 $ follows from Lempel's Theorem~\ref{lempel}, that
there exists a cycle in \BGr[2]{n-1} of length $\beta_n $, i.e. there
exists a $1$-regular subgraph in \BGr[2]{n-1} with $\beta_n $ vertices.
By Theorem \ref{piexisttheorem} (i) and (ii)
it follows for $1\le \beta_n < 2^{n-1} $,
that there exists a two level $\pi_2 (n+1 ,1) $-system
$\D \subseteq \A^n\cup\A^{n+1} $ with $|\D\cap\A^n |= \beta_n $
and $|\D\cap\A^{n+1} |= 2^n -2\beta_n =\beta_{n+1} $. Especially $\D $ is a
$\pi_2 (0,\ldots ,0,\beta_n ,\beta_{n+1}\, ;\, 1 ) $-system.
If $\beta_n =2^{n-1} $ then $\beta_{n+1} =0 $ and the cycle is a Hamilton
circuit in \BGr[2]{n-1}, i.e. a $1$-factor of \BGr[2]{n-1}. It follows
from Theorem \ref{piexisttheorem} (iii), that the edge set of the cycle
is a one-level $\pi_2 (n;1) $-system, i.e. a
$\pi_2 (0,\ldots ,0,\beta_n ,\beta_{n+1}\, ;\, 1 ) $-system.
Also for $\beta_{n+1}=2^n $ and $\beta_n =0 $ there exists a
$\pi_2 (0,\ldots ,0,\beta_n ,\beta_{n+1}\, ;\, 1 ) $-system. This follows
by the same argument,i.e
the edge set of a Hamilton circuit in \BGr[2]{n} is a
$\pi_2 (0,\ldots ,0,\beta_{n+1}\, ;\, 1 ) $-system.\qed\\

\smallskip
For $\A =\{ 0,1 \} $
from Construction \ref{piconstr1} follows, that we obtain a two level
$\pi_2 (n+1;1)$-system $\D\subseteq\A^n\cup\A^{n+1} $ with
$| \D\cup\A^n | =L $, if we choose $ \D\cap\A^n $ to be the edge set
of a cycle
in \BGr[2]{n-1} of length $L$ and $ \D\cap\A^{n+1} = 0\V^c0 \cup 1\V^c1 $.\footnote{
$\V^c $ is the set of vertices, which do not lay on the cycle.}
Furthermore every one level $\pi_2 (n ;1 ) $-system is
the edge set of a Hamilton circuit in \BGr[2]{n-1} and vice versa.\\

\smallskip
>From Theorem~\ref{piexisttheorem} and Theorem~\ref{piext} we obtain the following
generalization of Yekhanin's Theorem~\ref{yek} for arbitrary alphabets:\\

\begin{theorem}\label{pietheorem2}
Let $|\A | =q \ge 2\, ,\quad 1\le k < q $, $(\alpha_l )_{l\in\NN }$ be a
sequence of nonnegative integers
with $ \sum\limits_{l=1}^{\infty } \alpha_l q^{-l} \le \gamma_k $ ,where
$\gamma_k $ is chosen as in Theorem~\ref{piext}, and
$n\in\NN $ be the first integer with $\alpha_n\neq 0 $.\\
\begin{enumerate}
\renewcommand{\labelenumi}{(\roman{enumi})}
\item If $\frac{\alpha_n }{q^n } + \frac{\alpha_{n+1}}{q^{n+1}} \ge  \frac{k}{q} $
, $  \alpha_n =kL $ for some $1\le L < q^{n-1}  $ and there exists a
$k$-regular subgraph in \BGr{n-1} with $L$ vertices,
then there exists a fix-free code which
fits to $(\alpha_l )_{l\in\NN }$.

\item If $\frac{\alpha_n }{q^n } \ge \frac{k}{q} $ then there exists a fix-free
code which fits to $(\alpha_l )_{l\in\NN }$.

\end{enumerate}
\end{theorem}

{\bf Proof: } Let $\A , k  ,  (\alpha_l )_{l\in\NN } $ and $n$
be as in the theorem.
Furthermore let\\
$\frac{\alpha_n }{q^n } + \frac{\alpha_{n+1}}{q^{n+1}} \ge  \frac{k}{q} $
and $\alpha_n =kL $ for some $1\le L < q^{n-1}  $. Because of $\alpha_n < kq^{n-1} $
it follows, that there exists $1\le \beta \le \alpha_{n+1} $ with

\begin{equation}\label{yekgengl1}
\frac{\alpha_n }{q^n } + \frac{\beta }{q^{n+1}} =  \frac{k}{q} \, .
\end{equation}

We obtain $\beta = k(q^n-qL) $.\\

Let us assume that
there exists a $k$-regular subgraph with $L$ vertices in \BGr{n-1}. Then from
Theorem \ref{piexisttheorem} (ii) and (i) follows, that there exists
a two level $\pi_q (n+1;k) $-system $\D\subseteq\A^n\cup\A^{n+1} $ with
$|\D\cap\A^n | =kL =\alpha_n $ and
$|\D\cap\A^{n+1} | =k(q^n - q L ) =\beta $. $\D $ is
a $\pi_q (\alpha_1,\ldots \alpha_n,\beta ;k )$-system.
By (\ref{yekgengl1}) and Theorem~\ref{piexisttheorem} follows, that
there exists a fix-free extension of $\D $ which fits to $(\alpha_l )_{l\in\NN } $.
This shows (i). Part (ii) of the theorem has been shown already in
Proposition~\ref{1levelpiprop1}. Another proof is the following:\\
If $\frac{\alpha_n }{q^n } \ge \frac{k}{q} $ then there exists
$0 \le \beta \le \alpha_n $ such that
$\frac{\beta }{q^n } = \frac{k}{q} $. Then $\beta = kq^n $. Moreover
there exists a $k$-factor in \BGr{n-1}. Therefore from  Theorem
\ref{piexisttheorem} (iii)
follows, that there exists a one level $\pi_q (n;k)$-system $\D\subseteq \A^n $.
Obviously $\D $ is a $\pi_q (\alpha_1,\ldots ,\alpha_{n-1},\beta ;k ) $-system
and therefore from Theorem \ref{piext} follows that there exists a fix-free
extension of $\D $ which fits to $(\alpha_l )_{l\in\NN } $.\qed\\

\smallskip
As shown in the proof of Corollary~\ref{piext}, we have
$\gamma_{\lceil\frac{q}{2}\rceil } \ge \frac{3}{4} $ for all $q\ge 2 $,
therefore we obtain for $k= \lceil\frac{q}{2}\rceil $ the following corollary
of Theorem~\ref{pietheorem2} :
\begin{cor}\label{pietheorem3}
Let $|\A | =q \ge 2\, , \; (\alpha_l )_{l\in\NN }$ be a
sequence of nonnegative integers
with $ \sum\limits_{l=1}^{\infty } \alpha_l q^{-l} \le \frac{3}{4} $. Let
$n\in\NN $ be the first integer with $\alpha_n\neq 0 $.\\
\begin{enumerate}
\renewcommand{\labelenumi}{(\roman{enumi})}
\item If $\frac{\alpha_n }{q^n } + \frac{\alpha_{n+1}}{q^{n+1}} \ge
\lceil\frac{q}{2}\rceil \frac{1}{q} $
, $  \alpha_n =\lceil\frac{q}{2}\rceil L $ for some $1\le L < q^{n-1}  $
and there exists a
$\lceil\frac{q}{2}\rceil $-regular subgraph in \BGr{n-1} with L vertices
then there exists a fix-free code which
fits to $(\alpha_l )_{l\in\NN }$.

\item If $\frac{\alpha_n }{q^n } \ge \lceil\frac{q}{2}\rceil \frac{1}{q} $
then there exists a fix-free code which fits to $(\alpha_l )_{l\in\NN }$.

\end{enumerate}

\end{cor}

Let $\A =\{ 0,1 \} $. By Lempels Theorem~\ref{lempel} follows, that for
every $1\le L \le 2^{n-1} $ there exists a cycle of length $L$ in \BGr[2]{n-1},
i.e. there exists a $1$-regular subgraph in \BGr[2]{n-1} with $L$ vertices.
Therefore we obtain from Corollary~\ref{pietheorem3}, Yekhanin's Theorem~\ref{yek},
which was:

\begin{quote}
Let $ |\A |=2 $ ,$(\alpha_l)_{n\in\NN } $ be a sequence of nonnegative
integers with \\
$ \sum\limits_{l=1}^{\infty } \alpha_l q^{-l} \le \frac{3}{4} $.
Let $n\in\NN $ be the smallest integer with $\alpha_n\neq 0 $.\\
If $\frac{\alpha_n }{q^n } + \frac{\alpha_{n+1}}{q^{n+1}} \ge\frac{1}{2} $
then there exists a fix-free code which fits to $(\alpha_l)_{n\in\NN } $.\\
\end{quote}

\medskip
In the generalization of Theorem~\ref{yek} for arbitrary alphabets,
Theorem~\ref{pietheorem2} and Corollary~\ref{pietheorem3}, two extra
conditions occur. First $\alpha_n =kL$ for some
$1\le L\le q^{n-1} $, if $\alpha_n\le kq^{n-1} $ and secondly there
has to exists a $k$-regular subgraph in \BGr{n-1} with $L$ vertices.
One can ask if there is a generalization of Theorem~\ref{yek}, without
such extra conditions ?
However, if we take into account, that Theorem~\ref{piexisttheorem} gives us
a one-to-one correspondence between two level
$\pi $-systems $\D\subseteq\A^n\cup\A^{n+1} $
and regular subgraphs in de Bruin digraphs, it is obviously, that
Theorem~\ref{pietheorem2} and Corollary~\ref{pietheorem3} are the best
generalizations of Yekhanin's original Theorem \ref{yek}, which
can be obtained by using the technique of $\pi $-systems.
One can only try to replace the condition for the existence of regular subgraphs
in Theorem~\ref{pietheorem2} and Corollary~\ref{pietheorem3} by the values
of $L$, for which $k$-regular subgraphs with $L$ vertices in \BGr{n-1} exist.
It was shown in Chapter 3 Theorem~\ref{klnseq2}, there do not exist $k$-regular
subgraphs with $L$ vertices in \BGr{n-1}, if $L< k^{n-1} $ or
$k^{n-1} < L < k^{n-1}+k^{n-2} $. This means, that for these values of $L$
there does not exist a $\pi_q (n;k) $-system $\D\subseteq\A^n\cup\A^{n+1} $ with $kL$
codewords on the $n$-th level. Furthermore in Chapter 3 there are
several constructions of $k$-regular subgraphs
of \BGr{n} for certain values of $L$.\\

\chapter{The $\frac{3}{4}$-conjecture for binary fix-free codes}

In this chapter we examine the $\frac{3}{4} $-conjecture for the special case
$|\A| =2 $. There are some results which was shown only for this case.\\

In \cite{zeger} Kukorelly and Zeger have shown the following theorem.

\begin{theorem}[Kukorelly and Zeger \cite{zeger}]\label{zweiertheorem}
Let $|\A |=2 $ and $\alpha_1,\ldots ,\alpha_n \in\NN_0 $. If
$\sum\limits_{l=1}^n\alpha_l \left( \frac{1}{2}\right)^l \le \frac{3}{4} $
and $\alpha_l \le 2 $ for all $1\le l\le n $, then there exists a fix-free
set $\C\subseteq\A^* $ which fits to $(\alpha_1,\ldots ,\alpha_n )$.
\end{theorem}

To prove the theorem, Kukorelly and Zeger distinguish eight cases, where
the theorem is easy to show or follows from other theorems for
the first seven cases.
We show the theorem only for this seven easy cases, a proof
of the theorem for the last case can be found in \cite{zeger}.

{\bf Proof:}
Let $|\A | =2 $ and $(\alpha_1,\ldots ,\alpha_n )$
as in the theorem. It is sufficient to show that the theorem holds for all
$(\alpha_1,\ldots m,\alpha_n) $ with Kraftsum $\frac{3}{4} $.
We distinguish eight cases:
\begin{itemize}

\item[{\em Case 1:}] $\alpha_1 =1 $\\
In this case the theorem follows from Theorem \ref{yek}.

\item[{\em Case 2:}] $\alpha_1 =0 $ and $\alpha_2 =2 $ \\
Also in this case the theorem follows from Theorem \ref{yek}.

\item[{\em Case 3:}] $\alpha_1 =\alpha_2 =0  $\\
In this case with theorem \ref{qfall3} follows that the Theorem holds.

\item[{\em Case 4:}] $\alpha_1 =0\, ,\quad \alpha_2=1 $ and $\alpha_3\ge 2 $\\
In this case the theorem follows from Theorem \ref{yek}.

\item[{\em Case 5:}] $\alpha_1 =0\, ,\quad
\alpha_2 =1 \, ,\quad\alpha_3\le 1 $ and $n=3 $\\
In this case the theorem follows from Theorem \ref{qfall2}.

\item[{\em Case 6:}] $\alpha_1 =0\, ,\quad
\alpha_2 =1 \, ,\quad\alpha_3\le 1 $ and $n=4 $\\
While the Kraftsum of $(\alpha_1,\ldots ,\alpha_4)  $ is $\frac{3}{4} $ it follows,
that either $\alpha_4= 6 $ or $\alpha_4=8 $.
Two examples for such fix-free codes are listed below:\\
{\small
\[\begin{array}{|r|r|}
\hline
\mbox{\scriptsize
$\quad (\alpha_1,\ldots ,\alpha_4)$} &
\mbox{\scriptsize $\quad (\alpha_1,\ldots ,\alpha_4)$}  \\
\mbox{\scriptsize $=(\, 0\, ,\, 1 \, ,\, 1\, ,\, 6\, )$} &
\mbox{\scriptsize $=(\, 0\, ,\, 1 \, ,\, 0\, ,\, 8\, )$}\\ \hline
11 & 11\\
101 & 0000\\
0000 & 0010 \\
0010 & 0100 \\
0100 & 0110 \\
0110 & 1001 \\
1001 & 1000 \\
1000 & 1010 \\
\quad & 0101 \\ \hline
\end{array}\] }

\item[{\em Case 7:}] $\alpha_1 =0\, ,\quad
\alpha_2 =1 \, ,\quad\alpha_3\le 1 $ and $n=5 $\\
In this case there are six possibilities for $(\alpha_1,\ldots ,\alpha_5) $. For
each of them examples for fix-free codes are shown in the tabular below:\\

{\footnotesize  \[\begin{array}{|r|r|r|r|r|r|}\hline
\mbox{\scriptsize $(\alpha_1,\ldots ,\alpha_5)=$} &
\mbox{\scriptsize $(\alpha_1,\ldots ,\alpha_5)=$} &
\mbox{\scriptsize $(\alpha_1,\ldots ,\alpha_5)=$} &
\mbox{\scriptsize $(\alpha_1,\ldots ,\alpha_5)=$} &
\mbox{\scriptsize $(\alpha_1,\ldots ,\alpha_5)=$} &
\mbox{\scriptsize $(\alpha_1,\ldots ,\alpha_5)=$} \\
\mbox{\scriptsize $( 0\, ,\, 1 \, ,\, 1\, ,\, 2\, ,\, 8  )$} &
\mbox{\scriptsize $( 0\, ,\, 1 \, ,\, 1\, ,\, 1\, ,\, 10  )$} &
\mbox{\scriptsize $( 0\, ,\, 1 \, ,\, 1\, ,\, 0\, ,\, 12  )$} &
\mbox{\scriptsize $( 0\, ,\, 1 \, ,\, 0\, ,\, 2\, ,\, 12  )$} &
\mbox{\scriptsize $( 0\, ,\, 1 \, ,\, 0\, ,\, 1\, ,\, 14  )$} &
\mbox{\scriptsize $( 0\, ,\, 1 \, ,\, 1\, ,\, 0\, ,\, 16  )$} \\\hline
11 & 11 & 11 & 11 & 11 & 11 \\
101 & 101 & 101 & 1001 & 1001 & 00000 \\
1001 & 1001 & 00000 & 0110 & 00000 & 00010 \\
0110 & 00000 & 00010 & 00000 & 00010 & 00100 \\
00000 & 00010 & 00100 & 00010 & 00100 & 00110 \\
00010 & 00100 & 00110 & 00100 & 00110 & 01000 \\
00100 & 00110 & 01000 & 01000 & 01000 & 01010 \\
01000 & 01000 & 01010 & 01010 & 01010 & 01100 \\
01010 & 01010 & 01100 & 01110 & 01100 & 01110 \\
01110 & 01100 & 01110 & 10001 & 01110 & 10001 \\
10001 & 01110 & 10001 & 10000 & 10001 & 10000 \\
10000 & 10001 & 10000 & 00001 & 10000 & 00001 \\
\quad & 10000 & 10010 & 00101 & 00001 & 00101 \\
\quad & \quad & 01001 & 10100 & 00101 & 10100 \\
\quad & \quad & \quad & 10101 & 10100 & 10101 \\
\quad & \quad & \quad & \quad & 10101 & 10010 \\
\quad & \quad & \quad & \quad & \quad & 01001 \\ \hline
\end{array}\] }

\item[{\em Case 8:}] $\alpha_1 =0\, ,\quad
\alpha_2 =1 \, ,\quad\alpha_3\le 1 $ and $n\ge 6 $\\
For this case a proof of the theorem can be found in \cite{zeger}.\qed\\
\end{itemize}

\medskip
Ye and Yeung have shown in \cite{yeung} some results which are related
to the binary $\frac{3}{4}$-conjecture. Especially they prove a sufficient
and a necessary condition for the existence of binary fix-free codes,
where the conditions depends on the lengths sequence of a fix-free code.
Let us remind that a lengths sequence $\vec{l_n} =(l_1,\ldots ,l_n )$ is an
increasing finite sequence of natural numbers. A code $\C $ fits to
the lengths sequence $\vec{l_n} $, if the numbers $l_1,\ldots ,l_n $ are
the lengths of the codewords in $\C $.\\

We define for a number $x\in\RR $:
\[ x^+ := \left\{\begin{array}{ll} x & \mbox{ for }\; x> 0 \\
0 & \mbox{ for }\; x=0 \end{array}\right. \, .  \]

Let $\vec{l_n}=(l_1,\ldots ,l_n )\in\NN^n $ be a lengths sequence.
We define $su(\vec{l_n} ) $, $ne(\vec{l_n}) $ and $h(i) $ as:

\[\begin{array}{lcl}
h(i) & := & \min \{ j\, | \, l_j=l_{i+1} \}
\quad\mbox{ for all }\; 1\le i<n \, ,\\[2mm]
su(\vec{l_n} ) & := &
\prod\limits_{i=1}^{n-1}
\big( 1-2\sum\limits_{1\le j\le i} 2^{-l_i} +(\, i+1-h(i)\, )\cdot
2^{-l_i+1} +\hspace*{-8mm}
\sum\limits_{\begin{array}{l} \mbox{\tiny $1\le j,k\le h(i)-1$}\\
\mbox{\tiny s.t. $l_j+l_k \le l_i+1$}\end{array}}
\hspace*{-8mm}2^{-l_j-l_k} \big)^+ \, ,\\
ne(\vec{l_n} ) & := &
\prod\limits_{i=1}^{n-1}
\big( 1-2\sum\limits_{1\le j\le i} 2^{-l_i} +(\, i+1-h(i)\, )\cdot
2^{-l_i+1} +\hspace*{-5mm}
\sum\limits_{1\le j,k\le h(i)-1}\hspace*{-5mm}
2^{(l_{i+1}-l_j-l_k)^+-l_{i+1}} \big)^+ \, . \\
\end{array}\]

\begin{theorem}[Ye and Yeung]\label{mad}
Let $|\A | = 2 $ and $\vec{l_n}\in\NN^n $ be a lengths sequence.
\begin{enumerate}
\renewcommand{\labelenumi}{(\roman{enumi})}

\item (Sufficient Condition) If $su(\vec{l_n} ) > 0 $, then there exists a
fix-free code $\C\subseteq\A^+ $ which fits to $\vec{l_n} $.

\item (Necessary Condition) If $ne(\vec{l_n} ) = 0 $, then there does not
exist a fix-free code $\C\subseteq\A^+ $ which fits to $\vec{l_n} $.
\end{enumerate}
\end{theorem}

\smallskip
Furthermore Ye and Yeung have shown in \cite{yeung} the following corollary
of part (i) of the theorem above.

\begin{cor}[Ye and Yeung]\label{madcor}
Let $|\A | = 2 $ and $\vec{l_n}\in\NN^n $ be a lengths sequence.
If
\[ \sum\limits_{1\le j\le n} 2^{-l_i} <\frac{1}{2} +
\frac{n+2-h(n-1)}{2}\cdot 2^{-l_n } \, ,\]
then there exists a
fix-free code $\C\subseteq\A^+ $ which fits to $\vec{l}_n $.
\end{cor}

Proofs of Theorem~\ref{mad} and Corollary~\ref{madcor} can be found in
\cite{yeung}.\\

\medskip
Moreover Ye and Yeung have shown the following
proposition.

\begin{prop}[Ye and Yeung]\label{58yeung1}
Let $|\A | = 2 $ and $\alpha_1,\ldots ,\alpha_n \in\NN_0 $. If \\
$\alpha_1 =1 $
and $\sum\limits_{l=1}^n \alpha_l \left(\frac{1}{2}\right)^l \le \frac{5}{8} $,
then there exists a
fix-free code $\C\subseteq\A^+ $ which fits to $(\alpha_1,\ldots ,\alpha_n )$.
\end{prop}

Ye and Yeung gave in \cite{yeung} two different proofs of the proposition
above. The first proof works with Theorem~\ref{mad} and
the second proof use Lemma~\ref{einhalb}. With a proof of Yekhanin we will
show in the last section of this chapter, that the proposition also
holds for sequences with $\alpha_1=0 $. However the proposition above
follows also from Theorem~\ref{yek}. If $(\alpha_l)_{l\in\NN } $ is
a sequence of nonnegative integers with $\alpha_1 =1 $,
then $\frac{\alpha_1}{2}+\frac{\alpha_2}{2^2}\ge \frac{1}{2}$.
Therefore we obtain by Theorem~\ref{yek} the more general proposition:

\begin{prop}[Yekhanin]\label{58yeung1}
Let $|\A | = 2 $ and $(\alpha_l)_{l\in\NN}$ be a sequence
with \\
$\sum\limits_{l=1}^n \alpha_l \left(\frac{1}{2}\right)^l \le \frac{3}{4} $.
If $\alpha_1 =1 $ then there exists a
fix-free code $\C\subseteq\A^+ $ which fits to $(\alpha_l)_{l\in\NN}$.
\end{prop}

\medskip
The binary $\frac{3}{4}$-conjecture was verified by computer research
for several sequences. The
results are collected in the proposition below.

\begin{prop}\label{computer1}
Let $\A =\{ 0,1 \} $ and $\alpha_1,\ldots ,\alpha_n\in\NN_0 $  with
$\sum\limits_{l=1}^{n } \alpha_l2^{-l} \le \frac{3}{4} $.

\begin{enumerate}
\renewcommand{\labelenumi}{(\roman{enumi})}
\item {\bf Ye and Yeung \cite{yeung}}\\
If $n< 8 $,
then there exists a fix-free set $\C\subseteq\A^* $
which fits to $(\alpha_1,\ldots ,\alpha_n )$.

\item {\bf Yekhanin \cite{yek1}}\\
If $n< 9 $,
then there exists a fix-free set $\C\subseteq\A^* $
which fits to $(\alpha_1,\ldots ,\alpha_n )$.

\end{enumerate}
\end{prop}

\pagebreak
\section{Binary fix-free codes obtained from quaternary fix-free codes}
In Chapter 2 and Chapter 4, we gave a lot of results for the $q$-ary
case of the $\frac{3}{4}$-conjecture. By identifying the letters in
$\{ 0,1,2,3 \} $ with the the words of length 2 in $\{ 0,1 \}^2 $, it is possible
to obtain some new results for the binary $\frac{3}{4}$-conjecture
from the old $4$-ary results in Chapter 2 and Chapter 4.\\

In this section we denote with $\A $ and $\B $ the alphabets
$\A :=\{ 0,1 \} $ and\\
$\B :=\{ 0,1,2,3 \} $.
Let $\phi :\B \leftrightarrow \A^2 $ be a bijection. For example:
\[ \phi (0)= 00 ,\phi (1) =01 ,\phi (2 )= 10 \;\mbox{ and }\; \phi (3) = 11 \, .\]

Let $w=w_1\ldots w_{2n} \in\A^{2n} , v=v_1\ldots v_n \in\B^n $ with
$w_1,\ldots ,w_{2n} \in\A , v_1,\ldots ,v_n\in\B $ and
$\C\subseteq\bigcup\limits_{l=1}^{\infty } \A^{2l} \, , \, \D\subseteq\B^+ $.
We define $\phi (v) ,\phi^{-1}(w) ,\phi (\D ) $ and $\phi^{-1} (\C )$ as follows:
\[\begin{array}{lcl}
\phi (v) & := & \phi (v_1)\ldots \phi (v_n) \in\A^{2n} \, ,\\
\phi^{-1} (w) & := & \phi^{-1} (w_1w_2)\phi^{-1} (w_3w_4)\ldots
\phi^{-1} (w_{2n-1}w_{2n})\in\B^n \, ,\\
\phi (\C ) & := & \{ \phi (v)\in\A^+ \, |\, v\in\D \} \subseteq \B^+ \, ,\\
\phi^{-1} (\D ) & := & \{ \phi^{-1} (w) \, | \, w\in\D \} \subseteq
\bigcup\limits_{l=1}^{\infty }\A^{2l} \, ,\\
\phi (e) & := & e \;\mbox{ and }\; \phi^{-1} (e) := e \, .
\end{array}\]

Obviously the map $\phi $
is a one-to-one map from $\B^* $ onto $\bigcup\limits_{l=0}^{\infty }\A^{2l} $
with inverse map $\phi^{-1} $. Furthermore we obtain:

\begin{equation}\label{quat1}
\phi (\B^n ) = \A^{2n} \;\mbox{ for all }\; n\in\NN_0 \, .
\end{equation}

It is easy to verify, that the following equations hold:

\begin{equation}\label{quat2}\begin{array}{lcll}
\phi (uv) & =  & \phi (u) \phi(v)  &\mbox{for all }\; u,v\in\B^* \, ,\\
\phi^{-1} (u'v') & = & \phi^{-1} (u) \phi^{-1} (v)  &\mbox{for all }\;
u',v'\in\bigcup\limits_{l=0}^{\infty }\A^{2l} \, .
\end{array}\end{equation}

\begin{lemma}\label{quatlem}
Let $\A =\{ 0,1 \} ,\B =\{ 0,1,2,3 \} ,$ $\phi:\B\leftrightarrow \A^2 $
be a bijection and $\C\subseteq \A^+ ,\D\subseteq\B^+ $ such that
$\phi (\D )=\C $.
\begin{enumerate}
\renewcommand{\labelenumi}{(\roman{enumi})}
\item $ |\C\cap\A^{2l+1} |= 0 $ and $|\C\cap\A^{2l} |=|\D\cap\B^{l} | $ for all
$l\in\NN_0 $.
\item $\C $ is fix-free if and only if $\D $ is fix-free.
\item
$ \Kraft (\C ) =
\sum\limits_{l=0}^{\infty } |\C\cap\A^l |\left(\frac{1}{2}\right)^l =
\sum\limits_{l=0}^{\infty } |\D\cap\B^l |\left(\frac{1}{4}\right)^l =
\Kraft (\D ) $
\end{enumerate}
\end{lemma}

\pagebreak
{\bf Proof:}\\
\begin{enumerate}
\renewcommand{\labelenumi}{(\roman{enumi})}
\item $|\C\cap\A^{2l+1} |= 0 $ for all $l\in\NN_0 $, because $\phi (\D )=\C $ and
$\phi :\B \leftrightarrow \bigcup\limits_{l=0}^{\infty }\A^{2l } $ is
a bijection. Furthermore we have $\phi (B^l ) =\A^{2l} $ for all $l\in\NN_0 $.
Therefore we obtain for all $l\in\NN_0 $:
\[|\C\cap\A^{2l} |=|\phi (\D )\cap\A^{2l} | =
|\phi (\D\cap\B^{l}) | = |\D\cap\B^{l} | \, .\]

\item This follows by (\ref{quat2}).

\item Since $\phi_{\big| \B^l }:\B^l \leftrightarrow \A^{2l } $ is a bijection
and $\phi (\D )=\C \subseteq\bigcup\limits_{l=0}^{\infty }\A^{2l} $, we obtain:
\[\begin{array}{lclcl}
\Kraft (\C ) & = &
\sum\limits_{l=0}^{\infty }|\C\cap \A^l |\left(\frac{1}{2}\right)^l
& = &
\sum\limits_{l=0}^{\infty }|\C\cap \A^{2l} |\left(\frac{1}{2}\right)^{2l} \\
\quad & = &
\sum\limits_{l=0}^{\infty }|\phi (\D ) \cap \A^{2l} |\left(\frac{1}{2}\right)^{2l}
& = &
\sum\limits_{l=0}^{\infty }|\phi (\D\cap \B^l) |\left(\frac{1}{2^2}\right)^l \\
\quad & = &
\sum\limits_{l=0}^{\infty }|\D\cap \B^l |\left(\frac{1}{4}\right)^l
& = & \Kraft (\D ) \; .\mbox{\quad\qed}
\end{array}\]
\end{enumerate}

If we use the above lemma together with the theorems for the $q$-ary  case in
Chapter 2 and Chapter 4, we obtain the following proposition.

\begin{prop}\label{quatprop}
Let $\A :=\{ 0,1 \} $ and $(\alpha_l )_{l\in\NN } $ be a sequence of nonnegative
integers with
$\sum\limits_{l=1}^{\infty }\alpha_l \left(\frac{1}{2}\right)^l\le \frac{3}{4} $.

\begin{enumerate}
\renewcommand{\labelenumi}{(\roman{enumi})}

\item If there exists an $n\ge 2 $ such that
$\alpha_2 = \alpha_{2l+1}  =0 $ for all $l\in\NN_0 $,
$\alpha_{2l}=2^l $ for all $2\le l < n $, $\alpha_{2n} \ge 2^{n+1} $ and
$\alpha_{2l} \in\NN_0 $ for all $l>n $, then there exists a fix-free code $\C\subseteq\A^+ $
which fits to $(\alpha_l )_{l\in\NN } $.

\item If there exists an $n\ge 3 $ such that
$\alpha_2 = \alpha_4 =\alpha_{2l+1}  =0 $ for all $l\in\NN_0 $,
$\alpha_{2l}=2^{l+1} $ for all $2\le l < n $, $\alpha_{2n} \ge 2^{n+2} $ and
$\alpha_{2l} \in\NN_0 $ for all $l> n $, then there exists a fix-free code
$\C\subseteq\A^+ $ which fits to $(\alpha_l )_{l\in\NN } $.

\item If there exists an $n\in\NN $ such that
$\alpha_2 =\alpha_4 =\ldots =\alpha_{2n-2} =\alpha_{2l+1} =0 $ for all $l\in\NN_0 $,
$\alpha_{2n} $ is even,
$\frac{\alpha_{2n}}{2^{2n}} +\frac{\alpha_{2n+2}}{2^{2n+2}}\ge \frac{1}{2} $ and
there exists a $2$-regular subgraph of \BGr[4]{n-1} with $\frac{\alpha_{2n}}{2} $
vertices,
then there exists a fix-free code
$\C\subseteq\A^+ $ which fits to $(\alpha_l )_{l\in\NN } $.

\item If there exists an $n\in\NN $ such that
$\alpha_2 =\alpha_4 =\ldots =\alpha_{2n-2} =\alpha_{2l+1} =0 $ for all $l\in\NN_0 $
and
$\frac{\alpha_{2n}}{2^{2n}} \ge \frac{1}{2} $,
then there exists a fix-free code
$\C\subseteq\A^+ $ which fits to $(\alpha_l )_{l\in\NN } $.

\item Let $l_{min} :=\min \{ l \, | \, \alpha_l\neq 0 \} $ and
$l_{max} := \sup \{ l \, | \, \alpha_l\neq 0 \} $. If
$l_{max} <\infty $, $4\le l_{min} $ is even, $\alpha_{2l+1}=0 $ for all $l\in\NN_0 $
and $\alpha_{2l} \le 2^{\frac{l_{min}}{2}-2+l } $ for all $2l\neq l_{max} $,
then there exists a fix-free code
$\C\subseteq\A^+ $ which fits to $(\alpha_l )_{l\in\NN } $.
\end{enumerate}

\end{prop}

{\bf Proof: } Let $\B :=\{ 0,1,2,3 \} $ and $\phi :\B\leftrightarrow \A^2 $
be a bijection.
We define the sequence $(\beta_l )_{l\in\NN } $ as:

\[ \beta_l :=\alpha_{2l } \quad \mbox{ for all }\; l\in\NN \, . \]

In all cases of the proposition we have $\alpha_{2l+1} =0 $ for all $l\in\NN $.
Let us assume
that $\D \subseteq\B^+ $ is
a fix-free code which fits to $(\beta_l )_{l\in\NN } $.
By Lemma~\ref{quatlem} follows, that $\C := \phi (\D )\subseteq\A^+ $
is a fix-free code which fits to $(\alpha_l )_{l\in\NN } $.
Therefore it is for all cases of the proposition sufficient to show that
there exists a fix-free code $\D\subseteq\B^+  $ which fits to
$(\beta_l )_{l\in\NN } $.
We obtain for the Kraftsum of $(\beta_l )_{l\in\NN }$:
\[ \sum\limits_{l=0}^{\infty }\beta_l \left(\frac{1}{4}\right)^l
= \sum\limits_{l=0}^{\infty }\alpha_{2l} \left(\frac{1}{2}\right)^{2l} =
\sum\limits_{l=0}^{\infty }\alpha_l \left(\frac{1}{2}\right)^l\le \frac{3}{4} \, .\]

\begin{enumerate}
\renewcommand{\labelenumi}{(\roman{enumi})}

\item In this case we obtain for $(\beta_l )_{l\in\NN } $:
\[ \beta_1=0,
\beta_l =2^l =\left(\frac{4}{2}\right)^l \;\mbox{ for all }\; 2\le l< n
\;\mbox{ and }\; \beta_n \ge 2^{n+1}= 4\left(\frac{4}{2}\right)^{n-1} \, .   \]
By Proposition~\ref{ex1prop2} follows, that there exist a fix-free code
$\D\subseteq\B^+ $ which fits  to $(\beta_l )_{l\in\NN } $.

\item In this case from Proposition~\ref{ex3prop2} follows that there exists
a fix-free code $\D\subseteq\B^+ $ which fits  to $(\beta_l )_{l\in\NN } $.

\item In this case from Corollary~\ref{pitheorem3} (i) follows that there exists
a fix-free code $\D\subseteq\B^+ $ which fits  to $(\beta_l )_{l\in\NN } $.

\item In this case from Corollary~\ref{pitheorem3} (ii) follows that there exists
a fix-free code $\D\subseteq\B^+ $ which fits  to $(\beta_l )_{l\in\NN } $.

\item Let $l'_{min} :=\min\{ l \, | \, \beta_l \neq 0 \} $ and
$l'_{max} :=\sup\{ l \, | \, \beta_l \neq 0 \} $. It follows that
$l'_{max} =\frac{l_{max}}{2} <\infty $ and $ l'_{min} =\frac{l_{min}}{2}\ge 2 $.
Furthermore we obtain:
\[ \beta_l =\alpha_{2l} \le 2^{l'_{min}-2+l } =
4^{l'_{min}-2}\cdot 2^2 \cdot 2^{l-l'_{min}} \;\mbox{ for all }\; l\neq l_{max} \, . \]
By Theorem~\ref{qfall4} follows, that there exists
a fix-free code $\D\subseteq\B^+ $ which fits  to $(\beta_l )_{l\in\NN } $.\qed\\

\end{enumerate}

\pagebreak

\section{Binary fix-free codes with Kraftsum $\frac{5}{8} $ }
For $\C\subseteq \A^* $ and $a,b\in\A $ we define:

\begin{eqnarray*}
\,^a\C & := & \{aw\in\A^*|\, aw\in\C  \}=a\A^*\cap\C \\
\C^b & := & \{ wb\in\A^*|\, wb\in\C  \} = \A^*b\cap\C \\
\,^a\C^b & := &\,^a\C\cap\C^b= a\A^*b\cap\C = \{ awb\in\A^* |\, awb\in\C \}
\end{eqnarray*}
We show first the following proposition:
\begin{prop}\label{propbin1}
Let $|\A |=q \, ,\; a,b\in\A \, ,\; n\in\NN $
and $\C\subseteq\bigcup\limits_{l=1}^n\A^l $ be fix-free then:
\begin{enumerate}
\renewcommand{\labelenumi}{(\roman{enumi})}
\item \[ \big|a\A^{n-1}b-\bfs[n+1](\C ) \big| \ge \max \left\{ 0,
q^{n-1} -\big|\pfs[n](\,^a\C )\big| -\big|\sfs[n](\C^b)\big| \right\} \, ,\]

\item
\begin{eqnarray*}
\big|\pfs[n] (\,^a\C ) \big| & = & q^n\sum\limits_{c\in\A}\Kraft (\,^a\C^c )\, ,\\
\big|\sfs[n] (\C^b ) \big| & = & q^n\sum\limits_{c\in\A}\Kraft (\,^c\C^b ) \, .
\end{eqnarray*}
\end{enumerate}
\end{prop}

{\bf Proof: } \\
We show (i):
\begin{eqnarray*}
\big|a\A^{n-1}b-\bfs[n+1](\C ) \big| & = &
\big|\big(a\A^{n}\cap\A^{n}b\big) - \big(\pfs[n+1](\C )\cup \sfs[n+1](\C )\big)\big| \\
\quad & = & \big|\big( a\A^n -\pfs[n+1](\C )\big)\cap
\big(\A^nb-\sfs[n+1](\C ) \big)\big| \\
\quad & = & \big|\big( a\A^{n-1} -\pfs[n](\C )\big)\A\cap
\A\big(\A^{n-1}b-\sfs[n](\C ) \big)\big| \\
\mbox{\footnotesize (with lemma \ref{yeklemma1} (ii) )} & \ge &
\big|a\A^{n-1}-\pfs[n](\C ) \big| + \big|\A^{n-1}b-\sfs[n](\C )\big| -q^{n-1} \\
\quad & = & \big| a\A^{n-1} -\pfs[n] (\,^a\C )\big| +
\big| \A^{n-1}-\sfs[n](\C^b ) \big| -q^{n-1} \\
\quad & = & q^{n-1}-\big|\pfs[n](\,^a\C )\big|+
q^{n-1} -\big|\sfs[n](\C^b)\big| -q^{n-1} \\
\quad & = & q^{n-1} -\big|\pfs[n](\,^a\C )\big| -\big|\sfs[n](\C^b)\big| \, .
\end{eqnarray*}

\smallskip
We show (ii):
\begin{eqnarray*}
\big|\pfs[n] (\,^a\C ) \big| & = &
\big|\pfs[n] (\bigcup\limits_{c\in\A }\,^a\C^c )\big|
 =  \sum\limits_{l=1 }^n
\big| \A^l\cap\bigcup\limits_{c\in\A }\,^a\C^c )\big|\cdot q^{n-l}\\
\quad & = & q^n\sum\limits_{c\in\A}\sum\limits_{l=1 }^n
\big| \,^a\C^c \cap \A^l \big|\cdot q^{-l}
= q^n\sum\limits_{c\in\A}\Kraft (\,^a\C^c ) \, .
\end{eqnarray*}
The second part of (ii) follows the same way.\qed\\

The next theorem was shown by Yekhanin in \cite{yek2}.

\begin{theorem}[Yekhanin \cite{yek2}]\label{5to8theorem}
Let $|\A |=2 $ and $(\alpha_l )_{l\in\NN } $ a
sequence of nonnegative integers with
$\sum\limits_{l=1}^{\infty }\alpha_l \big(\frac{1}{2}\big)^l \le \frac{5}{8} $
then there exists a fix-free $\C\subseteq\A^* $ which fits
$(\alpha_l )_{l\in\NN } $. \end{theorem}

\smallskip
{\bf Proof: } Every sequence  $(\alpha_l )_{l\in\NN } $, with
Kraftsum smaller than $\frac{5}{8} $ can be extended to a sequence
$(\alpha'_l )_{l\in\NN } $ with $\alpha'_l\ge \alpha_l $ for all
$l\in\NN $ and Kraftsum equal to $\frac{5}{8}$. Therefore it is
sufficient to show the theorem for a sequence $(\alpha_l
)_{l\in\NN } $ of nonnegative integers with
$\sum\limits_{l=1}^{\infty }\alpha_l \big(\frac{1}{2}\big)^l = \frac{5}{8} $.\\

\smallskip

We distinguish three cases:\\

{\em Case 1:} $\alpha_1 $=1\\
Then $\frac{\alpha_1}{2}=\frac{1}{2} $ and by Theorem~\ref{yek} it follows,
that there exist a fix-free code which fits to $(\alpha_l )_{l\in\NN } $.\\

\smallskip
{\em Case 2:} $\alpha_1= 0 $ and $\alpha_2 =2 $\\
In this case $\frac{\alpha_2}{2^2}=\frac{1}{2} $ and by
Theorem~\ref{yek}, there exists a fix-free code which fits to
$(\alpha_l )_{l\in\NN } $.\\

{\em Case 3:} $\alpha_1 = 0 $ and $\alpha_2 < 2 $ \\
In this case we can find unique sequences of nonnegative integers
 $(\beta^{00}_l )_{l\in\NN }\, ,\; (\beta^{01}_l )_{l\in\NN }\,
,\; (\beta^{10}_l )_{l\in\NN } $ and $(\beta^{11}_l )_{l\in\NN } $
such that:
\[\begin{array}{l}
\sum\limits_{l=1}^{\infty }\beta^{00}_l \big(\frac{1}{2}\big)^l = \frac{1}{4} \, ,\\
\sum\limits_{l=1}^{\infty }\beta^{01}_l \big(\frac{1}{2}\big)^l =
\sum\limits_{l=1}^{\infty }\beta^{10}_l \big(\frac{1}{2}\big)^l =
\sum\limits_{l=1}^{\infty }\beta^{11}_l \big(\frac{1}{2}\big)^l = \frac{1}{8} \, ,\\
\alpha_l=\beta^{00}_l+\beta^{01}_l+\beta^{10}_l+\beta^{11}_l\quad
\forall\, l\in\NN \, ,\\
\beta^{00}_m> 0 \Rightarrow
\beta^{01}_l =\beta^{10}_l =\beta^{11}_l =0 \quad \forall\,
l\in\{ 1,\ldots ,m-1 \} \, ,\\
\beta^{01}_m> 0 \Rightarrow
\beta^{10}_l =\beta^{11}_l =0 \quad \forall\,
l\in\{ 1,\ldots ,m-1 \} \, ,\\
\beta^{10}_m> 0 \Rightarrow
\beta^{11}_l =0 \quad \forall\, l\in\{ 1,\ldots ,m-1 \} \, .\\
\end{array}\]

To give an example, let the first eight terms of
$(\alpha_l)_{l\in\NN },(\beta^{00}_l)_{l\in\NN },\ldots
,(\beta^{11}_l)_{l\in\NN }$ given by:
\[ \begin{array}{lc@{(}c@{\, ,\,}c@{\, ,\,}c@{\, ,\,}c@{\, ,\,}c@{\, ,\,}c@{\, ,\,}c@{\, ,\,}c@{)}}
(\alpha_1,\ldots ,\alpha_8) &  = &0&0&1&1&5&2&14&36 \, ,\\
(\beta^{00}_1,\ldots ,\beta^{00}_8) &  = &0&0&1&1&2&0&0&0 \, ,\\
(\beta^{01}_1,\ldots ,\beta^{01}_8) &  = &0&0&0&0&3&2&0&0 \, ,\\
(\beta^{10}_1,\ldots ,\beta^{10}_8) &  = &0&0&0&0&0&0&14&4 \, ,\\
(\beta^{11}_1,\ldots ,\beta^{11}_8) &  = &0&0&0&0&0&0&0&32 \, .\\

\end{array}\]
and $\alpha_l = \beta^{00}_l =\ldots = \beta^{11}_l =0 $ for $l > 8 $.
Then
$\sum\limits_{l=1}^{\infty } \alpha_l\cdot\big(\frac{1}{2}\big)^l =\frac{5}{8}$
and the $(\beta^{ab}_l)_{l\in\NN } $ are the unique sequences with the above
properties.\\

We construct  by induction a fix-free $\C \subseteq \A^* $ such that
$\big| \,^a\C^b\cap \A^l \big| =\beta^{ab}_l $ for all $l\in\NN $
, $a,b\in\{ 0,1\} $. $\C $ is a fix-free code which fits to
$(\alpha_l )_{l\in\NN } $, because $\C $ is the disjoint union of
$\,^0\C^0 \, ,\; \,^1\C^0 \, ,\; \,^0\C^1 $ and $ \,^1\C^1 $.\\
To construct a $\C $ with the above properties it is sufficient to find
a sequence $\C_1\subseteq C_2\subseteq \C_3\subseteq\ldots $ of fix-free sets
such that
$\C_n \subseteq\bigcup\limits_{l=1}^n \A^l$ and
$\big| \,^a\C_n^b\cap \A^l \big| =\beta^{ab}_l $ for all
$l\in\{ 1\ldots ,n \}  \, ,\quad a,b\in\{ 0,1\} $.
Then we obtain $\C $ by
$\C:=\bigcup\limits_{l=1}^{\infty }\C_n $.\\
Let $\C_1 := \emptyset $, then $\C_1 $ is fix-free and
$\big| \,^a\C^b\cap \A^l \big| =\beta^{ab}_1 =0 $
for all $a,b\in\{ 0,1\} $.\\

\smallskip
Let  $\C_n \subseteq\bigcup\limits_{l=1}^n \A^l$ be a fix-free set such that
$\big| \,^a\C^b_n\cap \A^l \big| =\beta^{ab}_l $ for all $l\in\{ 1,\ldots ,n\} $,
$a,b\in\{ 0,1\} $. Then we obtain:

\[
\begin{array}{l}
2^{-2}\ge \sum\limits_{l=1}^{n+1} \beta^{00}_{l}\big(\frac{1}{2}\big)^l
=2^{-n-1}\beta^{00}_{n+1} +\Kraft (\,^0\C^0_n ) \, ,\\
2^{-3} \ge \sum\limits_{l=1}^{n+1} \beta^{ab}_{l}\big(\frac{1}{2}\big)^l
=2^{-n-1}\beta^{ab}_{n+1} +\Kraft (\,^a\C^b_n )\quad \forall\,
ab\in\{ 01,10,11\} \, .
\end{array}
\]
>From the above follows :
\begin{equation}\label{bingl1}
\begin{array}{l}
\beta^{00}_{n+1} \le 2^{n-1} -2^{n+1}\Kraft (\,^0\C^0_n )  \, ,\\
\beta^{01}_{n+1} \le 2^{n-2} -2^{n+1}\Kraft (\,^0\C^1_n ) \, ,\\
\beta^{10}_{n+1} \le 2^{n-2} -2^{n+1}\Kraft (\,^1\C^0_n ) \, ,\\
\beta^{11}_{n+1} \le 2^{n-2} -2^{n+1}\Kraft (\,^1\C^1_n ) \, .\\
\end{array}
\end{equation}

By Proposition~(\ref{propbin1}) (i) we obtain:
\begin{equation}\label{bingl2}
\big| a\A^{n-1}b - \bfs[n+1](\C_n ) \big| \ge 2^{n-1}
-\big|\pfs[n](\,^a\C_n) \big| -\big|\sfs[n](\C^b_n) \big|\quad
\forall\, a,b\in\{ 0,1\}
\end{equation}
If $\beta^{ab}_{n+1}\le 2^{n-1}
-\big|\pfs[n](\,^a\C_n) \big| -\big|\sfs[n](\C^b_n) \big| $ for $a,b\in\{ 0,1\} $
then from  (\ref{bingl2}) follows that there are
$\beta^{ab}_{n+1} $ words in $a\A^{n-1} b $ which are not in the the
bifix-shadow of $\C_n $ by adding these codewords for every $a,b\in\{ 0,1 \} $
to $\C_n $ we obtain a new fix-free code
$\C_{n+1} \subseteq \bigcup\limits_{l=1}^{n+1}\A^l $ with
$\big| \C_{n+1}\cap\A^l \big| =\alpha_l $ and
$\big| \,^a\C^b_{n+1}\cap\A^l \big| =\beta^{ab}_l $ for all
$l\in\{ 1,\ldots ,n+1\} $, $a,b\in\{ 0,1\} $.\\

\smallskip
Therefore it is sufficient to show that :
\begin{equation}\label{bingl3} \beta^{ab}_{n+1}\le 2^{n-1}
-\big|\pfs[n](\,^a\C_n) \big| -\big|\sfs[n](\C^b_n) \big| \quad \forall\,
a,b\in\{ 0,1\} \mbox{ with } \beta^{ab}_{n+1}> 0
\end{equation}
By the definition of the $(\beta^{ab}_l )_{l\in\NN } $ we have to distinguish five
cases:
\begin{itemize}
\item[{\em Case 1:}]
\[\Kraft (\,^0\C_n^0) =\sum\limits_{l=1}^n\beta^{00}_l \bigg(\frac{1}{2}\bigg)^l
<\frac{1}{4} \mbox{ and }
\Kraft (\,^a\C_n^b) =\sum\limits_{l=1}^n\beta^{ab}_l \bigg(\frac{1}{2}\bigg)^l
=0 \quad\forall\, ac\in\{ 01,10,11 \} \]
With Proposition~\ref{propbin1} (ii) follows:
\[\begin{array}{l}
\big| \pfs[n] (\,^0\C_n )\big| =\big|\sfs[n](\C^0 ) \big| =
2^n\cdot\Kraft (\,^0\C^0 ) \\
\big|\pfs[n] (\,^1\C_n ) \big| =\big| \sfs[n] (\C_n^1 )\big| =0
\end{array} \]
We obtain that (\ref{bingl3} ) holds for all $a,b\in\{ 0,1 \} $,
since by (\ref{bingl1})  follows:
\[\begin{array}{lcl}
\beta^{00}_{n+1} & \le & 2^{n-1}-2\cdot 2^n\cdot\Kraft (\,^0\C_n^0 )=
2^{n-1} -\big|\pfs[n](\,^0\C_n) \big| -\big|\sfs[n](\C^0_n) \big| \, ,\\[3mm]
\beta^{01}_{n+1} & \le & 2^{n-2} =2^{n-1} -2^{n-2} < 2^{n-1}
-2^n\cdot\Kraft (\,^0\C_n^0 ) \, ,\\
\quad & = & 2^{n-1} -\big|\pfs[n](\,^0\C_n) \big| -\big|\sfs[n](\C^1_n)\big|\, ,\\[3mm]
\beta^{10}_{n+1} & \le & 2^{n-2} =2^{n-1}-2^{n-2} < 2^{n-1}
-2^n\cdot\Kraft (\,^0\C_n^0 ) \, ,\\
\quad & = & 2^{n-1} -\big|\pfs[n](\,^1\C_n) \big| -\big|\sfs[n](\C^0_n)\big|\, ,\\[3mm]
\beta^{11}_{n+1} & \le & 2^{n-2} < 2^{n-1} =
2^{n-1} -\big|\pfs[n](\,^1\C_n) \big| -\big|\sfs[n](\C^1_n)\big| \, .
\end{array}\]

\item[{\em Case 2:}]
\[ \begin{array}{l}
\Kraft (\,^0\C_n^0 ) =\sum\limits_{l=1}^n\beta^{00}_l \bigg(\frac{1}{2}\bigg)^l
= 2^{-2} \, ,\quad
\Kraft (\,^0\C_n^1 ) =\sum\limits_{l=1}^n\beta^{01}_l \bigg(\frac{1}{2}\bigg)^l
< \frac{1}{8} \;\mbox{ and }\\
\Kraft (\,^a\C_n^b ) =\sum\limits_{l=1}^n\beta^{ab}_l \bigg(\frac{1}{2}\bigg)^l
= 0 \quad\forall\, ab\in\{ 10,11 \}
\end{array}\]

In this case with Proposition \ref{propbin1} (ii) follows:

\[\begin{array}{l}
\big| \pfs[n] (\,^0\C_n )\big| =
2^n\cdot\Kraft (\,^0\C^0_n ) +2^n\cdot\Kraft (\,^0\C^1_n )=
2^{n-2}+2^n\cdot\Kraft (\,^0\C^1_n )  \, , \quad\\
\big| \sfs[n] (\C_n^0 )\big| =2^n\cdot\Kraft (\,^0\C^0_n )= 2^{n-2}  \, ,\\
\big| \pfs[n] (\,^1\C_n )\big| =0 \;\mbox{ and }\;
\big| \sfs[n] (\C^1_n ) \big| = 2^n\cdot\Kraft (\,^0\C_n^1 )
\end{array} \]

Once again in this case with (\ref{bingl1}) follows that
(\ref{bingl3}) holds, because:
\[\begin{array}{lcl}
\beta^{00}_{n+1} & = & 0 \\[3mm]

\beta^{01}_{n+1} & \le &  2^{n-2} -2^{n+1}\cdot\Kraft (\,^0\C_n^1 )=
2^{n-1}-2^{n-2} -2^n\cdot\Kraft (\,^0\C_n^1 ) -2^n\cdot\Kraft (\,^0\C_n^1 )\\
\quad & = & 2^{n-1} -\big|\pfs[n](\,^0\C_n) \big| -\big|\sfs[n](\C^1_n)\big|\\[3mm]

\beta^{10}_{n+1} & \le & 2^{n-2} =2^{n-1}-2^{n-2}
 =  2^{n-1} -\big|\pfs[n](\,^1\C_n) \big| -\big|\sfs[n](\C^0_n)\big|\\[3mm]

\beta^{11}_{n+1} & \le &  2^{n-2} < 2^{n-1}-2^{n-3}
< 2^{n-1} -2^n\cdot\Kraft (\,^0\C_n^1 )\\
\quad & = & 2^{n-1} -\big|\pfs[n](\,^1\C_n) \big| -\big|\sfs[n](\C^1_n)\big|
\end{array}\]

\item[{\em Case 3:}]
\[\begin{array}{l}
\Kraft (\,^0\C_n^0 ) =\sum\limits_{l=1}^n\beta^{00}_l \bigg(\frac{1}{2}\bigg)^l
= 2^{-2} \, ,\quad
\Kraft (\,^0\C_n^1 ) =\sum\limits_{l=1}^n\beta^{01}_l \bigg(\frac{1}{2}\bigg)^l
= 2^{-3} \, ,\\
\Kraft (\,^1\C_n^0 ) =\sum\limits_{l=1}^n\beta^{10}_l \bigg(\frac{1}{2}\bigg)^l
< 2^{-3} \;\mbox{ and }\;
\Kraft (\,^1\C_n^1 ) =\sum\limits_{l=1}^n\beta^{11}_l \bigg(\frac{1}{2}\bigg)^l
= 0 \
\end{array}\]

In this case with Proposition~\ref{propbin1} (ii) follows:

\[\begin{array}{l}
\big| \pfs[n] (\,^0\C_n )\big| =
2^n\cdot\Kraft (\,^0\C^0_n ) +2^n\cdot\Kraft (\,^0\C^1_n )=
2^{n-2}+2^{n-3}  \, , \quad\\
\big| \sfs[n] (\C_n^0 )\big| =2^n\cdot\Kraft (\,^0\C^0_n )
+2^n\cdot\Kraft (\,^1\C^0_n )= 2^{n-2}+2^n\cdot\Kraft (\,^1\C^0_n )  \, ,\\
\big| \pfs[n] (\,^1\C_n )\big| =2^n\cdot\Kraft (\,^1\C^0_n ) \;\mbox{ and }\;
\big| \sfs[n] (\C^1_n ) \big| = 2^n\cdot\Kraft (\,^0\C_n^1 ) =2^{n-3}
\end{array} \]

Also in this case (\ref{bingl3}) holds, because by (\ref{bingl1}) follows:
\[\begin{array}{lcl}
\beta^{00}_{n+1} & = &\beta^{01}_{n+1}= 0 \\[3mm]

\beta^{10}_{n+1} & \le & 2^{n-2}-2^{n+1}\cdot\Kraft (\,^1\C^0_n )
= 2^{n-1} -2^n\cdot\Kraft (\,^1\C^0_n )-2^{n-2}-2^n\cdot\Kraft (\,^1\C^0_n )\\
\quad & = & 2^{n-1} -\big|\pfs[n](\,^1\C_n) \big| -\big|\sfs[n](\C^0_n)\big|\\[3mm]

\beta^{11}_{n+1} & \le & 2^{n-2} = 2^{n-1} -2\cdot 2^{n-3}
<2^{n-1}- 2^n\cdot\Kraft (\,^1\C^0_n ) -2^{n-3}\\
\quad & = & 2^{n-1} -\big|\pfs[n](\,^1\C_n) \big| -\big|\sfs[n](\C^1_n)\big|
\end{array}\]

\item[{\em Case 4:}]
\[\begin{array}{l}
\Kraft (\,^0\C_n^0 ) =\sum\limits_{l=1}^n\beta^{00}_l \bigg(\frac{1}{2}\bigg)^l
= 2^{-2} \, ,\quad
\Kraft (\,^0\C_n^1 ) =\sum\limits_{l=1}^n\beta^{01}_l \bigg(\frac{1}{2}\bigg)^l
= 2^{-3} \, ,\\
\Kraft (\,^1\C_n^0 ) =\sum\limits_{l=1}^n\beta^{10}_l \bigg(\frac{1}{2}\bigg)^l
= 2^{-3} \;\mbox{ and }\;
\Kraft (\,^1\C_n^1 ) =\sum\limits_{l=1}^n\beta^{11}_l \bigg(\frac{1}{2}\bigg)^l
<2^{-3}
\end{array}\]

In this case with Proposition \ref{propbin1} (ii) follows:

\[\begin{array}{l}
\big| \pfs[n] (\,^0\C_n )\big| = \big| \sfs[n] (\C_n^0 )\big|=
2^n\cdot\Kraft (\,^0\C^0_n ) +2^n\cdot\Kraft (\,^0\C^1_n )=
2^{n-2}+2^{n-3}  \;\mbox{ and } \quad\\

\big| \pfs[n] (\,^1\C_n )\big| = \big| \sfs[n] (\C^1_n ) \big| =
2^n\cdot\Kraft (\,^1\C^0_n )
+2^n\cdot\Kraft (\,^1\C^1_n ) = 2^{n-3} +2^n\cdot\Kraft (\,^1\C^1_n )
\end{array} \]

Also in this case (\ref{bingl3}) holds because with (\ref{bingl1}) follows:
\[\begin{array}{lcl}
\beta^{00}_{n+1} & = & \beta^{01}_{n+1} = \beta^{10}_{n+1} = 0 \\[3mm]

\beta^{11}_{n+1} & \le &
2^{n-2} -2^{n+1}\cdot\Kraft (\,^1\C^1_n ) =
2^{n-1} -2\cdot\big( 2^{n-3} + 2^n\cdot\Kraft (\,^1\C^1_n )\big) \\
\quad & = & 2^{n-1} -\big|\pfs[n](\,^1\C_n) \big| -\big|\sfs[n](\C^1_n)\big|
\end{array}\]

\item[{\em Case 5:}]
\[\begin{array}{l}
\Kraft (\,^0\C_n^0 ) =\sum\limits_{l=1}^n\beta^{00}_l \bigg(\frac{1}{2}\bigg)^l
= 2^{-2} \;\mbox{ and }\;
\Kraft (\,^a\C_n^b ) =\sum\limits_{l=1}^n\beta^{ab}_l \bigg(\frac{1}{2}\bigg)^l
=2^{-3}
\; \mbox{ for }\; ab\neq 00
\end{array}\]
Since (\ref{bingl1}), we obtain $\beta^{ab}_{n+1} =0 $  for all
$ab\in\{ 00,01,10,11 \} $.\qed
\end{itemize}

\smallskip
One can try to generalize the above theorem for alphabets of arbitrary length.
Let $ \A  =\{ 0,\ldots ,q-1  \}$ for some $q\ge 2 $
and $(\alpha_l )_{l \in\NN }$ be a sequence of nonnegative sequence with
$\alpha_1 = 0 \, ,\; \alpha_2\le 2 $ and
$\sum\limits_{l=1}^{\infty } \alpha_l\cdot q^{-l } =
\frac{3q^2-q}{2q^3} $.
Let $\preceq $ be a linear ordering on $\A^2 $ with leats element $00\in\A^2 $.
It is easy to verify that there exists a unique set of  sequences
of nonnegative integers:
\[ \{ (\beta^{ab}_l)_{l\in\NN } \, | \, a,b\in\A  \} \]
with the properties:
\begin{equation}\label{betaprop2}
\begin{array}{ll}
\sum\limits_{l=1}^{\infty } \beta_l^{aa}\cdot q^{-l} =
\frac{q-a}{q^3} &\;\forall\, a\in\A \\
\sum\limits_{l=1}^{\infty } \beta_l^{ab}\cdot q^{-l} =
\frac{1}{q^3} &\;\forall\, a,b\in\A \, ,\; a\neq b\\
\sum\limits_{a,b\in\A }\beta^{ab}_l =\alpha_l &\;\forall\ l\in\NN \\
\beta^{ab}_m > 0 \Rightarrow \beta^{cd }_l  = 0
&\;\forall\, cd\succ ab \, ,  l < m
\end{array}
\end{equation}

For example let $\prec $ the lexicographic ordering on $\A $.
This means for $ab ,cd\in\A^2 $:
\[ ab \preceq cd \Leftrightarrow a\le c \;\mbox{ or }\; a=b , b\le d \, .\]

If $\A =\{ 0,1 \} $ then the sequences
$(\beta^{00}_l)_{l\in\NN }\, ,\; (\beta^{01}_l)_{l\in\NN } \, ,\;
(\beta^{10}_l)_{l\in\NN }\, ,\;(\beta^{11}_l)_{l\in\NN } $ in the proof
of Theorem~\ref{5to8theorem} have the above properties for the lexicograpic
ordering.\\

\smallskip
Let $| \A | = 3 $. If the first eight
terms of $( \alpha_l )_{l\in\NN } $ and the $(\beta^{ab}_l )_{l\in\NN } $ are
given by:
\[ \begin{array}{lc@{(}c@{\, ,\,}c@{\, ,\,}c@{\, ,\,}c@{\, ,\,}c@{\, ,\,}c@{\, ,\,}c@{\, ,\,}c@{)}}
(\alpha_1,\ldots ,\alpha_8) &  := &0&0&2&4&22&6&394&276 \\
(\beta^{00}_1,\ldots ,\beta^{00}_8) &  := &0&0&2&3&0&0&0&0 \\
(\beta^{01}_1,\ldots ,\beta^{01}_8) &  := &0&0&0&1&6&0&0&0 \\
(\beta^{02}_1,\ldots ,\beta^{02}_8) &  := &0&0&0&0&9&0&0&0 \\
(\beta^{10}_1,\ldots ,\beta^{10}_8) &  := &0&0&0&0&7&6&0&0 \\
(\beta^{11}_1,\ldots ,\beta^{11}_8) &  := &0&0&0&0&0&0&162&0 \\
(\beta^{12}_1,\ldots ,\beta^{12}_8) &  := &0&0&0&0&0&0&81&0 \\
(\beta^{20}_1,\ldots ,\beta^{20}_8) &  := &0&0&0&0&0&0&81&0 \\
(\beta^{21}_1,\ldots ,\beta^{21}_8) &  := &0&0&0&0&0&0&70&33 \\
(\beta^{22}_1,\ldots ,\beta^{22}_8) &  := &0&0&0&0&0&0&0&243 \\
\end{array}\]
and $\alpha_l = \beta^{00}_l =\ldots = \beta^{22}_l =0 $ for $l> 8 $.\\
\smallskip

Then
$\sum\limits_{l=1}^{\infty } \alpha_l \cdot 3^{-l} = \frac{12}{27}
=\frac{3\cdot 3^{2} -3}{2\cdot 3^3} $ and
$(\beta^{00}_l)_{l\in\NN },\ldots ,(\beta^{22}_l)_{l\in\NN } $
are the unique sequences which have the properties in (\ref{betaprop2}) for the
lexicographic ordering.\\

\begin{conj}
Let $q\ge 2\, ,\; \A =\{ 0,\ldots ,q-1 \} $ and $ (\alpha_l )_{l\in\NN } $ be a sequence
of nonnegative integers with $\alpha_1 =0 \, ,\;\alpha_2 \le 1 $ and
$\sum\limits_{l=1}^{\infty } \alpha_l\cdot q^{-l } =
\frac{3q^2-q}{2q^3} $.
\renewcommand{\labelenumi}{(\arabic{enumi})}
\begin{enumerate}
\item Then there exists a linear ordering $\preceq $ on $\A^2 $ with least element $00$ and
 a fix-free code $\C\subseteq \A^* $ with
\[ \big| \,^a\C^b \cap\A_l \big| =\beta^{ab}_l \quad\forall\, l\in\NN ,\]
where the $(\beta_l^{ab})_{l\in\NN } $ are the unique sequences which fulfill
(\ref{betaprop2}) for $\preceq $.

\item The first part of the conjecture holds for the lexicographic ordering of
$\A^2 $.
\end{enumerate}
\end{conj}

The conjecture above is a generalization of the idea of the proof of
Theorem~\ref{5to8theorem}

Furthermore the proof of theorem \ref{5to8theorem} shows that both part of the
conjecture
holds for $q=2 $.\\

\smallskip
If part (1) of the conjecture holds for some
$q\ge 2 $ then from the second property of the $(\beta_l^{ab})_{l\in\NN } $
follows that for every sequence $(\alpha_l)_{l\in\NN } $ with
$\alpha_1 =0 \, ,\; \alpha_2\le 1 $ and
$\sum\limits_{l=1}^{\infty } \alpha_l\cdot q^{-l } \le\frac{3q^2-q}{2q^3} $.
there exists a fix-free code which fits to $(\alpha_l)_{l\in\NN } $.
On the other hand, this gives nothing new for $q\ge 3 $, because
$\frac{3q^2-q}{2q^3} $ is a decreasing sequence for $q\in\NN $, as one
can easy verify, and $\frac{3\cdot 3^2-3}{2\cdot 3^3} =\frac{24}{54}<\frac{1}{2} $.
Indeed Theorem~\ref{einhalb} says already that for  every sequence
$(\alpha_l)_{l\in\NN } $ with Kraftsum smaller than or equal to $\frac{1}{2} $
there exist a fix-free Code $\C\subseteq\A^* $ which fits to $(\alpha_l)_{l\in\NN } $.
The only new, is the special form of the fix-free
code. Therefore we omit a full proof of the conjecture and finish this
section, by showing that both part of the conjecture holds for $q=3 $.\\

\medskip
{\bf Proof of the Conjecture for $q=3 $ :} Let $q=3 $,
$\A =\{ 0,1,2 \} $ and $\preceq $ the lexicographic ordering on $\A^2 $.
Let $ (\alpha_l )_{l\in\NN } $,
$(\beta^{00}_l )_{l\in\NN } ,\ldots , (\beta^{22}_l )_{l\in\NN } $
as in the conjecture.
The proof of the conjecture will
be similar to the proof of Theorem~\ref{5to8theorem}. This means,
we will construct by induction, fix-free sets $\C_1\subseteq \C_2\subseteq \C_3\subseteq\ldots $
with the property:\\

\begin{equation}\label{dreigl1}
\C_n\subseteq\bigcup\limits_{l=1}^{n}\A^l \;\mbox{ and }\;
\big| \,^a\C_n^b\cap\A^l  \big| =\beta^{ab}_l \quad \forall\,
l\in\{ 1,\ldots ,n\}\, ,\; a,b\in\A
\end{equation}
Then $\C :=\bigcup\limits_{l=1}^{\infty } \C_n $ is a fix-free code
for which the conditions of the conjecture holds.\\

If $\C_1 :=\emptyset $ then $\C_1 $ is a fix-free set for which
(\ref{dreigl1}) hold.\\
Let $\C_n $ be a fix-free set for which (\ref{dreigl1}) holds. Then we obtain with
(\ref{betaprop2}):
\[\begin{array}{lcl}
3^{-2} & = & 3\cdot 3^{-3} \ge \sum\limits_{l=1}^{n+1} \beta^{00}_l\cdot q^{-l}
=3^{-n-1}\beta_{n+1}^{00}+\Kraft (\,^0\C^0 ) \, ,\\
2\cdot 3^{-3} & \ge & \sum\limits_{l=1}^{n+1} \beta^{11}_l\cdot q^{-l}
=3^{-n-1}\beta_{n+1}^{11}+\Kraft (\,^1\C^1 ) \, \\
3^{-3} & \ge & \sum\limits_{l=1}^{n+1} \beta^{22}_l\cdot q^{-l}
=3^{-n-1}\beta_{n+1}^{22}+\Kraft (\,^2\C^2 )  \, \\
3^{-3} & \ge & \sum\limits_{l=1}^{n+1} \beta^{ab}_l\cdot q^{-l}
=3^{-n-1}\beta_{n+1}^{ab}+\Kraft (\,^a\C^b )\quad\forall\,
a,b\in\{ 0,1,2 \}\, ,\; a\neq b  \, .
\end{array}\]

With this follows:
\begin{equation}\label{dreigl2}\begin{array}{lcl}
\beta^{00}_{n+1} \le 3^{n-1}-3^{n+1}\cdot\Kraft (\,^0\C^0 ) & ; &
\beta^{01}_{n+1} \le 3^{n-2}-3^{n+1}\cdot\Kraft (\,^0\C^1 ) \; ;\\
\beta^{02}_{n+1} \le 3^{n-2}-3^{n+1}\cdot\Kraft (\,^0\C^2 ) & ; &
\beta^{10}_{n+1} \le 3^{n-2}-3^{n+1}\cdot\Kraft (\,^1\C^0 ) \; ;\\
\beta^{11}_{n+1} \le 2\cdot 3^{n-2}-3^{n+1}\cdot\Kraft (\,^1\C^1 ) & ; &
\beta^{12}_{n+1} \le 3^{n-2}-3^{n+1}\cdot\Kraft (\,^1\C^2 ) \; ;\\
\beta^{20}_{n+1} \le 3^{n-2}-3^{n+1}\cdot\Kraft (\,^2\C^0 ) & ; &
\beta^{21}_{n+1} \le 3^{n-2}-3^{n+1}\cdot\Kraft (\,^2\C^1 ) \;\mbox{ and }\\
\beta^{22}_{n+1} \le 3^{n-2}-3^{n+1}\cdot\Kraft (\,^2\C^2 ) &\quad &\quad
\end{array}\end{equation}

By Proposition \ref{propbin1} (i) we have:

\begin{equation}\label{dreigl3}
\big| a\A^{n-2} b -\bfs[n+1] (\C_n ) \big| \ge
\max\left\{ 0, 3^{n-1}-\big| \pfs[n] (\,^a\C_n ) \big|-\big| \sfs[n] (\C_n^b ) \big|
\right\}
\quad \forall\, a,b\in\{ 0,1,2 \}\end{equation}

Therefore it follows, that
for the existence of a fix-free  set $\C_{n+1}\supseteq \C_n $
with property (\ref{dreigl1}), it is sufficient to show that:

\begin{equation}\label{dreigl4}
\beta^{ab}_{n+1} \le
\max\left\{ 0, 3^{n-1}-\big| \pfs[n] (\,^a\C_n ) \big|-\big| \sfs[n] (\C_n^b ) \big|
\right\}
\quad
\forall\, a,b\in\{ 0,1,2 \} \;\mbox{ with }\; \beta^{ab}_{n+1}> 0
\end{equation}

If  \ref{dreigl4} holds, then by (\ref{dreigl3}) and \ref{dreigl4}) follows,
that for all $a,b\in\A $
there exist $\beta^{ab}_{n+1} $ codewords in $a\A^{n-1}b $ which are not
in the bifix-shadow of $\C_n $. By adding these codewords to $\C_n $
we obtain a fix-free code $\C_{n+1}\supseteq \C_n $ for which
(\ref{dreigl1}) holds. To show (\ref{dreigl4}) we have to distinguish
ten cases:

\begin{itemize}
\item[{\em Case 1:}]
\[ \Kraft (\,^0\C_n^0)<3^{-2}\;\mbox{ and }\;
\Kraft (\,^a\C_n^b)=0 \quad\forall\, a,b\in\A\, ,\; ab\neq 00 \]
By Proposition~\ref{propbin1} (ii) we obtain:
\[ \begin{array}{ll}
\big| \pfs[n](\,^0\C_n ) \big|=\big| \sfs[n](\C_n^0 ) \big|=
3^n\Kraft (\,^0\C_n^0) < 3^{n-2}\\
\big|\pfs[n] (\,^a\C_n )\big| =\big| \sfs[n] (\C_n^b )\big| =0
&\;\forall\, a,b\in\A\, ,\; a,b\neq 0
\end{array} \]
>From (\ref{dreigl2}) follows that
(\ref{dreigl4}) holds, because:

\[ \begin{array}{llcl}

\quad & \beta^{00}_{n+1} & \le & 3^{n-1}-3^{n+1}\cdot\Kraft (\,^0\C_n^0 ) <
3^{n-1}-2\cdot 3^n\cdot \Kraft (\,^0\C_n^0 ) \\
\quad &\quad & = & 3^{n-1}-\big|\pfs[n] (\,^0\C_n )\big| -\big| \sfs[n] (\C_n^0 ) \big|\\[3mm]

\forall\, b\neq 0\, : & \beta^{0b}_{n+1} & \le & 3^{n-2} < 3^{n-1}-3^{n-2}\le
3^{n-1}-3^n\cdot \Kraft (\,^0\C_n^0 ) \\
\quad & \quad & = &
3^{n-1}-\big|\pfs[n] (\,^0\C_n )\big| -\big| \sfs[n] (\C_n^b ) \big|\\[3mm]

\quad & \beta_{n+1}^{11} & \le &2\cdot 3^{n-2} < 3^{n-1}=
3^{n-1}-\big|\pfs[n] (\,^1\C_n )\big| -\big| \sfs[n] (\C_n^1 ) \big|\\[3mm]

\forall\, b\neq 1: & \beta_{n+1}^{1b} & \le & 3^{n-2} < 3^{n-1}-3^{n-2}\le
3^{n-1}-\big|\pfs[n] (\,^1\C_n )\big| -\big| \sfs[n] (\C_n^b ) \big|\\[3mm]

\forall\, b\in\A: & \beta_{n+1}^{2b} & \le & 3^{n-2} < 3^{n-1} -3^{n-2}\le
3^{n-1}-\big|\pfs[n] (\,^2\C_n )\big| -\big| \sfs[n] (\C_n^b ) \big|

\end{array}\]

\item[{\em Case 2:}]
\[ \begin{array}{l} \Kraft (\,^0\C_n^0)=3^{-2}\, ,\;\Kraft (\,^0\C_n^1)<3^{-3} \\
\Kraft (\,^a\C_n^b)=0 \quad\forall\, ab\in\A^2 -\{ 00,01 \}
\end{array}\]
In this case we obtain with proposition \ref{propbin1} (ii):
\[\begin{array}{lcl}
\big| \pfs[n](\,^0\C_n ) \big| & = & 3^{n-2}+3^n\cdot\Kraft (\,^0\C_n^1 )<
2\cdot 3^{n-2} \, ,\;  \big| \pfs[n] (\,^a\C_n )\big|  =  0 \quad\forall\,
a\ge 1 \, ,\\

\big| \sfs[n](\C_n^0 ) \big| & = & 3^{n-2} \, ,\;
\big|\sfs[n] (C_n^1) \big|  =  3^n\cdot\Kraft (\,^0\C_n^1) < 3^{n-3}
\;\mbox{ and }\;\big|\sfs[n] (\C_n^2 )\big|  = 0
\end{array}\]
By (\ref{dreigl2})  follows (\ref{dreigl4}):

\[\begin{array}{llcl}
\quad & \beta^{00}_{n+1} & = & 0  \\[3mm]

\quad & \beta^{01}_{n+1} & \le & 3^{n-2}-3^{n+1}\cdot\Kraft (\,^0\C_n^1) \le
3^{n-1}-3^{n-2}-2\cdot 3^n\cdot\Kraft (\,^0\C_n^1 ) \\
\quad &\quad & = &
3^{n-1}-\big|\pfs[n] (\,^0\C_n )\big|-\big|\sfs[n](\C_n^1)\big| \\[3mm]

\quad & \beta^{02}_{n+1} & \le  & 3^{n-2} = 3^{n-1}-2\cdot 3^{n-2}
\le 3^{n-1}-3^{n-2}-3^n\cdot \Kraft (\,^0\C_n^2 ) \\
\quad & \quad & = &
3^{n-1}-\big| \pfs[n] (\,^0\C_n )\big|-\big| \sfs[n] (\C_n^2 )\big|\\[3mm]

\quad  & \beta^{11}_{n+1} & \le & 2\cdot 3^{n-2} < 3^{n-1} -3^{n-3} \le
 3^{n-1}-\big| \pfs[n] (\,^1\C_n )\big|-\big| \sfs[n] (\C_n^1 )\big|\\[3mm]

\quad  & \beta^{12}_{n+1} & \le & 3^{n-2} < 3^{n-1}
= 3^{n-1}-\big| \pfs[n] (\,^1\C_n )\big|-\big| \sfs[n] (\C_n^2 )\big|\\[3mm]

\forall \, b\in \A  & \beta^{2b}_{n+1} & \le & 3^{n-2} < 3^{n-1} -3^{n-2}
\le 3^{n-1}-\big| \pfs[n] (\,^2\C_n )\big|-\big| \sfs[n] (\C_n^b )\big|

\end{array}\]

\item[{\em Case 3:}]
\[ \begin{array}{l} \Kraft (\,^0\C_n^0)=3^{-2}\, ,\;\Kraft (\,^0\C_n^1)=3^{-3}
\, ,\;\Kraft (\,^0\C_n^2)< 3^{-3} \, ,\;\\
 \Kraft (\,^a\C_n^b )=0 \quad\forall\, a,b\in\A -\{ 00,01,02 \}
\end{array}\]

In this case we obtain by proposition \ref{propbin1} (ii):
\[\begin{array}{lcl}
\big| \pfs[n](\,^0\C_n ) \big| & = &
3^{n-2}+3^{n-3} + 3^n\cdot\Kraft (\,^0\C_n^2 )< 5\cdot 3^{n-3}
\, ,\;  \big| \pfs[n] (\,^a\C_n )\big|  =  0 \quad\forall\, a\ge 1 \, ,\\

\big| \sfs[n](\C_n^0 ) \big| & = & 3^{n-2} \, ,\;
\big|\sfs[n] (\C_n^1) \big|  =  3^{n-3}
\;\mbox{ and }\; \big|\sfs[n] (\C_n^2 )\big|  =
3^{n}\cdot\Kraft (\,^0\C_n^2 ) < 3^{n-3}
\end{array}\]

Once again with (\ref{dreigl2}) follows (\ref{dreigl4}):

\[\begin{array}{llcl}
\quad & \beta^{00}_{n+1} & = &\beta^{01}_{n+1}= 0  \\[3mm]

\quad & \beta^{02}_{n+1} & \le & 3^{n-2}-3^{n+1}\cdot\Kraft (\,^0\C_n^2) \\
\quad &\quad & < &
3^{n-1}- 3^{n-2} -3^{n-3} -3^n\cdot\Kraft (\,^0\C_n^2 )-3^n\cdot\Kraft (\,^0\C_n^2) \\
\quad &\quad & = &
3^{n-1}-\big|\pfs[n] (\,^0\C_n )\big|-\big|\sfs[n](\C_n^1)\big| \\[3mm]

\quad  & \beta^{11}_{n+1} & \le & 2\cdot 3^{n-2} <
8\cdot 3^{n-3} = 3^{n-1} - 3^{n-3}\\
\quad &\quad & < &
3^{n-1}-\big| \pfs[n] (\,^1\C_n )\big|-\big| \sfs[n] (\C_n^1 )\big| \\[3mm]

\quad  & \beta^{12}_{n+1} & \le & 3^{n-2} <  3^{n-1} - 3^{n-3}\\
\quad &\quad & < & 3^{n-1}-\big| \pfs[n] (\,^1\C_n )\big|-\big| \sfs[n] (\C_n^2 )\big|
\\[3mm]

\quad \forall\, b\in\A \, : & \beta^{2b}_{n+1} & \le & 3^{n-2} <  3^{n-1} - 3^{n-2}\\
\quad &\quad & < & 3^{n-1}-\big| \pfs[n] (\,^2\C_n )\big|-\big| \sfs[n] (\C_n^b )\big|

\end{array}\]

\item[{\em Case 4:}]
\[ \begin{array}{l} \Kraft (\,^0\C_n^0)=3^{-2}\, ,\;\Kraft (\,^0\C_n^1)=
\Kraft (\,^0\C_n^2)= 3^{-3} \, ,\;\\
\Kraft (\,^1\C_n^0 )< 3^{-3}\, ,\;
\Kraft (\,^a\C_n^b )=0 \quad\forall\, ab\in\A -\{ 00,01,02,10 \}
\end{array}\]

In this case we obtain with proposition \ref{propbin1} (ii):

\[\begin{array}{lcl}
\big| \pfs[n](\,^0\C_n ) \big| & = &
3^{n-2}+2\cdot 3^{n-3} =4\cdot 3^{n-3}

\, ,\;  \big| \pfs[n] (\,^1\C_n )\big|  =  3^n\Kraft (\,^1\C_n^0 ) < 3^{n-3}

\, ,\;\\
\big| \pfs[n] (\,^2\C_n )\big|  & = & 0 \, , \;

\big| \sfs[n](\C_n^0 ) \big|  =
3^{n-2}+3^n\cdot\Kraft (\,^1\C_n^0 )<4\cdot3^{n-3}  \, ,\;

\big|\sfs[n] (\C_n^1 ) \big|  =  3^{n-3} \, \\
\big|\sfs[n] (\C_n^2 )\big|  & = &  3^{n-3}
\end{array}\]

Once again with (\ref{dreigl2}) follows (\ref{dreigl4}):

\[\begin{array}{llcl}
\quad & \beta^{00}_{n+1} & = & \beta^{01}_{n+1}=\beta^{02}_{n+1} = 0  \\[3mm]

\quad & \beta^{10}_{n+1} & \le & 3^{n-2}-3^{n+1}\cdot\Kraft (\,^1\C_n^0 )\\
\quad & \quad & < &
3^{n-1} -3^{n}\cdot\Kraft (\,^1\C_n^0 ) -3^{n-2}-3^n\cdot\Kraft (\,^1\C_n^0 ) \\
\quad &\quad & = &
3^{n-1}-\big|\pfs[n] (\,^1\C_n )\big|-\big|\sfs[n](\C_n^0)\big| \\[3mm]

\quad &\beta^{11}_{n+1} & \le & 2\cdot 3^{n-2} < 7\cdot 3^{n-3}
=3^{n-1}-2\cdot 3^{n-3}\\
\quad &\quad & < &
3^{n-1}-\big| \pfs[n] (\,^1\C_n )\big|-\big| \sfs[n] (\C_n^1 )\big|\\[3mm]

\quad &\beta^{12}_{n+1} & \le & 3^{n-2} < 7\cdot 3^{n-3} = 3^{n-1}-2\cdot 3^{n-3}\\
\quad &\quad & < &
3^{n-1}-\big| \pfs[n] (\,^1\C_n )\big|-\big| \sfs[n] (\C_n^2 )\big|\\[3mm]

\quad & \beta^{20}_{n+1} &\le & 3^{n-2} <5\cdot 3^{n-3} = 3^{n-1}-4\cdot 3^{n-3}\\
\quad &\quad & < &
3^{n-1}-\big| \pfs[n] (\,^2\C_n )\big|-\big| \sfs[n] (\C_n^0 )\big|\\[3mm]

\forall\, b\ge 1 & \beta^{2b}_{n+1} &\le & 3^{n-2} < 8\cdot 3^{n-3}
= 3^{n-1} -3^{n-3}
= 3^{n-1}-\big| \pfs[n] (\,^2\C_n )\big|-\big| \sfs[n] (\C_n^b )\big|
\end{array}\]

\item[{\em Case 5:}]
\[ \begin{array}{l} \Kraft (\,^0\C_n^0)=3^{-2}
\, ,\;\Kraft (\,^0\C_n^1)=
\Kraft (\,^0\C_n^2)=\Kraft (\,^1\C_n^0 )= 3^{-3} \, ,\;\\
\Kraft (\,^1\C_n^1)<2\cdot 3^{-3} \, ,\;
\Kraft (\,^a\C_n^b )=0 \quad\forall\, ab\in\{ 12,20,21,22 \}
\end{array}\]

In this case we obtain with proposition \ref{propbin1} (ii):

\[\begin{array}{lclclcl}
\big| \pfs[n](\,^0\C_n ) \big| & = &
3^{n-2}+2\cdot 3^{n-3} =5\cdot 3^{n-3}
& , &
\big| \sfs[n](\C_n^0 ) \big| & = &
3^{n-2}+3^{n-3} =4\cdot3^{n-3}  \\

\big| \pfs[n] (\,^1\C_n )\big|
& = & 3^{n-3}+  3^n\cdot\Kraft (\,^1\C_n^1 ) < 3^{n-2}
& , &
\big|\sfs[n] (\C_n^1 ) \big|  & = &  3^{n-3} +3^n
\cdot\Kraft (\,^1\C_n^1 )<3^{n-2}\, ,\;\\

\big| \pfs[n] (\,^2\C_n )\big|  & = & 0
& , &
\big|\sfs[n] (\C_n^2 )\big|  & = &  3^{n-3}
\end{array}\]

Also in this case follows (\ref{dreigl4}) with (\ref{dreigl2}):

\[\begin{array}{llcl}
\quad & \beta^{00}_{n+1} & = & \beta^{01}_{n+1}=\beta^{02}_{n+1} =
\beta^{10}_{n+1} = 0  \\[3mm]

\quad & \beta^{11}_{n+1} & \le &
2\cdot 3^{n-2}-3^{n+1}\cdot\Kraft (\,^1\C_n^1 )
<3^{n-1}-2\cdot 3^{n-3} -2\cdot 3^n\Kraft (\,^1\C_n^1) \\
\quad &\quad & = &
3^{n-1}-\big|\pfs[n] (\,^1\C_n )\big|-\big|\sfs[n](\C_n^1)\big| \\[3mm]

\quad &\beta^{12}_{n+1} & \le & 3^{n-2} < ^5\cdot 3^{n-3}
=3^{n-1}-4\cdot 3^{n-3}\\
\quad &\quad & < &
3^{n-1}-\big| \pfs[n] (\,^1\C_n )\big|-\big| \sfs[n] (\C_n^2 )\big|\\[3mm]

\forall\, b\in\A & \beta^{2b}_{n+1} &\le & 3^{n-2} <5\cdot 3^{n-3}
= 3^{n-1}-4\cdot 3^{n-3}\\
\quad &\quad & \le &
3^{n-1}-\big| \pfs[n] (\,^2\C_n )\big|-\big| \sfs[n] (\C_n^b )\big|

\end{array}\]

\item[{\em Case 6:}]
\[ \begin{array}{l} \Kraft (\,^0\C_n^0)=3^{-2} \, ,\;
\Kraft (\,^1\C_n^1)=2\cdot 3^{-3} \, ,\;
\Kraft (\,^0\C_n^1)=\Kraft (\,^0\C_n^2)=\Kraft (\,^1\C_n^0 )= 3^{-3} \, ,\;\\
\Kraft (\,^1\C_n^2)<3^{-3} \, ,\;
\Kraft (\,^2\C_n^b )=0 \quad\forall\, b\in\A
\end{array}\]

In this case we obtain with proposition \ref{propbin1} (ii):

\[\begin{array}{lcl}
\big| \pfs[n](\,^0\C_n ) \big| & = &
3^{n-2}+2\cdot 3^{n-3} =5\cdot 3^{n-3} \\

\big| \pfs[n] (\,^1\C_n )\big|
& = & 3\cdot 3^{n-3}+  3^n\cdot\Kraft (\,^1\C_n^2 ) < 4\cdot 3^{n-3}\\
\big| \pfs[n] (\,^2\C_n )\big|   & = &  0\\

\big| \sfs[n](\C_n^0 ) \big| & = &
3^{n-2}+3^{n-3} =4\cdot3^{n-3}  \\

\big|\sfs[n] (\C_n^1 ) \big|  & = &  3 \cdot 3^{n-3} = 3^{n-2}\\

\big|\sfs[n] (\C_n^2 )\big|  & = &
3^{n-3}+3^n\cdot\Kraft (\,^1\C_n^2)<2\cdot 3^{n-3}

\end{array}\]

Now (\ref{dreigl4}) follows with (\ref{dreigl2}):

\[\begin{array}{llcl}
\quad & \beta^{00}_{n+1} & = & \beta^{01}_{n+1}=\beta^{02}_{n+1} =
\beta^{10}_{n+1} =\beta^{11}_{n+1} = 0  \\[3mm]

\quad & \beta^{12}_{n+1} & \le &
3^{n-2}-3^{n+1}\cdot\Kraft (\,^1\C_n^2 )
<5\cdot 3^{n-3} -2\cdot 3^n\Kraft (\,^1\C_n^2) \\
\quad &\quad & = &
3^{n-1}-\big|\pfs[n] (\,^1\C_n )\big|-\big|\sfs[n](\C_n^2)\big| \\[3mm]

\forall\, b\in\A & \beta^{2b}_{n+1} &\le & 3^{n-2} <5\cdot 3^{n-3}
= 3^{n-1}-4\cdot 3^{n-3}\\
\quad &\quad & \le &
3^{n-1}-\big| \pfs[n] (\,^2\C_n )\big|-\big| \sfs[n] (\C_n^b )\big|

\end{array}\]

\item[{\em Case 7:}]
\[ \begin{array}{l} \Kraft (\,^0\C_n^0)=3^{-2} \, ,\;
\Kraft (\,^1\C_n^1)=2\cdot 3^{-3} \, ,\;
\Kraft (\,^0\C_n^1)=\Kraft (\,^0\C_n^2)=\Kraft (\,^1\C_n^0 )=
\Kraft (\,^1\C_n^2 )=3^{-3} \, ,\;\\
\Kraft (\,^2\C_n^0)<3^{-3} \, ,\;
\Kraft (\,^2\C_n^b )=0 \quad\forall\, b\in\{ 1,2 \}
\end{array}\]

In this case we obtain with proposition \ref{propbin1} (ii):

\[\begin{array}{lcl}
\big| \pfs[n](\,^0\C_n ) \big| & = &
3^{n-2}+2\cdot 3^{n-3} =5\cdot 3^{n-3} \\

\big| \pfs[n] (\,^1\C_n )\big|
& = &  4\cdot 3^{n-3}\\
\big| \pfs[n] (\,^2\C_n )\big|   & = &  3^n\cdot\Kraft (\,^2\C_n^0 )<3^{n-3}\\

\big| \sfs[n](\C_n^0 ) \big| & = &
4\cdot 3^{n-3} +3^n\cdot\Kraft (\,^2\C_n^0 ) < 5\cdot 3^{n-3}  \\

\big|\sfs[n] (\C_n^1 ) \big|  & = &  3 \cdot 3^{n-3} = 3^{n-2}\\

\big|\sfs[n] (\C_n^2 )\big|  & = & 2\cdot 3^{n-3}

\end{array}\]

Now (\ref{dreigl4}) follows with (\ref{dreigl2}):

\[\begin{array}{llcl}
\quad & \beta^{00}_{n+1} & = & \beta^{01}_{n+1}=\beta^{02}_{n+1} =
\beta^{10}_{n+1} =\beta^{11}_{n+1} \beta^{12}_{n+1}= 0  \\[3mm]

\quad & \beta^{20}_{n+1} & \le &
3^{n-2}-3^{n+1}\cdot\Kraft (\,^2\C_n^0 )
<5\cdot 3^{n-3} -2\cdot 3^n\Kraft (\,^2\C_n^0) \\
\quad &\quad & = &
3^{n-1}-\big|\pfs[n] (\,^2\C_n )\big|-\big|\sfs[n](\C_n^0)\big| \\[3mm]

\quad & \beta^{21}_{n+1} &\le & 3^{n-2} <5\cdot 3^{n-3}
= 3^{n-1}-4\cdot 3^{n-3}\\
\quad &\quad & < &
3^{n-1}-\big| \pfs[n] (\,^2\C_n )\big|-\big| \sfs[n] (\C_n^1 )\big|\\[3mm]

\quad & \beta^{22}_{n+1} &\le & 3^{n-2} <5\cdot 3^{n-3}
= 3^{n-1}-4\cdot 3^{n-3}\\
\quad &\quad & < &
3^{n-1}-\big| \pfs[n] (\,^2\C_n )\big|-\big| \sfs[n] (\C_n^2 )\big|

\end{array}\]

\item[{\em Case 8:}]
\[ \begin{array}{l} \Kraft (\,^0\C_n^0)=3^{-2} \, ,\;
\Kraft (\,^1\C_n^1)=2\cdot 3^{-3} \, ,\; \Kraft (\,^2\C_n^2 )=0 \, ,\;
\Kraft (\,^2\C_n^1)<3^{-3} \\
\Kraft (\,^0\C_n^1)=\Kraft (\,^0\C_n^2)=\Kraft (\,^1\C_n^0 )=
\Kraft (\,^1\C_n^2 )=\Kraft (\,^2\C_n^0)=3^{-3}
\end{array}\]

In this case we obtain with proposition \ref{propbin1} (ii):

\[\begin{array}{lcl}
\big| \pfs[n](\,^0\C_n ) \big| & = &
3^{n-2}+2\cdot 3^{n-3} =5\cdot 3^{n-3} \\

\big| \pfs[n] (\,^1\C_n )\big|
& = &  4\cdot 3^{n-3}\\

\big| \pfs[n] (\,^2\C_n )\big|   & = &
3^{n-3}+3^n\cdot\Kraft (\,^2\C_n^1 )<2\cdot 3^{n-3}\\

\big| \sfs[n](\C_n^0 ) \big| & = &  5\cdot 3^{n-3}  \\

\big|\sfs[n] (\C_n^1 ) \big|  & = &
3 \cdot 3^{n-3} +3^n\cdot\Kraft (\,^2\C_n^1 )<4\cdot 3^{n-3} \\

\big|\sfs[n] (\C_n^2 )\big|  & = & 2\cdot 3^{n-3}

\end{array}\]

Now (\ref{dreigl4}) follows with (\ref{dreigl2}):

\[\begin{array}{llcl}
\quad & \beta^{00}_{n+1} & = & \beta^{01}_{n+1}=\beta^{02}_{n+1} =
\beta^{10}_{n+1} =\beta^{11}_{n+1} \beta^{12}_{n+1}=
\beta^{20}_{n+1} =0  \\[3mm]

\quad & \beta^{21}_{n+1} & \le &
3^{n-2}-3^{n+1}\cdot\Kraft (\,^2\C_n^1 )
<5\cdot 3^{n-3} -2\cdot 3^n\cdot\Kraft (\,^2\C_n^1) \\
\quad &\quad & = &
3^{n-1}-\big|\pfs[n] (\,^2\C_n )\big|-\big|\sfs[n](\C_n^1)\big| \\[3mm]

\quad & \beta^{22}_{n+1} &\le & 3^{n-2} <5\cdot 3^{n-3} <
3^{n-1}-\big| \pfs[n] (\,^2\C_n )\big|-\big| \sfs[n] (\C_n^2 ) \big|

\end{array}\]

\item[{\em Case 9:}]
\[ \begin{array}{l} \Kraft (\,^0\C_n^0)=3^{-2} \, ,\;
\Kraft (\,^1\C_n^1)=2\cdot 3^{-3} \, ,\; \Kraft (\,^2\C_n^2 )<3^{-3} \, , \\
\Kraft (\,^0\C_n^1)=\Kraft (\,^0\C_n^2)=\Kraft (\,^1\C_n^0 )=
\Kraft (\,^1\C_n^2 )=\Kraft (\,^2\C_n^0)=\Kraft (\,^2\C_n^1)=3^{-3}
\end{array}\]

In this case we obtain with proposition \ref{propbin1} (ii):

\[\begin{array}{lcl}
\big| \pfs[n](\,^0\C_n ) \big| & = &
3^{n-2}+2\cdot 3^{n-3} =5\cdot 3^{n-3} \\

\big| \pfs[n] (\,^1\C_n )\big|
& = &  4\cdot 3^{n-3}\\

\big| \pfs[n] (\,^2\C_n )\big|   & = &
2\cdot 3^{n-3}+3^n\cdot\Kraft (\,^2\C_n^2 )< 3^{n-2}\\

\big| \sfs[n](\C_n^0 ) \big| & = &  5\cdot 3^{n-3}  \\

\big|\sfs[n] (\C_n^1 ) \big|  & = & 4\cdot 3^{n-3} \\

\big|\sfs[n] (\C_n^2 )\big|  & = &
2\cdot 3^{n-3} +3^n\cdot\Kraft (\,^2\C_n^2 ) < 3^{n-2}

\end{array}\]

Also in this case (\ref{dreigl4}) follows with (\ref{dreigl2}):

\[\begin{array}{llcl}
\quad & \beta^{00}_{n+1} & = & \beta^{01}_{n+1}=\beta^{02}_{n+1} =
\beta^{10}_{n+1} =\beta^{11}_{n+1} \beta^{12}_{n+1}=
\beta^{20}_{n+1} =\beta^{21}_{n+1}=0  \\[3mm]

\quad & \beta^{22}_{n+1} & \le &
3^{n-2}-3^{n+1}\cdot\Kraft (\,^2\C_n^2 )
<5\cdot 3^{n-3} -2\cdot 3^n\cdot\Kraft (\,^2\C_n^2) \\
\quad &\quad & = &
3^{n-1}-\big|\pfs[n] (\,^2\C_n )\big|-\big|\sfs[n](\C_n^1)\big|

\end{array}\]
\item[{\em Case 10:}]
\[ \begin{array}{l} \Kraft (\,^0\C_n^0)=3^{-2} \, ,\;
\Kraft (\,^1\C_n^1)=2\cdot 3^{-3} \, , \\
\Kraft (\,^0\C_n^1)=\Kraft (\,^0\C_n^2)=\Kraft (\,^1\C_n^0 )=
\Kraft (\,^1\C_n^2 )=\Kraft (\,^2\C_n^0)=\Kraft (\,^2\C_n^1)=
\Kraft (\,^2\C_n^2 )=3^{-3}
\end{array}\]
In this case we obtain $\beta^{ab}_{n+1} =0 $ for all $a,b\in\A $.\qed
\end{itemize}

\begin{appendix}
\chapter{Overview of known results about the $\frac{3}{4}$-conjecture}
In the appendix we give a collection of all known results about the
$\frac{3}{4}$-conjecture up to now. Throughout the appendix we denote with
$\A $ a finite alphabet with $|\A |\ge 2 $ and $(\alpha_l)_{l\in\NN } $
should be a finite sequence of nonnegative integers. We write a set
$\C\subseteq\A^+ $ fits to $(\alpha_l)_{l\in\NN } $, if
$|\C\cap \A^l | =\alpha_l $ for all $l\in\NN $.

\begin{theorem}[Kraft and McMillan \cite{kraft}]\label{ap1}
If $\C \subseteq \A^+ $ is a code which fits to $(\alpha_l)_{l\in\NN } $, then
$\sum\limits_{l=1}^{\infty }\alpha_l q^{-l} \le 1 $.
\end{theorem}
For prefix-free codes also the other direction of the theorem above holds.

\begin{theorem}[Kraft and McMillan \cite{kraft}]\label{ap2}
$\sum\limits_{l=1}^{\infty }\alpha_l q^{-l} \le 1 $ if and only
if there exists a prefix-free code which fits to $(\alpha_l)_{l\in\NN } $.
\end{theorem}

Krafts theorem holds also for suffix-free codes.\\

The $\frac{3}{4} $-conjecture is a possible generalization
of the second implication in Krafts theorem for fix-free codes. The first
implication is Theorem~\ref{ap1}, which holds for all codes.

\begin{conj}[ Ahlswede, Balkenhol and Khachatrian]
If $\sum\limits_{l=1}^{\infty }\alpha_l q^{-l} \le \frac{3}{4} $, then
there exists a fix-free code which fits to $(\alpha_l)_{l\in\NN } $.
\end{conj}

We distinguish theorems for the binary case $|\A |=2 $ and the general
case $|\A | = q $ for some $q\ge 2 $. The next theorem shows, that for sequences with
Kraftsum bigger than  $\frac{3}{4} $ the conjecture can not holds, but
that the conjecture holds for sequences with Kraftsum smaller than
or equal to $\frac{1}{2} $.

\begin{theorem}\quad\\
{\bf
\begin{tabular}[t]{lcl}(Binary case &:&
Ahlswede, Balkenhol and Khachatrian \cite{ahlswede}\\
General case &:& Harada and Kobayashi \cite{harada})\\
\end{tabular} }
\begin{enumerate}
\renewcommand{\labelenumi}{(\roman{enumi})}

\item If $\sum\limits_{l=1}^{\infty }\alpha_l q^{-l} \le \frac{1}{2} $, then
there exists a fix-free code which fits to $\fo{\alpha }$.

\item For every $\gamma > \frac{3}{4} $ there exists a sequence
$\fo{\alpha } $ with
\[ \frac{3}{4}<\sum\limits_{l=1}^{\infty }\alpha_l q^{-l} \le \gamma \, ,\]
such that there doesn't exist a  fix-free code which fits to $\fo{\alpha }$.
\end{enumerate}
\end{theorem}

The next theorem shows, for which sequences the $\frac{3}{4} $-conjecture
is already proven.

\begin{theorem}
Let $\sum\limits_{l=1}^{\infty } \alpha_l q^{-l} \le \frac{3}{4}$.
\begin{enumerate}
\renewcommand{\labelenumi}{(\roman{enumi})}
\item {\bf\begin{tabular}[t]{lcl}(Binary case &:&
Ahlswede, Balkenhol and Khachatrian \cite{ahlswede}\\
General case &:& Harada and Kobayashi \cite{harada})
\end{tabular} }\\
If
$\, 2k\le \inf \{ l\, |\, \alpha_l\neq 0 \, , l>k\}
\;\mbox{ for all }\; k\in\NN \;\mbox{ with }\; \alpha_k\neq 0\, ,$
then there exists a fix-free code which fits to $\fo{\alpha }$.

\item {\bf \begin{tabular}[t]{lcl} (General case  & : &
Harada and Kobayashi \cite{harada} )\end{tabular} }\\
If there exists $n,m\in\NN $ such that $\alpha_l =0 $ for all $l\not\in\{ n,m\} $,
then there exists a fix-free code which fits to $\fo{\alpha }$.

\item {\bf\begin{tabular}[t]{lcl}(Binary case &:&
Kukorelly and Zeger \cite{zeger})\\
General case &:& This survey, Chapter 2
\end{tabular} }\\
Let $l_{min} :=\min \{ l  | \alpha_l > 0 \} \, $ and
$\, l_{max} :=\mbox{sup}\, \{ l \in\NN | \alpha_l > 0 \} \le \infty $.
If $\; l_{min} \ge 2 $, $l_{max} <\infty $ and
$ \alpha_l\le q^{l_{min}-2} \big\lfloor \frac{q}{2} \big\rfloor^2 \big\lceil
\frac{q}{2} \big\rceil^{l-l_{min}} $ for all
$l\neq l_{max} $,
then there exists a fix-free code which fits to $\fo{\alpha }$.

\item {\bf\begin{tabular}[t]{lcl}(Binary case &:&
Yekhanin \cite{yek1} without a full proof. \\
\quad & & A full proof is in this survey Chapter 4.)
\end{tabular} }\\
Let $|\A | =q=2 $ and $n:=\min\{ l \, | \,\alpha_l \neq 0 \} $.
If $\frac{\alpha_n}{2^n} + \frac{\alpha_{n+1}}{2^{n+1}}\ge \frac{1}{2} $,
then there exists a fix-free code which fits to $\fo{\alpha }$.

\item {\bf\begin{tabular}[t]{lcl}(Binary case &:& See (iv). \\
General case  & : & This survey, Chapter 4.)
\end{tabular} }\\
Let $\frac{\alpha_n }{q^n } + \frac{\alpha_{n+1}}{q^{n+1}} \ge
\lceil\frac{q}{2}\rceil \frac{1}{q} $.\\
If $\alpha_n\ge \lceil\frac{q}{2}\rceil\cdot q^{n-1} $ or
$  \alpha_n =\lceil\frac{q}{2}\rceil L $ for some $1\le L < q^{n-1}  $
and there exists a $\lceil\frac{q}{2}\rceil$-regular subgraph in
$\BGr{n-1} $ with $L$ vertices,
then there exists a fix-free code which fits to $\fo{\alpha }$.

\item {\bf\begin{tabular}[t]{lcl}(Binary case &:&
Yekhanin \cite{yek1}, without a full proof.\\
General case &:& This survey, Chapter 4.)
\end{tabular} }\\
Let $n:=\min\{ l \, | \,\alpha_l \neq 0 \} $. If
$\frac{\alpha_n}{q^n}\ge \left\lceil\frac{q}{2}\right\rceil\cdot q^{-1} $,
then there exists a fix-free code which fits to $\fo{\alpha }$.

\item {\bf\begin{tabular}[t]{lcl}(Binary case &:&
Kukorelly and Zeger \cite{zeger}.)
\end{tabular} }\\
Let $|\A |=q=2 $.
If $\alpha_l \le 2 $ for  all $l\in\NN $ and
$\sup\{ l\, |\, \alpha_l\neq 0 \} < \infty $,
then there exists a fix-free code which fits to $\fo{\alpha }$.

\item {\bf\begin{tabular}[t]{lcl}(General case &:&
This survey, Chapter 4 only for even $q$.)
\end{tabular} }\\
Let $|\A | = q $ with $q$ even.
If there exists an $n\ge 2$, with
$\alpha_1 =0 $, $\alpha_l =\big(\frac{q}{2}\big)^l $ for $2\le l < n $ and
$\alpha_n \ge q\cdot \big(\frac{q}{2}\big)^{n-1} $,
then there exists a fix-free Code which fits to
$(\alpha_l )_{l\in\NN } $.

\item {\bf\begin{tabular}[t]{lcl}(General case &:&
This survey, Chapter 4 only for even $q$.)
\end{tabular} }\\
Let $|\A | = q $ with $q$ even.
If there exists an $n\ge 3 $, with
$\alpha_1 =\alpha_2 = 0 $,
$\alpha_l =2\cdot \big(\frac{q}{2}\big)^l $ for $3\le l < n $ and
$\alpha_n \ge 2q \big(\frac{q}{2}\big)^{n-1} $,
then there exists a fix-free Code which fits to
$(\alpha_l )_{l\in\NN } $.

\item {\bf\begin{tabular}[t]{lcl}(Binary case &:&
This survey, Chapter 5.)
\end{tabular} }\\
Let $|\A |=q=2 $.
If there exists an $n\ge 2 $ such that
$\alpha_2 = \alpha_{2l+1}  =0 $ for all $l\in\NN_0 $,
$\alpha_{2l}=2^l $ for all $2\le l < n $, $\alpha_{2n} \ge 2^{n+1} $ and
$\alpha_{2l} \in\NN_0 $ for all $l>n $,
then there exists a fix-free Code which fits to
$(\alpha_l )_{l\in\NN } $.

\item {\bf\begin{tabular}[t]{lcl}(Binary case &:&
This survey, Chapter 5.)
\end{tabular} }\\
Let $|\A |=q=2 $.
If there exists an $n\ge 3 $ such that
$\alpha_2 = \alpha_4 =\alpha_{2l+1}  =0 $ for all $l\in\NN_0 $,
$\alpha_{2l}=2^{l+1} $ for all $2\le l < n $, $\alpha_{2n} \ge 2^{n+2} $ and
$\alpha_{2l} \in\NN_0 $ for all $l> n $, then there exists a fix-free code
which fits to $(\alpha_l )_{l\in\NN } $.

\item {\bf\begin{tabular}[t]{lcl}(Binary case &:&
This survey, Chapter 5.)
\end{tabular} }\\
Let $|\A |=q=2 $.\\
If there exists an $n\in\NN $ such that
$\alpha_2 =\alpha_4 =\ldots =\alpha_{2n-2} =\alpha_{2l+1} =0 $ for all $l\in\NN_0 $,
$\alpha_{2n} $ is even,
$\frac{\alpha_{2n}}{2^{2n}} +\frac{\alpha_{2n+2}}{2^{2n+2}}\ge \frac{1}{2} $ and
there exists a $2$-regular subgraph of \BGr[4]{n-1} with $\frac{\alpha_{2n}}{2} $
vertices,
then there exists a fix-free code
$\C\subseteq\A^+ $ which fits to $(\alpha_l )_{l\in\NN } $.

\item {\bf\begin{tabular}[t]{lcl}(Binary case &:&
This survey, Chapter 5.)
\end{tabular} }\\
Let $|\A |=q=2 $.\\
If there exists an $n\in\NN $ such that
$\alpha_2 =\alpha_4 =\ldots =\alpha_{2n-2} =\alpha_{2l+1} =0 $ for all $l\in\NN_0 $
and
$\frac{\alpha_{2n}}{2^{2n}} \ge \frac{1}{2} $,
then there exists a fix-free code
$\C\subseteq\A^+ $ which fits to $(\alpha_l )_{l\in\NN } $.

\item {\bf\begin{tabular}[t]{lcl}(Binary case &:&
This survey, Chapter 5.)
\end{tabular} }\\
Let $|\A |=q=2 $,
$l_{min} :=\min \{ l \, | \, \alpha_l\neq 0 \} $ and
$l_{max} := \sup \{ l \, | \, \alpha_l\neq 0 \} $. If
$l_{max} <\infty $, $4\le l_{min} $ is even, $\alpha_{2l+1}=0 $ for all $l\in\NN_0 $
and $\alpha_{2l} \le 2^{\frac{l_{min}}{2}-2+l } $ for all $2l\neq l_{max} $,
then there exists a fix-free code
which fits to $(\alpha_l )_{l\in\NN } $.

\item {\bf\begin{tabular}[t]{lcl}(Binary case &:&
Ye and Yeung \cite{yeung}, by computer research.)
\end{tabular} }\\
Let $|\A |=q=2 $.
If $\alpha_l =0 $ for all $l\ge 8 $,
then there exists a fix-free code
which fits to $(\alpha_l )_{l\in\NN } $.

\item {\bf\begin{tabular}[t]{lcl}(Binary case &:&
Yekhanin \cite{yek1}, by computer research.)
\end{tabular} }\\
Let $|\A |=q=2$.
If $\alpha_l =0 $ for all $l\ge 8 $,
then there exists a fix-free code
which fits to $(\alpha_l )_{l\in\NN } $.

\end{enumerate}
\end{theorem}

The next theorem shows results which are related to the
binary $\frac{3}{4}$-conjecture.
\begin{theorem}
\quad\\
\begin{enumerate}
\renewcommand{\labelenumi}{(\roman{enumi})}

\item {\bf\begin{tabular}[t]{lcl}(Binary case &:&
Ye and Yeung \cite{yeung}.)
\end{tabular} }\\
Let $|\A |=q =2 $.
If $\sup\{ l\, | \, \alpha_l\neq 0 \} <\infty $,
$\alpha_1 =1 $
and $\sum\limits_{l=1}^n \alpha_l \left(\frac{1}{2}\right)^l \le \frac{5}{8} $,
then there exists a fix-free code
which fits to $(\alpha_l )_{l\in\NN } $.

\item {\bf\begin{tabular}[t]{lcl}(Binary case &:&
Yekhanin \cite{yek2}.)
\end{tabular} }\\
Let $|\A |=q =2 $.
If $\sum\limits_{l=1}^n \alpha_l \left(\frac{1}{2}\right)^l \le \frac{5}{8} $,
then there exists a fix-free code
which fits to $(\alpha_l )_{l\in\NN } $.

\item {\bf\begin{tabular}[t]{lcl}(Binary case &:&
Ye and Yeung \cite{yeung}.)
\end{tabular} }\\
Let $|\A |=q =2 $.
Let $\vec{l}_n=(l_1,\ldots ,l_n )\in\NN^n $ be a lengths sequence and
$ h(i)  :=  \min \{ j\, | \, l_j=l_{i+1} \}
\quad\mbox{ for all }\; 1\le i<n $. If\\
{\footnotesize $\prod\limits_{i=1}^{n-1}
\big( 1-2\sum\limits_{1\le j\le i} 2^{-l_i} +(\, i+1-h(i)\, )\cdot
2^{-l_i+1} +\hspace*{-8mm}
\sum\limits_{\begin{array}{l} \mbox{\tiny $1\le j,k\le h(i)-1$}\\
\mbox{\tiny s.t. $l_j+l_k \le l_i+1$}\end{array}}
\hspace*{-8mm}2^{-l_k-l_k} \big)^+ > 0 \, $},
then there exists a fix-free code which fits to $ \vec{l}_n $.

\item {\bf\begin{tabular}[t]{lcl}(Binary case &:&
Ye and Yeung \cite{yeung}.)
\end{tabular} }\\
Let $|\A |=q =2 $.
Let $\vec{l}_n=(l_1,\ldots ,l_n )\in\NN^n $ be a lengths sequence and
$ h(i)  :=  \min \{ j\, | \, l_j=l_{i+1} \}
\quad\mbox{ for all }\; 1\le i<n $. If\\
{\footnotesize $ \prod\limits_{i=1}^{n-1}
\big( 1-2\sum\limits_{1\le j\le i} 2^{-l_i} +(\, i+1-h(i)\, )\cdot
2^{-l_i+1} +\hspace*{-5mm}
\sum\limits_{1\le j,k\le h(i)-1}\hspace*{-5mm}
2^{(l_{i+1}-l_j-l_k)^+-l_{i+1}} \big)^+ = 0 \, $},
then there doesn't
exists a fix-free code which fits to $\vec{l_n} $.

\item {\bf\begin{tabular}[t]{lcl}(Binary case &:&
Ye and Yeung \cite{yeung}.)
\end{tabular} }\\
Let $|\A |=q =2 $.
Let $\vec{l}_n=(l_1,\ldots ,l_n )\in\NN^n $ be a lengths sequence and
$ h(i)  :=  \min \{ j\, | \, l_j=l_{i+1} \}
\quad\mbox{ for all }\; 1\le i<n $. If\\
$ \sum\limits_{1\le j\le n} 2^{-l_i} <\frac{1}{2} +
\frac{n+2-h(n-1)}{2}\cdot 2^{-l_n } \, $ ,
then there exists a fix-free code which fits to $ \vec{l}_n $.
\end{enumerate}
\end{theorem}
\end{appendix}

{\quad\quad}
\end{document}